\documentclass[12pt]{article}
\usepackage{a41} %
\usepackage{epsfig} %
\usepackage{float} %
\usepackage{floatflt} %
\usepackage{amssymb} %
\usepackage{amsmath} %
\usepackage{verbatim} %
\usepackage{cite} %
\usepackage{color} %
\allowdisplaybreaks[4]
\usepackage{eufrak}

\newcommand{\LqPS}{\ensuremath{L_{q,2}^{\sf PS}}}
\newcommand{\LgS}{\ensuremath{L_{g,2}^{\sf S}}}
\newcommand{\HgS}{\ensuremath{H_{g,2}^{\sf S}}}

\newcommand{\HqPS}{\ensuremath{H_{q,2}^{\sf PS}}}

\newcommand{\AQg}{\ensuremath{A_{Qg}}}
\newcommand{\Agg}{\ensuremath{A_{gg,Q}}}
\newcommand{\AQqPS}{\ensuremath{A_{Qq}^{\sf PS}}}
\newcommand{\AqqNS}{\ensuremath{A_{qq,Q}^{\sf NS}}}
\newcommand{\Agq}{\ensuremath{A_{gq,Q}}}
\newcommand{\Aqg}{\ensuremath{A_{qg,Q}}}
\newcommand{\AqqPS}{\ensuremath{A_{qq,Q}^{\sf PS}}}

\newcommand{\Ahat}{\hat{A}}
\newcommand{\N}{\nonumber}
\newcommand{\ep}{\varepsilon}

\newcommand{\Ctil}{\tilde{C}}

\newcommand{\Mvec}{\mbox{\rm\bf M}}

\newcommand{\beq}{\begin{equation}}
\newcommand{\eeq}{\end{equation}}
\newcommand{\bea}{\begin{eqnarray}}
\newcommand{\eea}{\end{eqnarray}}

\newcommand{\GeV}{${\rm GeV}$}

\newcommand{\gsim}{\raisebox{-0.07cm}{$\, \stackrel{>}{{\scriptstyle
\sim}}\, $}}

\newcounter{lin}

\begin{document}
\begin{titlepage}

\begin{flushleft}
DESY 11--144   \hfill   
\\
DO-TH 13/33
\\
SFB-CPP/14--002
\\
LPN 14--002
\\
Higgstools 14--003
\\
March 2014 
\\
\end{flushleft}

\vspace{1cm}
\noindent
\begin{center}
{\LARGE\bf The Logarithmic Contributions to the}

\vspace*{2mm} 
{\LARGE\bf
\boldmath{$O(\alpha_s^3)$} Asymptotic Massive Wilson 
Coefficients}

\vspace*{2mm}
{\LARGE\bf 
and Operator Matrix Elements in} 

\vspace*{2mm}
{\LARGE\bf Deeply
Inelastic Scattering} \\
\end{center}
\begin{center}

\vspace{2cm}

{\large A.~Behring$^a$, I. Bierenbaum$^b$, J. Bl\"umlein$^a$, A. De Freitas$^a$,  S. Klein$^c$, 

\vspace*{2mm}
 and F. Wi\ss{}brock$^{b,d,}$\footnote{Present address: IHES, 35 Route de Chartres, 91440 
Bures-sur-Yvette,
France.}}

\vspace*{2mm}
{\it $^a$~Deutsches Elektronen Synchrotron, DESY,\\
Platanenallee 6, D--15738 Zeuthen, Germany}\\

\vspace*{2mm}
{\it $^b$~II. Institut f\"ur Theoretische Physik, Universit\"at Hamburg,
\\
Luruper Chaussee 149, D-22761 Hamburg, Germany}

\vspace*{2mm}
{\it $^c$~Institut f\"ur Theoretische Teilchenphysik und Kosmologie, \\
                 RWTH Aachen University, D--52056 Aachen, Germany}

\vspace*{2mm}
{\it $^d$~Research Institute for Symbolic Computation (RISC),\\
                          Johannes Kepler University, Altenbergerstra\ss{}e 69,
                          A--4040, Linz, Austria}

\vspace{3cm}
\end{center}

\begin{abstract}
\noindent
We calculate the logarithmic contributions to the massive Wilson coefficients for deep-inelastic 
scattering in the asymptotic region $Q^2 \gg m^2$ to 3-loop order in the fixed-flavor number scheme
and present the corresponding expressions for the massive operator matrix elements needed in the 
variable flavor number scheme. Explicit expressions are given both in Mellin-$N$ space and 
$z$-space. 
\end{abstract}
\end{titlepage}

\newpage
\sloppy

\section{Introduction}
\label{sec:intro}

\vspace{1mm}
\noindent
The heavy flavor corrections to deep-inelastic structure functions amount to
sizeable contributions, in particular in the region of small values of the Bjorken variable $x$.
Starting from lower values of the virtuality, over a rather wide kinematic range, their scaling violations
are very different from those of the massless contributions. Currently the heavy flavor 
corrections are known in semi-analytic form to 2--loop (NLO) order~\cite{NLO}. The present 
accuracy of the deep-inelastic data reaches the order of 1\%~\cite{H1Z:2009wt}. It therefore
requires the next-to-next-to-leading order (NNLO) corrections for precision determinations
of both the strong coupling constant $\alpha_s(M_Z^2)$ \cite{Bethke:2011tr} and the parton distribution 
functions (PDFs)~\cite{PDF,Alekhin:2013nda}, as well as the detailed understanding of the heavy flavor 
production cross
sections in lepton--nucleon scattering~\cite{EXP}. The precise knowledge of these quantities is 
also of central importance for the interpretation of the physics results at the Large Hadron 
Collider, 
LHC,~\cite{HERALHC}. 

In the kinematic region at HERA, where the twist-2 contributions to the deep-inelastic scattering 
(DIS) dominate cf.~\cite{Blumlein:2008kz}\footnote{For higher order corrections to the 
gluonic contributions in the threshold region, cf.~\cite{Kawamura:2012cr}.}, i.e. 
$Q^2/m^2~\gsim~10$, 
with $m = m_c$ the charm quark mass, it has been proven in Ref.~\cite{Buza:1995ie} that the heavy 
flavor 
Wilson coefficients factorize into massive operator matrix elements (OMEs) and the massless 
Wilson coefficients. The massless Wilson coefficients for the structure function $F_2(x,Q^2)$ are
known to 3-loop order \cite{Vermaseren:2005qc}. In the region $Q^2 \gg m^2$, where $Q^2=-q^2$, 
with $q$ the space-like 
4--momentum transfer and $m$ the heavy quark mass, the power corrections $O((m^2/Q^2)^k), k \geq 1$ 
to the heavy quark structure functions become very small. 

In Ref.~\cite{Bierenbaum:2009mv} a series of fixed Mellin moments $N$ up to $N = 10,...,14$, 
depending on the respective transition, has been calculated for all the OMEs at 3--loop 
order\footnote{For the corresponding contributions in case of transversity see 
\cite{Blumlein:2009rg}.}. Also the moments of the transition coefficients needed in the variable 
flavor scheme (VFNS) have been calculated. Here, the massive OMEs for given total spin $N$ were 
mapped
onto massive tadpoles which have been computed using {\tt MATAD}  \cite{Steinhauser:2000ry}. 

In the present paper, we calculate the logarithmic contributions to the unpolarized massive Wilson 
coefficients in the asymptotic region $Q^2 \gg m^2$ to 3--loop order and the massive OMEs needed 
in the VFNS. These include the logarithmic terms $\log(Q^2/m^2)$. In the following, we set
the factorization and renormalization scales equal $\mu_F = \mu_R \equiv \mu$ and exhibit 
the $\log(m^2/\mu^2)$ dependence on the Wilson coefficients, besides their dependence on 
the virtuality $Q^2$. The 
logarithmic contributions are determined by the lower order massive OMEs 
\cite{Buza:1996wv,TOL1b,TOL2a,TOL2b,Bierenbaum:2009zt}, 
the mass- and coupling constant renormalization constants, and the anomalous dimensions 
\cite{ANDIM2a,ANDIM2b}, as has been worked out in Ref.~\cite{Bierenbaum:2009mv}. For the 
structure function $F_L(x,Q^2)$ the asymptotic heavy flavor Wilson coefficients at $O(\alpha_s^3)$ 
were calculated in~\cite{Blumlein:2006mh}. They are also presented here, for inclusive hadronic final states. 
In this case the corrections, however, become effective only at much higher scales of $Q^2$ \cite{Buza:1995ie} 
compared to the case of $F_2(x,Q^2)$. 
We first choose the fixed flavor number scheme to express the heavy flavor contributions to the structure 
functions $F_2(x,Q^2)$ and $F_L(x,Q^2)$. This scheme has to be considered as the genuine scheme in 
quantum field theoretic calculations since the initial states, the twist--2 {\it massless} partons can, 
at least to a good approximation, be considered as LSZ-states. The representations in the VFNS can be 
obtained using the respective transition coefficients within the appropriate regions, where one single
heavy quark flavor  becomes effectively massless. Here, appropriate matching scales have to be 
applied,
which vary in dependence on the observable considered, cf.~\cite{Blumlein:1998sh}.

Two of the OMEs, $A_{qq,Q}^{\sf PS}(N)$ and $A_{qg,Q}(N)$, have already been calculated 
completely including the constant contribution in Ref.~\cite{Ablinger:2010ty}. They and the 
corresponding massive Wilson coefficients contribute first at 2-- and 3--loop order, respectively. For these quantities
we also derive numerical results. The quantities being presented in the present paper 
derive from OMEs which were computed in terms of generalized hypergeometric functions \cite{HYPER} and 
sums thereof, prior to the expansion in the dimensional variable $\varepsilon = D -4$,
cf.~\cite{Blumlein:2009ta,Blumlein:2010zv,Bytev:2009kb}. Finally, they are represented in terms of 
nested sums over products of hypergeometric terms and harmonic sums, which can be calculated using 
modern summation techniques \cite{SIGMA,Ablinger:2010pb,Blumlein:2012hg,Schneider2013,ROUND}. 
They are based on a refined difference field of \cite{KARR} and generalize the summation paradigms 
presented in \cite{AEB} to multi-summation. The results of this computation can be expressed 
in terms of nested harmonic 
sums~\cite{HSUM1,HSUM2}. The corresponding representations in $z$-space are obtained in terms of 
harmonic 
polylogarithms \cite{HPL}. Here, the variable $z$ denotes the partonic momentum fraction. The 
results in Mellin $N$-space can 
be continued to complex values of $N$ as has been described in Refs.~\cite{ANCONT,Blumlein:2009ta}.

It is the aim of the present paper to provide a detailed documentation of formulae both in $N$- and 
$z$-space for all logarithmic contributions to the heavy flavor Wilson coefficients of the structure 
functions $F_2(x,Q^2)$ and $F_L(x,Q^2)$ and the massive OMEs needed in the variable flavor number 
scheme up to $O(\alpha_s^3)$. Here, we refer to a minimal representation, i.e. we use all the 
algebraic 
relations between the harmonic sums and the harmonic polylogarithms, respectively, leading to a minimal 
number of basic functions. Based on the known Mellin moments \cite{Bierenbaum:2009mv} we also perform 
numerical comparisons between the different contributions to the Wilson coefficients and massive OMEs  
at $O(\alpha_s^3)$ referring to the parton distributions \cite{Alekhin:2013nda}.

The paper is organized as follows. In Section~\ref{sec:forma}, we summarize the basic formalism. 
The Wilson coefficients  $L_{q,2}^{\sf PS}$ and $L_{g,2}^{\sf S}$ are discussed in 
Section~\ref{sec:3}. As they are known in complete form we also present numerical results.
In Section~\ref{sec:4}, the logarithmic contributions to the Wilson coefficients $H_{q,2}^{\sf 
PS}$ 
and  $H_{g,2}^{\sf S}$\footnote{The expressions for the non-singlet Wilson-coefficient, are 
presented elsewhere together with the OME for transversity \cite{NS2014}.} are derived. 
The corresponding Wilson coefficients for the  longitudinal structure function $F_L(x,Q^2)$ in 
the asymptotic region are presented in Section~\ref{sec:5}. In Section~\ref{sec:6}, we compare the 
different 
loop contributions to the massive Wilson coefficients and OMEs for a series of Mellin moments in 
dependence on 
the virtuality $Q^2$. Section~\ref{sec:7} contains the conclusions. 
In Appendix~\ref{app:A} the massive OMEs needed in the VFNS are given in Mellin $N$-space. The 
asymptotic heavy flavor Wilson coefficients contributing to the structure function 
$F_2(x,Q^2)$ are presented in $z$-space in 
Appendix~\ref{app:B}, retaining all contributions except for the 3-loop constant part of the 
unrenormalized OMEs $a_{ij}^{(3)}$ being not yet known. Likewise, 
in Appendix~\ref{app:C} and \ref{app:D}, the asymptotic heavy flavor Wilson coefficients for the 
structure function $F_L(x,Q^2)$  and the massive OMEs are given in $z$-space.
\section{The heavy flavor Wilson coefficients in the asymptotic region}
\label{sec:forma}

\vspace*{1mm}
\noindent
We consider the heavy flavor contributions to the inclusive unpolarized structure functions 
$F_2(x,Q^2)$ and $F_L(x,Q^2)$ in deep-inelastic scattering, cf. \cite{Arbuzov:1995id,Blumlein:2012bf}, 
in case of single electro-weak gauge-boson exchange at large virtualities $Q^2$. At higher orders in 
the strong coupling constant these corrections receive both contributions from massive
and massless partons in the hadronic final state, which is summed over completely.
In the latter case, the heavy flavor corrections are also due to virtual contributions.
We consider the situation in which the contributions to the twist-2 operators dominate 
in the Bjorken limit. Here, no transverse momentum effects of the initial state contribute.
In the present paper, we consider only heavy flavor contributions due to $N_F$ massless
and one massive flavor of mass $m$.\footnote{At 3--loop order there are also contributions
by graphs carrying heavy quark lines of different mass. These are dealt with elsewhere
\cite{Ablinger:2011pb}.} The Wilson coefficients are calculable perturbatively and are denoted by
\begin{eqnarray}
{\cal C}^{\sf S,PS,NS}_{i, (2,L)} \Bigl(x,N_F+1,\frac{Q^2}{\mu^2},\frac{m^2}{\mu^2}\Bigr)~. 
\label{Calldef}
\end{eqnarray}
Here, $x$ denotes the Bjorken variable, the index $i$ refers to the respective initial state 
on-shell parton $i=q,g$ being a quark or gluon, and  ${\sf S,PS,NS}$ label the flavor 
singlet, pure--singlet and non--singlet contributions, respectively. In the twist-2 approximation
the Bjorken variable $x$ and the parton momentum fraction $z$ are identical. Representations
in momentum fraction space are therefore also called $z$-space representation in what follows.
 
The massless flavor 
contributions in (\ref{Calldef}) may be identified and separated in the Wilson coefficients 
into a purely light part $C_{i,(2,L)}$, and a heavy part by~:
\begin{eqnarray}
\label{eqLH}
{\cal C}^{\sf S,PS,NS}_{i,(2,L)}
               \left(x,N_F+1,\frac{Q^2}{\mu^2},\frac{m^2}{\mu^2}\right) 
      = 
       && C_{i,(2,L)}^{\sf S,PS,NS}\left(x,N_F,\frac{Q^2}{\mu^2}\right)
\N\\ &&\hspace{-25mm}
      + H_{i,(2,L)}^{\sf S,PS}
               \left(x,N_F+1,\frac{Q^2}{\mu^2},\frac{m^2}{\mu^2}\right)
      + L_{i,(2,L)}^{\sf S,PS,NS}
             \left(x,N_F+1,\frac{Q^2}{\mu^2},\frac{m^2}{\mu^2}\right)~.\N\\
       \label{Callsplit}
\end{eqnarray}
The heavy flavor Wilson coefficients are defined by $L_{i,j}$ and $H_{i,j}$, depending on whether the 
exchanged electro-weak gauge boson couples to a light $(L)$ or heavy $(H)$ quark line. 
From this it follows that the light flavor Wilson coefficients $C_{i,j}$ depend on $N_F$ light 
flavors only, whereas $H_{i,j}$ and $L_{i,j}$ may contain light flavors in addition to the heavy
quark, indicated by the argument $N_F+1$. The perturbative series of the heavy flavor Wilson 
coefficients read
\begin{eqnarray}
H_{g,(2,L)}^{\sf S}
          \left(x,N_F+1,\frac{Q^2}{\mu^2},\frac{m^2}{\mu^2}\right)&=&
           \sum_{i=1}^{\infty}a_s^i
             H_{g,(2,L)}^{(i), \sf S}
          \left(x,N_F+1,\frac{Q^2}{\mu^2},\frac{m^2}{\mu^2}\right)~,
\label{Hg2Lpert}
\\
      H_{q,(2,L)}^{\sf PS}
          \left(x,N_F+1,\frac{Q^2}{\mu^2},\frac{m^2}{\mu^2}\right)&=&
           \sum_{i=2}^{\infty}a_s^i
             H_{q,(2,L)}^{(i), \sf PS}
          \left(x,N_F+1,\frac{Q^2}{\mu^2},\frac{m^2}{\mu^2}\right)~,
\label{Hq2LPSpert}
\\
      L_{g,(2,L)}^{\sf S}
          \left(x,N_F+1,\frac{Q^2}{\mu^2},\frac{m^2}{\mu^2}\right)&=&
           \sum_{i=2}^{\infty}a_s^i
             L_{g,(2,L)}^{(i), \sf S}
          \left(x,N_F+1,\frac{Q^2}{\mu^2},\frac{m^2}{\mu^2}\right)~,
\label{Lg2Lpert}
\\
      L_{q,(2,L)}^{\sf S}
          \left(x,N_F+1,\frac{Q^2}{\mu^2},\frac{m^2}{\mu^2}\right)&=&
           \sum_{i=2}^{\infty}a_s^i
             L_{q,(2,L)}^{(i), \sf S}
          \left(x,N_F+1,\frac{Q^2}{\mu^2},\frac{m^2}{\mu^2}\right)~.
\label{Lq2LSpert}
\end{eqnarray}
Here, we defined $a_s = \alpha_s/(4\pi)$.
At leading order, only the term $H_{g,(2,L)}$ contributes via the 
photon--gluon fusion process, 
\cite{Witten:1975bh,Babcock:1977fi,Shifman:1977yb,Leveille:1978px,Gluck:1979aw,Gluck:1980cp},
\begin{eqnarray}
\gamma^*+g\rightarrow~Q+\overline{Q}~. 
\label{VBfusion}
\end{eqnarray}
At $O(a_s^2)$, the terms $H_{q,(2,L)}^{\sf PS}$, $L_{q,(2,L)}^{\sf S}$ and  $L_{g,(2,L)}^{\sf S}$ contribute as well. 
They result from the processes 
\begin{eqnarray}
\gamma^*+q(\overline{q})\rightarrow~q(\overline{q})+X~,\\
\gamma^*+g\rightarrow~q(\overline{q})+X~,
\end{eqnarray}
where $X$ may contain heavy flavor contributions. $L_{q,(2,L)}^{\sf S}$ can be split into 
the flavor non-singlet and pure-singlet contributions
\begin{eqnarray}
L_{q,(2,L)}^{\sf S}=L_{q,(2,L)}^{\sf NS}+L_{q,(2,L)}^{\sf PS}, 
\end{eqnarray}
and at $O(a_s^2)$ only the non-singlet term contributes. The pure-singlet term emerges at 3--loop 
order.

The heavy quark contribution to the structure functions $F_{(2,L)}(x,Q^2)$
for one heavy quark of mass $m$ and $N_F$ light flavors is then given by,
cf. \cite{Buza:1996wv}, in case of pure photon exchange\footnote{For the heavy flavor 
corrections in case of $W^\pm$-boson exchange up to $O(\alpha_s^2)$ 
see~\cite{Gluck:1996ve,Blumlein:2011zu,Buza:1997mg,BHP2014}.} 
\begin{eqnarray}
\label{eqF2}
    \frac{1}{x}   F_{(2,L)}^{Q\overline{Q}}(x,N_F\!\!\!&+&\!\!\!1,Q^2,m^2) =\N\\
       &&\sum_{k=1}^{N_F}e_k^2\Biggl\{ 
                   L_{q,(2,L)}^{\sf NS}\left(x,N_F+1,\frac{Q^2}{\mu^2}
                                                ,\frac{m^2}{\mu^2}\right)
                \otimes 
                   \Bigl[f_k(x,\mu^2,N_F)+f_{\overline{k}}(x,\mu^2,N_F)\Bigr]
\N\\ &&\hspace{14mm}
               +\frac{1}{N_F}L_{q,(2,L)}^{\sf PS}\left(x,N_F+1,\frac{Q^2}{\mu^2}
                                                ,\frac{m^2}{\mu^2}\right) 
                \otimes 
                   \Sigma(x,\mu^2,N_F)
\N\\ &&\hspace{14mm}
               +\frac{1}{N_F}L_{g,(2,L)}^{\sf S}\left(x,N_F+1,\frac{Q^2}{\mu^2}
                                                 ,\frac{m^2}{\mu^2}\right)
                \otimes 
                   G(x,\mu^2,N_F) 
                             \Biggr\}
\N\\
       &+&e_Q^2\Biggl[
                   H_{q,(2,L)}^{\sf PS}\left(x,N_F+1,\frac{Q^2}{\mu^2}
                                        ,\frac{m^2}{\mu^2}\right) 
                \otimes 
                   \Sigma(x,\mu^2,N_F)
\N\\ &&\hspace{7mm}
                  +H_{g,(2,L)}^{\sf S}\left(x,N_F+1,\frac{Q^2}{\mu^2}
                                           ,\frac{m^2}{\mu^2}\right)
                \otimes 
                   G(x,\mu^2,N_F)
                                  \Biggr]~,
\end{eqnarray}
The meaning of the argument $(N_F +1)$ in Eqs.~(\ref{eqF2}) in the massive Wilson coefficients
shall be interpreted as $N_F$ {\it massless} and one {\it massive flavor}. $N_F$ denotes the 
number of massless flavors.
The symbol $\otimes$ denotes the Mellin convolution,\footnote{Note that the
heavy flavor threshold in the limit $Q^2 \gg m^2$ is again $x$ and not $x(1+4m^2/Q^2)$, which is 
the case retaining also power corrections.}
\begin{eqnarray}
[A \otimes B](x) = 
\int_0^1 \int_0^1 dx_1 dx_2~\delta(x - x_1 x_2) A(x_1) B(x_2)~.
\end{eqnarray}
The charges of the light quarks are denoted by $e_k$ and that of the heavy quark by $e_Q$.
The scale $\mu^2$ is the factorization scale, and $f_k, f_{\overline{k}}, \Sigma$ and $G$ are the 
quark, 
anti-quark, flavor singlet and gluon distribution functions, with
\begin{eqnarray}
\Sigma(x,\mu^2,N_F) = \sum_{k=1}^{N_F} \left[ f_k(x,\mu^2,N_F) 
                           + f_{\overline{k}}(x,\mu^2,N_F)\right]~.
\end{eqnarray}

An important part of the kinematic region in case of heavy flavor production in DIS is located 
at larger values of $Q^2$, cf. e.g. \cite{Gluck:1987uk,Ingelman:1988qn}. As has 
been shown in Ref.~\cite{Buza:1995ie}, the heavy flavor Wilson coefficients $H_{i,j},~L_{i,j}$ 
factorize in the limit $Q^2 \gg m^2$ into massive operator matrix elements $A_{ki}$ and the 
massless Wilson coefficients $C_{i,j}$, if one heavy quark flavor and $N_F$ light flavors are 
considered. The massive OMEs are process independent quantities and contain all the mass dependence 
except for the power corrections $\propto~(m^2/Q^2)^k,~k\ge~1$. The process dependence is implied
by the massless Wilson coefficients. This allows the analytic calculation of the NLO heavy 
flavor Wilson coefficients, \cite{Buza:1995ie,TOL2a}. Comparing these asymptotic 
expressions with the exact LO and NLO results obtained in 
Refs.~\cite{Witten:1975bh,Babcock:1977fi,Shifman:1977yb,Leveille:1978px,Gluck:1980cp} 
and \cite{NLO}, respectively, one finds that this 
approximation becomes valid in case of $F_2^{Q\overline{Q}}$ for $Q^2/m^2 \gsim 10$. These scales are 
sufficiently low and match with the region analyzed in deeply inelastic scattering for precision 
measurements. In case of $F_L^{Q\overline{Q}}$, this approximation is only valid for $Q^2/m^2 \gsim 
800$, \cite{Buza:1995ie}. For the latter case, the 3--loop corrections were calculated in 
Ref.~\cite{Blumlein:2006mh}. This difference is due to the emergence of terms $\propto (m^2/Q^2) 
\ln(m^2/Q^2)$, which only vanish slowly in the limit $Q^2/m^2 \rightarrow \infty$. 

In order to derive the factorization formula, one considers the inclusive Wilson coefficients 
${\cal C}^{\sf S,PS,NS}_{i,j}$, which have been defined in Eq.~(\ref{Calldef}). After applying 
the light cone expansion (LCE) \cite{LCE}
to the partonic tensor, or the forward Compton amplitude, corresponding to the respective
Wilson coefficients, one arrives at the factorization relation,
\begin{eqnarray}
{\cal C}^{{\sf S,PS,NS}, \small{{\sf \sf asymp}}}_{j,(2,L)}
\Bigl(N,N_F+1,\frac{Q^2}{\mu^2},\frac{m^2}{\mu^2}\Bigr) 
&=& \N\\ && \hspace{-55mm}
\sum_{i} A^{\sf S,PS,NS}_{ij}\Bigl(N,N_F+1,\frac{m^2}{\mu^2}\Bigr)
                    C^{\sf S,PS,NS}_{i,(2,L)}
                      \Bigl(N,N_F+1,\frac{Q^2}{\mu^2}\Bigr)
           +O\Bigl(\frac{m^2}{Q^2}\Bigr)~. 
\label{CallFAC}
\end{eqnarray}
Here, $\mu$ refers to the factorization scale between the heavy and light contributions in 
${\cal {C}}_{j,i}$ and {\sf 'asymp'} denotes the limit $Q^2 \gg m^2$. The $C_{i,j}$ are the 
light Wilson coefficients, cf.~\cite{Vermaseren:2005qc}, taken at $N_F+1$ flavors. This can be inferred from 
the fact that in the LCE the Wilson coefficients describe the singularities for very large 
values of $Q^2$, which can not depend on the presence of a quark mass. The mass dependence is 
given by the OMEs $A_{ij}$, between partonic states. Eq.~(\ref{CallFAC}) accounts for all mass 
effects but corrections which are power suppressed, $(m^2/Q^2)^k, k\ge~1$. This factorization 
is only valid if the heavy quark coefficient functions are defined in such a way that all 
radiative corrections containing heavy quark loops are included. Otherwise, (\ref{CallFAC}) 
would not show the correct asymptotic $Q^2$--behavior, \cite{Bierenbaum:2009zt,Buza:1996wv}. 
An equivalent way of describing Eq.~(\ref{CallFAC}) is obtained by considering the calculation 
of the massless Wilson coefficients. Here, the initial state collinear singularities are given 
by evaluating the massless OMEs between off--shell partons, leading to transition functions 
$\Gamma_{ij}$. The $\Gamma_{ij}$ are given in terms of the anomalous dimensions of the twist--$2$ 
operators and transfer the initial state singularities to the bare parton--densities due to 
mass factorization, cf. e.g. \cite{Buza:1995ie,Buza:1996wv}. In the case at hand, something 
similar happens: The initial state collinear singularities are transferred to the parton densities
except for those which are regulated by the quark mass and described by the OMEs. Instead of 
absorbing these terms into the parton densities as well, they are used to reconstruct the  
asymptotic behavior of the heavy flavor Wilson coefficients. Here,
\begin{eqnarray}
     \label{eqAij}
      A_{ij}^{\sf S,NS}\Bigl(N,N_F+1,\frac{m^2}{\mu^2}\Bigr)
               = \langle j| O_i^{\sf S,NS}|j \rangle 
               =\delta_{ij}+\sum_{i=k}^{\infty}a_s^k A_{ij}^{(k),{\sf S,NS}}
                \label{pertomeren}
    \end{eqnarray}
are the operator matrix elements of the local twist--2 operators between on--shell partonic states 
$|j\rangle,~~j = q, g$. 

Let us now derive the explicit expressions for the massive Wilson coefficients in the asymptotic
region. One may split  Eq.~(\ref{CallFAC}) into parts by considering the different $N_F$ 
contributions. We define
\begin{eqnarray}
       {\tilde{f}}(N_F)&\equiv&\frac{f(N_F)}{N_F}~. 
\label{gammapres2}
\end{eqnarray}
This is necessary in order to separate the different types of contributions in Eq.~(\ref{eqF2}), 
weighted by the electric charges of the light and heavy flavors, respectively. Since we would like 
to derive the heavy flavor part, we define as well for later use     
\begin{eqnarray}
\hat{f}(N_F)&\equiv&f(N_F+1)-f(N_F)~, 
\label{gammapres1}
\end{eqnarray}
where $\hat{\hspace*{-1mm}{\tilde{f}}}(N_F) \equiv \widehat{[{\tilde{f}}(N_F)]}$.
The following Eqs. (\ref{LNSFAC})--(\ref{HgFAC}) are the same as Eqs.~(2.31)--(2.35) 
in Ref.~\cite{Buza:1996wv}. We present these terms here again, however, since Ref.~\cite{Buza:1996wv} 
contains a few inconsistencies regarding the $\tilde{f}$--description. Contrary to the latter 
reference, the argument corresponding to the number of flavors stands for all flavors, light or heavy. 
The separation for the ${\sf NS}$--term is obtained by
\begin{eqnarray}
   C_{q,(2,L)}^{\sf NS}\Bigl(N,N_F,\frac{Q^2}{\mu^2}\Bigr)
     + L_{q,(2,L)}^{\sf NS}
          \Bigl(N,N_F+1,\frac{Q^2}{\mu^2},\frac{m^2}{\mu^2}\Bigr)
     &=&
\N\\ &&
       \hspace{-50mm}
        A_{qq,Q}^{\sf NS}\Bigl(N,N_F+1,\frac{m^2}{\mu^2}\Bigr)
        C_{q,(2,L)}^{\sf NS}\Bigl(N,N_F+1,\frac{Q^2}{\mu^2}\Bigr)~.
        \label{LNSFAC}
\end{eqnarray}
Here and in the following, we omit the index $"{\sf asymp}"$ to denote the asymptotic heavy flavor 
Wilson coefficients. For the remaining terms, we suppress the arguments $N$, $Q^2/\mu^2$ and 
$m^2/\mu^2$ for brevity, all of which can be inferred from Eqs. (\ref{Callsplit}, \ref{CallFAC}). 
Additionally, we will suppress from now on the index ${\sf S}$ and label only the ${\sf NS}$
and ${\sf PS}$ terms explicitly. 
The contributions to $L_{i,j}$ read
\begin{eqnarray}
     C_{q,(2,L)}^{\sf PS}(N_F)
       +L_{q,(2,L)}^{\sf PS}
            (N_F+1)
     &=&
        \Bigl[
               A_{qq,Q}^{\sf NS}(N_F+1)
              +A_{qq,Q}^{\sf PS}(N_F+1)
              +A_{Qq}^{\sf PS}(N_F+1)
         \Bigr]
\N\\ &&
         \times
         N_F \tilde{C}_{q,(2,L)}^{\sf PS}(N_F+1)
        +A_{qq,Q}^{\sf PS}(N_F+1)
         C_{q,(2,L)}^{\sf NS}(N_F+1)
\N\\ &&        
+A_{gq,Q}(N_F+1)
         N_F \tilde{C}_{g,(2,L)}(N_F+1)~, \N\\ 
                 \label{LPSFAC} \\
      C_{g,(2,L)}(N_F)
     +L_{g,(2,L)}(N_F+1)
    &=&
           A_{gg,Q}(N_F+1)
           N_F \tilde{C}_{g,(2,L)}(N_F+1)
\N\\ &&
         + A_{qg,Q}(N_F+1)
           C_{q,(2,L)}^{\sf NS}(N_F+1)
\N\\ &&         
+\Bigl[
                A_{qg,Q}(N_F+1)
               +A_{Qg}(N_F+1)
          \Bigr]
        N_F\tilde{C}_{q,(2,L)}^{\sf PS}(N_F+1)~.\N\\
        \label{LgFAC}
    \end{eqnarray}
    The terms $H_{i,j}$ are given by
    \begin{eqnarray}
     H_{q,(2,L)}^{\sf PS}
          (N_F+1)
     &=&  
        A_{Qq}^{\sf PS}(N_F+1)
           \Bigl[ 
                 C_{q,(2,L)}^{\sf NS}(N_F+1)
                +\tilde C_{q,(2,L)}^{\sf PS}
                         (N_F+1)
          \Bigr]
\N\\ &&
          +\Bigl[ 
                A_{qq,Q}^{\sf NS}(N_F+1)
               +A_{qq,Q}^{\sf PS}(N_F+1)
         \Bigr]
        \tilde{C}_{q,(2,L)}^{\sf PS}(N_F+1)
\N\\ &&
       +A_{gq,Q}(N_F+1)
       \tilde{C}_{g,(2,L)}(N_F+1)~,         \label{HPSFAC} \\
     H_{g,(2,L)}(N_F+1)
      &=&
         A_{gg,Q}(N_F+1)
         \tilde{C}_{g,(2,L)}(N_F+1)
        +A_{qg,Q}(N_F+1)
          \tilde{C}_{q,(2,L)}^{\sf PS}(N_F+1)
\N\\ &&
        + A_{Qg}(N_F+1)
          \Bigl[ C_{q,(2,L)}^{\sf NS}(N_F+1)
            +\tilde{C}_{q,(2,L)}^{\sf PS}(N_F+1)
             \Bigr]~.         \label{HgFAC}
\end{eqnarray}
Expanding the above relations up to $O(a_s^3)$, we obtain, using Eqs. (\ref{gammapres2}, 
\ref{gammapres1}), the heavy flavor Wilson coefficients in the asymptotic limit, 
cf.~\cite{Bierenbaum:2009mv}~: 
\begin{eqnarray}
     \label{eqWIL1}
     L_{q,(2,L)}^{\sf NS}(N_F+1) &=& 
     a_s^2 \Bigl[A_{qq,Q}^{(2), {\sf NS}}(N_F+1)~\delta_2 +
     \hat{C}^{(2), {\sf NS}}_{q,(2,L)}(N_F)\Bigr]
     \N\\
     &+&
     a_s^3 \Bigl[A_{qq,Q}^{(3), {\sf NS}}(N_F+1)~\delta_2
     +  A_{qq,Q}^{(2), {\sf NS}}(N_F+1) C_{q,(2,L)}^{(1), {\sf NS}}(N_F+1)
       \N \\
     && \hspace*{5mm}
     + \hat{C}^{(3), {\sf NS}}_{q,(2,L)}(N_F)\Bigr]~,  \\
      \label{eqWIL2}
      L_{q,(2,L)}^{\sf PS}(N_F+1) &=& 
     a_s^3 \Bigl[~A_{qq,Q}^{(3), {\sf PS}}(N_F+1)~\delta_2
     +  A_{gq,Q}^{(2)}(N_F+1)~~N_F\Ctil_{g,(2,L)}^{(1)}(N_F+1) \N \\
     && \hspace*{5mm}
     + N_F \hat{\Ctil}^{(3), {\sf PS}}_{q,(2,L)}(N_F)\Bigr]~,
     \\
     \label{eqWIL3}
      L_{g,(2,L)}^{\sf S}(N_F+1) &=& 
     a_s^2 A_{gg,Q}^{(1)}(N_F+1)N_F \Ctil_{g,(2,L)}^{(1)}(N_F+1)
     \N\\ &+&
      a_s^3 \Bigl[~A_{qg,Q}^{(3)}(N_F+1)~\delta_2 
     +  A_{gg,Q}^{(1)}(N_F+1)~~N_F\Ctil_{g,(2,L)}^{(2)}(N_F+1)
     \N\\ && \hspace*{5mm}
     +  A_{gg,Q}^{(2)}(N_F+1)~~N_F\Ctil_{g,(2,L)}^{(1)}(N_F+1)
     \N\\ && \hspace*{5mm}
     +  ~A^{(1)}_{Qg}(N_F+1)~~N_F\Ctil_{q,(2,L)}^{(2), {\sf PS}}(N_F+1)
     + N_F \hat{\Ctil}^{(3)}_{g,(2,L)}(N_F)\Bigr]~,
 \\ \N \\
\label{eq:WILPS}
     H_{q,(2,L)}^{\sf PS}(N_F+1)
     &=& a_s^2 \Bigl[~A_{Qq}^{(2), {\sf PS}}(N_F+1)~\delta_2
     +~\Ctil_{q,(2,L)}^{(2), {\sf PS}}(N_F+1)\Bigr]
     \\
     &+& a_s^3 \Bigl[~A_{Qq}^{(3), {\sf PS}}(N_F+1)~\delta_2
     +~\Ctil_{q,(2,L)}^{(3), {\sf PS}}(N_F+1) \N\\
 && 
     + A_{gq,Q}^{(2)}(N_F+1)~\Ctil_{g,(2,L)}^{(1)}(N_F+1) 
     + A_{Qq}^{(2), {\sf PS}}(N_F+1)~C_{q,(2,L)}^{(1), {\sf NS}}(N_F+1) 
        \Bigr]~,       \label{eqWIL4}
         \N\\ 
\label{eq:WILS}
     H_{g,(2,L)}^{\sf S}(N_F+1) &=& a_s \Bigl[~A_{Qg}^{(1)}(N_F+1)~\delta_2
     +~\Ctil^{(1)}_{g,(2,L)}(N_F+1) \Bigr] \N\\
     &+& a_s^2 \Bigl[~A_{Qg}^{(2)}(N_F+1)~\delta_2
     +~A_{Qg}^{(1)}(N_F+1)~C^{(1), {\sf NS}}_{q,(2,L)}(N_F+1)\N\\ && 
     \hspace*{5mm}
     +~A_{gg,Q}^{(1)}(N_F+1)~\Ctil^{(1)}_{g,(2,L)}(N_F+1) 
     +~\Ctil^{(2)}_{g,(2,L)}(N_F+1) \Bigr]
     \N\\ &+&
     a_s^3 \Bigl[~A_{Qg}^{(3)}(N_F+1)~\delta_2
     +~A_{Qg}^{(2)}(N_F+1)~C^{(1), {\sf NS}}_{q,(2,L)}(N_F+1)
     \N\\ &&
     \hspace*{5mm}
     +~A_{gg,Q}^{(2)}(N_F+1)~\Ctil^{(1)}_{g,(2,L)}(N_F+1)
     \N\\ && \hspace*{5mm}
     +~A_{Qg}^{(1)}(N_F+1)\Bigl\{
     C^{(2), {\sf NS}}_{q,(2,L)}(N_F+1)
     +~\Ctil^{(2), {\sf PS}}_{q,(2,L)}(N_F+1)\Bigr\}
     \N\\ && \hspace*{5mm}
     +~A_{gg,Q}^{(1)}(N_F+1)~\Ctil^{(2)}_{g,(2,L)}(N_F+1)
     +~\Ctil^{(3)}_{g,(2,L)}(N_F+1) \Bigr]~, 
\label{eqWIL5}
\end{eqnarray}
with $\delta_2 =1$ for $F_2$ and  $\delta_2 =0$ for $F_L$. Again, the argument $(N_F+1)$ in the 
massive OMEs signals that these functions depend on $N_F$ massless and one massive flavor, while
the setting of $N_F$ in the massless Wilson coefficients is a functional one.
The above equations include radiative 
corrections due to heavy quark 
loops to the Wilson coefficients. Therefore, in order to compare e.g. with the calculation in 
Refs.~\cite{NLO}, these terms still have to be subtracted. Since the light flavor Wilson 
coefficients were calculated in the $\overline{\sf MS}$--scheme, the {\sf same} scheme has 
to be used for the massive OMEs. It should also be thoroughly used for renormalization to derive 
consistent results in QCD analyses of deep-inelastic scattering data and to be able to compare to 
other analyses of hard scattering data directly. This requests special attendance w.r.t. the 
choice of the scheme in which $a_s$  is defined, cf.~\cite{Bierenbaum:2009mv}.

The renormalized massive OMEs depend on the ratio $m^2/\mu^2$, while the scale ratio in the 
massless Wilson coefficients is $\mu^2/Q^2$. The latter are pure functions of the momentum 
fraction $z$, or the Mellin variable $N$, if one sets $\mu^2 = Q^2$. The mass dependence on
the heavy flavor Wilson coefficients in the asymptotic region derives from the unrenormalized 
massive OMEs
\begin{eqnarray}
\Ahat^{(3)}_{ij}(\ep)  = \frac{1}{\ep^3} \hat{a}^{(3),3}_{ij} 
+\frac{1}{\ep^2} \hat{a}^{(3),2}_{ij} +\frac{1}{\ep} \hat{a}^{(3),1}_{ij} +\hat{a}^{(3),0}_{ij}~,
\label{eq:unep}
\end{eqnarray}
applying mass, coupling constant, and operator-renormalization, as well as mass factorization, 
cf.~Ref.~\cite{Bierenbaum:2009mv}. The renormalized massive OMEs obey then the general structure
\begin{eqnarray}
      A^{(3)}_{ij}\left(\frac{m^2}{Q^2}\right)  = 
        a^{(3),3}_{ij} \ln^3\left(\frac{m^2}{Q^2}\right)
      + a^{(3),2}_{ij} \ln^2\left(\frac{m^2}{Q^2}\right)
      + a^{(3),1}_{ij} \ln\left(\frac{m^2}{Q^2}\right)
      + a^{(3),0}_{ij}~.
\label{eq:log}
\end{eqnarray}
The subsequent calculations will be performed in the $\overline{\rm MS}$ scheme for the coupling 
constant and the on-shell scheme for the heavy quark mass $m$. The transition to the scheme in 
which $m$ is renormalized in the $\overline{\rm MS}$-scheme is described in 
Ref.~\cite{Bierenbaum:2009mv}.
The strong coupling constant is obtained as the {\it perturbative} solution of the equation
\begin{eqnarray}
\frac{d a_s(\mu^2)}{d \ln(\mu^2)} = - \sum_{l=0}^\infty \beta_l a_s^{l+2}(\mu^2)
\label{eq:as}
\end{eqnarray}
to 3--loop order, where $\beta_k$ are the expansion coefficients of the QCD $\beta$--function and 
$\mu^2$ denotes the renormalization scale. For simplicity we identify the factorization $(\mu_F)$ 
and renormalization $(\mu_R)$ scales from now on. In the subsequent sections we present explicit 
expressions of the asymptotic heavy flavor Wilson coefficients in Mellin-$N$ space. They depend on 
the logarithms
\begin{eqnarray}
L_Q = \ln\left(\frac{Q^2}{\mu^2}\right)~~~\text{and}~~~L_M = \ln\left(\frac{m^2}{\mu^2}\right)~,
\end{eqnarray}
where $\mu \equiv \mu_F = \mu_R$.

Besides the Wilson coefficients (\ref{eqWIL1}--\ref{eqWIL5}) the massive OMEs are important 
themselves to establish the matching conditions in the variable flavor number scheme 
in describing the process of a single massive quark becoming massless\footnote{For the VFNS in 
case of both the bottom and charm quarks transmuting into massless states, see 
\cite{Ablinger:2011pb}.}  
at large enough scales $\mu^2$,~\cite{Buza:1996wv,Bierenbaum:2009mv}. Here, the PDFs for
$N_F+1$ massless quarks are related to the former $N_F$ massless quarks process independently. 
The corresponding relations to 3--loop order read, cf. also \cite{Buza:1996wv}\footnote{Here, we 
have 
corrected some typographical errors in (\ref{eq:VFNS3}--\ref{eq:VFNS2}) in \cite{Buza:1996wv}, in 
accordance with the appendix of Ref.~\cite{Buza:1996wv}.}
~:
{\small
\begin{eqnarray}
\label{eq:VFNS1}
f_k(N_F+1, \mu^2) + f_{\overline{k}}(N_F+1, \mu^2) 
&=& A_{qq,Q}^{\sf NS}\Big(N_F, \frac{\mu^2}{m^2}\Big)
\otimes \left[f_k(N_F, \mu^2) + f_{\overline{k}}(N_F, \mu^2)\right] 
\nonumber\\
&& 
+ {\tilde A_{qq,Q}^{\sf PS}\Big(N_F, \frac{\mu^2}{m^2}\Big)}\otimes
  {\Sigma(N_F, \mu^2)} 
\nonumber\\
&& + {\tilde A_{qg,Q}^{\sf  S}\Big(N_F, \frac{\mu^2}{m^2}\Big)}\otimes
  {G(N_F, \mu^2)}
\\
\label{eq:VFNS3}
  {f_{Q+\bar Q}(N_F+1, \mu^2)}
   &=& {A_{Qq}^{\sf PS}\Big(N_F, \frac{\mu^2}{m^2}\Big)}\otimes
       {\Sigma(N_F, \mu^2)}
   + {A_{Qg}^{\sf S}\Big(N_F, \frac{\mu^2}{m^2}\Big)} \otimes
    {G(N_F, \mu^2)}
\nonumber\\
\\
 {G(N_F+1, \mu^2)} &=&  {A_{gq,Q}^{\sf S}\Bigl(N_F,\frac{\mu^2}{m^2}\Bigr)}
\otimes {\Sigma(N_F,\mu^2)}
+ {A_{gg,Q}^S\Bigl(N_F,\frac{\mu^2}{m^2}\Bigr)} \otimes {G(N_F,\mu^2)}~.
\nonumber\\
\\
{\Sigma(N_F+1,\mu^2)} 
&=& \left[ {A_{qq,Q}^{\sf NS}\left(N_F, \frac{\mu^2}{m^2}\right)} +
N_F
{\tilde{A}_{qq,Q}^{\sf PS} \left(N_F, \frac{\mu^2}{m^2}\right)} +
{{A}_{Qq}^{\sf PS} \left(N_F, \frac{\mu^2}{m^2}\right)} \right]
\nonumber\\ &&
\otimes {\Sigma(N_F,\mu^2)} \nonumber\\ &&
+\left[N_F
{\tilde{A}^{\sf S}_{qg,Q}\left(N_F, \frac{\mu^2}{m^2}\right)}
+ {{A}^{\sf S}_{Qg}\left(N_F, \frac{\mu^2}{m^2}\right)} \right]
\otimes {G(N_F,\mu^2)}
\label{eq:VFNS2}
\end{eqnarray}
}

\normalsize \noindent
Here, the $N_F$-dependence of the OMEs is understood as functional and $\mu^2$ denotes the 
matching scale, which for the heavy-to-light transitions is normally 
much larger than mass scale $m^2$, \cite{Blumlein:1998sh}. We will present the corresponding 
OMEs in Appendix~\ref{app:A}. The results of 
the calculations being presented in the subsequent sections have been obtained making mutual
use of the packages {\tt HarmonicSums.m} \cite{HARMSUM} and {\tt Sigma.m} \cite{SIGMA}.

\section{The Wilson Coefficients \boldmath $L_{q,2}^{\sf PS}$ and $L_{g,2}^{\sf S}$}
\label{sec:3}

\vspace*{1mm}
\noindent
The OMEs for these Wilson coefficients have been calculated in \cite{Ablinger:2010ty}.
They contribute for the first time at 3-- and 2--loop order, respectively, and stem from processes 
in which the virtual electro-weak gauge boson couples to a massless quark.
As a shorthand notation we also define the function
\begin{eqnarray}
\tilde{\gamma}_{qg}^{0} = -4 \frac{N^2 + N + 2}{N (N+1) (N+2)}
\label{eq:gqgtil}
\end{eqnarray}
denoting the kinetic part of the leading order anomalous dimensions separating off the
corresponding color factor.

In Mellin-$N$ space the Wilson coefficient $L_{q,2}^{\sf PS}$ reads~:
{\small
\begin{eqnarray}
\lefteqn{L_{q,2}^{\sf PS} = \tfrac{1}{2}\left[1 + (-1)^N\right]} \nonumber\\ &&  
\times \textcolor{blue}{a_s^3} \Biggl\{
\textcolor{blue}{C_F} \textcolor{blue}{N_F} \textcolor{blue}{T_F^2} \Biggl[
-\frac{32 P_{4} L_Q^2}{9 (N-1) N^3 (N+1)^3 (N+2)^2}
+L_Q \Biggl[\frac{64 P_{6}}{27 (N-1) N^4 (N+1)^4 (N+2)^3}
\N\\&&
-\frac{256 P_{1} (-1)^N}{9 (N-1) N^2 (N+1)^3 (N+2)^3}
+\frac{2 (\tilde{\gamma}_{qg}^{0})^2 (N+2)
L_M^2}{3 (N-1)}
\N\\&&
+\Biggl[
\frac{64 \big(N^2+N+2\big) \big(8 N^3+13 N^2+27 N+16\big)}{9 (N-1) N^2 (N+1)^3 (N+2)}
-\frac{64 \big(N^2+N+2\big)^2 S_1}{3 (N-1) N^2 (N+1)^2 (N+2)}\Biggr] L_M
\N\\&&
+\frac{512 S_{-2}}{3 (N-1) N (N+1) (N+2)}\Biggr]
-\frac{32 P_{4} L_M^2}{9 (N-1) N^3 (N+1)^3 (N+2)^2}
\N\\&&
+\Biggl[
-\frac{32 P_{7}}{27 (N-1) N^4 (N+1)^4 (N+2)^3}
+\frac{64 P_{2} S_1}{3 (N-1) N^3 (N+1)^3 (N+2)^2}
\N\\&&
+\frac{\big(N^2+N+2\big)^2}{(N-1) N^2 (N+1)^2 (N+2)} \frac{32}{3} \left(S_1^2
- S_2\right)\Biggr] L_M
\N\\&&
-\frac{32 P_{9}}{243 (N-1) N^5 (N+1)^5 (N+2)^4}
+\frac{32 P_{8} S_1}{81 (N-1) N^4 (N+1)^4 (N+2)^3}
\N\\&&
-\frac{16 P_{3} S_1^2}{27 (N-1) N^3 (N+1)^3 (N+2)^2}
-\frac{16 P_{5} S_2}{27 (N-1) N^3 (N+1)^3 (N+2)^2}
\N\\&&
+\frac{32 L_Q^3 \big(N^2+N+2\big)^2}{9 (N-1) N^2 (N+1)^2 (N+2)}
-\frac{32 \big(N^2+N+2\big)^2 L_M^3}{9 (N-1) N^2 (N+1)^2 (N+2)}
\N\\&&
+\frac{\big(N^2+N+2\big)^2}{(N-1) N^2 (N+1)^2 (N+2)} 
\Bigl[-\frac{64}{27} S_1^3
+\frac{32}{9} S_2 S_1
+\frac{160 S_3}{27}
+\frac{256 \zeta_3}{9}\Bigr]\Biggr]
\N\\&&
+\textcolor{blue}{N_F} \hat{\tilde{C}}_{2,q}^{{\sf PS},(3)}({N_F})\Biggr\}~,
\end{eqnarray}

}
with the polynomials
{\small
\begin{eqnarray}
P_{1}&=&4 N^6+22 N^5+48 N^4+53 N^3+45 N^2+36 N+8
    \\
P_{2}&=&N^7-15 N^5-58 N^4-92 N^3-76 N^2-48 N-16
    \\
P_{3}&=&N^7-37 N^6-248 N^5-799 N^4-1183 N^3-970 N^2-580 N-168
    \\
P_{4}&=&11 N^7+37 N^6+53 N^5+7 N^4-68 N^3-56 N^2-80 N-48
    \\
P_{5}&=&49 N^7+185 N^6+340 N^5+287 N^4+65 N^3+62 N^2-196 N-168
 \\
P_{6}&=&85 N^{10}+530 N^9+1458 N^8+2112 N^7+1744 N^6+2016 N^5+3399 N^4+2968 N^3
\N\\&&
+1864 N^2+1248 N+432
 \\
P_{7}&=&143 N^{10}+838 N^9+1995 N^8+1833 N^7-1609 N^6-5961 N^5-7503 N^4-6928 N^3
\N\\&&
-4024 N^2-816 N+144
 \\
P_{8}&=&176 N^{10}+973 N^9+1824 N^8-948 N^7-10192 N^6-19173 N^5-20424 N^4-16036 N^3
\N\\&&
-7816 N^2-1248 N+288
 \\
P_{9}&=&1717 N^{13}+16037 N^{12}+66983 N^{11}+161797 N^{10}+241447 N^9+216696 N^8+86480 N^7
\N\\&&
-67484 N^6-170003 N^5-165454 N^4-81976 N^3-15792 N^2-1008 N-864~.
\end{eqnarray}

}

\noindent
For the massless 3-loop Wilson coefficients $C_{i,j}^{k}$ we refer to 
Ref.~\cite{Vermaseren:2005qc}. Here and in the following, their expression will be kept 
symbolically.
The corresponding $z$-space expressions are given in Appendix~\ref{app:B}.
  
Likewise the Wilson coefficient $L_{g,2}^{\sf S}$ is given by~:
{\small
\begin{eqnarray}
\label{eq:L2gS}
\lefteqn{L_{g,2}^{\sf S} = \tfrac{1}{2} \left[1 + (-1)^N\right] 
\Biggl\{\textcolor{blue}{a_s^2 T_F^2} \textcolor{blue}{N_F} \Biggl\{
L_M \Biggl[\frac{4}{3} \tilde{\gamma}_{qg}^{0} S_1
-\frac{16 \big(N^3-4 N^2-N-2\big)}{3 N^2 (N+1) (N+2)}\Biggr]}
\N\\&&
-\frac{4}{3} \tilde{\gamma}_{qg}^{0} L_Q L_M\Biggr\} 
+\textcolor{blue}{a_s^3}
\Biggl\{
\textcolor{blue}{N_F}  \textcolor{blue}{T_F^3}
\Biggl[
L_M^2 \Biggl[\frac{16}{9} \tilde{\gamma}_{qg}^{0} S_1
-\frac{64 \big(N^3-4 N^2-N-2\big)}{9 N^2 (N+1) (N+2)}\Biggr]
-\frac{16}{9} \tilde{\gamma}_{qg}^{0} L_Q L_M^2\Biggr]
\N\\&&
+\textcolor{blue}{C_A} \textcolor{blue}{N_F}  \textcolor{blue}{T_F^2} \Biggl[
\Biggl[
\frac{64 \big(N^2+N+1\big) \big(N^2+N+2\big)}{9 (N-1) N^2 (N+1)^2 (N+2)^2}
+\frac{8}{9} \tilde{\gamma}_{qg}^{0} S_1\Biggr] L_Q^3
+\Biggl[
-\frac{64 (-1)^N \big(N^3+4 N^2+7 N+5\big)}{3 (N+1)^3 (N+2)^3}
\N\\&&
+\frac{8 P_{25}}{9 (N-1) N^3 (N+1)^3 (N+2)^3}
+\frac{32 \big(8 N^4-7 N^3+5 N^2-17 N-13\big) S_1}{9 (N-1) N (N+1)^2 (N+2)}
\N\\&&
+L_M \Biggl[
\frac{64 \big(N^2+N+1\big) \big(N^2+N+2\big)}{3 (N-1) N^2 (N+1)^2 (N+2)^2}
+\frac{8}{3} \tilde{\gamma}_{qg}^{0} S_1\Biggr]
+\tilde{\gamma}_{qg}^{0} \Bigl[
-\frac{4}{3} S_1^2
+\frac{4 S_2}{3}
+\frac{8}{3} S_{-2}\Bigr]\Biggr] L_Q^2
\N\\&&
+\Biggl[
-\frac{32 \big(8 N^4-7 N^3+5 N^2-17 N-13\big) S_1^2}{9 (N-1) N (N+1)^2 (N+2)}
+\frac{128 (-1)^N \big(N^3+4 N^2+7 N+5\big) S_1}{3 (N+1)^3 (N+2)^3}
\N\\&&
-\frac{32 P_{24} S_1}{27 (N-1) N^2 (N+1)^3 (N+2)^3}
+\frac{64 (-1)^N P_{18}}{9 (N-1) N^2 (N+1)^4 (N+2)^4}
\N\\&&
-\frac{16 P_{32}}{27 (N-1) N^3 (N+1)^4 (N+2)^4}
+L_M^2 \Biggl[
\frac{64 \big(N^2+N+1\big) \big(N^2+N+2\big)}{3 (N-1) N^2 (N+1)^2 (N+2)^2}
\N\\&&
+\frac{8}{3} \tilde{\gamma}_{qg}^{0} S_1\Biggr]
+\frac{32 \big(8 N^4+13 N^3-22 N^2-9 N-26\big) S_2}{9 (N-1) N (N+1) (N+2)^2}
+\frac{128 \big(N^2+N-1\big) S_3}{9 N (N+1) (N+2)}
\N\\&&
+\frac{64 \big(8 N^5+15 N^4+6 N^3+11 N^2+16 N+16\big) S_{-2}}{9 (N-1) N (N+1)^2 (N+2)^2}+L_M \Biggl[
\frac{32 P_{26}}{9 (N-1) N^3 (N+1)^3 (N+2)^3}
\N\\&&
-\frac{128 (-1)^N \big(N^3+4 N^2+7 N+5\big)}{3 (N+1)^3 (N+2)^3}
-\frac{64 (2 N-1) \big(N^3+9 N^2+7 N+7\big) S_1}{9 (N-1) N (N+1)^2 (N+2)}
\N\\&&
+\tilde{\gamma}_{qg}^{0} \Bigl[
-\frac{8}{3} S_1^2
+\frac{8 S_2}{3}
+\frac{16}{3} S_{-2}\Bigr]\Biggr]
-\frac{128 \big(N^2+N+3\big) S_{-3}}{3 N (N+1) (N+2)}+\tilde{\gamma}_{qg}^{0} \Bigl[\frac{8}{9} S_1^3-8 S_2 S_1
+\frac{32}{3} S_{2,1}\Bigr]
\N\\&&
+\frac{256 S_{-2,1}}{3 N (N+1) (N+2)}+\frac{(N-1) \Bigl[\frac{64}{3} S_{-2} S_1-32 \zeta_3\Bigr]}{N (N+1)}\Biggr] L_Q
+\frac{16 P_{12} S_1^2}{81 N (N+1)^3 (N+2)^3}
\N\\&&
+\frac{8 P_{39}}{243 (N-1) N^5 (N+1)^5 (N+2)^5}
+\frac{512}{9} \frac{\big(N^2+N+1\big) \big(N^2+N+2\big)}{(N-1) N^2 (N+1)^2 (N+2)^2} \zeta_3
\N\\&&
+\frac{8 P_{36} S_1}{243 (N-1) N^4 (N+1)^4 (N+2)^4}+L_M^3 \Biggl[
-\frac{64 \big(N^2+N+1\big) \big(N^2+N+2\big)}{9 (N-1) N^2 (N+1)^2 (N+2)^2}
\N\\&&
-\frac{8}{9} \tilde{\gamma}_{qg}^{0} S_1\Biggr]
-\frac{16 P_{13} S_2}{81 N (N+1)^3 (N+2)^3}
+\frac{64 \big(5 N^4+38 N^3+59 N^2+31 N+20\big) S_3}{81 N (N+1)^2 (N+2)^2}
\N\\&&
-\frac{32 \big(121 N^3+293 N^2+414 N+224\big) S_{-2}}{81 N (N+1)^2 (N+2)}
+L_M^2 \Biggl[
-\frac{64 (-1)^N \big(N^3+4 N^2+7 N+5\big)}{3 (N+1)^3 (N+2)^3}
\N\\&&
+\frac{8 P_{25}}{9 (N-1) N^3 (N+1)^3 (N+2)^3}
+\frac{32 \big(8 N^4-7 N^3+5 N^2-17 N-13\big) S_1}{9 (N-1) N (N+1)^2 (N+2)}+\tilde{\gamma}_{qg}^{0} \Bigl[
-\frac{4}{3} S_1^2
\N
\\
&&
+\frac{4 S_2}{3}
+\frac{8}{3} S_{-2}\Bigr]\Biggr]
+\frac{128 \big(5 N^2+8 N+10\big) S_{-3}}{27 N (N+1) (N+2)}
\N\\&&
+\frac{\big(5 N^4+20 N^3+41 N^2+49 N+20\big) \Bigl[\frac{32}{81} S_1^3
-\frac{32}{27} S_2 S_1
+\frac{128}{27} S_{2,1}\Bigr]}{N (N+1)^2 (N+2)^2}
\N\\&&
+L_M \Biggl[\frac{32 \big(2 N^5+21 N^4+27 N^3+11 N^2+25 N-14\big) S_1^2}{9 (N-1) N (N+1)^2 (N+2)^2}
+\frac{16 P_{27} S_1}{27 (N-1) N^3 (N+1)^3 (N+2)^3}
\N\\&&
+\frac{128 (-1)^N \big(N^3+4 N^2+7 N+5\big) S_1}{3 (N+1)^3 (N+2)^3}
-\frac{16}{3} \tilde{\gamma}_{qg}^{0} S_2 S_1
-\frac{64 (-1)^N P_{14}}{9 (N-1) N^2 (N+1)^3 (N+2)^4}
\N\\&&
+\frac{16 P_{34}}{27 (N-1) N^4 (N+1)^4 (N+2)^4}
-\frac{32 \big(2 N^5+21 N^4+51 N^3+23 N^2-11 N-14\big) S_2}{9 (N-1) N (N+1)^2 (N+2)^2}
\N\\&&
+\frac{64 S_3}{3 (N+2)}
-\frac{64 \big(2 N^5+21 N^4+36 N^3-7 N^2-68 N-56\big) S_{-2}}{9 (N-1) N (N+1)^2 (N+2)^2}+\frac{\frac{256}{3} S_{-2,1}
-\frac{128}{3} S_{-3}}{N (N+1) (N+2)}
\N\\&&
+\frac{(N-1) \big(\frac{64}{3} S_{-2} S_1-32 \zeta_3\big)}{N (N+1)}\Biggr]+\tilde{\gamma}_{qg}^{0} \Bigl[\frac{1}{27} S_1^4
-\frac{2}{9} S_2 S_1^2+\Bigl[\frac{16}{9} S_{2,1}
-\frac{40 S_3}{27}\Bigr] S_1
\N\\&&
+\frac{64}{9} \zeta_3 S_1
+\frac{1}{9} S_2^2
+\frac{14 S_4}{9}
+\frac{32}{9} S_{-4}
+\frac{32}{9} S_{3,1}
-\frac{16}{9} S_{2,1,1}\Bigr]\Biggr]
\N\\&&
+\textcolor{blue}{C_F} \textcolor{blue}{N_F}  \textcolor{blue}{T_F^2} \Biggl[
\Biggl[\frac{16 \big(N^2+N+1\big) \big(N^2+N+2\big) \big(3 N^4+6 N^3-N^2-4 N+12\big)}{9 (N-1) N^3 (N+1)^3 (N+2)^2}
+\frac{16}{9} \tilde{\gamma}_{qg}^{0} S_1\Biggr] L_Q^3
\N\\&&
+\Biggl[
-\frac{4 P_{31}}{9 (N-1) N^4 (N+1)^4 (N+2)^3}
+\frac{16 P_{21} S_1}{9 (N-1) N^3 (N+1)^3 (N+2)^2}
+\tilde{\gamma}_{qg}^{0} \Bigl[\frac{20 S_2}{3}-4 S_1^2\Bigr]
\N\\&&
+L_M \Biggl[\frac{8 \big(N^2+N+2\big) P_{10}}{3 (N-1) N^3 (N+1)^3 (N+2)^2}
+\frac{8}{3} \tilde{\gamma}_{qg}^{0} S_1\Biggr]
\Biggr] L_Q^2
+\Biggl[
\frac{16 L_M^2 \big(N^2+N+2\big)^3}{(N-1) N^3 (N+1)^3 (N+2)^2}
\N\\&&
-\frac{16 P_{22} S_1^2}{9 (N-1) N^3 (N+1)^3 (N+2)^2}
+\frac{64 (-1)^N P_{37}}{45 (N-2) (N-1)^2 N^3 (N+1)^4 (N+2)^4 (N+3)^3}
\N\\&&
+\frac{4 P_{42}}{45 (N-1)^2 N^5 (N+1)^5 (N+2)^4 (N+3)^3}
-\frac{8 P_{30} S_1}{9 (N-1) N^4 (N+1)^4 (N+2)^3}
\N\\&&
+\frac{16 P_{23} S_2}{9 (N-1) N^3 (N+1)^3 (N+2)^2}+L_M \Biggl[
-\frac{16 P_{28}}{3 (N-1) N^4 (N+1)^4 (N+2)^3}
\N\\&&
+\frac{16 P_{17} S_1}{3 (N-1) N^3 (N+1)^3 (N+2)^2}+\tilde{\gamma}_{qg}^{0} \big(\frac{16 S_2}{3}
-\frac{16}{3} S_1^2\big)\Biggr]
-\frac{256 \big(N^2+N+1\big) S_3}{3 N (N+1) (N+2)}
\N\\&&
+\frac{64 P_{16} S_{-2}}{3 (N-2) (N-1) N^2 (N+1)^2 (N+2)^2 (N+3)}+\tilde{\gamma}_{qg}^{0} \Bigl[\frac{8}{3} S_1^3-8 S_2 S_1
-\frac{32}{3} S_{2,1}\Bigr]
\N\\&&
+\frac{\frac{512}{3} S_1 S_{-2}
+\frac{256}{3} S_{-3}
-\frac{512}{3} S_{-2,1}}{N (N+1) (N+2)}
+\frac{64 (N-1) \zeta_3}{N (N+1)}\Biggr] L_Q
-\frac{64}{9} \frac{\big(N^2+N+2\big) P_{10} \zeta_3}{(N-1) N^3 (N+1)^3 (N+2)^2} 
\N\\&&
+\frac{8 \big(215 N^4+481 N^3+930 N^2+748 N+120\big) S_1^2}{81 N^2 (N+1)^2 (N+2)}
+\frac{P_{40}}{243 (N-1) N^6 (N+1)^6 (N+2)^5}
\N\\&&
-\frac{4 P_{35} S_1}{243 (N-1) N^5 (N+1)^5 (N+2)^2}
+L_M^3 \Biggl[\frac{8 \big(N^2+N+2\big) P_{10}}{9 (N-1) N^3 (N+1)^3 (N+2)^2}
+\frac{8}{9} \tilde{\gamma}_{qg}^{0} S_1\Biggr]
\N\\&&
+L_M^2 \Biggl[\frac{4 P_{29}}{9 (N-1) N^4 (N+1)^4 (N+2)^3}
-\frac{16 P_{20} S_1}{9 (N-1) N^3 (N+1)^3 (N+2)^2}
+\tilde{\gamma}_{qg}^{0} \Bigl[
-\frac{4}{3} S_1^2
-\frac{4 S_2}{3}\Bigr]\Biggr]
\N\\&&
+\frac{8 \big(109 N^4+291 N^3+478 N^2+324 N+40\big) S_2}{27 N^2 (N+1)^2 (N+2)}
+\frac{\big(10 N^3+13 N^2+29 N+6\big) \Bigl[
-\frac{16}{81} S_1^3
-\frac{16}{27} S_2 S_1\Bigr]}{N^2 (N+1) (N+2)}
\N\\&&
+\frac{32 \big(5 N^3-16 N^2+N-6\big) S_3}{81 N^2 (N+1) (N+2)}
+\tilde{\gamma}_{qg}^{0} \Bigl[
-\frac{1}{27} S_1^4
-\frac{2}{9} S_2 S_1^2
-\frac{8}{27} S_3 S_1
-\frac{64}{9} \zeta_3 S_1
-\frac{1}{9} S_2^2
+\frac{14 S_4}{9}\Bigr]
\N\\&&
+L_M \Biggl[
-\frac{8 P_{19} S_1^2}{9 (N-1) N^3 (N+1)^3 (N+2)^2}
-\frac{16 P_{33} S_1}{27 (N-1) N^4 (N+1)^4 (N+2)^3}
\N\\&&
+\frac{64 (-1)^N P_{38}}{45 (N-2) (N-1)^2 N^3 (N+1)^4 (N+2)^4 (N+3)^3}
+\frac{8 \big(N^2+N+2\big) P_{11} S_2}{9 (N-1) N^3 (N+1)^3 (N+2)^2}
\N\\&&
+\frac{4 P_{41}}{135 (N-1)^2 N^5 (N+1)^5 (N+2)^4 (N+3)^3}
+\frac{64 P_{15} S_{-2}}{3 (N-2) (N-1) N^2 (N+1)^2 (N+2)^2 (N+3)}
\N\\&&
+\tilde{\gamma}_{qg}^{0} \big(\frac{8}{3} S_1^3
-\frac{8}{3} S_2 S_1
-\frac{16}{3} S_{2,1}\big)+\frac{\frac{512}{3} S_1 S_{-2}
+\frac{256}{3} S_{-3}
-\frac{512}{3} S_{-2,1}}{N (N+1) (N+2)}+\frac{(N-1) \big(64 \zeta_3
-\frac{64 S_3}{3}\big)}{N (N+1)}\Biggr]\Bigg]
\N\\&&
+\textcolor{blue}{N_F} \hat{\tilde{C}}_{2,g}^{{\sf S},(3)}({N_F})\Biggr\}\Biggr\}~,
\end{eqnarray}

}
where
{\small
\begin{eqnarray}
P_{10}&=&3 N^6+9 N^5-N^4-17 N^3-38 N^2-28 N-24
    \\
P_{11}&=&47 N^6+141 N^5+59 N^4-117 N^3+2 N^2+84 N+72
    \\
P_{12}&=&65 N^6+455 N^5+1218 N^4+1820 N^3+1968 N^2+1460 N+448
    \\
P_{13}&=&139 N^6+1093 N^5+3438 N^4+5776 N^3+5724 N^2+3220 N+752
    \\
P_{14}&=&9 N^7+71 N^6+214 N^5+320 N^4+275 N^3+215 N^2+160 N+32
    \\
P_{15}&=&N^8+8 N^7-2 N^6-60 N^5-23 N^4+108 N^3+96 N^2+16 N+48
    \\
P_{16}&=&N^8+8 N^7-2 N^6-60 N^5+N^4+156 N^3+24 N^2-80 N-240
    \\
P_{17}&=&3 N^8+8 N^7-2 N^6-24 N^5+15 N^4+88 N^3+152 N^2+96 N+48
    \\
P_{18}&=&5 N^8-8 N^7-137 N^6-436 N^5-713 N^4-672 N^3-407 N^2-192 N-32
    \\
P_{19}&=&7 N^8+4 N^7-90 N^6-224 N^5-21 N^4+388 N^3+608 N^2+336 N+144
    \\
P_{20}&=&10 N^8+46 N^7+105 N^6+139 N^5+87 N^4-17 N^3+50 N^2+84 N+72
    \\
P_{21}&=&19 N^8+70 N^7+63 N^6-41 N^5-192 N^4-221 N^3-142 N^2-60 N-72
    \\
P_{22}&=&38 N^8+146 N^7+177 N^6+35 N^5-249 N^4-373 N^3-218 N^2-60 N-72
    \\
P_{23}&=&56 N^8+194 N^7+213 N^6+83 N^5-231 N^4-469 N^3-290 N^2-60 N-72
    \\
P_{24}&=&113 N^8+348 N^7+109 N^6-289 N^5-272 N^4-859 N^3-778 N^2-172 N+72
    \\
P_{25}&=&9 N^9+54 N^8+56 N^7-110 N^6-381 N^5-568 N^4-364 N^3-72 N^2+128 N+96
 \\
P_{26}&=&9 N^9+54 N^8+167 N^7+397 N^6+780 N^5+1241 N^4+1448 N^3+1200 N^2+608 N+144
 \\
P_{27}&=&55 N^9+336 N^8+218 N^7-2180 N^6-6529 N^5-9764 N^4-9368 N^3-6032 N^2
\N
\\
&&
-2448 N-576
 \\
P_{28}&=&N^{11}-56 N^9-236 N^8-373 N^7+82 N^6+1244 N^5+2330 N^4+2560 N^3+1712 N^2
\N\\&&
+896 N+288
 \\
P_{29}&=&33 N^{11}+231 N^{10}+662 N^9+1254 N^8+1801 N^7+2759 N^6+5440 N^5+9884 N^4
\N\\&&
+12512 N^3+9200 N^2+5184 N+1728
 \\
P_{30}&=&45 N^{11}+383 N^{10}+958 N^9+526 N^8-763 N^7+1375 N^6+7808 N^5+13028 N^4
\N\\&&
+12976 N^3+8016 N^2+4608 N+1728
 \\
P_{31}&=&81 N^{11}+483 N^{10}+1142 N^9+1086 N^8-767 N^7-4645 N^6-8936 N^5-11980 N^4
\N\\&&
-12352 N^3-8272 N^2-4800 N-1728
 \\
P_{32}&=&120 N^{11}+1017 N^{10}+2737 N^9+1292 N^8-8086 N^7-20743 N^6-24563 N^5-16702 N^4
\N\\&&
-6840 N^3+120 N^2+2432 N+960
 \\
P_{33}&=&121 N^{11}+988 N^{10}+3554 N^9+6972 N^8+7131 N^7-846 N^6-14806 N^5-25354 N^4
\N\\&&
-26096 N^3-16752 N^2-8352 N-2592
 \\
P_{34}&=&27 N^{12}+441 N^{11}+2206 N^{10}+5360 N^9+7445 N^8+8555 N^7+18766 N^6+44852 N^5
\N\\&&
+67572 N^4
+63960 N^3+39632 N^2+15648 N+2880
 \\
P_{35}&=&2447 N^{12}+16902 N^{11}+59649 N^{10}+125860 N^9+128761 N^8-36530 N^7-248341 N^6
\N\\&&
-304460 N^5-162188 N^4
-11724 N^3+29160 N^2+19440 N+7776
 \\
P_{36}&=&3361 N^{12}+23769 N^{11}+62338 N^{10}+59992 N^9-63303 N^8-317823 N^7-585520 N^6
\N\\&&
-640602 N^5-430132 N^4-167536 N^3-27648 N^2+9504 N+5184
 \\
P_{37}&=&76 N^{14}+802 N^{13}+2979 N^{12}+1847 N^{11}-19377 N^{10}-58253 N^9-26543 N^8+170601 N^7
\N\\&&
+362177 N^6+225119 N^5-103240 N^4-193092 N^3-137160 N^2-117072 N-25920
 \\
P_{38}&=&76 N^{14}+1042 N^{13}+5979 N^{12}+16367 N^{11}+11883 N^{10}-47693 N^9-125723 N^8-86079 N^7
\N\\&&
+36437 N^6+22559 N^5-51700 N^4+24828 N^3+132840 N^2+116208 N+25920
 \\
P_{39}&=&3180 N^{15}+38835 N^{14}+188728 N^{13}+456665 N^{12}+460954 N^{11}-406761 N^{10}-1972948 N^9
\N\\&&
-2827653 N^8-1857970 N^7+109786 N^6+1302824 N^5+1092456 N^4
\N\\&&
+265888 N^3-227616 N^2-194688 N-44928
 \\
P_{40}&=&28503 N^{17}+297639 N^{16}+1232041 N^{15}+2461407 N^{14}+2169615 N^{13}+662941 N^{12}
\N\\&&
+2110979 N^{11}+5346653 N^{10}+2021366 N^9-7290864 N^8-11721384 N^7-3689680 N^6
\N\\&&
+15676192 N^5+32276800 N^4+31869312 N^3+18809856 N^2+6856704 N+1244160
 \\
P_{41}&=&75 N^{18}+3330 N^{17}+35497 N^{16}+175010 N^{15}+486862 N^{14}+966996 N^{13}+2037362 N^{12}
\N\\&&
+3604404 N^{11}-1625689 N^{10}-29506022 N^9-78753403 N^8-107977014 N^7-71548880 N^6
\N\\&&
+18344016 N^5+89016048 N^4+92657952 N^3+58942080 N^2+25505280 N+5598720
 \\
P_{42}&=&325 N^{18}+4280 N^{17}+17759 N^{16}-14880 N^{15}-412326 N^{14}-1696848 N^{13}-3216546 N^{12}
\N\\&&
-1169232 N^{11}+8956857 N^{10}+23914216 N^9+31536899 N^8+25361392 N^7+9982840 N^6
\N\\&&
-10154128 N^5-26098704 N^4-26761536 N^3-17642880 N^2-8087040 N-1866240~.
\end{eqnarray}

}

\noindent
In all the representations of the massive Wilson coefficients and OMEs in $N$--space we apply algebraic reduction
\cite{Blumlein:2003gb}.
The 2--loop term in (\ref{eq:L2gS}) is purely multiplicative and induced by renormalization only,
while the 3--loop contributions require the calculation of massive OMEs. The above Wilson 
coefficients depend on the harmonic sums 
\begin{eqnarray}
S_1, S_{-2}, S_2, S_{-3}, S_3, S_{-4}, S_4, S_{-2,1}, S_{2,1}, S_{3,1}, S_{2,1,1},
\end{eqnarray}
apart of those defining the massless 3--loop Wilson coefficients \cite{Vermaseren:2005qc}\footnote{For the 
algebraically reduced representations see \cite{Blumlein:2009tj}.}. The harmonic sums are defined 
recursively by, cf.~\cite{HSUM1,HSUM2}, 
\begin{eqnarray}
S_{b,\vec{a}}(N) = \sum_{k=1}^N \frac{({\rm sign}(b))^k}{k^{|b|}} S_{\vec{a}}(k),~~~
b, a_i \in \mathbb{Z} \backslash \{0\}, N \in \mathbb{N}, N \geq 1, S_\emptyset = 1~.
\end{eqnarray}
In the above $\zeta_l = \sum_{k=1}^\infty 1/k^l, l \in \mathbb{N}, l \geq 2$ denote the 
Riemann  
$\zeta$-values, which are convergent harmonic sums in the limit $N \rightarrow \infty$. In the 
constant part of the other Wilson coefficients it is expected that more complicated multiple zeta 
values emerge, which have been dealt with in \cite{Blumlein:2009cf}.

In Eq.~(\ref{eq:L2gS}) denominator terms $\propto 1/(N-2)$ occur. They cancel in the complete expression
and the rightmost singularity is located at $N = 1$ as expected for this Wilson coefficient.
Let us now consider both the small- and large-$x$ dominant terms for both Wilson coefficients.
Those of the massless parts were given in \cite{Vermaseren:2005qc} before. Both Wilson coefficients
contain terms $\propto 1/(N-1)$. For simplicity we consider the choice of scale  $Q^2 = \mu^2$ here.
The expansion of the heavy flavor contribution, subtracting 
the massless 3--loop Wilson coefficients, denoted by $\widehat{L}_i$,  around $N = 1 (x 
\rightarrow 0)$ and in the limit $N 
\rightarrow \infty (x \rightarrow 1)$, setting $Q^2 = \mu^2$,
are given by
\begin{eqnarray}
\widehat{L}_{q,2}^{\sf PS}(N \rightarrow 1) &\propto& 
\frac{1}{N-1} C_F T_F^2 N_F \Biggl\{ \frac{1024}{27} \zeta_3
-\frac{64}{729} \Biggl[
  54 L_M^3
- 81 L_M^2
+342 L_M
+500 \Biggr] \Biggr\}
\\
\widehat{L}_{g,2}^{\sf S}(N \rightarrow 1) &\propto& \frac{1}{N-1} \Biggl\{C_A T_F^2 N_F \Biggl\{
 \frac{512}{27} \zeta_3 
-\frac{16}{729} \Biggl[108 L_M^3 + 540 L_M^2 + 54 L_M +3091
\Biggr]\Biggr\}
\nonumber\\ 
&&
+ C_F T_F^2 N_F \Biggl\{\frac{1024}{27} \zeta_3 
- \frac{32}{729} \Biggl[108 L_M^3
-864 L_M^2
+1314 L_M
-1091 \Biggr] \Biggr\}\Biggr\}
\\
\widehat{L}_{q,2}^{\sf PS}(N \rightarrow \infty) &\propto& - \frac{64}{27} C_F T_F^2 N_F 
\frac{\ln^3(\bar{N})}{N^2}
\\
\widehat{L}_{g,2}^{\sf S}(N \rightarrow \infty) &\propto&
-\frac{4}{27} \frac{(N-2) \ln^4(\bar{N})}{N^2} (C_A - C_F) T_F^2 N_F~.
\end{eqnarray}
The corresponding limits for the contributions of the massless Wilson coefficients behave like 
\begin{eqnarray}
N_F \hat{\tilde{C}}_{2,q}^{{\sf PS},(3)}({N_F})(N \rightarrow 1) &\propto& \frac{4}{N-1} C_F T_F^2 
N_F
\left[\frac{22112}{729} - \frac{32}{9} \zeta_2 + \frac{128}{27} \zeta_3 \right]~\text{[\citen{Vermaseren:2005qc}]}
\\
N_F \hat{\tilde{C}}_{2,g}^{{\sf S},(3)}({N_F}) (N \rightarrow 1) &\propto& \frac{4}{N-1} \Biggl\{
C_A T_F^2 N_F \left[ -\frac{572}{729} + \frac{160}{27} \zeta_2 + \frac{64}{27} \zeta_3 \right]
\nonumber\\ &&
+ C_F T_F^2 N_F \left[ \frac{45468}{729} - \frac{512}{27} \zeta_2 + \frac{128}{27} \zeta_3 \right]
\Biggr\}~\text{[\citen{Vermaseren:2005qc}]}
\\
N_F \hat{\tilde{C}}_{2,q}^{{\sf PS},(3)}({N_F})(N \rightarrow \infty) &\propto& 
\frac{\ln^3(\bar{N})}{N^3} C_F T_F^2 N_F \frac{64}{27} 
\\
N_F \hat{\tilde{C}}_{2,g}^{{\sf S},(3)}({N_F})(N \rightarrow \infty)  &\propto& 
\frac{\ln^4(\bar{N})}{N} \left[\frac{68}{27} C_F T_F^2 N_F + \frac{28}{27} C_A T_F^2 N_F\right],
\end{eqnarray}
where $\bar{N} = N \exp(\gamma_E)$ and $\gamma_E$ denotes the Euler-Mascheroni constant.

\begin{figure}[H]
\begin{center}
\includegraphics[scale=0.8]{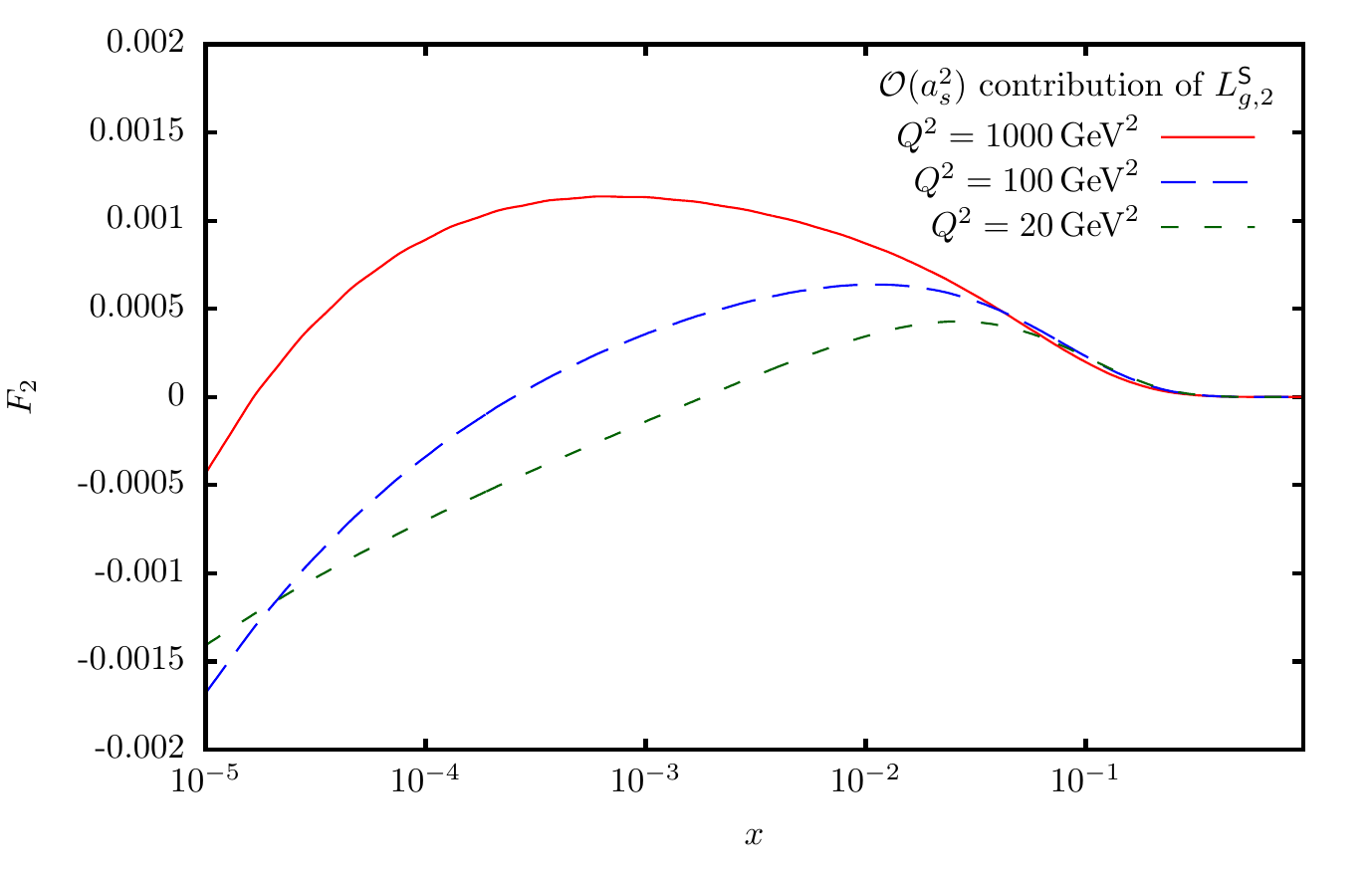}
\end{center}
\caption[]{\sf The $O(a_s^2)$ contribution by $L_{g,2}^{\sf S}$ to the structure function $F_2(x,Q^2)$. \label{fig1}}
\end{figure}

\begin{figure}[H]
\begin{center}
\includegraphics[scale=0.8]{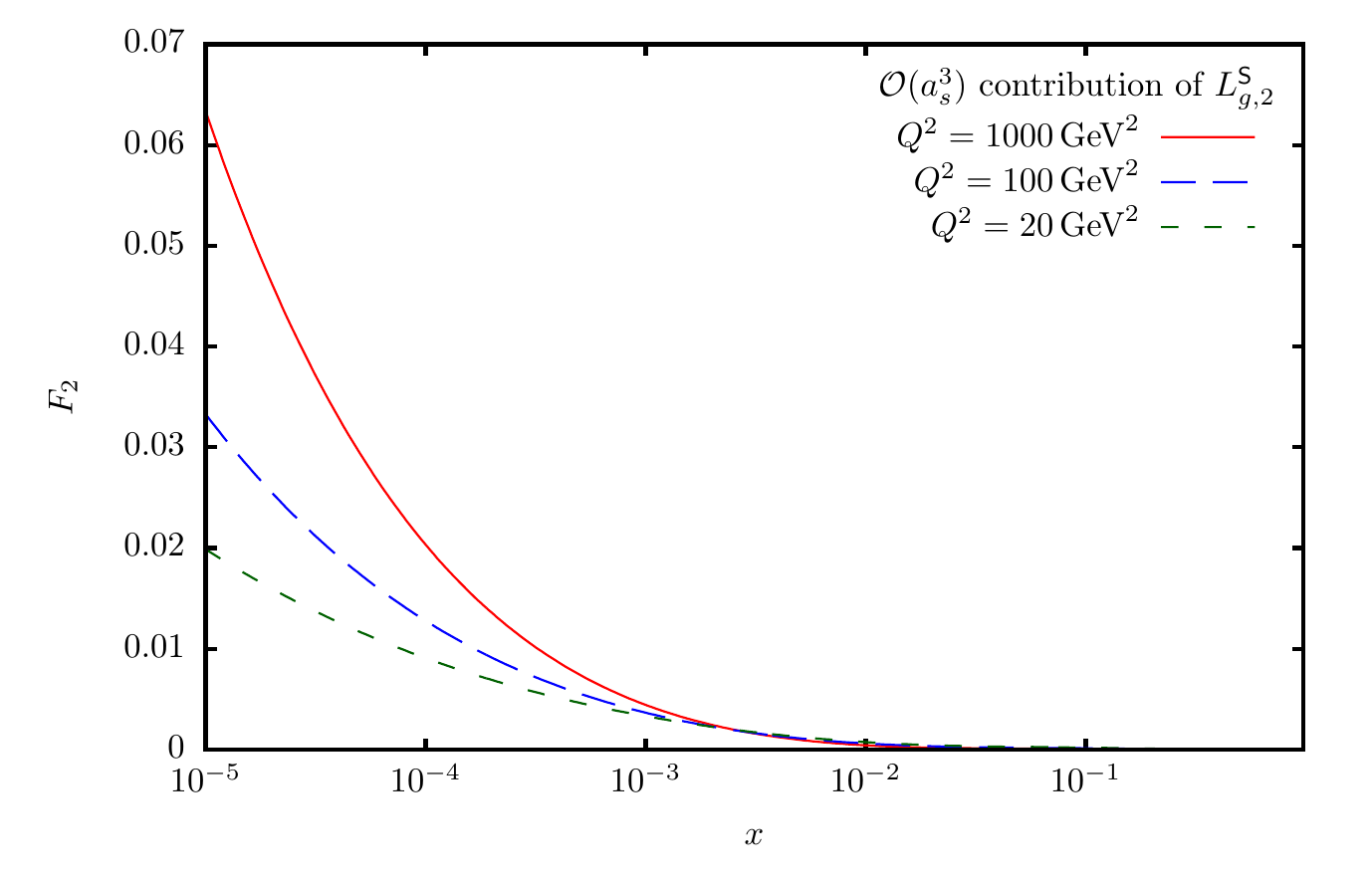}
\end{center}
\caption[]{\sf The $O(a_s^3)$ contribution by $L_{g,2}^{\sf S}$ to the structure function $F_2(x,Q^2)$. \label{fig2}}
\end{figure}
While the expression for $L_{q,2}^{\sf PS}$ is the same in the $\overline{\rm MS}$-- and on-mass-shell scheme to $O(a_s^3)$,
$L_{g,2}^{\sf S}$, in its 3-loop contribution, changes by the term
\begin{eqnarray}
L_{g,2}^{{\sf S}, \rm (3),\overline{\rm MS}}(N) &=& L_{g,2}^{{\sf S},\rm (3),\rm OMS}(N)
+ a_s^3 \frac{32}{3} C_F T_F^2 N_F \left[3L_M - 4 \right]
\nonumber\\ && \times
\left[\frac{\big(N^2+N+2\big)}{N (N+1) (N+2)} S_1
+\frac{\big(N^3-4 N^2-N-2\big)}{N^2 (N+1) (N+2)}\right]
\end{eqnarray}
setting $Q^2 = \mu^2$. Here, we have identified the logarithms $L_M$ in both schemes symbolically. 
In applications,
either the on-shell or the $\overline{\rm MS}$ mass has to be used here. The 
corresponding expression in $z$--space reads
\begin{eqnarray}
L_{g,2}^{{\sf S}, \rm (3),\overline{\rm MS}}(z) &=& L_{g,2}^{{\sf S}, \rm (3),\rm OMS}(z)
+ 
a_s^3 \frac{32}{3} C_F T_F^2 N_F \left[3 L_M-4 \right]
\nonumber\\ && \times
\Biggl[\big(2 z^2-2 z+1\big) \big[H_0(z)+H_1(z)\big]+8 z^2-8 z+1\Biggr],
\end{eqnarray}
with $H_{\vec{a}}(z)$ harmonic polylogarithms, see~Eq.~(\ref{eq:HPL}).

\begin{figure}[H]
\begin{center}
\includegraphics[scale=0.8]{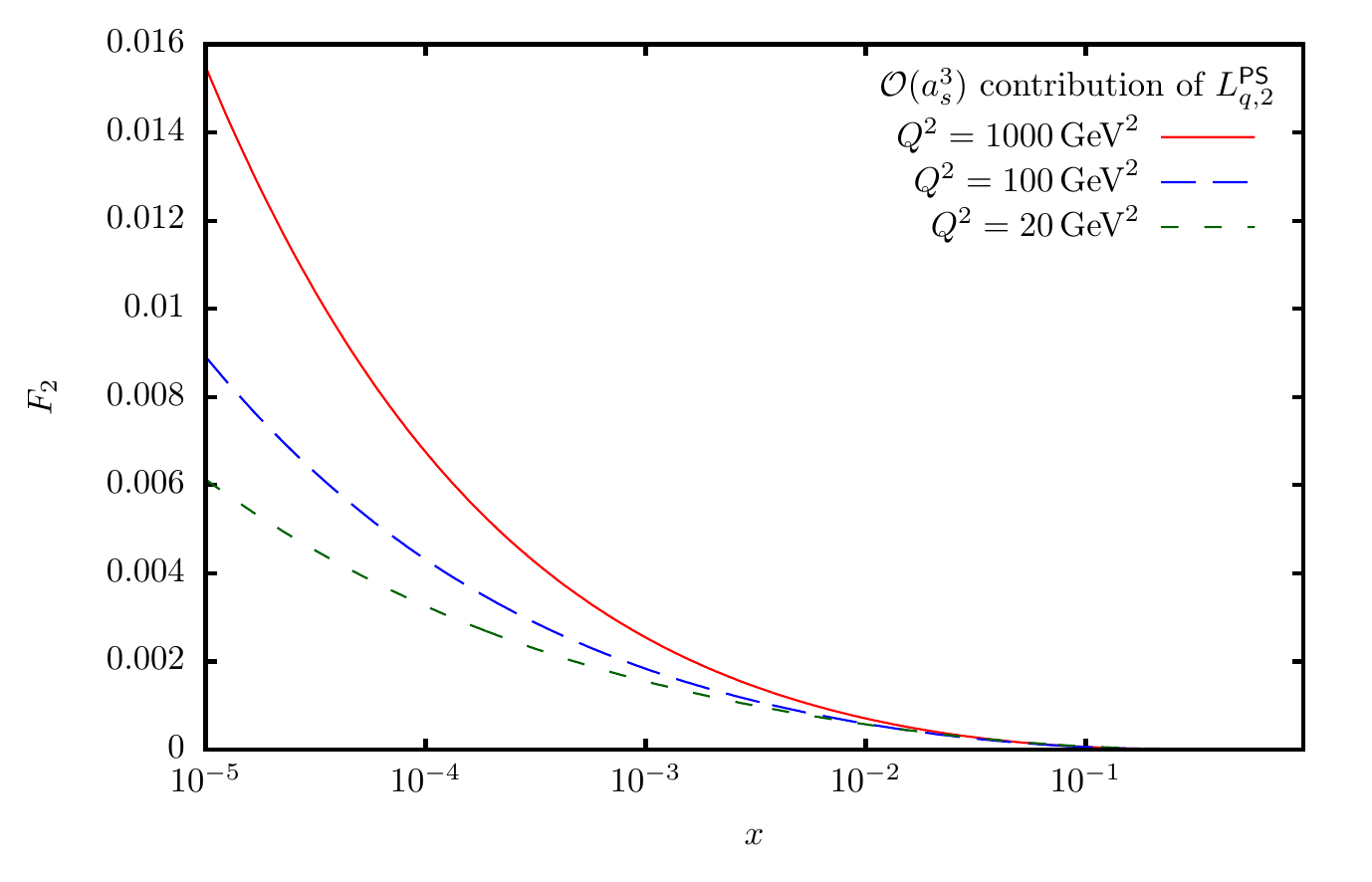}
\end{center}
\caption[]{\sf The $O(a_s^3)$ contribution by $L_{q,2}^{\sf PS}$ to the structure function $F_2(x,Q^2)$. \label{fig3}}
\end{figure}
\renewcommand{\arraystretch}{1.3}
\begin{center}
\begin{table}[H]\centering
\begin{tabular}{|l||r|r|r|}
\hline
$x$   
&  $Q^2 = 20   \GeV^2$
&  $Q^2 = 100  \GeV^2$
&  $Q^2 = 1000 \GeV^2$
\\
\hline
$10^{-4}$  & 1.946 &  3.200 &  5.340 \\
$10^{-3}$  & 1.141 &  1.702 &  2.526 \\
$10^{-2}$  & 0.641 &  0.825 &  1.040 \\
$10^{-1}$  & 0.400 &  0.409 &  0.412 \\
\hline
\end{tabular}
\caption[]{\sf Values of the structure function $F_2(x,Q^2)$ in the low $x$ region using the PDF-parameterization 
\cite{Alekhin:2013nda}. \label{tab0}}
\end{table}
\renewcommand{\arraystretch}{1.0}
\end{center}
In Figure~\ref{fig1} we illustrate the contribution of $L_{q,2}^{\sf PS}$ to the structure function 
$F_2(x,Q^2)$ using the PDFs of Ref.~\cite{Alekhin:2013nda}, cf.~Eq.~(\ref{eqF2}). 
Likewise Figures~\ref{fig2} and \ref{fig3} show the corresponding contributions by $L_{g,2}^{\sf S}$
at $O(a_s^2)$ and $O(a_s^3)$, respectively. Note that the  $O(a_s^2)$-terms, cf. 
also Ref.~\cite{Bierenbaum:2009zt} are smaller than those
at $O(a_s^3)$, which is caused by terms $\propto 1/z$ in the 3-loop contribution to $L_{g,2}^{\sf S}$,
which are absent at 2-loop order.

These contributions emerging on the 2- and 3-loop level are minor compared to the values of the structure
function $F_2(x,Q^2)$, for which typical values are given in Table~\ref{tab0}.\footnote{Note that the kinematic region
at small $x$ probed at HERA is limited to values $x \geq Q^2/(yS)$, with $S \simeq 10^5 \GeV^2$ and $y \in [0,1]$.} 
A global comparison 
of all heavy flavor contributions up to 3-loop order can presently only be performed using the known
number of Mellin moments, cf.~\cite{Bierenbaum:2009mv}, given in Section~\ref{sec:6}.

\section{The Logarithmic Contributions to \boldmath $H_{q,2}^{\sf PS}$ and  $H_{g,2}^{\sf S}$ to 
$O(a_s^3)$}
\label{sec:4}

\vspace*{1mm}
\noindent
The pure-singlet Wilson coefficient $H_{q,2}^{\sf PS}$, except for the constant part 
$a_{Qq}^{{\sf PS},\rm (3)}$ of the unrenormalized operator matrix element in the on-shell
scheme can be expressed by harmonic sums and rational functions in $N$ only. As before we reduce 
to a basis eliminating the algebraic relations \cite{Blumlein:2003gb}. It is given by~:
{\small
\begin{eqnarray}
\label{eq:HQ2PS}
\lefteqn{H_{q,2}^{\sf PS} = \tfrac{1}{2}[1 + (-1)^N]}\nonumber\\
&&
\times \Biggl\{
\textcolor{blue}{a_s^2} \Biggl\{
\textcolor{blue}{C_F} \textcolor{blue}{T_F} \Biggl[
-\frac{4 L_M^2 \big(N^2+N+2\big)^2}{(N-1) N^2 (N+1)^2 (N+2)}
+\frac{\big(4 S_1^2-12 S_2\big) \big(N^2+N+2\big)^2}{(N-1) N^2 (N+1)^2 (N+2)}
\N\\&&
+\frac{4 L_Q^2 \big(N^2+N+2\big)^2}{(N-1) N^2 (N+1)^2 (N+2)}
-\frac{32 (-1)^N P_{45}}{3 (N-1) N^2 (N+1)^3 (N+2)^3}
\N\\&&
+\frac{8 P_{75}}{3 (N-1) N^4 (N+1)^4 (N+2)^3}
+\frac{8 P_{57} S_1}{(N-1) N^3 (N+1)^3 (N+2)^2}
\N\\&&
+L_Q \Biggl[
-\frac{8 S_1 \big(N^2+N+2\big)^2}{(N-1) N^2 (N+1)^2 (N+2)}
-\frac{8 P_{57}}{(N-1) N^3 (N+1)^3 (N+2)^2}\Biggr]
\N\\&&
+\frac{64 S_{-2}}{(N-1) N (N+1) (N+2)}
-\frac{8 \big(N^2+5 N+2\big) \big(5 N^3+7 N^2+4 N+4\big) L_M}{(N-1) N^3 (N+1)^3 (N+2)^2}
\Biggr]
\Biggr\}
\N\\&&
+\textcolor{blue}{a_s^3} \Biggl\{
\textcolor{blue}{C_F^2} \textcolor{blue}{T_F} \Biggl[
 L_Q^3 \Biggl[\frac{8 \big(N^2+N+2\big)^2 \big(3 N^2+3 N+2\big)}{3 (N-1) N^3 (N+1)^3 (N+2)}
-\frac{32 \big(N^2+N+2\big)^2 S_1}{3 (N-1) N^2 (N+1)^2 (N+2)}\Biggr]
\N\\&&
+ L_Q^2
\Biggl[\frac{\big(24 S_1^2-24 S_2\big) \big(N^2+N+2\big)^2}{(N-1) N^2 (N+1)^2 (N+2)}
-\frac{4 P_{73}}{(N-1) N^4 (N+1)^4 (N+2)^2}
\N\\&&
+\frac{8 P_{62} S_1}{(N-1) N^3 (N+1)^3 (N+2)^2}\Biggr]
+L_Q \Biggl[\frac{\big(104 S_1 S_2
-\frac{56}{3} S_1^3\big) \big(N^2+N+2\big)^2}{(N-1) N^2 (N+1)^2 (N+2)}
\N\\&&
-\frac{16 \big(N^2+N-22\big) S_3 \big(N^2+N+2\big)}{3 (N-1) N^2 (N+1)^2 (N+2)}
+\frac{\big(128 S_{-3}-256 S_{-2,1}-384 \zeta_3\big) \big(N^2+N+2\big)}{(N-1) N^2 (N+1)^2 (N+2)}
\N\\&&
-\frac{4 P_{67} S_1^2}{(N-1) N^3 (N+1)^3 (N+2)^2}
-\frac{64 (-1)^N P_{98}}{15 (N-2) (N-1)^3 N^3 (N+1)^5 (N+2)^4 (N+3)^3}
\N\\&&
+\frac{4 P_{100}}{15 (N-1)^3 N^5 (N+1)^5 (N+2)^4 (N+3)^3}
+\frac{128 (-1)^N P_{45} S_1}{3 (N-1) N^2 (N+1)^3 (N+2)^3}
\N\\&&
-\frac{8 P_{79} S_1}{3 (N-1) N^4 (N+1)^4 (N+2)^3}
+\frac{512 S_{-2} S_1}{(N-1) N^2 (N+1)^2 (N+2)}
+\frac{4 P_{65} S_2}{(N-1) N^3 (N+1)^3 (N+2)^2}
\N\\&&
+L_M^2 \Biggl[
\frac{16 \big(N^2+N+2\big)^2 S_1}{(N-1) N^2 (N+1)^2 (N+2)}
-\frac{4 \big(N^2+N+2\big)^2 \big(3 N^2+3 N+2\big)}{(N-1) N^3 (N+1)^3 (N+2)}\Biggr]
\N\\&&
+L_M \Biggl[
\frac{32 \big(N^2+5 N+2\big) \big(5 N^3+7 N^2+4 N+4\big) S_1}{(N-1)
  N^3 (N+1)^3 (N+2)^2}
\N\\&&
-\frac{8 \big(N^2+5 N+2\big) \big(3 N^2+3 N+2\big) \big(5 N^3+7 N^2+4
  N+4\big)}{(N-1) N^4 (N+1)^4 (N+2)^2}\Biggr]
\N\\&&
+\frac{32 P_{69} S_{-2}}{(N-2) (N-1) N^3 (N+1)^3 (N+2) (N+3)}\Biggr]
-\frac{8 (3 N+2) \big(N^2+N+2\big) S_1^3}{3 (N-1) N^3 (N+1) (N+2)}
\N\\&&
+\frac{4 P_{74} S_1^2}{(N-1) N^4 (N+1)^4 (N+2)^3}-2  \frac{\big(N^2+N+2\big) \zeta_2}{(N-1) N^4 (N+1)^4 (N+2)} P_{55}
-\frac{4 P_{94}}{(N-1) N^5 (N+1)^6 (N+2)^3}
\N\\&&
-\frac{4}{3}  \frac{\big(N^2+N+2\big)^2 \big(3 N^2+3 N+2\big) \zeta_3}{(N-1) N^3 (N+1)^3 (N+2)}
+4  \frac{\big(N^2+N+2\big) \big(5 N^4+4 N^3+N^2-10 N-8\big) \zeta_2}{(N-1) N^3 (N+1)^3 (N+2)} S_1
\N\\&&
+\frac{4 P_{91} S_1}{(N-1) N^5 (N+1)^5 (N+2)^3}
+L_M^3 \Biggl[\frac{4 \big(N^2+N+2\big)^2 \big(3 N^2+3 N+2\big)}{3 (N-1) N^3 (N+1)^3 (N+2)}
\N\\&&
-\frac{16 \big(N^2+N+2\big)^2 S_1}{3 (N-1) N^2 (N+1)^2 (N+2)}\Biggr]
-\frac{8 \big(N^2+N+2\big) \big(3 N^4+9 N^3+15 N^2+11 N-2\big) S_1 S_2}{(N-1) N^3 (N+1)^3 (N+2)}
\N\\&&
-\frac{4 P_{76} S_2}{(N-1) N^4 (N+1)^4 (N+2)^3}
+L_M^2 \Biggl[
-\frac{4 \big(13 N^2+5 N-6\big) S_1 \big(N^2+N+2\big)^2}{(N-1) N^3 (N+1)^3 (N+2)}
\N\\&&
+\frac{\big(24 S_2-8 S_1^2\big) \big(N^2+N+2\big)^2}{(N-1) N^2 (N+1)^2 (N+2)}
+\frac{4 P_{51} \big(N^2+N+2\big)}{(N-1) N^4 (N+1)^4 (N+2)}\Biggr]
\N\\&&
-\frac{8 \big(N^2+N+2\big) \big(3 N^4+48 N^3+43 N^2-22 N-8\big) S_3}{3 (N-1) N^3 (N+1)^3 (N+2)}
+\frac{32 \big(N^2-3 N-2\big) \big(N^2+N+2\big) S_{2,1}}{(N-1) N^3 (N+1)^2 (N+2)}
\N\\&&
+\frac{\big(N^2+N+2\big)^2 }{(N-1) N^2 (N+1)^2 (N+2)}
\Big[\frac{2}{3} S_1^4-12 S_2 S_1^2
+\Bigl[\frac{16 S_3}{3}+32 S_{2,1}\Bigr] S_1
+\frac{16}{3} \zeta_3 S_1+18 S_2^2-12 S_4
\N\\&&
+32 S_{3,1}-64 S_{2,1,1}+\big(4 S_1^2-12 S_2\big) \zeta_2\Bigr]
+L_M \Biggl[\frac{ \big(N^2+N+2\big)^2}{(N-1) N^2 (N+1)^2 (N+2)} \Biggl[\frac{8}{3} S_1^3-24 S_2 S_1
\N\\&&
-\frac{80 S_3}{3}+32 S_{2,1}+96 \zeta_3\Biggr]
-\frac{4 P_{60} S_1^2}{(N-1) N^3 (N+1)^3 (N+2)^2}
-\frac{4 P_{92}}{(N-1) N^5 (N+1)^5 (N+2)^3}
\N\\&&
+\frac{8 P_{77} S_1}{(N-1) N^4 (N+1)^4 (N+2)^3}
-\frac{4 P_{59} S_2}{(N-1) N^3 (N+1)^3 (N+2)^2}\Biggr]\Biggr]
\N\\&&
+\textcolor{blue}{C_F} \textcolor{blue}{T_F^2} \Biggl[
\Biggl[\frac{32 \big(N^2+N+2\big)^2 L_Q^3}{9 (N-1) N^2 (N+1)^2 (N+2)}
-\frac{32 P_{66} L_Q^2}{9 (N-1) N^3 (N+1)^3 (N+2)^2}
\N\\&&
+   L_Q \Biggl[
\frac{32 \big(N^2+N+2\big)^2 L_M^2}{3 (N-1) N^2 (N+1)^2 (N+2)}
+\Biggl[\frac{64 \big(N^2+N+2\big) \big(8 N^3+13 N^2+27 N+16\big)}{9 (N-1) N^2 (N+1)^3 (N+2)}
\N\\&&
-\frac{64 \big(N^2+N+2\big)^2 S_1}{3 (N-1) N^2 (N+1)^2 (N+2)}\Biggr] L_M
-\frac{256 (-1)^N P_{45}}{9 (N-1) N^2 (N+1)^3 (N+2)^3}
\N\\&&
+\frac{64 P_{82}}{27 (N-1) N^4 (N+1)^4 (N+2)^3}
+\frac{512 S_{-2}}{3 (N-1) N (N+1) (N+2)}\Biggr]
\N\\&&
-\frac{128 \big(N^2+N+2\big)^2 L_M^3}{9 (N-1) N^2 (N+1)^2 (N+2)}
+\frac{16 \big(N^2+N+2\big) \big(7 N^4+16 N^3+32 N^2+19 N+2\big) S_1^2}{3 (N-1) N^3 (N+1)^3 (N+2)}
\N\\&&
-\frac{32 \big(11 N^5+26 N^4+57 N^3+142 N^2+84 N+88\big) L_M^2}{9 (N-1) N^2 (N+1)^2 (N+2)^2}
+\frac{32 \zeta_2 P_{48}}{9 (N-1) N^3 (N+1)^2 (N+2)^2} 
\N\\&&
+\frac{32 P_{95}}{81 (N-1) N^5 (N+1)^5 (N+2)^4}
-\frac{32 P_{68} S_1}{27 (N-1) N^3 (N+1)^4 (N+2)}
\N\\&&
+L_M \Biggl[
\frac{\Bigl[\frac{32}{3} S_1^2-32 S_2\Bigr] \big(N^2+N+2\big)^2}{(N-1) N^2 (N+1)^2 (N+2)}
-\frac{64 P_{80}}{27 (N-1) N^4 (N+1)^4 (N+2)^3}
\N\\&&
+\frac{64 P_{57} S_1}{3 (N-1) N^3 (N+1)^3 (N+2)^2}\Biggr]
+\frac{16 P_{61} S_2}{9 (N-1) N^3 (N+1)^3 (N+2)^2}
\N\\&&
+\frac{\big(N^2+N+2\big)^2}{(N-1) N^2 (N+1)^2 (N+2)} \Bigl[
-\frac{64}{9} S_1^3
-\frac{32}{3} S_2 S_1
-\frac{32}{3} \zeta_2 S_1
+\frac{160 S_3}{9}
+\frac{128 \zeta_3}{9}\Bigr]\Biggr]\Biggr]
\N\\&&
+\textcolor{blue}{N_F} \textcolor{blue}{T_F^2} \textcolor{blue}{C_F}\Biggl[
\frac{32 \big(N^2+N+2\big)^2 L_Q^3}{9 (N-1) N^2 (N+1)^2 (N+2)}
-\frac{32 P_{66} L_Q^2}{9 (N-1) N^3 (N+1)^3 (N+2)^2}
\N\\&&
+\Biggl[\frac{\Bigl[
-\frac{16}{3} S_1^2
-\frac{16 S_2}{3}\Bigr] \big(N^2+N+2\big)^2}{(N-1) N^2 (N+1)^2 (N+2)}
+\frac{32 \big(8 N^3+13 N^2+27 N+16\big) S_1 \big(N^2+N+2\big)}{9 (N-1) N^2 (N+1)^3 (N+2)}
\N\\&&
-\frac{256 (-1)^N P_{45}}{9 (N-1) N^2 (N+1)^3 (N+2)^3}
+\frac{32 P_{84}}{27 (N-1) N^4 (N+1)^4 (N+2)^3}
\N\\&&
+\frac{512 S_{-2}}{3 (N-1) N (N+1) (N+2)}\Biggr] L_Q
-\frac{32 \big(N^2+N+2\big)^2 L_M^3}{9 (N-1) N^2 (N+1)^2 (N+2)}
\N\\&&
-\frac{16}{9} \frac{\zeta_2}{(N-1) N^3 (N+1)^3 (N+2)^2} P_{63}
-\frac{32 P_{87}}{3 (N-1) N^5 (N+1)^5 (N+2)^4}
\N\\&&
+L_M^2 \Biggl[\frac{32 P_{64}}{9 (N-1) N^3 (N+1)^3 (N+2)^2}
-\frac{32 \big(N^2+N+2\big)^2 S_1}{3 (N-1) N^2 (N+1)^2 (N+2)}\Biggr]
\N\\&&
+L_M \Biggl[
\frac{\Bigl[
-\frac{16}{3} S_1^2
-\frac{80 S_2}{3}\Bigr] \big(N^2+N+2\big)^2}{(N-1) N^2 (N+1)^2 (N+2)}
-\frac{32 P_{78}}{27 (N-1) N^4 (N+1)^4 (N+2)^3}
\N
\\
&&
+\frac{32 P_{63} S_1}{9 (N-1) N^3 (N+1)^3 (N+2)^2}\Biggr]
+\frac{64 \big(N^2+5 N+2\big) \big(5 N^3+7 N^2+4 N+4\big) S_2}{3 (N-1) N^3 (N+1)^3 (N+2)^2}
\N\\&&
+\frac{\big(N^2+N+2\big)^2 \Bigl[\frac{64 S_3}{3}
+\frac{16}{3} S_1 \zeta_2
+\frac{32 \zeta_3}{9}\Bigr]}{(N-1) N^2 (N+1)^2 (N+2)}\Biggr]
\N\\&&
+\textcolor{blue}{C_A} \textcolor{blue}{C_F} \textcolor{blue}{T_F} \Biggl[
 L_Q^3
\Biggl[
-\frac{16 S_1 \big(N^2+N+2\big)^2}{3 (N-1) N^2 (N+1)^2 (N+2)}
\N\\&&
-\frac{8 \big(11 N^4+22 N^3-23 N^2-34 N-12\big) \big(N^2+N+2\big)^2}{9 (N-1)^2 N^3 (N+1)^3 (N+2)^2}
\Biggr]
\N\\&&
+ L_Q^2 \Biggl[
-\frac{16 \big(5 N^2-1\big) S_1 \big(N^2+N+2\big)^2}{(N-1)^2 N^3 (N+1)^3 (N+2)}
+\frac{\big(16 S_1^2-16 S_2-32 S_{-2}\big) \big(N^2+N+2\big)^2}{(N-1) N^2 (N+1)^2 (N+2)}
\N\\&&
+\frac{16 (-1)^N \big(N^5+9 N^4+24 N^3+36 N^2+32 N+8\big) \big(N^2+N+2\big)}{(N-1) N^3 (N+1)^4 (N+2)^3}
\N\\&&
+\frac{8 P_{85}}{9 (N-1)^2 N^4 (N+1)^3 (N+2)^3}\Biggr]
+L_Q
\Biggl[
\frac{ \big(N^2+N+2\big)^2}{(N-1) N^2 (N+1)^2 (N+2)}\Bigl[
-\frac{40}{3} S_1^3 + 40 S_2 S_1
\N\\&&
-144 S_{-3} + 96 S_{-2,1}\Bigr]
+\frac{4 P_{47} S_1^2 \big(N^2+N+2\big)}{3 (N-1)^2 N^3 (N+1)^3 (N+2)^2}
+48  \frac{\big(N^2+N+6\big) \big(N^2+N+2\big) \zeta_3}{(N-1) N^2 (N+1)^2 (N+2)} 
\N\\&&
+\frac{32 S_{-2} S_1 \big(N^2+N+2\big)}{N^2 (N+1)^2}
+\frac{4 P_{52} S_2 \big(N^2+N+2\big)}{3 (N-1)^2 N^3 (N+1)^3 (N+2)^2}
+\frac{32 (-1)^N P_{93}}{9 (N-1)^2 N^4 (N+1)^5 (N+2)^4}
\N\\&&
-\frac{8 \big(13 N^2+13 N+62\big) S_3 \big(N^2+N+2\big)}{3 (N-1) N^2 (N+1)^2 (N+2)}
-\frac{8 P_{97}}{27 (N-1)^2 N^5 (N+1)^5 (N+2)^4}
\N\\&&
-\frac{32 (-1)^N P_{72} S_1}{3 (N-1) N^3 (N+1)^4 (N+2)^3}
-\frac{8 P_{88} S_1}{9 (N-1)^2 N^4 (N+1)^4 (N+2)^3}
\N\\&&
+\frac{16 P_{71} S_{-2}}{3 (N-1)^2 N^3 (N+1)^3 (N+2)^2}\Biggr]
+\frac{8 \big(N^2+N+2\big) \big(N^3+8 N^2+11 N+2\big) S_1^3}{3 (N-1) N^2 (N+1)^3 (N+2)^2}
\N\\&&
+\frac{4 \big(N^2+N+2\big) P_{43} S_1^2}{(N-1) N^2 (N+1)^4 (N+2)^3}
+8 \big(N^2+N+2\big) \frac{(-1)^N \zeta_2}{(N-1) N^3 (N+1)^4 (N+2)^3} P_{44}
\N\\&&
-\frac{8}{9}  \frac{\big(N^2+N+2\big) \zeta_3}{(N-1)^2 N^3 (N+1)^3 (N+2)^2} P_{49}
+\frac{4}{9} \frac{\zeta_2}{(N-1)^2 N^4 (N+1)^4 (N+2)^3} P_{89}
\N\\&&
+\frac{8 P_{99}}{3 (N-1)^2 N^6 (N+1)^6 (N+2)^5}
-\frac{4}{3}  \frac{\big(N^2+N+2\big) \zeta_2}{(N-1)^2 N^3 (N+1)^3 (N+2)^2} P_{53} S_1
\N\\&&
-\frac{8 \big(N^2+N+2\big) P_{70} S_1}{(N-1) N^2 (N+1)^5 (N+2)^4}
+L_M^3 \Biggl[
\frac{16 S_1 \big(N^2+N+2\big)^2}{3 (N-1) N^2 (N+1)^2 (N+2)}
\N\\&&
+\frac{8 \big(11 N^4+22 N^3-23 N^2-34 N-12\big) \big(N^2+N+2\big)^2}{9 (N-1)^2 N^3 (N+1)^3 (N+2)^2}
\Biggr]
+\frac{4 P_{86} S_2}{3 (N-1)^2 N^4 (N+1)^4 (N+2)^3}
\N\\&&
-\frac{8 \big(N^2+N+2\big) \big(3 N^3-12 N^2-27 N-2\big) S_1 S_2}{(N-1) N^2 (N+1)^3 (N+2)^2}
-\frac{16 \big(N^2+N+2\big) P_{50} S_3}{3 (N-1)^2 N^3 (N+1)^3 (N+2)^2}
\N\\&&
+\frac{32 (-1)^N \big(N^2+N+2\big) \big(N^4+2 N^3+7 N^2+22 N+20\big) S_{-2}}{(N-1) N (N+1)^4 (N+2)^3}
\N\\&&
+L_M^2 \Biggl[\frac{\big(16 S_2+32 S_{-2}\big) \big(N^2+N+2\big)^2}{(N-1) N^2 (N+1)^2 (N+2)}
+\frac{8 P_{54} S_1 \big(N^2+N+2\big)}{3 (N-1)^2 N^3 (N+1)^3 (N+2)^2}
\N\\ &&
-\frac{16 (-1)^N \big(N^5+9 N^4+24 N^3+36 N^2+32 N+8\big) \big(N^2+N+2\big)}{(N-1) N^3 (N+1)^4 (N+2)^3}
\N\\&&
-\frac{8 P_{83}}{9 (N-1)^2 N^4 (N+1)^3 (N+2)^3}\Biggr]
+\frac{\big(N^2-N-4\big) \big(N^2+N+2\big) }{(N-1) N (N+1)^3 (N+2)^2} \Bigl[-64 (-1)^N S_1 S_{-2}
\N\\&&
-32 (-1)^N S_{-3}+64 S_{-2,1}-32 (-1)^N S_1 \zeta_2-24 (-1)^N \zeta_3\Bigr]
\N\\&&
+\frac{\big(N^2+N+2\big)^2}{(N-1) N^2 (N+1)^2 (N+2)}
\Bigl[
-\frac{2}{3} S_1^4-20 S_2 S_1^2-32 (-1)^N S_{-3} S_1+\Bigl(64 S_{-2,1} 
-\frac{160 S_3}{3}\Bigr) S_1
\N\\&& 
-\frac{8}{3} \big(-7+9 (-1)^N\big) \zeta_3 S_1-2 S_2^2
+S_{-2} \big(-32 (-1)^N S_1^2-32 (-1)^N S_2\big)
-36 S_4-16 (-1)^N S_{-4}
\N\\&&
+16 S_{3,1}+32 S_{-2,2}
+32 S_{-3,1}+16 S_{2,1,1}
-64 S_{-2,1,1}+\big(-4 \big(-3+4 (-1)^N\big) S_1^2
\N\\&&
-4 \big(-1+4 (-1)^N\big) S_2
-8 \big(1+2 (-1)^N\big) S_{-2}\big) \zeta_2\Bigr]
+L_M \Biggl[
\frac{ \big(N^2+N+2\big)^2}{(N-1) N^2 (N+1)^2 (N+2)}
\Bigl[
-\frac{8}{3} S_1^3
\N\\&&
+40 S_2 S_1+32 \big(1+(-1)^N\big) S_{-2} S_1
+16 (-1)^N S_{-3}-32 S_{2,1} + 12 \big(-9+(-1)^N\big) \zeta_3\Bigr]
\N\\&&
+\frac{4 \big(17 N^4-6 N^3+41 N^2-16 N-12\big) S_1^2 \big(N^2+N+2\big)}{3 (N-1)^2 N^3 (N+1)^2 (N+2)}
+\frac{4 P_{56} S_2 \big(N^2+N+2\big)}{3 (N-1)^2 N^3 (N+1)^3 (N+2)^2}
\N\\&&
+\frac{8 \big(31 N^2+31 N+74\big) S_3 \big(N^2+N+2\big)}{3 (N-1) N^2 (N+1)^2 (N+2)}
+\frac{16 \big(7 N^2+7 N+10\big) S_{-3} \big(N^2+N+2\big)}{(N-1) N^2 (N+1)^2 (N+2)}
\N\\&&
-\frac{128 \big(N^2+N+1\big) S_{-2,1} \big(N^2+N+2\big)}{(N-1) N^2 (N+1)^2 (N+2)}
+\frac{\big(N^2-N-4\big) \big(N^2+N+2\big)
32 (-1)^N S_{-2}
}{(N-1) N (N+1)^3 (N+2)^2}
\N\\&&
-\frac{64 (-1)^N P_{81}}{9 (N-1) N^3 (N+1)^5 (N+2)^4}
+\frac{8 P_{96}}{27 (N-1)^2 N^5 (N+1)^5 (N+2)^4}
\N\\&&
+\frac{64 (-1)^N P_{46} S_1}{3 (N-1) N^2 (N+1)^3 (N+2)^3}
-\frac{8 P_{90} S_1}{9 (N-1)^2 N^4 (N+1)^4 (N+2)^3}
\N\\&&
+\frac{16 P_{58} S_{-2}}{(N-1) N^3 (N+1)^3 (N+2)^2}\Biggr]
\Biggr]
+a_{Qq}^{{\sf PS},(3)} 
+\tilde{C}_{2,q}^{{\sf PS},(3)}({N_F}+1)
\Biggr\}\Biggr\}~,
\end{eqnarray}

}
with the polynomials
{\small
\begin{eqnarray}
P_{43}&=&N^6+6 N^5+7 N^4+4 N^3+18 N^2+16 N-8
    \\
P_{44}&=&2 N^6+7 N^5+31 N^4+82 N^3+86 N^2+32 N+8
    \\
P_{45}&=&4 N^6+22 N^5+48 N^4+53 N^3+45 N^2+36 N+8
    \\
P_{46}&=&5 N^6+29 N^5+78 N^4+118 N^3+114 N^2+72 N+16
    \\
P_{47}&=&5 N^6+135 N^5+327 N^4+329 N^3+220 N^2-176 N-120
    \\
P_{48}&=&8 N^6+29 N^5+84 N^4+193 N^3+162 N^2+124 N+24
    \\
P_{49}&=&11 N^6+6 N^5+75 N^4+68 N^3-200 N^2-80 N-24
    \\
P_{50}&=&11 N^6+29 N^5-7 N^4-25 N^3-56 N^2-72 N-24
    \\
P_{51}&=&16 N^6+35 N^5+33 N^4-11 N^3-41 N^2-36 N-12
    \\
P_{52}&=&17 N^6-57 N^5-213 N^4-175 N^3-140 N^2+64 N+72
    \\
P_{53}&=&17 N^6+27 N^5+75 N^4+149 N^3-20 N^2-80 N-24
    \\
P_{54}&=&17 N^6+51 N^5+51 N^4+89 N^3+40 N^2-80 N-24
    \\
P_{55}&=&38 N^6+108 N^5+151 N^4+106 N^3+21 N^2-28 N-12
    \\
P_{56}&=&73 N^6+189 N^5+45 N^4+31 N^3-238 N^2-412 N-120
    \\
P_{57}&=&N^7-15 N^5-58 N^4-92 N^3-76 N^2-48 N-16
    \\
P_{58}&=&2 N^7+14 N^6+37 N^5+102 N^4+155 N^3+158 N^2+132 N+40
    \\
P_{59}&=&3 N^7-15 N^6-153 N^5-577 N^4-854 N^3-652 N^2-408 N-128
    \\
P_{60}&=&5 N^7+19 N^6+61 N^5+197 N^4+266 N^3+212 N^2+136 N+32
    \\
P_{61}&=&5 N^7+37 N^6+188 N^5+643 N^4+925 N^3+742 N^2+460 N+120
    \\
P_{62}&=&7 N^7+21 N^6+5 N^5-117 N^4-244 N^3-232 N^2-192 N-80
    \\
P_{63}&=&8 N^7+37 N^6+68 N^5-11 N^4-86 N^3-56 N^2-104 N-48
    \\
P_{64}&=&8 N^7+37 N^6+83 N^5+85 N^4+61 N^3+58 N^2-20 N-24
    \\
P_{65}&=&9 N^7+15 N^6-103 N^5-575 N^4-998 N^3-948 N^2-696 N-256
    \\
P_{66}&=&11 N^7+37 N^6+53 N^5+7 N^4-68 N^3-56 N^2-80 N-48
    \\
P_{67}&=&25 N^7+91 N^6+101 N^5-195 N^4-546 N^3-556 N^2-520 N-224
    \\
P_{68}&=&62 N^7+329 N^6+986 N^5+1790 N^4+2242 N^3+1653 N^2+650 N+96
    \\
P_{69}&=&N^8+8 N^7+8 N^6-14 N^5-53 N^4-82 N^3+60 N^2+104 N+96
    \\
P_{70}&=&2 N^8+22 N^7+117 N^6+386 N^5+759 N^4+810 N^3+396 N^2+72 N+32
    \\
P_{71}&=&6 N^8-42 N^7-241 N^6-579 N^5-307 N^4+477 N^3+602 N^2+492 N+168
    \\
P_{72}&=&10 N^8+71 N^7+244 N^6+497 N^5+698 N^4+720 N^3+512 N^2+248 N+48
 \\
P_{73}&=&19 N^9+86 N^8+144 N^7-38 N^6-535 N^5-1016 N^4-1180 N^3-872 N^2
\N\\&&
-416 N-96
 \\
P_{74}&=&N^{10}+15 N^9+105 N^8+361 N^7+660 N^6+828 N^5+814 N^4+384 N^3
\N\\&&
-112 N^2-128 N-32
 \\
P_{75}&=&6 N^{10}+49 N^9+197 N^8+472 N^7+833 N^6+1469 N^5+2142 N^4+1904 N^3
\N\\&&
+1040 N^2+432 N+96
 \\
P_{76}&=&11 N^{10}+123 N^9+541 N^8+1273 N^7+1806 N^6+1672 N^5+1006 N^4+320 N^3
\N\\&&
-16 N^2-64 N-32
 \\
P_{77}&=&19 N^{10}+143 N^9+412 N^8+426 N^7-N^6+159 N^5+1066 N^4+1552 N^3
\N\\&&
+1456 N^2+848 N+224
 \\
P_{78}&=&43 N^{10}+320 N^9+939 N^8+912 N^7-218 N^6-510 N^5-654 N^4-1232 N^3
\N\\&&
+16 N^2+672 N+288
 \\
P_{79}&=&60 N^{10}+397 N^9+1073 N^8+1111 N^7+623 N^6+4328 N^5+12432 N^4+15944 N^3
\N\\&&
+12704 N^2+6816 N+1728
 \\
P_{80}&=&67 N^{10}+383 N^9+867 N^8+696 N^7-755 N^6-2391 N^5-3027 N^4-2744 N^3
\N\\&&
-1256 N^2-48 N+144
 \\
P_{81}&=&77 N^{10}+646 N^9+2553 N^8+6903 N^7+14498 N^6+22898 N^5+24861 N^4
\N\\&&
+17068 N^3+7040 N^2+1760 N+192
 \\
P_{82}&=&85 N^{10}+530 N^9+1458 N^8+2112 N^7+1744 N^6+2016 N^5+3399 N^4+2968 N^3
\N\\&&
+1864 N^2+1248 N+432
 \\
P_{83}&=&118 N^{10}+675 N^9+1588 N^8+1652 N^7+326 N^6+357 N^5+876 N^4
\N\\&&
+1672 N^3+3440 N^2+2544 N+576
 \\
P_{84}&=&127 N^{10}+740 N^9+1737 N^8+1308 N^7-1592 N^6-2226 N^5+1386 N^4
\N\\&&
+3064 N^3+3040 N^2+2496 N+864
 \\
P_{85}&=&151 N^{10}+708 N^9+1156 N^8+464 N^7-967 N^6+372 N^5+3672 N^4
\N\\&&
+5236 N^3+6152 N^2+3792 N+864
 \\
P_{86}&=&3 N^{11}+66 N^{10}+104 N^9-1152 N^8-3801 N^7-2510 N^6+3318 N^5+8076 N^4
\N\\&&
+9608 N^3+6512 N^2+2432 N+384
 \\
P_{87}&=&5 N^{11}+62 N^{10}+252 N^9+374 N^8+38 N^7-400 N^6-473 N^5-682 N^4
\N\\&&
-904 N^3-592 N^2-208 N-32
 \\
P_{88}&=&118 N^{11}+529 N^{10}+1264 N^9+3846 N^8+11353 N^7+23684 N^6+32793 N^5
\N\\&&
+31801 N^4+22836 N^3+10448 N^2+2592 N+432
 \\
P_{89}&=&127 N^{11}+820 N^{10}+2197 N^9+1890 N^8-1847 N^7-1960 N^6+3843 N^5
\N\\&&
+9730 N^4+13632 N^3+10688 N^2+4944 N+864
 \\
P_{90}&=&136 N^{11}+1039 N^{10}+3100 N^9+3534 N^8-1295 N^7-6352 N^6-8421 N^5
\N\\&&
-11729 N^4-7644 N^3+1376 N^2+1920 N+144
 \\
P_{91}&=&7 N^{12}+47 N^{11}+123 N^{10}+76 N^9-598 N^8-2178 N^7-3626 N^6
\N\\&&
-3933 N^5-3254 N^4-1608 N^3-144 N^2+112 N+32
 \\
P_{92}&=&37 N^{12}+305 N^{11}+1017 N^{10}+1462 N^9+592 N^8+408 N^7+4064 N^6
\N\\&&
+9645 N^5+12222 N^4+10280 N^3+6064 N^2+2192 N+352
 \\
P_{93}&=&242 N^{12}+1853 N^{11}+6173 N^{10}+12711 N^9+18608 N^8+17040 N^7
-302 N^6
\N\\&&
-24986 N^5-32225 N^4-20010 N^3-7904 N^2-2016 N-288
 \\
P_{94}&=&5 N^{13}+27 N^{12}-97 N^{11}-1410 N^{10}-5754 N^9-12428 N^8-16530 N^7
\N\\&&
-14531 N^6-7956 N^5-1038 N^4+2176 N^3+1632 N^2+448 N+32
 \\
P_{95}&=&119 N^{13}+1897 N^{12}+12595 N^{11}+48221 N^{10}+124877 N^9+239946 N^8
\N\\&&
+345670 N^7+356234 N^6+253043 N^5+129982 N^4+55768 N^3+20112 N^2
\N\\&&
+5616 N+864
 \\
P_{96}&=&686 N^{14}+8408 N^{13}+39228 N^{12}+89257 N^{11}+113445 N^{10}+109336 N^9
\N\\&&
+76360 N^8-109649 N^7-393915 N^6-482272 N^5-376932 N^4-263440 N^3
\N\\&&
-155472 N^2
\nonumber\\ &&
-56448 N-8640
 \\
P_{97}&=&1790 N^{14}+15938 N^{13}+56250 N^{12}+90805 N^{11}+43917 N^{10}-38450 N^9-42314 N^8
\N\\&&
-169217 N^7-616623 N^6-992860 N^5-964980 N^4-697072 N^3-376464 N^2
\N\\&&
-127872 N-19008
 \\
P_{98}&=&30 N^{16}+397 N^{15}+1996 N^{14}+3786 N^{13}-3905 N^{12}-30084 N^{11}-44372 N^{10}
\N\\&&
+5100 N^9+71344 N^8+27709 N^7-104744 N^6-146534 N^5-30293 N^4+77346 N^3
\N\\&&
+33768 N^2-23544 N-3888
 \\
P_{99}&=&12 N^{17}+162 N^{16}+1030 N^{15}+4188 N^{14}+11527 N^{13}+19051 N^{12}+11176 N^{11}
\N\\&&
-17182 N^{10}-36527 N^9-27469 N^8-11770 N^7+5554 N^6+32640 N^5+46456 N^4
\N\\&&
+34528 N^3+14816 N^2+3584 N+384
 \\
P_{100}&=&1245 N^{18}+19980 N^{17}+133282 N^{16}+461805 N^{15}+787161 N^{14}+185392 N^{13}
\N\\&&
-1368400 N^{12}-225082 N^{11}+6978631 N^{10}+13143336 N^9+5808466 N^8
\N\\&&
-11433627 N^7-19928573 N^6
-12013164 N^5+1462668 N^4+8209584 N^3
\N\\&&
+6906384 N^2+2980800 N+544320~. 
\end{eqnarray}

}

The Wilson coefficient $H_{g,2}^{\sf S}$, except for the constant contribution 
$a_{Qg}^{(3)}$, has a similar structure. It is given by~:
{\small


}
with the polynomials
{\small
\begin{eqnarray}
P_{101}&=&N^6-81 N^5-264 N^4-185 N^3-307 N^2-256 N-204
    \\
P_{102}&=&N^6+6 N^5+7 N^4+4 N^3+18 N^2+16 N-8
    \\
P_{103}&=&N^6+7 N^5-7 N^4-39 N^3+14 N^2+40 N+48
    \\
P_{104}&=&N^6+21 N^5+57 N^4+31 N^3+26 N^2+20 N+24
    \\
P_{105}&=&2 N^6-7 N^5-41 N^4-31 N^3-29 N^2-22 N-16
    \\
P_{106}&=&2 N^6-7 N^5-24 N^4-35 N^3-44 N^2-44 N-16
    \\
P_{107}&=&3 N^6+5 N^5+27 N^4+35 N^3+6 N^2+12 N+8
    \\
P_{108}&=&3 N^6+9 N^5-N^4-17 N^3-38 N^2-28 N-24
    \\
P_{109}&=&3 N^6+9 N^5+2 N^4-11 N^3-23 N^2-16 N-12
    \\
P_{110}&=&3 N^6+30 N^5+15 N^4-64 N^3-56 N^2-20 N-8
    \\
P_{111}&=&4 N^6+5 N^5-10 N^4-39 N^3-40 N^2-24 N-8
    \\
P_{112}&=&6 N^6-12 N^5+17 N^4+106 N^3+127 N^2+104 N+84
    \\
P_{113}&=&6 N^6+18 N^5+7 N^4-16 N^3-31 N^2-20 N-12
    \\
P_{114}&=&7 N^6-93 N^5-327 N^4-287 N^3-316 N^2-112 N-24
    \\
P_{115}&=&7 N^6-20 N^5-176 N^4-335 N^3-276 N^2-116 N-16
\\
P_{116}&=&7 N^6-19 N^5-171 N^4-325 N^3-264 N^2-108 N-16
    \\
P_{117}&=&7 N^6+21 N^5+5 N^4-25 N^3-204 N^2-188 N-192
    \\
P_{118}&=&8 N^6+13 N^5-111 N^4-193 N^3-89 N^2-56 N-20
    \\
P_{119}&=&9 N^6+21 N^5+11 N^4-5 N^3-104 N^2-76 N-144
    \\
P_{120}&=&9 N^6+39 N^5+53 N^4+25 N^3+94 N^2+44 N+312
\\
P_{121}&=&10 N^6+18 N^5-111 N^4-164 N^3-61 N^2-16 N+36
    \\
P_{122}&=&10 N^6+63 N^5+105 N^4+31 N^3+17 N^2+14 N+48
    \\
P_{123}&=&11 N^6-15 N^5-327 N^4-181 N^3+292 N^2-20 N-48
    \\
P_{124}&=&11 N^6+15 N^5-285 N^4-319 N^3-254 N^2-368 N-240
    \\
P_{125}&=&11 N^6+33 N^5-189 N^4-361 N^3-194 N^2-92 N-72
    \\
P_{126}&=&11 N^6+33 N^5-114 N^4-247 N^3-263 N^2-176 N-108
    \\
P_{127}&=&11 N^6+33 N^5-87 N^4-85 N^3+4 N^2-116 N-48
    \\
P_{128}&=&11 N^6+35 N^5+59 N^4+57 N^3-38 N^2-68 N+40
    \\
P_{129}&=&11 N^6+47 N^5+7 N^4+9 N^3+90 N^2+28 N+96
    \\
P_{130}&=&11 N^6+57 N^5-39 N^4-109 N^3-44 N^2-116 N-48
    \\
P_{131}&=&11 N^6+81 N^5+9 N^4-133 N^3-92 N^2-116 N-48
    \\
P_{132}&=&13 N^6+36 N^5+39 N^4+8 N^3-21 N^2-29 N-10
    \\
P_{133}&=&16 N^6+78 N^5-23 N^4-228 N^3-503 N^2-408 N-228
    \\
P_{134}&=&17 N^6+111 N^5+234 N^4+203 N^3-89 N^2-296 N-36
    \\
P_{135}&=&22 N^6+69 N^5+71 N^4+23 N^3-57 N^2-68 N+84
    \\
P_{136}&=&23 N^6-7 N^5-237 N^4-593 N^3-678 N^2-548 N-200
    \\
P_{137}&=&23 N^6+9 N^5-71 N^4-53 N^3-184 N^2-92 N-16
    \\
P_{138}&=&25 N^6+35 N^5-55 N^4-243 N^3-286 N^2-204 N-72
    \\
P_{139}&=&29 N^6+91 N^5+235 N^4+405 N^3+272 N^2+288 N+120
    \\
P_{140}&=&29 N^6+176 N^5+777 N^4+1820 N^3+1878 N^2+776 N+232
    \\
P_{141}&=&35 N^6-15 N^5-183 N^4-133 N^3-356 N^2-164 N-48
    \\
P_{142}&=&35 N^6-15 N^5-101 N^4+31 N^3+54 N^2+164 N+120
    \\
P_{143}&=&44 N^6+96 N^5+369 N^4+290 N^3-695 N^2-428 N-108
    \\
P_{144}&=&55 N^6+141 N^5-195 N^4-401 N^3-772 N^2-748 N-384
    \\
P_{145}&=&55 N^6+165 N^5-420 N^4-899 N^3-1561 N^2-1336 N-1188
    \\
P_{146}&=&57 N^6+161 N^5-25 N^4-193 N^3-172 N^2-36 N+48
    \\
P_{147}&=&65 N^6+199 N^5+197 N^4-143 N^3-330 N^2-316 N-120
    \\
P_{148}&=&77 N^6+339 N^5-105 N^4-487 N^3-356 N^2-668 N-240
    \\
P_{149}&=&80 N^6+60 N^5+9 N^4+230 N^3+901 N^2+988 N+1188
    \\
P_{150}&=&81 N^6+211 N^5-23 N^4-355 N^3-334 N^2-4 N-344
    \\
P_{151}&=&83 N^6+249 N^5-111 N^4-637 N^3-956 N^2-596 N-624
    \\
P_{152}&=&130 N^6+865 N^5+2316 N^4+3811 N^3+4434 N^2+2884 N+536
    \\
P_{153}&=&133 N^6+699 N^5+1395 N^4+217 N^3-880 N^2+164 N+288
    \\
P_{154}&=&155 N^6+369 N^5+211 N^4-65 N^3-1002 N^2-556 N-1416
    \\
P_{155}&=&215 N^6+429 N^5+891 N^4+491 N^3-2486 N^2-1436 N-408
    \\
P_{156}&=&3 N^7+28 N^6+66 N^5+90 N^4+107 N^3+78 N^2+36 N+8
    \\
P_{157}&=&9 N^7+71 N^6+214 N^5+320 N^4+275 N^3+215 N^2+160 N+32
    \\
P_{158}&=&21 N^7+120 N^6-128 N^5-1038 N^4-89 N^3+2382 N^2+1636 N-600
\\
P_{159}&=&81 N^7+247 N^6+291 N^5+277 N^4+108 N^3-56 N^2+20 N+24
    \\
P_{160}&=&N^8+5 N^7+10 N^6+27 N^5+65 N^4+112 N^3+124 N^2+80 N+32
    \\
P_{161}&=&N^8+5 N^7+14 N^6+23 N^5+25 N^4+52 N^3+56 N^2+48 N+16
    \\
P_{162}&=&N^8+8 N^7-2 N^6-60 N^5-23 N^4+108 N^3+96 N^2+16 N+48
    \\
P_{163}&=&N^8+8 N^7-2 N^6-60 N^5+N^4+156 N^3+24 N^2-80 N-240
    \\
P_{164}&=&N^8+22 N^7+111 N^6+211 N^5+42 N^4-281 N^3-406 N^2-204 N-72
    \\
P_{165}&=&2 N^8+N^7-6 N^6+26 N^5+64 N^4+51 N^3+54 N^2+28 N+8
    \\
P_{166}&=&2 N^8+22 N^7+117 N^6+386 N^5+759 N^4+810 N^3+396 N^2+72 N+32
    \\
P_{167}&=&2 N^8+44 N^7+211 N^6+485 N^5+654 N^4+581 N^3+391 N^2+192 N+32
    \\
P_{168}&=&3 N^8+41 N^7+136 N^6+233 N^5+331 N^4+360 N^3+208 N^2+80 N+16
    \\
P_{169}&=&3 N^8+54 N^7+118 N^6-44 N^5-353 N^4-314 N^3-272 N^2-200 N-144
    \\
P_{170}&=&5 N^8-8 N^7-137 N^6-436 N^5-713 N^4-672 N^3-407 N^2-192 N-32
    \\
P_{171}&=&7 N^8+40 N^7+110 N^6+193 N^5+261 N^4+313 N^3+260 N^2+96 N+16
    \\
P_{172}&=&9 N^8+54 N^7+80 N^6-110 N^5-645 N^4-1168 N^3-1132 N^2-672 N-160
    \\
P_{173}&=&10 N^8+46 N^7+87 N^6+85 N^5-75 N^4-251 N^3-274 N^2-132 N-72
    \\
P_{174}&=&11 N^8+74 N^7+213 N^6+281 N^5-30 N^4-427 N^3-446 N^2-180 N-72
    \\
P_{175}&=&15 N^8+36 N^7+50 N^6-252 N^5-357 N^4+152 N^3-68 N^2+88 N+48
    \\
P_{176}&=&18 N^8+101 N^7+128 N^6+208 N^5+190 N^4-769 N^3-1200 N^2-212 N-48
    \\
P_{177}&=&19 N^8+70 N^7+63 N^6-41 N^5-192 N^4-221 N^3-142 N^2-60 N-72
    \\
P_{178}&=&21 N^8+42 N^7-38 N^6-360 N^5-631 N^4-730 N^3-472 N^2-216 N-48
    \\
P_{179}&=&23 N^8+2 N^7-135 N^6+29 N^5+210 N^4-151 N^3-350 N^2-132 N-72
    \\
P_{180}&=&27 N^8-36 N^7-956 N^6-1724 N^5+187 N^4+1288 N^3+70 N^2-224 N-72
    \\
P_{181}&=&38 N^8+146 N^7+177 N^6+35 N^5-249 N^4-373 N^3-218 N^2-60 N-72
    \\
P_{182}&=&41 N^8+5 N^7-195 N^6-97 N^5+326 N^4+424 N^3+208 N^2+72 N+16
    \\
P_{183}&=&56 N^8+194 N^7+213 N^6+83 N^5-231 N^4-469 N^3-290 N^2-60 N-72
    \\
P_{184}&=&79 N^8+196 N^7+132 N^6+274 N^5+465 N^4+82 N^3+332 N^2+456 N+288
 \\
P_{185}&=&105 N^8+978 N^7+1688 N^6-1330 N^5-5245 N^4-4672 N^3-2212 N^2-544 N-288
    \\
P_{186}&=&113 N^8+348 N^7+109 N^6-289 N^5-272 N^4-859 N^3-778 N^2-172 N+72
    \\
P_{187}&=&170 N^8+369 N^7-521 N^6-1393 N^5-761 N^4-952 N^3-544 N^2+32 N+144
    \\
P_{188}&=&264 N^8+1407 N^7+2246 N^6+1746 N^5+804 N^4-1069 N^3-674 N^2-92 N-24
    \\
P_{189}&=&283 N^8+838 N^7+1482 N^6+628 N^5-1497 N^4-1130 N^3-772 N^2+456 N+288
 \\
P_{190}&=&633 N^8+2532 N^7+5036 N^6+6142 N^5+4275 N^4+1118 N^3-176 N^2-184 N-48
    \\
P_{191}&=&N^9+21 N^8+85 N^7+105 N^6+42 N^5+290 N^4+600 N^3+456 N^2+256 N+64
    \\
P_{192}&=&4 N^9+53 N^8+193 N^7+233 N^6+87 N^5+554 N^4+1172 N^3+904 N^2+512 N+128
 \\
P_{193}&=&6 N^9+93 N^8+576 N^7+1296 N^6+586 N^5+359 N^4+2000 N^3+1996 N^2
\nonumber\\ &&
+1488 N+384
    \\
P_{194}&=&9 N^9+54 N^8+56 N^7-110 N^6-381 N^5-568 N^4-364 N^3-72 N^2+128 N+96
 \\
P_{195}&=&9 N^9+54 N^8+167 N^7+397 N^6+780 N^5+1241 N^4+1448 N^3+1200 N^2+608 N+144
 \\
P_{196}&=&11 N^9+78 N^8+214 N^7+335 N^6+383 N^5+571 N^4+916 N^3+876 N^2+480 N+96
 \\
P_{197}&=&35 N^9+150 N^8+232 N^7+137 N^6+119 N^5+661 N^4+1174 N^3+876 N^2+480 N+96
 \\
P_{198}&=&37 N^9+210 N^8-52 N^7-2738 N^6-7249 N^5-9368 N^4-8216 N^3-5888 N^2
\nonumber\\ &&
-2448 N-576
 \\
P_{199}&=&45 N^9+270 N^8+820 N^7+1478 N^6+1683 N^5+1996 N^4+2356 N^3+2328 N^2
\N\\&&
+1408 N+288
 \\
P_{200}&=&57 N^9+624 N^8+1756 N^7+1092 N^6-1803 N^5-1512 N^4+966 N^3+1116 N^2
\N\\&&
+920 N+528
    \\
P_{201}&=&69 N^9+366 N^8+1124 N^7+1966 N^6+2523 N^5+5228 N^4+7340 N^3+5352 N^2
\N\\&&
+3008 N+672
 \\
P_{202}&=&94 N^9+597 N^8+1616 N^7+2410 N^6+1841 N^5+1165 N^4+2191 N^3+3802 N^2
\N\\&&
+2916 N+648
 \\
P_{203}&=&121 N^9+696 N^8+1535 N^7+1585 N^6+416 N^5-749 N^4-836 N^3+16 N^2
\N\\&&
+528 N+144
 \\
P_{204}&=&197 N^9+1242 N^8+2938 N^7+3524 N^6+2713 N^5+2234 N^4+3680 N^3+6176 N^2
\N\\&&
+4080 N
+864
 \\
P_{205}&=&439 N^9+2634 N^8+6008 N^7+6694 N^6+3545 N^5+736 N^4+2008 N^3+6208 N^2
\N\\&&
+5136 N
+1152
 \\
P_{206}&=&538 N^9+3333 N^8+7802 N^7+7630 N^6+458 N^5-1415 N^4+7786 N^3+12340 N^2
\N\\&&
+5592 N
+864
 \\
P_{207}&=&664 N^9+3861 N^8+9038 N^7+11830 N^6+9344 N^5+3793 N^4+3874 N^3+11044 N^2
\N\\&&
+9624 N
+2592
 \\
P_{208}&=&891 N^9+4455 N^8+16078 N^7+28774 N^6+37047 N^5+45835 N^4+42192 N^3+28888 N^2
\N\\&&
+10640 N+1776
 \\
P_{209}&=&923 N^9+5208 N^8+11824 N^7+12854 N^6+2185 N^5-7030 N^4+1436 N^3+15032 N^2
\N\\&&
+12864 N+3456
 \\
P_{210}&=&965 N^9+4884 N^8+10816 N^7+20810 N^6+36895 N^5+40442 N^4+27692 N^3+22712 N^2
\N\\&&
+14496 N+3456
 \\
P_{211}&=&2 N^{10}-46 N^9-98 N^8+282 N^7+1063 N^6+1569 N^5+1275 N^4+403 N^3-94 N^2
\N\\&&
-108 N-24
 \\
P_{212}&=&2 N^{10}+12 N^9+24 N^8+11 N^7-48 N^6-151 N^5-282 N^4-480 N^3-664 N^2
\N\\&&
-576 N-288
 \\
P_{213}&=&11 N^{10}+44 N^9+74 N^8+196 N^7+31 N^6-1426 N^5-3044 N^4-2762 N^3-1476 N^2
\N\\&&
-480 N
-96
 \\
P_{214}&=&11 N^{10}+76 N^9+138 N^8-204 N^7-1041 N^6-988 N^5+752 N^4+1896 N^3+944 N^2
\N\\&&
-384 N
-576
 \\
P_{215}&=&37 N^{10}+392 N^9+2106 N^8+6514 N^7+9211 N^6+1258 N^5-9218 N^4-6116 N^3-72 N^2
\N\\&&
-752 N-192
    \\
P_{216}&=&85 N^{10}+425 N^9+902 N^8+932 N^7-521 N^6-685 N^5+2022 N^4+2928 N^3+968 N^2
\N\\&&
-1296 N-576
 \\
P_{217}&=&103 N^{10}+575 N^9+1124 N^8-334 N^7-1505 N^6+3755 N^5+4926 N^4+36 N^3-472 N^2
\N\\&&
-2160 N-864
 \\
P_{218}&=&118 N^{10}+425 N^9+197 N^8+86 N^7+1240 N^6+2489 N^5+4401 N^4+3480 N^3+524 N^2
\N\\&&
-1728 N-864
 \\
P_{219}&=&118 N^{10}+557 N^9+461 N^8-94 N^7+1300 N^6+3521 N^5+4509 N^4+1920 N^3
\N\\&&
-1132 N^2
-2376 N-1008
 \\
P_{220}&=&127 N^{10}+536 N^9+611 N^8+602 N^7+1474 N^6+2099 N^5+798 N^4-2301 N^3
\N\\&&
-4486 N^2
-3708 N-936
 \\
P_{221}&=&170 N^{10}+883 N^9+2041 N^8+2998 N^7-448 N^6-5465 N^5+129 N^4+6624 N^3
\N\\&&
+1132 N^2
-2016 N-864
    \\
P_{222}&=&170 N^{10}+1213 N^9+3235 N^8+2794 N^7-2692 N^6-3767 N^5-1293 N^4
\N\\&&
-1632 N^3
-5324 N^2-6240 N-2016
 \\
P_{223}&=&226 N^{10}+317 N^9-811 N^8+662 N^7+4552 N^6+3857 N^5+3933 N^4+2364 N^3
 \N\\&&
+236 N^2-1656 N-720
 \\
P_{224}&=&489 N^{10}+2934 N^9+9364 N^8+18830 N^7+18627 N^6+124 N^5-19856 N^4-19296 N^3
 \N\\&&
-10640 N^2-2880 N-1152
 \\
P_{225}&=&3 N^{11}+42 N^{10}+144 N^9+74 N^8-459 N^7-1060 N^6-1152 N^5-1424 N^4-1688 N^3
 \N\\&&
-1232 N^2-736 N-192
 \\
P_{226}&=&11 N^{11}+37 N^{10}-27 N^9-118 N^8+21 N^7-249 N^6-1097 N^5-1138 N^4+552 N^3
 \N\\&&
+3448 N^2+3456 N+2016
 \\
P_{227}&=&21 N^{11}+231 N^{10}+1334 N^9+4086 N^8+6277 N^7+1775 N^6-9488 N^5-18076 N^4
 \N\\&&
-18208 N^3
-11344 N^2-5568 N-1728
 \\
P_{228}&=&33 N^{11}+231 N^{10}+698 N^9+1290 N^8+1513 N^7+1463 N^6+2236 N^5+5096 N^4
 \N\\&&
+7328 N^3
+5456 N^2+3456 N+1152
 \\
P_{229}&=&45 N^{11}+383 N^{10}+958 N^9+526 N^8-763 N^7+1375 N^6+7808 N^5+13028 N^4
 \N\\&&
+12976 N^3+8016 N^2+4608 N+1728
 \\
P_{230}&=&51 N^{11}+269 N^{10}+46 N^9-1934 N^8-3973 N^7-875 N^6+7364 N^5+14972 N^4
 \N\\&&
+16768 N^3+10896 N^2+5376 N+1728
 \\
P_{231}&=&51 N^{11}+357 N^{10}+1238 N^9+2586 N^8+2755 N^7-1435 N^6-9212 N^5-15028 N^4
 \N\\&&
-15280 N^3-9808 N^2-5184 N-1728
 \\
P_{232}&=&81 N^{11}+483 N^{10}+1142 N^9+1086 N^8-767 N^7-4645 N^6-8936 N^5-11980 N^4
 \N\\&&
-12352 N^3-8272 N^2-4800 N-1728
 \\
P_{233}&=&120 N^{11}+1017 N^{10}+2737 N^9+1292 N^8-8086 N^7-20743 N^6-24563 N^5-16702 N^4
 \N\\&&
-6840 N^3+120 N^2+2432 N+960
 \\
P_{234}&=&243 N^{11}+1701 N^{10}+5378 N^9+10350 N^8+11479 N^7+1193 N^6-14684 N^5-20572 N^4
 \N\\&&
-16288 N^3-8944 N^2-4992 N-1728
 \\
P_{235}&=&333 N^{11}+2331 N^{10}+6556 N^9+9270 N^8+5081 N^7-6701 N^6-17554 N^5-20036 N^4
 \N\\&&
-15680 N^3-9200 N^2-5664 N-1728
 \\
P_{236}&=&753 N^{11}+4809 N^{10}+13174 N^9+20466 N^8+17717 N^7+6829 N^6+3908 N^5
 \N\\&&
+15304 N^4+25408 N^3+20272 N^2+8448 N+1152
 \\
P_{237}&=&837 N^{11}+7757 N^{10}+30120 N^9+68575 N^8+119176 N^7+191350 N^6+262979 N^5
 \N\\&&
+258308 N^4+163106 N^3+63360 N^2+14848 N+1536
 \\
P_{238}&=&1017 N^{11}+6195 N^{10}+14050 N^9+12738 N^8-2023 N^7-5093 N^6+27548 N^5
 \N\\&&
+69760 N^4+80752 N^3+54064 N^2+20928 N+3456
 \\
P_{239}&=&3 N^{12}+21 N^{11}+17 N^{10}-202 N^9-842 N^8-1924 N^7-3378 N^6-5059 N^5
 \N\\&&
-6008 N^4-4860 N^3-2536 N^2-960 N-192
 \\
P_{240}&=&9 N^{12}+63 N^{11}+38 N^{10}-414 N^9-1035 N^8-1341 N^7-1511 N^6-2972 N^5
 \N\\&&
-6011 N^4-8038 N^3-6892 N^2-3432 N-864
 \\
P_{241}&=&9 N^{12}+63 N^{11}+71 N^{10}-381 N^9-1536 N^8-2529 N^7-1946 N^6-1331 N^5
 \N\\&&
-2096 N^4-4036 N^3-4144 N^2-2304 N-576
 \\
P_{242}&=&39 N^{12}+585 N^{11}+2938 N^{10}+7136 N^9+9083 N^8+7745 N^7+14668 N^6+38246 N^5
 \N\\&&
+59856 N^4+55560 N^3+32144 N^2+12480 N+2304
 \\
P_{243}&=&48 N^{12}+459 N^{11}+2322 N^{10}+8290 N^9+20159 N^8+30862 N^7+28247 N^6+16109 N^5
 \N\\&&
+9312 N^4+7488 N^3+4064 N^2+1328 N+192
 \\
P_{244}&=&61 N^{12}+302 N^{11}+531 N^{10}+348 N^9-349 N^8-786 N^7+457 N^6+2524 N^5
 \N\\&&
+2012 N^4+204 N^3-360 N^2-240 N-96
 \\
P_{245}&=&92 N^{12}+796 N^{11}+3089 N^{10}+7550 N^9+10547 N^8+1029 N^7-19496 N^6
 \N\\&&
-24199 N^5-8960 N^4+736 N^3+1744 N^2+816 N+192
 \\
P_{246}&=&201 N^{12}+1845 N^{11}+6910 N^{10}+12854 N^9+8915 N^8-7741 N^7-17126 N^6
 \N\\&&
-4294 N^5+16260 N^4+22080 N^3+12416 N^2+4128 N+576
 \\
P_{247}&=&239 N^{12}+1338 N^{11}+3137 N^{10}+3164 N^9-983 N^8-6640 N^7-8123 N^6
 \N\\&&
-4526 N^5-342 N^4+1232 N^3+848 N^2+256 N+32
 \\
P_{248}&=&255 N^{12}+2169 N^{11}+6496 N^{10}+7694 N^9-127 N^8-6973 N^7+4132 N^6
 \N\\&&
+25502 N^5+31956 N^4+22656 N^3+9632 N^2+864 N-576
 \\
P_{249}&=&581 N^{12}+7035 N^{11}+37826 N^{10}+112904 N^9+190293 N^8+174327 N^7+92032 N^6
 \N\\&&
+69438 N^5+78364 N^4+44464 N^3+11520 N^2-3168 N-1728
 \\
P_{250}&=&825 N^{12}+7363 N^{11}+25396 N^{10}+40686 N^9+26213 N^8-12749 N^7-55498 N^6
 \N\\&&
-89796 N^5-110552 N^4-134960 N^3-127584 N^2-64704 N-12672
 \\
P_{251}&=&69 N^{13}+420 N^{12}+794 N^{11}-1357 N^{10}-10401 N^9-15678 N^8+532 N^7+239 N^6
 \N\\&&
-40018 N^5-69432 N^4-69152 N^3-43792 N^2-18336 N-3456
 \\
P_{252}&=&76 N^{13}+922 N^{12}+4479 N^{11}+9107 N^{10}-3747 N^9-52973 N^8-76133 N^7+42261 N^6
 \N\\&&
+199307 N^5+123839 N^4-77470 N^3-84132 N^2-2160 N-432
 \\
P_{253}&=&295 N^{13}+2387 N^{12}+8005 N^{11}+13687 N^{10}+10883 N^9+389 N^8-2641 N^7+6029 N^6
 \N\\&&
+11034 N^5+6644 N^4+1384 N^3+80 N^2+128 N+64
 \\
P_{254}&=&296 N^{13}+2368 N^{12}+10916 N^{11}+27006 N^{10}+23644 N^9-19764 N^8-61931 N^7
 \N\\&&
-63733 N^6
-52001 N^5-56865 N^4-38104 N^3+2664 N^2+7344 N+432
 \\
P_{255}&=&377 N^{13}+4649 N^{12}+21813 N^{11}+38539 N^{10}-39339 N^9-272611 N^8-332971 N^7
 \N\\&&
+220377 N^6
+801934 N^5+384958 N^4-362030 N^3-297864 N^2-1080 N-864
 \\
P_{256}&=&859 N^{13}+7376 N^{12}+25294 N^{11}+47088 N^{10}+63868 N^9+80876 N^8+63648 N^7
 \N\\&&
-35856 N^6
-146697 N^5-157168 N^4-91320 N^3-34800 N^2-8640 N-1152
 \\
P_{257}&=&1211 N^{13}+5680 N^{12}+3338 N^{11}-17355 N^{10}-31517 N^9-48486 N^8-139667 N^7
 \N\\&&
-278026 N^6
-340745 N^5-269457 N^4-138568 N^3-34632 N^2+9072 N+3888
 \\
P_{258}&=&70 N^{14}+555 N^{13}+1599 N^{12}+1192 N^{11}-4430 N^{10}-13305 N^9-11835 N^8+8440 N^7
 \N\\&&
+35816 N^6+57126 N^5+60340 N^4+44464 N^3+27808 N^2+12768 N+2880
 \\
P_{259}&=&76 N^{14}+802 N^{13}+2979 N^{12}+1847 N^{11}-19377 N^{10}-58253 N^9-26543 N^8+170601 N^7
 \N\\&&
+362177 N^6+225119 N^5-103240 N^4-193092 N^3-137160 N^2-117072 N-25920
 \\
P_{260}&=&76 N^{14}+1042 N^{13}+5979 N^{12}+16367 N^{11}+11883 N^{10}-47693 N^9-125723 N^8-86079 N^7
 \N\\&&
+36437 N^6+22559 N^5-51700 N^4+24828 N^3+132840 N^2+116208 N+25920
 \\
P_{261}&=&4 N^{15}+50 N^{14}+267 N^{13}+765 N^{12}+1183 N^{11}+682 N^{10}-826 N^9-1858 N^8-1116 N^7
 \N\\&&
+457 N^6+1500 N^5+2268 N^4+2400 N^3+1392 N^2+448 N+64
    \\
P_{262}&=&26 N^{15}+314 N^{14}+1503 N^{13}+3222 N^{12}+2510 N^{11}+1996 N^{10}+15041 N^9+40728 N^8
 \N\\&&
+54008 N^7
+44956 N^6+31936 N^5+30416 N^4+29568 N^3+16704 N^2+5376 N+768
 \\
P_{263}&=&101 N^{15}+1234 N^{14}+6867 N^{13}+21904 N^{12}+40098 N^{11}+32226 N^{10}-22057 N^9
 \N\\&&
-86972 N^8
-114557 N^7-111416 N^6-89204 N^5-37312 N^4+13392 N^3+23040 N^2
 \N\\&&
+9792 N+1536
 \\
P_{264}&=&390 N^{15}+5121 N^{14}+30556 N^{13}+114173 N^{12}+321958 N^{11}+771597 N^{10}
 \N\\&&
+1583594 N^9+2637549 N^8
+3381542 N^7+3199120 N^6+2183360 N^5+1123200 N^4
 \N\\&&
+489952 N^3+178176 N^2+48384 N+6912
 \\
P_{265}&=&75 N^{16}+1245 N^{15}+8291 N^{14}+27609 N^{13}+43437 N^{12}+14221 N^{11}-5995 N^{10}
 \N\\&&
+182937 N^9+488696 N^8+296818 N^7-452292 N^6-730430 N^5-186180 N^4
 \N\\&&
+259728 N^3+241056 N^2+116640 N+25920
 \\
P_{266}&=&115 N^{16}+1838 N^{15}+11829 N^{14}+36114 N^{13}+30900 N^{12}-133946 N^{11}-454068 N^{10}
 \N\\&&
-457420 N^9+249211 N^8+864716 N^7+312979 N^6-634466 N^5-587862 N^4
 \N\\&&
-19556 N^3+104832 N^2+9504 N+1728
 \\
P_{267}&=&185 N^{16}+2988 N^{15}+19694 N^{14}+62954 N^{13}+64470 N^{12}-207876 N^{11}-792388 N^{10}
 \N\\&&
-861230 N^9+437231 N^8+1750616 N^7+869954 N^6-1016136 N^5-1130122 N^4
 \N\\&&
-96596 N^3+199872 N^2+31104 N+1728
 \\
P_{268}&=&939 N^{16}+10527 N^{15}+37207 N^{14}+18679 N^{13}-202006 N^{12}-617170 N^{11}-930025 N^{10}
 \N\\&&
-882917 N^9-157123 N^8+1388549 N^7+2739376 N^6+2837500 N^5+2088640 N^4
 \N\\&&
+1259696 N^3+622464 N^2+211392 N+34560
 \\
P_{269}&=&1155 N^{16}+12417 N^{15}+37693 N^{14}-12293 N^{13}-285754 N^{12}-613900 N^{11}-571735 N^{10}
 \N\\&&
-134309 N^9+778901 N^8+2698745 N^7+4995724 N^6+5915740 N^5+4978144 N^4
 \N\\&&
+3161840 N^3+1498752 N^2+479808 N+76032
 \\
P_{270}&=&1665 N^{16}+33005 N^{15}+287646 N^{14}+1402624 N^{13}+4031902 N^{12}+6199846 N^{11}
 \N\\&&
+1054640 N^{10}-16668628 N^9-37272559 N^8-38892027 N^7-17387942 N^6+3962700 N^5
 \N\\&&
+15625800 N^4+26960688 N^3+27379296 N^2+12985920 N+2332800
 \\
P_{271}&=&87 N^{17}+1099 N^{16}+6055 N^{15}+19019 N^{14}+37119 N^{13}+45159 N^{12}+29583 N^{11}-2639 N^{10}
 \N\\&&
-30218 N^9-40778 N^8-39994 N^7-35844 N^6-30808 N^5-30384 N^4-28256 N^3
 \N\\&&
-16064 N^2-5248 N-768
 \\
P_{272}&=&829 N^{17}+13413 N^{16}+83461 N^{15}+226391 N^{14}+55508 N^{13}-1239070 N^{12}-2862466 N^{11}
 \N\\&&
-1217372 N^{10}+3372689 N^9+2779147 N^8-2705687 N^7+171733 N^6
 \N\\&&
+8617302 N^5+5817902 N^4-3127236 N^3-3652560 N^2-336096 N-25920
 \\
P_{273}&=&1407 N^{17}+18107 N^{16}+103463 N^{15}+347083 N^{14}+760095 N^{13}+1142715 N^{12}+1220067 N^{11}
 \N
\\
&&
+983393 N^{10}+702746 N^9+533822 N^8+337702 N^7-3552 N^6
 \N\\&&
-300296 N^5-332160 N^4-188128 N^3-63232 N^2-13184 N-1536
 \\
P_{274}&=&95 N^{18}+3940 N^{17}+48989 N^{16}+308380 N^{15}+1166094 N^{14}+2843192 N^{13}+4428234 N^{12}
 \N\\&&
+3171928 N^{11}-4692053 N^{10}-19875244 N^9-34305831 N^8-34774388 N^7-16392680 N^6
 \N\\&&
+11584912 N^5+30493776 N^4+29700864 N^3+18783360 N^2+8294400 N+1866240
 \\
P_{275}&=&325 N^{18}+4280 N^{17}+17759 N^{16}-14880 N^{15}-412326 N^{14}-1696848 N^{13}-3216546 N^{12}
 \N\\&&
-1169232 N^{11}+8956857 N^{10}+23914216 N^9+31536899 N^8+25361392 N^7+9982840 N^6
 \N\\&&
-10154128 N^5-26098704 N^4-26761536 N^3-17642880 N^2-8087040 N-1866240
 \\
P_{276}&=&500 N^{18}+8215 N^{17}+56287 N^{16}+201810 N^{15}+361782 N^{14}+98826 N^{13}-759348 N^{12}
 \N\\&&
-495786 N^{11}+3942186 N^{10}+11896133 N^9+16709737 N^8+13315736 N^7+3779660 N^6
 \N\\&&
-7306454 N^5-14232852 N^4-13254768 N^3-8367840 N^2-3771360 N-855360
 \\
P_{277}&=&150 N^{19}+2815 N^{18}+24285 N^{17}+131358 N^{16}+511310 N^{15}+1515954 N^{14}
 \N\\&&
+3372978 N^{13}+5213980 N^{12}
+4715522 N^{11}+980739 N^{10}-2709391 N^9-3741506 N^8
 \N\\&&
-4630558 N^7-5623132 N^6
-2333736 N^5+3419632 N^4+5238496 N^3+3231936 N^2
 \N\\&&
+1123200 N+172800
 \\
P_{278}&=&5410 N^{19}+98215 N^{18}+764965 N^{17}+3280996 N^{16}+8031920 N^{15}+8939378 N^{14}
 \N\\&&
-7608074 N^{13}-44964380 N^{12}
-74768226 N^{11}-57879177 N^{10}-5243187 N^9+13745888 N^8
 \N\\&&
-28158216 N^7-49672024 N^6
+14757808 N^5+94650144 N^4+100507392 N^3
 \N\\&&
+53764992 N^2+15655680 N+1866240
 \\
P_{279}&=&7060 N^{20}+123495 N^{19}+898682 N^{18}+3394183 N^{17}+6222824 N^{16}+376386 N^{15}
 \N\\&&
-22032204 N^{14}
-39912378 N^{13}-13976964 N^{12}
+31985011 N^{11}+4994394 N^{10}
 \N\\&&
-91499501 N^9
-97243208 N^8
+54501988 N^7+183103272 N^6
+127073120 N^5
 \N\\&&
-20272608 N^4
-88410816 N^3
-62225280 N^2
-21772800 N-3110400~. 
\end{eqnarray}

}
The corresponding expressions in $z$-space are given in Appendix~\ref{app:B}.

Note that our result for $H_{g,2}^{\sf S}$ differs from the one given in Eq.~(B.7) in $z$-space in 
Ref.~\cite{Kawamura:2012cr} by the term
\begin{eqnarray}
C_F T_F^2 N_F \frac{4(N^2+N+2)}{N(N+1)(N+2)} \left[28 \zeta_2 -69\right]
\end{eqnarray}
in $N$-space. This result of \cite{Kawamura:2012cr} is based on the calculation carried out in 
Ref.~\cite{Bierenbaum:2009mv}, including the renormalization formulae derived there.
We have checked, however, that our result Eq.~(\ref{eq:H2S}) is in full agreement with 
Eq.~(\ref{eq:WILS}) and the moments having been calculated by part of the present authors in 
Ref.~\cite{Bierenbaum:2009mv}. The corresponding expression in $z$-space is presented in Appendix~B.
\section{The Asymptotic Wilson Coefficients for the Longitudinal Structure Function}
\label{sec:5}

\vspace*{1mm}
\noindent
The Wilson coefficients have been calculated in Ref.~\cite{Blumlein:2006mh} for exclusive 
heavy flavor production, retaining three contributions only. In total also here five Wilson 
coefficients contribute and the expressions are slightly modified in the 
inclusive case of the complete structure function $F_L(x,Q^2)$, 
cf.~\cite{Bierenbaum:2009mv}. In Mellin-$N$ space they read~:
{\small
\begin{eqnarray}
\lefteqn{L_{q,L}^{{\sf PS},(3)} = \tfrac{1}{2}[1 + (-1)^N]} \nonumber\\ 
&&
\times \Biggl\{
\textcolor{blue}{a_s^3} \Biggl\{
\textcolor{blue}{C_F} \textcolor{blue}{N_F} \textcolor{blue}{T_F^2} \Biggl[
\frac{128 L_Q^2 \big(N^2+N+2\big)}{3 (N-1) N (N+1)^2 (N+2)}
+\frac{128 \big(N^2+N+2\big) L_M^2}{3 (N-1) N (N+1)^2 (N+2)}
\N\\&&
-\frac{256 L_Q \big(11 N^5+35 N^4+59 N^3+55 N^2-4 N-12\big)}{9 (N-1) N^2 (N+1)^3 (N+2)^2}
+\Biggl[
\frac{256 \big(8 N^3+13 N^2+27 N+16\big)}{9 (N-1) N (N+1)^3 (N+2)}
\N\\&&
-\frac{256 \big(N^2+N+2\big) S_1}{3 (N-1) N (N+1)^2 (N+2)}\Biggr] L_M
+\frac{64 \big(N^2+N+2\big) \big[S_1^2+ S_2\big]}{3(N-1) N (N+1)^2 (N+2)}
\N\\&&
-\frac{128 \big(8 N^3+13 N^2+27 N+16\big) S_1}{9 (N-1) N (N+1)^3 (N+2)}
+\frac{128 \big(43 N^4+105 N^3+224 N^2+230 N+86\big)}{27 (N-1) N (N+1)^4 (N+2)}\Biggr]
\N\\&&
+\textcolor{blue}{N_F} \hat{\tilde{C}}_{L,q}^{{\sf PS},(3)}({N_F})\Biggr\} \Biggr\},
\end{eqnarray}

}
{\small
\begin{eqnarray}
\lefteqn{L_{g,L}^{\sf S} = \tfrac{1}{2}[1 + (-1)^N]} \nonumber\\  
&&
\times \Biggl\{
\textcolor{blue}{a_s^2} \frac{64 \textcolor{blue}{N_F} \textcolor{blue}{T_F^2} L_M }{3 (N+1) (N+2)}
+\textcolor{blue}{a_s^3} \Biggl\{
\textcolor{blue}{N_F} \textcolor{blue}{T_F^3} \frac{256  L_M^2 }{9 (N+1) (N+2)}
+\textcolor{blue}{C_A} \textcolor{blue}{N_F} \textcolor{blue}{T_F^2} \Biggl[
\N\\&&
\Biggl[\frac{256 \big(N^2+N+1\big)}{3 (N-1) N (N+1)^2 (N+2)^2}
-\frac{128 S_1}{3 (N+1) (N+2)}\Biggr] L_Q^2
+\Biggl[\frac{256 (-1)^N \big(N^3+4 N^2+7 N+5\big)}{3 (N+1)^3 (N+2)^3}
\N\\&&
+\frac{64 Q_2}{9 (N-1) N (N+1)^3 (N+2)^3}
+\frac{256 \big(11 N^3-6 N^2-8 N-3\big) S_1}{9 (N-1) N (N+1)^2 (N+2)}
\N\\&&
+L_M \Biggl[\frac{512 \big(N^2+N+1\big)}{3 (N-1) N (N+1)^2 (N+2)^2}
-\frac{256 S_1}{3 (N+1) (N+2)}\Biggr]
\N\\&&
+\frac{1}{(N+1) (N+2)} \Bigl[\frac{128}{3} S_1^2
-\frac{128 S_2}{3}
-\frac{256}{3} S_{-2}\Bigr]\Biggr] L_Q
+\frac{32 Q_5}{27 (N-1) N^3 (N+1)^4 (N+2)^2}
\N\\&&
-\frac{64 (56 N+47) S_1}{27 (N+1)^2 (N+2)}
+L_M^2 \Biggl[
\frac{256 \big(N^2+N+1\big)}{3 (N-1) N (N+1)^2 (N+2)^2}
-\frac{128 S_1}{3 (N+1) (N+2)}\Biggr]
\N\\&&
+L_M \Biggl[
\frac{256 (-1)^N \big(N^3+4 N^2+7 N+5\big)}{3 (N+1)^3 (N+2)^3}
+\frac{128 Q_4}{9 (N-1) N^2 (N+1)^3 (N+2)^3}
\N\\&&
+\frac{256 \big(N^3-6 N^2+2 N-3\big) S_1}{9 (N-1) N (N+1)^2 (N+2)}
+\frac{\frac{128}{3} S_1^2
-\frac{128 S_2}{3}
-\frac{256}{3} S_{-2}}{(N+1) (N+2)}\Biggr]\Biggr] 
\N\\&&
+\textcolor{blue}{C_F} \textcolor{blue}{T_F^2} \textcolor{blue}{N_F} \Biggl[
\frac{64 \big(N^2+N+2\big) \big(N^4+2 N^3+2 N^2+N+6\big) L_Q^2}{3 (N-1) N^2 (N+1)^3 (N+2)^2}
\N\\&&
+\Biggl[\frac{256 (-1)^N Q_6}{45 (N-2) (N-1)^2 N^2 (N+1)^3 (N+2)^2 (N+3)^3}
\N\\&&
-\frac{32 Q_9}{45 (N-1)^2 N^3 (N+1)^4 (N+2)^3 (N+3)^3}
-\frac{128 Q_1 S_1}{3 (N-1) N^2 (N+1)^3 (N+2)^2}
\N\\&&
+\frac{256 (N-1) S_{-2}}{3 (N-2) (N+1) (N+3)}
+\frac{64 \big(N^2+N+2\big) \big(N^4+2 N^3-7 N^2-8 N-12\big) L_M}{3 (N-1) N^2 (N+1)^3 (N+2)^2}\Biggr] L_Q
\N\\&&
+\frac{64 \big(N^2+N+2\big)^2 L_M^2}{(N-1) N^2 (N+1)^3 (N+2)^2}
-\frac{16 Q_7}{(N-1) N^4 (N+1)^5 (N+2)^2}
\N\\&&
+L_M \Biggl[\frac{256 (-1)^N Q_6}{45 (N-2) (N-1)^2 N^2 (N+1)^3 (N+2)^2 (N+3)^3}
\N\\&&
+\frac{64 Q_8}{45 (N-1)^2 N^3 (N+1)^4 (N+2)^3 (N+3)^3}
-\frac{64 Q_3 S_1}{3 (N-1) N^2 (N+1)^3 (N+2)^2}
\N\\&&
+\frac{256 (N-1) S_{-2}}{3 (N-2) (N+1) (N+3)}\Biggr]\Biggr] 
+\textcolor{blue}{N_F} \hat{\tilde{C}}_{L,g}^{{\sf S},(3)}({N_F})\Biggr\}\Biggr\},
\end{eqnarray}

}
with
{\small
\begin{eqnarray}
Q_1 &=&2 N^6+6 N^5+7 N^4+4 N^3+9 N^2+8 N+12
    \\
Q_2 &=&3 N^6+3 N^5-121 N^4-391 N^3-474 N^2-308 N-80
    \\
Q_3 &=&3 N^6+9 N^5-N^4-17 N^3-38 N^2-28 N-24
    \\
Q_4 &=&6 N^7+24 N^6+47 N^5+104 N^4+219 N^3+316 N^2+208 N+48
    \\
Q_5 &=&15 N^8+60 N^7+572 N^6+1470 N^5+2135 N^4+1794 N^3+722 N^2-24 N-72
    \\
Q_6 &=&N^{10}-13 N^9-39 N^8+222 N^7+1132 N^6+1787 N^5+913 N^4+392 N^3+645 N^2
\N\\&&
-324 N-108
 \\
Q_7 &=&15 N^{10}+75 N^9+112 N^8+14 N^7-61 N^6+107 N^5+170 N^4+36 N^3-36 N^2
\nonumber\\ &&
-32 N-16
 \\
Q_8 &=&45 N^{13}+656 N^{12}+4397 N^{11}+17513 N^{10}+43665 N^9+63005 N^8+27977 N^7-71993 N^6
\N\\&&
-140386 N^5-78985 N^4+25350 N^3+80460 N^2+100008 N+38880
 \\
Q_9 &=&95 N^{13}+1218 N^{12}+6096 N^{11}+14484 N^{10}+11570 N^9-28440 N^8-117844 N^7
\N\\&&
-225884 N^6-238953 N^5-83290 N^4+57660 N^3+122040 N^2+182304 N+77760, 
\end{eqnarray}

}
{\small
\begin{eqnarray}
\lefteqn{L_{q,L}^{\sf NS} = \tfrac{1}{2}[1 + (-1)^N]} \nonumber\\ 
&&
\times \Biggl\{
\textcolor{blue}{a_s^2} \textcolor{blue}{C_F} \textcolor{blue}{T_F} \Biggl\{
\frac{16 L_Q}{3 (N+1)}
-\frac{8 \big(19 N^2+7 N-6\big)}{9 N (N+1)^2}
-\frac{16 S_1}{3 (N+1)}\Biggr\} 
\N\\&&
+\textcolor{blue}{a_s^3} \Biggl\{
\textcolor{blue}{C_F^2} \textcolor{blue}{T_F} \Biggl[
\Biggl[\frac{8 \big(3 N^2+3 N+2\big)}{N (N+1)^2}
-\frac{32 S_1}{N+1}\Biggr] L_Q^2
\N\\&&
+\Biggl[\frac{256 (-1)^N Q_{11}}{15 (N-2) (N-1)^2 N^2 (N+1)^4 (N+2)^2 (N+3)^3}
\N\\&&
-\frac{32 Q_{12}}{45 (N-1)^2 N^2 (N+1)^4 (N+2)^2 (N+3)^3}
-\frac{16 (N+10) (5 N+3) S_1}{9 N (N+1)^2}
\N\\&&
+\frac{512 \big(N^4+2 N^3-N^2-2 N-6\big) S_{-2}}{3 (N-2) N (N+1)^2 (N+3)}
+\frac{1}{N+1} \Bigl[\frac{128}{3} S_1^2
-\frac{512}{3} S_{-2} S_1
-\frac{128 S_2}{3}
-\frac{256 S_3}{3}
\N\\&&
-\frac{256}{3} S_{-3}
+\frac{512}{3} S_{-2,1}+256 \zeta_3\Bigr]\Biggr] L_Q
+\frac{2 Q_{10}}{27 N^3 (N+1)^4}
+L_M^2 \Biggl[\frac{8 \big(3 N^2+3 N+2\big)}{3 N (N+1)^2}
\N\\&&
-\frac{32 S_1}{3 (N+1)}\Biggr]
+L_M \Biggl[
\frac{8 \big(3 N^4+6 N^3+47 N^2+20 N-12\big)}{9 N^2 (N+1)^3}
+\frac{\frac{64 S_2}{3}
-\frac{320 S_1}{9}}{N+1}\Biggr]
\N\\&&
+\frac{
-\frac{896}{27} S_1
+\frac{160 S_2}{9}
-\frac{32 S_3}{3}}{N+1}\Biggr] 
+\textcolor{blue}{C_F} \textcolor{blue}{T_F^2} \Biggl[
\frac{64 L_Q^2}{9 (N+1)}+\Biggl[
-\frac{64 \big(19 N^2+7 N-6\big)}{27 N (N+1)^2}
-\frac{128 S_1}{9 (N+1)}\Biggr] L_Q \Biggr] 
\N\\&&
+ \textcolor{blue}{C_F} \textcolor{blue}{T_F^2} \textcolor{blue}{N_F} \Biggl[
\frac{128 L_Q^2}{9 (N+1)}
+\Biggl[
-\frac{128 \big(19 N^2+7 N-6\big)}{27 N (N+1)^2}
-\frac{256 S_1}{9 (N+1)}\Biggr] L_Q\Biggr]
+\textcolor{blue}{C_F} \textcolor{blue}{C_A} \textcolor{blue}{T_F} \Biggl[L_Q \Biggl[
\N\\&&
-\frac{128 (-1)^N Q_{11}}{15 (N-2) (N-1)^2 N^2 (N+1)^4 (N+2)^2 (N+3)^3}
-\frac{256 \big(N^4+2 N^3-N^2-2 N-6\big) S_{-2}}{3 (N-2) N (N+1)^2 (N+3)}
\N\\&&
+\frac{16 Q_{13}}{135 (N-1)^2 N^2 (N+1)^4 (N+2)^2 (N+3)^3}
+\frac{1}{N+1}\Bigl[\frac{256}{3} S_{-2} S_1
+\frac{1088 S_1}{9}
+\frac{128 S_3}{3}
\N\\&&
+\frac{128}{3} S_{-3}
-\frac{256}{3} S_{-2,1}-128 \zeta_3\Bigr]\Biggr]
-\frac{352 L_Q^2}{9 (N+1)}\Biggr] 
+\hat{C}_{L,q}^{{\sf NS},(3)}(N_F)\Biggr\} \Biggr\}~,
\end{eqnarray}

}
with
{\small
\begin{eqnarray}
Q_{10}&=&219 N^6+657 N^5+1193 N^4+763 N^3-40 N^2-48 N+72
 \\
Q_{11}&=&2 N^{11}+41 N^{10}+226 N^9+556 N^8+963 N^7+2733 N^6+7160 N^5+8610 N^4+1969 N^3
\N\\&&
-2748 N^2-864 N-216
 \\
Q_{12}&=&180 N^{12}+2385 N^{11}+11798 N^{10}+23030 N^9-10466 N^8-131068 N^7-245294 N^6
\N\\&&
-196786 N^5-22282 N^4+86571 N^3+50688 N^2-7236 N-3888
 \\
Q_{13}&=&2345 N^{12}+31510 N^{11}+163614 N^{10}+380250 N^9+208092 N^8-794874 N^7-1604762 N^6
\N\\&&
-833938 N^5+451419 N^4+584028 N^3+113724 N^2-36288 N+7776, 
\end{eqnarray}

}
{\small
\begin{eqnarray}
\lefteqn{H_{q,L}^{\sf PS} = \tfrac{1}{2}[1 + (-1)^N]} \nonumber\\ 
&&
\times \Biggl\{
\textcolor{blue}{a_s^2} \textcolor{blue}{C_F} \textcolor{blue}{T_F} \Biggl\{
-\frac{32 S_1 \big(N^2+N+2\big)}{(N-1) N (N+1)^2 (N+2)}
+\frac{32 L_Q \big(N^2+N+2\big)}{(N-1) N (N+1)^2 (N+2)}
\N\\&&
-\frac{32 \big(N^5+2 N^4+2 N^3-5 N^2-12 N-4\big)}{(N-1) N^2 (N+1)^3 (N+2)^2}\Biggr\} 
+\textcolor{blue}{a_s^3} \Biggl\{
\textcolor{blue}{C_F^2} \textcolor{blue}{T_F} 
\Biggl[\Biggl[
\frac{64 \big(N^2+N+1\big) \big(N^2+N+2\big)}{(N-1) N^2 (N+1)^3 (N+2)}
\N\\&&
-\frac{64 \big(N^2+N+2\big) S_1}{(N-1) N (N+1)^2 (N+2)}\Biggr] L_Q^2
+\Biggl[
\frac{128 (-1)^N \big(N^2+N+2\big) Q_{16}}{15 (N-2) (N-1)^3 N^3 (N+1)^4 (N+2)^2 (N+3)^3}
\N\\&&
-\frac{32 Q_{18}}{15 (N-1)^3 N^3 (N+1)^4 (N+2)^2 (N+3)^3}
+\frac{128 \big(2 N^5+5 N^4+7 N^3+2 N^2-12 N-8\big) S_1}{(N-1) N^2 (N+1)^3 (N+2)^2}
\N\\&&
+\frac{\big(N^2+N+2\big) \big(64 S_1^2-64 S_2\big)}{(N-1) N (N+1)^2 (N+2)}
+\frac{128 \big(N^2+N+2\big) S_{-2}}{(N-2) N (N+1)^2 (N+3)}\Biggr] L_Q
\N\\&&
-\frac{16 \big(N^2+N+2\big)^2 L_M^2}{(N-1) N^2 (N+1)^3 (N+2)}
+\frac{16 Q_{17}}{(N-1) N^4 (N+1)^5 (N+2)^3}
-\frac{32 \big(N^2+N+2\big)^2 S_2}{(N-1) N^2 (N+1)^3 (N+2)}
\N\\&&
-\frac{32 \big(N^2+5 N+2\big) \big(5 N^3+7 N^2+4 N+4\big) L_M}{(N-1) N^3 (N+1)^4 (N+2)^2}
\Biggr] 
 + \textcolor{blue}{C_F}\textcolor{blue}{T_F^2} \textcolor{blue}{N_F}\Biggl[
 \frac{128 L_Q^2 \big(N^2+N+2\big)}{3 (N-1) N (N+1)^2 (N+2)}
 \N\\&&
 -\frac{256 L_Q \big(11 N^5+35 N^4+59 N^3+55 N^2-4 N-12\big)}{9 (N-1) N^2 (N+1)^3 (N+2)^2}\Biggr] 
+ \textcolor{blue}{C_F} \textcolor{blue}{T_F^2} \Biggl[
\frac{128 \big(N^2+N+2\big) L_Q^2}{3 (N-1) N (N+1)^2 (N+2)}
\N\\&&
-\frac{256 \big(11 N^5+35 N^4+59 N^3+55 N^2-4 N-12\big) L_Q}{9 (N-1) N^2 (N+1)^3 (N+2)^2}
+\frac{128 \big(N^2+N+2\big) L_M^2}{3 (N-1)
 N (N+1)^2 (N+2)}
 \N\\&&
 +\frac{128 \big(43 N^4+105 N^3+224 N^2+230 N+86\big)}{27 (N-1) N (N+1)^4 (N+2)}
 -\frac{128 \big(8 N^3+13 N^2+27 N+16\big) S_1}{9 (N-1) N (N+1)^3 (N+2)}
 \N\\&&
 +L_M \Biggl[\frac{256 \big(8 N^3+13 N^2+27 N+16\big)}{9 (N-1) N (N+1)^3 (N+2)}
 -\frac{256 \big(N^2+N+2\big) S_1}{3 (N-1) N (N+1)^2 (N+2)}\Biggr]
 \N\\&&
 +\frac{\big(N^2+N+2\big) \Bigl[\frac{64}{3} S_1^2+\frac{64 S_2}{3}\Bigr]}{(N-1) N (N+1)^2 (N+2)}\Biggr]
 \N\\&&
 +\textcolor{blue}{C_F} \textcolor{blue}{C_A} \textcolor{blue}{T_F}\Biggl[\Biggl[
 -\frac{32 \big(N^2+N+2\big) \big(11 N^4+22 N^3-23 N^2-34 N-12\big)}{3 (N-1)^2 N^2 (N+1)^3 (N+2)^2}
 \N\\&&
 -\frac{64 \big(N^2+N+2\big) S_1}{(N-1) N (N+1)^2 (N+2)}\Biggr] L_Q^2
 +\Biggl[
 \frac{128 (-1)^N Q_{14}}{(N-1) N^2 (N+1)^4 (N+2)^3}
 \N\\&&
 +\frac{64 Q_{15}}{9 (N-1)^2 N^3 (N+1)^3 (N+2)^3}
 +\frac{128 \big(N^4-N^3-4 N^2-11 N-1\big) S_1}{(N-1)^2 N (N+1)^3 (N+2)}
 \N\\&&
 +\frac{\big(N^2+N+2\big) \Bigl[128 S_1^2-128 S_
2-256 S_{-2}\Bigr]}{(N-1) N (N+1)^2 (N+2)}\Biggr] L_Q\Biggr]
+\tilde{C}_{L,q}^{{\sf PS},(3)}\big({N_F}+1)\Biggr\} \Biggr\},
\end{eqnarray}

}
with
{\small
\begin{eqnarray}
Q_{14} &=&N^6+8 N^5+30 N^4+58 N^3+65 N^2+42 N+8
    \\
Q_{15} &=&142 N^8+593 N^7+801 N^6+199 N^5-1067 N^4-900 N^3+976 N^2+1128 N
\N\\ &&
+288
    \\
Q_{16} &=&N^{10}-13 N^9-39 N^8+222 N^7+1132 N^6+1787 N^5+913 N^4+392 N^3+645 N^2
\N\\&&
-324 N-108
 \\
Q_{17} &=&N^{10}+8 N^9+29 N^8+49 N^7-11 N^6-131 N^5-161 N^4-160 N^3-168 N^2
\N\\&&
-80 N-16
 \\
Q_{18} &=&225 N^{12}+2494 N^{11}+9980 N^{10}+14480 N^9-11602 N^8-68380 N^7-86828 N^6
\N\\&&
-15080 N^5+67401 N^4+60334 N^3-312 N^2-33912 N-12528, 
\end{eqnarray}

}
and
{\small
\begin{eqnarray}
\lefteqn{H_{g,L}^{\sf S} = \tfrac{1}{2}[1 + (-1)^N]} \nonumber\\
&&
\times\Biggl\{
\textcolor{blue}{a_s} \textcolor{blue}{T_F} \frac{16 }{(N+1)(N+2)}
+\textcolor{blue}{a_s^2} \Biggl\{
\frac{64 L_M \textcolor{blue}{T_F^2}}{3 (N+1) (N+2)}
+\textcolor{blue}{C_A} \textcolor{blue}{T_F} \Biggl[
\frac{64 (-1)^N \big(N^3+4 N^2+7 N+5\big)}{(N+1)^3 (N+2)^3}
\N\\&&
-\frac{32 \big(2 N^5+9 N^4+5 N^3-12 N^2-20 N-8\big)}{(N-1) N^2 (N+1)^2 (N+2)^3}
+\frac{64 \big(2 N^3-2 N^2-N-1\big) S_1}{(N-1) N (N+1)^2 (N+2)}
\N\\&&
+L_Q \Biggl[
\frac{128 \big(N^2+N+1\big)}{(N-1) N (N+1)^2 (N+2)^2}
-\frac{64 S_1}{(N+1) (N+2)}\Biggr]
+\frac{32 S_1^2-32 S_2-64 S_{-2}}{(N+1) (N+2)}\Biggr]
\N\\&&
+\textcolor{blue}{C_F} \textcolor{blue}{T_F}\Biggl[
-\frac{16 L_M \big(N^2+N+2\big)}{N (N+1)^2 (N+2)}
+\frac{16 L_Q \big(N^2+N+2\big)}{N (N+1)^2 (N+2)}
\N\\&&
+\frac{16 Q_{29}}{15 (N-1)^2 N^2 (N+1)^3 (N+2)^2 (N+3)^3}
+\frac{64 (-1)^N Q_{30}}{15 (N-2) (N-1)^2 N^2 (N+1)^3 (N+2)^2 (N+3)^3}
\N\\&&
-\frac{16 \big(3 N^2+3 N+2\big) S_1}{N (N+1)^2 (N+2)}
+\frac{64 (N-1) S_{-2}}{(N-2) (N+1) (N+3)}\Biggr] \Biggr\} 
\N\\&&
+\textcolor{blue}{a_s^3} \Biggl\{
\frac{256 L_M^2 \textcolor{blue}{T_F^3}}{9 (N+1) (N+2)}
+\textcolor{blue}{C_A} \textcolor{blue}{T_F^2}\Biggl[
\Biggl[\frac{256 \big(N^2+N+1\big)}{3 (N-1) N (N+1)^2 (N+2)^2}
-\frac{128 S_1}{3 (N+1) (N+2)}\Biggr] L_Q^2
\N\\&&
+\Biggl[
\frac{256 (-1)^N \big(N^3+4 N^2+7 N+5\big)}{3 (N+1)^3 (N+2)^3}
+\frac{64 Q_{22}}{9 (N-1) N (N+1)^3 (N+2)^3}
\N\\&&
+\frac{256 \big(11 N^3-6 N^2-8 N-3\big) S_1}{9 (N-1) N (N+1)^2 (N+2)}
+L_M \Biggl[
\frac{512 \big(N^2+N+1\big)}{3 (N-1) N (N+1)^2 (N+2)^2}
-\frac{256 S_1}{3 (N+1) (N+2)}\Biggr]
\N\\&&
+\frac{\frac{128}{3} S_1^2-\frac{128 S_2}{3}-\frac{256}{3} S_{-2}}{(N+1) (N+2)}\Biggr] L_Q
+\frac{32 Q_{27}}{27 (N-1) N^3 (N+1)^4 (N+2)^2}
-\frac{64 (56 N+47) S_1}{27 (N+1)^2 (N+2)}
\N\\&&
+L_M^2 \Biggl[
\frac{256 \big(N^2+N+1\big)}{3 (N-1) N (N+1)^2 (N+2)^2}
-\frac{128 S_1}{3 (N+1) (N+2)}\Biggr]
+L_M \Biggl[\frac{256 (-1)^N \big(N^3+4 N^2+7 N+5\big)}{3 (N+1)^3 (N+2)^3}
\N\\&&
+\frac{128 Q_{24}}{9 (N-1) N^2 (N+1)^3 (N+2)^3}
+\frac{256 \big(N^3-6 N^2+2 N-3\big) S_1}{9 (N-1) N (N+1)^2 (N+2)}
+\frac{\frac{128}{3} S_1^2-\frac{128 S_2}{3}-\frac{256}{3} S_{-2}}{(N+1) (N+2)}\Biggr]\Biggr] 
\N\\&&
+\textcolor{blue}{C_A} \textcolor{blue}{T_F^2} \textcolor{blue}{N_F} \Biggl[
\Biggl[
\frac{256 \big(N^2+N+1\big)}{3 (N-1) N (N+1)^2 (N+2)^2}
-\frac{128 S_1}{3 (N+1) (N+2)}\Biggr] L_Q^2
\N\\&&
+\Biggl[
\frac{256 (-1)^N \big(N^3+4 N^2+7 N+5\big)}{3 (N+1)^3 (N+2)^3}
+\frac{64 Q_{22}}{9 (N-1) N (N+1)^3 (N+2)^3}
\N\\&&
+\frac{256 \big(11 N^3-6 N^2-8 N-3\big) S_1}{9 (N-1) N (N+1)^2 (N+2)}
+\frac{\frac{128}{3} S_1^2-\frac{128 S_2}{3}-\frac{256}{3} S_{-2}}{(N+1) (N+2)}\Biggr] L_Q \Biggr]
%
\N\\&&
+\textcolor{blue}{C_A^2} \textcolor{blue}{T_F}\Biggl[
\Biggl[\frac{128 S_1^2}{(N+1) (N+2)}
+\frac{32 \big(11 N^4+22 N^3-59 N^2-70 N-48\big) S_1}{3 (N-1) N (N+1)^2 (N+2)^2}
\N\\&&
-\frac{64 \big(N^2+N+1\big) \big(11 N^4+22 N^3-35 N^2-46 N-24\big)}{3 (N-1)^2 N^2 (N+1)^3 (N+2)^3}\Biggr] L_Q^2
\N\\&&
+\Biggl[
-\frac{32 \big(59 N^4+70 N^3-155 N^2-118 N-72\big) S_1^2}{3 (N-1) N (N+1)^2 (N+2)^2}
-\frac{256 (-1)^N \big(N^3+4 N^2+7 N+5\big) S_1}{(N+1)^3 (N+2)^3}
\N\\&&
-\frac{64 Q_{28} S_1}{9 (N-1)^2 N^2 (N+1)^3 (N+2)^3}
-\frac{64 (-1)^N Q_{25}}{3 (N-1) N^2 (N+1)^4 (N+2)^4}
\N\\&&
-\frac{32 Q_{31}}{9 (N-1)^2 N^3 (N+1)^3 (N+2)^4}
+\frac{32 \big(11 N^4+22 N^3-83 N^2-94 N-72\big) S_2}{3 (N-1) N (N+1)^2 (N+2)^2}
\N\\&&
+\frac{64 \big(11 N^4+22 N^3-59 N^2-70 N-48\big) S_{-2}}{3 (N-1) N (N+1)^2 (N+2)^2}
+\frac{1 }{(N+1) (N+2)} \Bigl[-128 S_1^3
\N\\&&
+384 S_2 S_1+512 S_{-2} S_1+128 S_3
+128 S_{-3}-256 S_{-2,1}\Bigr]\Biggr] L_Q\Biggr] 
\N\\&&
+\textcolor{blue}{C_F^2} \textcolor{blue}{T_F} \Biggl[
\Biggl[
\frac{8 \big(N^2+N+2\big) \big(3 N^2+3 N+2\big)}{N^2 (N+1)^3 (N+2)}
-\frac{32 \big(N^2+N+2\big) S_1}{N (N+1)^2 (N+2)}\Biggr] L_Q^2
\N\\&&
+\Biggl[\frac{128 (-1)^N \big(N^2+N+2\big) Q_{34}}{5 (N-2) (N-1)^2 N^3 (N+1)^5 (N+2)^3 (N+3)^3}
-\frac{8 Q_{39}}{5 (N-1)^2 N^3 (N+1)^5 (N+2)^3 (N+3)^3}
\N\\&&
-\frac{16 \big(9 N^4+26 N^3+49 N^2+48 N+12\big) S_1}{N^2 (N+1)^3 (N+2)}
+L_M \Biggl[
\frac{64 \big(N^2+N+2\big) S_1}{N (N+1)^2 (N+2)}
\N\\&&
-\frac{16 \big(N^2+N+2\big) \big(3 N^2+3 N+2\big)}{N^2 (N+1)^3 (N+2)}\Biggr]
+\frac{256 \big(N^2+N+2\big) \big(N^4+2 N^3-N^2-2 N-6\big) S_{-2}}{(N-2) N^2 (N+1)^3 (N+2) (N+3)}
\N\\&&
+\frac{\big(N^2+N+2\big) }{N (N+1)^2 (N+2)} \Bigl[64 S_1^2-256 S_{-2} S_1-64 S_2-128 S_3-128 S_{-3}+256 S_{-2,1}+384 \zeta_3\Bigr]\Biggr] L_Q
\N\\&&
+\frac{16 (3 N+2) S_1^2}{N^2 (N+1) (N+2)}
+\frac{8 Q_{26}}{N^4 (N+1)^5 (N+2)}
+\frac{16 \big(N^4-N^3-20 N^2-10 N-4\big) S_1}{N^2 (N+1)^3 (N+2)}
\N\\&&
+L_M^2 \Biggl[\frac{8 \big(N^2+N+2\big) \big(3 N^2+3 N+2\big)}{N^2 (N+1)^3 (N+2)}
-\frac{32 \big(N^2+N+2\big) S_1}{N (N+1)^2 (N+2)}\Biggr]
\N\\&&
+\frac{16 \big(N^4+17 N^3+17 N^2-5 N-2\big) S_2}{N^2 (N+1)^3 (N+2)}
+\frac{\big(N^2+N+2\big) \Bigl[-\frac{16}{3} S_1^3-16 S_2 S_1+\frac{64 S_3}{3}\Bigr]}{N (N+1)^2 (N+2)}
\N\\&&
+L_M \Biggl[
-\frac{128 (-1)^N \big(N^2+N+2\big) Q_{34}}{5 (N-2) (N-1)^2 N^3 (N+1)^5 (N+2)^3 (N+3)^3}
\N\\&&
+\frac{8 Q_{39}}{5 (N-1)^2 N^3 (N+1)^5 (N+2)^3 (N+3)^3}
+\frac{16 \big(9 N^4+26 N^3+49 N^2+48 N+12\big) S_1}{N^2 (N+1)^3 (N+2)}
\N\\&&
-\frac{256 \big(N^2+N+2\big) \big(N^4+2 N^3-N^2-2 N-6\big) S_{-2}}{(N-2) N^2 (N+1)^3 (N+2) (N+3)}
+\frac{\big(N^2+N+2\big) }{N (N+1)^2 (N+2)} \Bigl[-64 S_1^2
\N\\&&
+256 S_{-2} S_1+64 S_2+128 S_3
+128 S_{-3}-256 S_{-2,1}-384 \zeta_3\Bigr]\Biggr]\Biggr]
\N\\&&
+\textcolor{blue}{C_F} \textcolor{blue}{T_F^2} \Biggl[
\frac{64 \big(N^2+N+2\big) \big(N^4+2 N^3+2 N^2+N+6\big) L_Q^2}{3 (N-1) N^2 (N+1)^3 (N+2)^2}
+\Biggl[
-\frac{128 L_M \big(N^2+N+2\big)^2}{(N-1) N^2 (N+1)^3 (N+2)^2}
\N\\&&
+\frac{256 (-1)^N Q_{30}}{45 (N-2) (N-1)^2 N^2 (N+1)^3 (N+2)^2 (N+3)^3}
-\frac{32 Q_{37}}{45 (N-1)^2 N^3 (N+1)^4 (N+2)^3 (N+3)^3}
\N\\&&
-\frac{128 Q_{21} S_1}{3 (N-1) N^2 (N+1)^3 (N+2)^2}
+\frac{256 (N-1) S_{-2}}{3 (N-2) (N+1) (N+3)}\Biggr] L_Q
\N\\&&
-\frac{64 (N-2) (N+3) \big(N^2+N+1\big) \big(N^2+N+2\big) L_M^2}{3 (N-1) N^2 (N+1)^3 (N+2)^2}
-\frac{16 Q_{32}}{(N-1) N^4 (N+1)^5 (N+2)^2}
\N\\&&
+L_M \Biggl[
\frac{256 (-1)^N Q_{30}}{45 (N-2) (N-1)^2 N^2 (N+1)^3 (N+2)^2 (N+3)^3}
+\frac{256 (N-1) S_{-2}}{3 (N-2) (N+1) (N+3)}
\N\\&&
+\frac{32 Q_{38}}{45 (N-1)^2 N^3 (N+1)^4 (N+2)^3 (N+3)^3}
-\frac{128 Q_{19} S_1}{3 (N-1) N^2 (N+1)^3 (N+2)^2}
\Biggr]\Biggr]
\N\\&&
+\textcolor{blue}{C_F} \textcolor{blue}{T_F^2} \textcolor{blue}{N_F} \Biggl[
\frac{64 \big(N^2+N+2\big) \big(N^4+2 N^3+2 N^2+N+6\big) L_Q^2}{3 (N-1) N^2 (N+1)^3 (N+2)^2}
\N\\&&
+\Biggl[
\frac{256 (-1)^N Q_{30}}{45 (N-2) (N-1)^2 N^2 (N+1)^3 (N+2)^2 (N+3)^3}
-\frac{128 Q_{21} S_1}{3 (N-1) N^2 (N+1)^3 (N+2)^2}
\N\\&&
-\frac{32 Q_{37}}{45 (N-1)^2 N^3 (N+1)^4 (N+2)^3 (N+3)^3}
+\frac{256 (N-1) S_{-2}}{3 (N-2) (N+1) (N+3)}
-\frac{64 \big(N^2+N+2\big) L_M}{3 N (N+1)^2 (N+2)}\Biggr] L_Q
\N\\&&
+L_M \Biggl[
\frac{32 \big(N^2+N+2\big) \big(19 N^2+7 N-6\big)}{9 N^2 (N+1)^3 (N+2)}
+\frac{64 \big(N^2+N+2\big) S_1}{3 N (N+1)^2 (N+2)}\Biggr]\Biggr]
\N\\&&
+\textcolor{blue}{C_F} \textcolor{blue}{C_A} \textcolor{blue}{T_F} \Biggl[
\Biggl[
-\frac{16 \big(N^2+N+2\big) \big(11 N^4+22 N^3-23 N^2-34 N-12\big)}{3 (N-1) N^2 (N+1)^3 (N+2)^2}
-\frac{32 \big(N^2+N+2\big) S_1}{N (N+1)^2 (N+2)}\Biggr] L_Q^2
\N\\&&
+\Biggl[
\frac{32 \big(5 N^2+5 N+2\big) S_1^2}{N (N+1)^2 (N+2)}
-\frac{256 (-1)^N Q_{30} S_1}{15 (N-2) (N-1)^2 N^2 (N+1)^3 (N+2)^2 (N+3)^3}
\N\\&&
+\frac{128 Q_{33} S_1}{15 (N-1)^2 N^2 (N+1)^3 (N+2)^2 (N+3)^3}
-\frac{128 \big(N^4+2 N^3+N^2+12\big) S_{-2} S_1}{(N-2) N (N+1)^2 (N+2) (N+3)}
\N\\&&
-\frac{64 (-1)^N Q_{41}}{45 (N-2) (N-1)^3 N^3 (N+1)^5 (N+2)^3 (N+3)^3}
+\frac{8 Q_{42}}{45 (N-1)^3 N^3 (N+1)^5 (N+2)^3 (N+3)^3}
\N\\&&
-\frac{128 Q_{23} S_{-2}}{3 (N-2) N^2 (N+1)^3 (N+2) (N+3)}
+\frac{176 \big(N^2+N+2\big) L_M}{3 N (N+1)^2 (N+2)}
\N\\&&
+\frac{\big(N^2+N+2\big) \Bigl[-32 S_2+64 S_3+64 S_{-3}-128 S_{-2,1}-192 \zeta_3\Bigr]}{N (N+1)^2 (N+2)}\Biggr] L_Q
\N\\&&
-\frac{16 \big(N^3+8 N^2+11 N+2\big) S_1^2}{N (N+1)^3 (N+2)^2}
+\frac{16 Q_{35}}{(N-1) N^4 (N+1)^5 (N+2)^4}
-\frac{16 Q_{20} S_1}{N (N+1)^4 (N+2)^3}
\N\\&&
+L_M^2 \Biggl[
\frac{32 \big(N^2+N+2\big) S_1}{N (N+1)^2 (N+2)}
-\frac{64 \big(N^2+N+1\big) \big(N^2+N+2\big)}{(N-1) N^2 (N+1)^3 (N+2)^2}\Biggr]
\N\\&&
-\frac{16 \big(7 N^5+21 N^4+13 N^3+21 N^2+18 N+16\big) S_2}{(N-1) N^2 (N+1)^3 (N+2)^2}
+\frac{\big(N^2-N-4\big) 
64 (-1)^N S_{-2}
}{(N+1)^3 (N+2)^2}
\N\\&&
+\frac{\big(N^2+N+2\big) }{N (N+1)^2 (N+2)} \Bigl[\frac{16}{3} S_1^3
+48 S_2 S_1+64 (-1)^N S_{-2} S_1
+\frac{128 S_3}{3}+32 (-1)^N S_{-3}-64 S_{-2,1}
\Bigr]
\N\\&&
+L_M \Biggl[
\frac{64 (-1)^N Q_{36}}{5 (N-2) (N-1)^2 N^3 (N+1)^5 (N+2)^3 (N+3)^3}
\N\\&&
-\frac{8 Q_{40}}{45 (N-1)^2 N^3 (N+1)^5 (N+2)^3 (N+3)^3}
-\frac{16 \big(23 N^4+92 N^3+209 N^2+256 N+92\big) S_1}{3 N (N+1)^3 (N+2)^2}
\N\\&&
+\frac{64 \big(N^2+N+2\big) \big(3 N^4+6 N^3-7 N^2-10 N-12\big) S_{-2}}{(N-2) N^2 (N+1)^3 (N+2) (N+3)}
\N\\&&
+\frac{\big(N^2+N+2\big) \Bigl[32 S_1^2-128 S_{-2} S_1+32 S_2
-64 S_3-64 S_{-3}+128 S_{-2,1}+192 \zeta_3\Bigr]}{N (N+1)^2 (N+2)}\Biggr]\Biggr]
\N\\&&
+\tilde{C}_{L,g}^{{\sf S},(3)}(N_F+1)
\Biggr\} \Biggr\}, 
\end{eqnarray}

}
with
{\small
\begin{eqnarray}
Q_{19}&=&N^6+3 N^5-2 N^4-9 N^3-17 N^2-12 N-12
    \\
Q_{20}&=&N^6+8 N^5+23 N^4+54 N^3+94 N^2+72 N+8
    \\
Q_{21}&=&2 N^6+6 N^5+7 N^4+4 N^3+9 N^2+8 N+12
    \\
Q_{22}&=&3 N^6+3 N^5-121 N^4-391 N^3-474 N^2-308 N-80
    \\
Q_{23}&=&10 N^6+30 N^5+N^4-48 N^3-89 N^2-60 N-36
    \\
Q_{24}&=&6 N^7+24 N^6+47 N^5+104 N^4+219 N^3+316 N^2+208 N+48
    \\
Q_{25}&=&11 N^8+66 N^7+106 N^6-121 N^5-775 N^4-1325 N^3-1130 N^2-552 N-96
    \\
Q_{26}&=&12 N^8+52 N^7+132 N^6+216 N^5+191 N^4+54 N^3-25 N^2-20 N-4
    \\
Q_{27}&=&15 N^8+60 N^7+572 N^6+1470 N^5+2135 N^4+1794 N^3+722 N^2-24 N-72
    \\
Q_{28}&=&133 N^8+430 N^7-271 N^6-1361 N^5+110 N^4+2023 N^3+1684 N^2-12 N-144
 \\
Q_{29}&=&26 N^9+539 N^8+3244 N^7+8465 N^6+9342 N^5+841 N^4-5720 N^3-2193 N^2
\N\\&&
+2484 N+1404
 \\
Q_{30}&=&N^{10}-13 N^9-39 N^8+222 N^7+1132 N^6+1787 N^5+913 N^4+392 N^3
\N\\&&
+645 N^2-324 N-108
 \\
Q_{31}&=&3 N^{10}-48 N^9-856 N^8-2702 N^7-1961 N^6+2142 N^5+3122 N^4-1924 N^3
\N\\&&
-5552 N^2-4032 N-1152
 \\
Q_{32}&=&15 N^{10}+75 N^9+112 N^8+14 N^7-61 N^6+107 N^5+170 N^4+36 N^3
\nonumber\\ &&
-36 N^2-32 N-16
 \\
Q_{33}&=&35 N^{10}+372 N^9+1263 N^8+673 N^7-5090 N^6-11596 N^5-8413 N^4+2305 N^3
\N\\&&
+8049 N^2
+3078 N+108
 \\
Q_{34}&=&2 N^{11}+41 N^{10}+226 N^9+556 N^8+963 N^7+2733 N^6+7160 N^5+8610 N^4+1969 N^3
\N\\&&
-2748 N^2-864 N-216
 \\
Q_{35}&=&2 N^{12}+20 N^{11}+86 N^{10}+192 N^9+199 N^8-N^7-297 N^6-495 N^5-514 N^4-488 N^3
\N\\&&
-416 N^2-176 N-32
 \\
Q_{36}&=&12 N^{13}+143 N^{12}+591 N^{11}+954 N^{10}+371 N^9+1658 N^8+11559 N^7+26626 N^6
\N\\&&
+29129 N^5
+14011 N^4-2374 N^3-6576 N^2-1944 N-432
 \\
Q_{37}&=&95 N^{13}+1218 N^{12}+6096 N^{11}+14484 N^{10}+11570 N^9-28440 N^8-117844 N^7
\N\\&&
-225884 N^6
-238953 N^5-83290 N^4+57660 N^3+122040 N^2+182304 N
\nonumber\\ &&
+77760
 \\
Q_{38}&=&185 N^{13}+2582 N^{12}+15584 N^{11}+53036 N^{10}+109190 N^9+124040 N^8+12604 N^7
\N\\&&
-200836 N^6
-294247 N^5-116270 N^4+85260 N^3+158760 N^2+193536 N
\nonumber\\ &&
+77760
 \\
Q_{39}&=&35 N^{14}+465 N^{13}+1962 N^{12}-348 N^{11}-32130 N^{10}-131686 N^9-280396 N^8
\N\\&&
-363984 N^7
-290209 N^6-122547 N^5+6730 N^4+47316 N^3+11928 N^2
\nonumber\\ &&
-21600 N-5184
 \\
Q_{40}&=&1255 N^{14}+18165 N^{13}+107824 N^{12}+331744 N^{11}+515430 N^{10}+132498 N^9
\N\\&&
-1057432 
N^8
-2202648 N^7-1979173 N^6-534079 N^5+350880 N^4-29088 N^3
\N\\&&
-519264 N^2-382320 N
-62208
 \\
Q_{41}&=&11 N^{15}-2 N^{14}+308 N^{13}+5275 N^{12}+24535 N^{11}+52925 N^{10}+50941 N^9-5977 N^8
\N\\&&
-85550 N^7-191059 N^6-294877 N^5-248414 N^4-64728 N^3+57636 N^2+28944 N
\N\\&&
+6480
 \\
Q_{42}&=&1255 N^{15}+16338 N^{14}+76085 N^{13}+117654 N^{12}-198422 N^{11}-971844 N^{10}
\N\\&&
-1002678 N^9+1019372 N^8+3525323 N^7+3236906 N^6+272625 N^5-1523746 N^4
\N\\&&
-632844 N^3+606888 N^2+635904 N+129600~. 
\end{eqnarray}

}
The expressions in $z$-space are presented in Appendix~\ref{app:C}. 

As has been outlined for the 2--loop results in Ref.~\cite{Buza:1995ie} already, the 
scales at which the asymptotic expressions are dominating are estimated to be $Q^2/m^2 \gsim 800$.
They are far outside the kinematic region in which the structure function $F_L(x,Q^2)$ can 
presently be measured in deep-inelastic scattering. The corresponding expressions are therefore
of merely theoretical character and cannot be used in current phenomenological analyses.
\section{Comparison of Mellin Moments for the Wilson Coefficients and OMEs}
\label{sec:6}

\vspace*{1mm}
\noindent
In order to compare the relative impact of the different Wilson coefficients on the structure function $F_2(x,Q^2)$ 
we will consider the Mellin moments for $N = 2$ to 10 in the following, folded with the moments of the respective
parton distribution functions in the flavor singlet case, i.e. the gluon $G(x,Q^2)$ and quark--singlet density 
$\Sigma(x,Q^2)$ for $N_F = 3$ and characteristic values of $Q^2$. Since only a series of Mellin moments has been 
calculated at large momentum transfer $Q^2$ in Ref.~\cite{Bierenbaum:2009mv}, a detailed numerical 
comparison is only possible 
in this way at the moment. The numerical results for the moments of the contributing parton densities are given in 
Table~\ref{tab1}. Note that for $N \geq 2$ the moments for the singlet-distribution are mostly larger than those of 
the gluon. 
We apply these parton densities to study the relative contributions of the different Wilson coefficients,
normalizing to $\HgS$ within the respective order in $a_s$ using the following ratios:
\begin{align*}
        R\left(\LgS,\HgS\right)  &= \frac{c_{N_F} \, \LgS  \, G     }{c_Q \, \HgS \, G} \\
        R\left(\LqPS,\HgS\right) &= \frac{c_{N_F} \, \LqPS \, \Sigma}{c_Q \, \HgS \, G} \\
        R\left(\HqPS,\HgS\right) &= \frac{c_Q     \, \HqPS \, \Sigma}{c_Q \, \HgS \, G},
\end{align*}
where
\begin{eqnarray}
        c_{N_F} = \frac{1}{N_F} \sum\limits_{k=0}^{N_F} e_k^2,
        ~~~~~~~~~c_Q     = e_Q^2.
\end{eqnarray}
In the numerical examples we set $e_Q = e_c = 2/3$. 
\renewcommand{\arraystretch}{1.3}
\begin{center}
\begin{table}[H]\centering
\begin{tabular}{|l||r|r|r|r|r|}
\hline
$Q^2$                             & \multicolumn{5}{c|}{$20\,\text{GeV}^2$} \\
\hline
$N$                               & $2$        & $4$         & $6$          & $8$           & $10$                   
\\
\hline
$G$                               & $0.4583$ & $0.0044$ & $0.0003$ & $3.62\times10^{-5}$ & $7.78\times10^{-6}$ \\
$\Sigma$                          & $0.5417$ & $0.0353$ & $0.0070$ & $2.10\times10^{-3}$ & $8.01\times10^{-4}$ \\
\hline
$Q^2$                             & \multicolumn{5}{c|}{$100\,\text{GeV}^2$} \\
\hline
$N$                               & $2$        & $4$         & $6$          & $8$           & $10$                   
\\
\hline
$G$                               & $0.4819$ & $0.0038$ & $0.0002$ & $3.60\times10^{-5}$ & $8.55\times10^{-6}$ \\
$\Sigma$                          & $0.5181$ & $0.0296$ & $0.0056$ & $1.61\times10^{-3}$ & $5.97\times10^{-4}$ \\
\hline
$Q^2$                             & \multicolumn{5}{c|}{$1000\,\text{GeV}^2$} \\
\hline
$N$                               & $2$        & $4$         & $6$          & $8$           & $10$                   
\\
\hline
$G$                               & $0.5042$ & $0.0032$ & $0.0002$ & $3.14\times10^{-5}$ & $7.60\times10^{-6}$ \\
$\Sigma$                          & $0.4958$ & $0.0244$ & $0.0043$ & $1.20\times10^{-3}$ & $4.32\times10^{-3}$ \\
\hline
\end{tabular}
\caption[]{\sf The moments $N =2, ..., 10$ of the gluon and quark-singlet momentum density using the parton distribution functions
\cite{Alekhin:2013nda}. \label{tab1}}
\end{table}
\renewcommand{\arraystretch}{1.0}
\end{center}
\begin{table}[H]\centering
\begin{tabular}{|ll||r|r|r|r|r|}
\hline \hline
$Q^2$  &                               & \multicolumn{5}{c|}{$20\,\text{GeV}^2$} \\
\hline
$N$    &                               & $2$         & $4$         & $6$         & $8$         & $10$        \\
\hline
$\Sigma/G$    &               & $ 1.1821$ & $ 7.9967$ & $25.847$  & $ 57.965$  & $ 103.06$     \\
\hline
$O(a_s^2):$ & $R(\LgS,\HgS)$  & $ 0.0387$ & $ 0.1349$ & $1.7000$  & $-0.2592$  & $-0.1384$ \\
& $R(\HqPS,\HgS)$             & $-0.2588$ & $-0.3018$ & $1.9946$  & $-1.6153$  & $-1.7222$ \\
\hline
$O(a_s^3):$ & $R(\LgS,\HgS)$  & $ 0.0829$ & $ 0.1983$ & $0.6628$  & $-2.5018$  & $-0.5957$ \\
& $R(\LqPS,\HgS)$             & $ 0.0438$ & $ 0.1042$ & $0.7483$  & $-5.7476$  & $-2.3762$ \\
& $R(\HqPS,\HgS)$             & $-0.2259$ & $-0.3472$ & $0.3387$  & $-9.3371$  & $-4.5870$ \\
\hline
$Q^2$              &                   & \multicolumn{5}{c|}{$100\,\text{GeV}^2$} \\
\hline
$\Sigma/G$         &          & $ 1.0753$  & $7.7514$  & $ 22.797$ & $ 44.660$ & $ 69.888$ \\
\hline
$O(a_s^2)$ & $R(\LgS,\HgS)$   & $ 0.0313$  & $ 0.0687$ & $ 0.1071$ & $ 0.1587$ & $ 0.2418$ \\
& $R(\HqPS,\HgS)$             & $-0.2496$  & $-0.3753$ & $-0.5429$ & $-0.6531$ & $-0.6743$ \\
\hline
$O(a_s^3)$ & $R(\LgS,\HgS)$   & $ 0.0533$  & $ 0.0853$ & $ 0.1449$ & $ 0.2186$ & $ 0.3195$ \\
& $R(\LqPS,\HgS)$             & $ 0.0340$  & $ 0.0378$ & $ 0.1006$ & $ 0.2600$ & $ 0.5828$ \\
& $R(\HqPS,\HgS)$             & $-0.3062$  & $-0.5165$ & $-0.7070$ & $-0.7471$ & $-0.5637$ \\
\hline
$Q^2$               &                  & \multicolumn{5}{c|}{$1000\,\text{GeV}^2$} \\
\hline
$\Sigma/G$     &                       & $ 0.9833$ & $ 7.5948$ & $ 20.958$ & $ 38.236$ & $ 56.876$ \\
\hline
$O(a_s^2)$ &
$R(\LgS,\HgS)$                         & $ 0.0243$ & $ 0.0420$ & $ 0.0531$ & $ 0.0615$ & $ 0.0687$ \\
& $R(\HqPS,\HgS)$                      & $-0.2837$ & $-0.4085$ & $-0.5690$ & $-0.6707$ & $-0.7237$ \\
\hline
$O(a_s^3)$ &$R(\LgS,\HgS)$             & $ 0.0337$ & $ 0.0392$ & $ 0.0597$ & $ 0.0764$ & $ 0.0907$ \\
& $R(\LqPS,\HgS)$                      & $ 0.0313$ & $ 0.0209$ & $ 0.0297$ & $ 0.0505$ & $ 0.0828$ \\
& $R(\HqPS,\HgS)$                      & $-0.3825$ & $-0.5903$ & $-0.8058$ & $-0.9253$ & $-0.9679$ \\
\hline
\end{tabular}
\caption[]{\sf Relative impact of the moments $N = 2,..., 10$ of the individual massive Wilson coefficients, weighted by 
moments 
of the
corresponding parton distributions \cite{Alekhin:2013nda}, at $O(a_s^2)$ and $O(a_s^3)$
normalized to the contribution
to $\HgS$ for $Q^2 = 20, 100$ and $1000~\GeV^2$. \label{tab2}}
\end{table}

Before we discuss quantitative results, a remark on the contributions by the color factor $d_{abc} 
d_{abc}$ to the massless Wilson coefficients  Refs.~\cite{Larin:1993vu,Larin:1996wd,Retey:2000nq} 
and \cite{Blumlein:2004xt,Vermaseren:2005qc} used in the present analysis, is in order.
For $SU(N)$ one obtains
\begin{eqnarray}
d_{abc} d_{abc} = \frac{(N^2-1)(N^2-4)}{N}~.
\end{eqnarray}
It emerges weighted by $1/N_c$ and $1/N_A$ for external quark and gluon lines, respectively, with $N_c = N$ and
$N_A = N^2 -1$. In Refs.~\cite{Larin:1993vu,Larin:1996wd,Retey:2000nq} this group-theoretic 
expression has been used, while in \cite{Blumlein:2004xt,Vermaseren:2005qc} a factor of $16$ has been taken out and was 
absorbed into the 
Lorentz structure of the corresponding contribution to the Wilson coefficient. We agree with the 
$N_F$-dependence as given 
in Refs.~\cite{Larin:1993vu,Larin:1996wd,Retey:2000nq}.
Furthermore, we 
note a typographical error in Eq.~(4.13) of \cite{Vermaseren:2005qc}. Here, the corresponding term 
reads
correctly\footnote{The expression in 
the parameterization
given at {\tt http://www.liv.ac.uk/{$\sim$}avogt/ } is correct, however.}
\begin{eqnarray}
c_{2,g}^{(3)}(x) \simeq  - 932.089 N_F \frac{L_0}{x} ...,~~~~~\text{with}~~L_0 = \ln(x)~.
\end{eqnarray}
Also in the pure-singlet case the massless Wilson coefficients  contain terms 
$\propto d_{abc} d_{abc}$, although with 
a generally different charge-weight factor, cf.~\cite{Larin:1993vu,Larin:1996wd,Retey:2000nq}.
\noindent
{\footnotesize
\begin{table}[H]\centering
\begin{tabular}{|lc||r|r|r|r|r|}
\hline
$Q^2$         &                         & \multicolumn{5}{c|}{$20\,\text{GeV}^2$} \\
\hline
$N$           &                         & $2$          & $4$          & $6$          & $8$          & $10$         \\
\hline
$\Sigma/G$    &                         & $1.1821$    & $7.9967$     & $25.847$     & $57.965$     & $103.06$     \\
\hline
$O(a_s):$ & $R(\Agg  ,\AQg)$   & $-1.0000$ & $-1.8182$ & $-2.5455$  & $-3.2432$   & $-3.9286$   \\
\hline
$O(a_s^2):$ & $R(\Agg  ,\AQg)$ & $-1.0000$  & $-1.6395$  & $-2.3808$  & $-3.1781$  & $-4.0262$    \\
& $R(\AQqPS,\AQg)$ & $-0.1259$ & $-0.3656$  & $-0.7822$  & $-1.3339$  & $-1.9352$    \\
& $R(\AqqNS,\AQg)$ & $-0.0584$ & $-1.1306$  & $-6.3206$  & $-20.735$  & $-49.508$    \\
& $R(\Agq  ,\AQg)$ & $0.1843$ & $ 1.1422$  & $ 3.9956$  & $ 9.8073$  & $ 18.995$    \\ 
\hline
$O(a_s^3):$ & $R(\Agg  ,\AQg)$  & $-1.0051$ & $-1.3397$  & $-1.8466$  & $-2.4306$ & $-3.0890$  \\
 & $R(\AQqPS,\AQg)$             & $-0.1604$ & $-0.4838$  & $-0.9635$  & $-1.5449$ & $-2.1295$  \\
 & $R(\AqqNS,\AQg)$             & $-0.0404$ & $-0.5832$  & $-2.9406$  & $-9.1817$ & $-21.375$  \\
 & $R(\Agq  ,\AQg)$             & $ 0.1473$ & $ 1.1265$  & $ 3.7972$  & $ 8.9925$ & $ 16.961$  \\
 & $R(\Aqg  ,\AQg)$             & $ 0.0051$ & $-0.0202$  & $-0.0326$  & $-0.0445$ & $-0.0567$  \\
 & $R(\AqqPS,\AQg)$             & $ 0.0534$ & $ 0.1093$  & $ 0.2460$  & $ 0.4678$ & $ 0.7619$  \\ 
\hline
\hline
$Q^2$              &                    & \multicolumn{5}{c|}{$100\,\text{GeV}^2$} \\
\hline
$\Sigma/G$         &                    & $1.0753$    & $7.7514$     & $22.797$     & $44.660$     & $69.888$     \\
\hline
$O(a_s):$ & $R(\Agg  ,\AQg)$ & $-1.0000$         & $-1.8182$   & $-2.5455$   & $-3.2432$   & $-3.9286$   \\
\hline
$O(a_s^2):$ & $R(\Agg  ,\AQg)$ & $-1.0000$  & $-1.7746$  & $-2.6448$  & $-3.5805$  & $-4.5771$    \\
& $R(\AQqPS,\AQg)$             & $-0.1884$  & $-0.4246$  & $-0.7627$  & $-1.0887$  & $-1.3514$    \\
& $R(\AqqNS,\AQg)$             & $-0.1247$  & $-2.4385$  & $-12.377$  & $-35.460$  & $-74.510$    \\
& $R(\Agq  ,\AQg)$             & $ 0.3131$  & $ 1.3950$  & $ 3.9075$  & $ 7.7692$  & $ 12.560$    \\ 
\hline
$O(a_s^3):$ & $R(\Agg  ,\AQg)$ & $-1.0048$  & $-1.6120$  & $-2.3201$  & $-3.0734$  & $-3.8667$ \\
& $R(\AQqPS,\AQg)$             & $-0.2799$  & $-0.5808$  & $-0.9473$  & $-1.2540$  & $-1.4615$ \\
& $R(\AqqNS,\AQg)$             & $-0.1772$  & $-2.8698$  & $-13.694$  & $-37.928$  & $-77.874$ \\
& $R(\Agq  ,\AQg)$             & $ 0.3924$  & $ 1.4140$  & $ 3.4078$  & $ 6.0520$  & $ 8.9253$ \\
& $R(\Aqg  ,\AQg)$             & $ 0.0048$  & $-0.0375$  & $-0.0491$  & $-0.0580$  & $-0.0657$ \\
& $R(\AqqPS,\AQg)$             & $ 0.0647$  & $ 0.0984$  & $ 0.1726$  & $ 0.2553$  & $ 0.3319$ \\ 
\hline \hline
\end{tabular}
\caption[]{\sf Relative impact of the moments $N = 2,..., 10$ of the individual massive OMEs, weighted by moments of the 
corresponding parton distributions \cite{Alekhin:2013nda}, at the different orders in $a_s$ 
normalized to the contribution
to $A_{Qg}$ for $Q^2 = 20$ and 100 GeV$^2$. \label{tab3}}
\end{table}

\normalsize
Let us now consider the relative impact of the individual massive Wilson coefficients.
The ratios at $O(a_s^2)$ and $O(a_s^3)$ for different values of $Q^2$ and the moments $N=2$ to 10
are given in Table~\ref{tab2}.
One first notes that at low values of $Q^2$ the moments of $L_{g,2}^{\sf S}$ change sign, 
which is also the case for $H_{q,2}^{\sf PS}$ in the whole region up to $Q^2 = 1000~\GeV^2$.
At $O(a_s^2)$  $L_{g,2}^{\sf S}$ is small for low moments and grows $24\%$ for $N=10$ compared to $\HgS$ 
at $Q^2 = 100~\GeV^2$, with lower values at higher $Q^2$. A comparable tendency is observed at $O(a_s^2)$.
The fraction $|R(H_{q,2}^{\sf PS},\HgS)|$ moves between $25\%$ and $170\%$ comparing the moments $N=2$ to $10$
at $Q^2 = 20~\GeV^2$ and upper values of $\sim 70\%$ at $Q^2 = 1000 \GeV^2$. 

In the case of the comparison of the massive OMEs we normalize to $\AQg$
with PDFs according to their appearance in the singlet and gluon transitions from
$N_F \rightarrow N_F + 1$ massless flavors in

\normalsize
\noindent
the variable flavor number scheme, 
cf.~Eqs.~(\ref{eq:VFNS3}--\ref{eq:VFNS2}):
\begin{align*}
        R(\Agg  ,\AQg) &= \frac{\Agg   \, G     }{\AQg \, G} &
        R(\AQqPS,\AQg) &= \frac{\AQqPS \, \Sigma}{\AQg \, G} \\
        R(\AqqNS,\AQg) &= \frac{\AqqNS \, \Sigma}{\AQg \, G} &
        R(\Agq  ,\AQg) &= \frac{\Agq   \, \Sigma}{\AQg \, G} \\
        R(\Aqg  ,\AQg) &= \frac{\Aqg   \, G     }{\AQg \, G} &
        R(\AqqPS,\AQg) &= \frac{\AqqPS \, \Sigma}{\AQg \, G}~.
\end{align*}
These ratios describe the relative impact, within the corresponding order in $a_s$, of the massive OMEs
in the variable flavor number scheme for the flavor singlet contributions.
\begin{table}[H]\centering
\begin{tabular}{|lc||r|r|r|r|r|}
\hline
$Q^2$         &                         & \multicolumn{5}{c|}{$1000\,\text{GeV}^2$} \\
\hline
$N$           &                         & $2$          & $4$          & $6$          & $8$          & $10$         \\
\hline
$\Sigma/G$    &                         & $0.9833$   & $7.5948$     & $20.958$     & $38.236$     & $56.876$     \\
\hline
$O(a_s):$ & $R(\Agg  ,\AQg)$ & $-1.0000$ & $-1.8182$ & $-2.5455$ & $-3.2432$ & $-3.9286$   \\
\hline
$O(a_s^2)$ &
$R(\Agg  ,\AQg)$             & $-1.0000$ & $-2.0101$ & $-3.0170$ & $-4.0596$ & $-5.1403$   \\
& $R(\AQqPS,\AQg)$           & $-0.2555$ & $-0.4521$ & $-0.6997$ & $-0.8850$ & $-1.0072$   \\
& $R(\AqqNS,\AQg)$           & $-0.2048$ & $-3.7111$ & $-16.978$ & $-44.037$ & $-85.945$   \\
& $R(\Agq  ,\AQg)$           & $ 0.4603$ & $ 1.5427$ & $ 3.5816$ & $ 6.1302$ & $ 8.8857$   \\
\hline
$O(a_s^3):$ &
$R(\Agg  ,\AQg)$             & $-1.0054$ & $-1.7515$ & $-2.4980$ & $-3.2560$ & $-4.0291$ \\
& $R(\AQqPS,\AQg)$           & $-0.3731$ & $-0.6275$ & $-0.9067$ & $-1.0906$ & $-1.1928$ \\
& $R(\AqqNS,\AQg)$           & $-0.2930$ & $-4.1599$ & $-17.532$ & $-43.424$ & $-82.110$ \\
& $R(\Agq  ,\AQg)$           & $ 0.5934$ & $ 1.5060$ & $ 2.9692$ & $ 4.5100$ & $ 5.9397$ \\
& $R(\Aqg  ,\AQg)$           & $ 0.0054$ & $-0.0469$ & $-0.0561$ & $-0.0624$ & $-0.0675$ \\
& $R(\AqqPS,\AQg)$           & $ 0.0727$ & $ 0.0902$ & $ 0.1341$ & $ 0.1721$ & $ 0.2011$ \\ 
\hline \hline
\end{tabular}
\caption[]{\sf The same as Table~\ref{tab3} for $Q^2 = 1000 \GeV^2$. \label{tab4}}
\end{table}

\normalsize
\noindent
The numerical values for different scales of $Q^2$ are given in Tables~\ref{tab3} and \ref{tab4}. 
$|A_{gg,Q}/A_{Qg}|$ rises from about 1 to higher values from $N = 2$ to 10, irrespectively of the values of $Q^2$
and the order in $a_s$. The smallest contributions are $|A_{qg,Q}|$ and $A_{qq,Q}^{\sf PS}$ contributing the ratios $R$
by $\sim 0.5\%$ to 5\% and form $\sim 5$ to $\sim 10\%$, respectively, for $N = 2$ and 4, i.e. in the region dominated
by lower values of the Bjorken variable $x$. The OMEs $|A_{Q,q}^{\sf PS}|$ and $A_{gq,Q}$ have contributions
of 16--62\% and 14--150\%, respectively, for $N = 2$ and $4$. Also the flavor non-singlet Wilson coefficient
$|A_{qq,Q}^{\sf NS}|$ contributes in the flavor singlet transitions and is weighted by the distribution $\Sigma$ here.
Its relative impact rises with $Q^2$ and amounts from $\sim 4\%$ to 370\% for the $R$-ratio considering the lower moments 
$N = 2$ and $N=4$ only.

Right after having obtained a series of moments for the massive OMEs at 3--loops in \cite{Bierenbaum:2009mv},
it became clear that the logarithmic contributions are of comparable order to the constant term.
Moreover, there is a strong functional dependence w.r.t. $N$, as displayed in Tables~\ref{tab1}--\ref{tab4}. To obtain 
a definite answer, the calculation of the constant parts of the unrenormalized OMEs $a_{ij}^{(3)}$ as a function of 
$N \in \mathbb{C}$ is necessary. In particular predictions for the range of small values $x \simeq 
10^{-4}$ appear 
to be rather difficult otherwise.
\section{Conclusions}
\label{sec:7}

\vspace*{1mm}
\noindent
\normalsize
We have derived the contributions of the massive Wilson coefficients to the structure 
functions $F_2(x,Q^2)$ and $F_L(x,Q^2)$ in deep-inelastic scattering and the corresponding massive OMEs 
to 3--loop order in the asymptotic region $Q^2 \gg m^2$ both in Mellin--$N$ and $z$--space 
except for the constant parts 
$a_{ij}$ of the unrenormalized OMEs, which are not known for all quantities yet. Here, we retained
both the scale-dependence due to the virtuality $Q^2$ and the factorization and renormalization scales $\mu^2$,
which were set equal. This allows for scale variation studies in applications.
Two of the Wilson coefficients, $L_{q,2}^{\sf PS}$ and $L_{g,2}^{\sf S}$, are known in complete form, and 
the corresponding results for $L_{q,2}^{\sf NS}$ will be given in \cite{NS2014}. In the variable 
flavor number scheme being applied to describe the process through which an initially massive 
quark transmutes into a massless one at high momentum scales, moreover, the matching coefficients 
$A_{ij}$ are needed. Here, $A_{qq,Q}^{\sf PS}$ and $A_{qg,Q}$ are known in complete form
to 3--loop order and the results for $A_{gq}$ and $A_{qq,Q}^{\sf NS}$ are given in \cite{GQ2014} 
and \cite{NS2014}, respectively.

We have given numerical results for the Wilson coefficients $L_{q,2}^{\sf PS}$ and $L_{g,2}^{\sf S}$.
Using the available Mellin moments we have performed a numerical comparison of 
the different Wilson coefficients
and operator matrix elements inside the respective order in the coupling constant for the moments $N = 2$ to 10
and in  the $Q^2$ range between 20 and 1000~GeV$^2$. While some of the quantities studied are of minor importance,
several others of the Wilson coefficients and OMEs are of similar size, which is 
varying in the kinematic range
of experimental interest for present and future precision measurements. Even in case of the charm-quark 
contributions the logarithmic terms are not dominant over the constant contributions in wide kinematic ranges, as. e.g. 
at HERA.

The expression which were derived in the present paper are available in form of computer-readable files on request
via e-mail to {\tt Johannes.Bluemlein@desy.de}. 
  
\appendix
\section{The massive operator matrix elements in \boldmath $N$-space}
\label{app:A}

\vspace*{1mm}
\noindent
In this appendix we present the massive OMEs in Mellin--space to be used in the matching 
coefficients
in the variable flavor number scheme Eqs.~(\ref{eq:VFNS1}--\ref{eq:VFNS2}). The corresponding 
representations in $z$--space are given in Appendix~\ref{app:D}. Thus far the OMEs 
$A_{qq,Q}^{{\sf PS}}$ and $A_{qg,Q}$ are known completely. The other OMEs are presented except 
for the 
3--loop constant part $a_{ij}$ in the unrenormalized OMEs. The OMEs $A_{qq,Q}^{\sf {NS}}$ and
$A_{gq,Q}^{\sf  {S}}$ are presented elsewhere \cite{NS2014,GQ2014}.

The transition matrix elements are given by $A_{qq,Q}^{{\sf PS}}$ and $A_{qg,Q}$~: 
{\small
\begin{eqnarray}
\lefteqn{A_{qq,Q}^{\sf PS} =  \tfrac{1}{2}[1 + (-1)^N]} \nonumber\\
&&
\times \Biggl\{
\textcolor{blue}{a_s^3} \textcolor{blue}{C_F} \textcolor{blue}{N_F} 
\textcolor{blue}{T_F^2} 
\Biggl\{
 L_M^2 \Biggl[
\frac{32 \big(N^2+N+2\big)^2 S_1}{3 (N-1) N^2  (N+1)^2 (N+2)}
-\frac{32 P_{280}}{9 (N-1) N^3 (N+1)^3 (N+2)^2}
\Biggr]
\nonumber\\&&
+\Biggl[
-\frac{32 P_{282}}{27 (N-1) N^4 (N+1)^4 (N+2)^3}+\frac{64 P_{280} S_1}{9 (N-1)
  N^3 (N+1)^3 (N+2)^2}
\nonumber\\&&
+\frac{\big(N^2+N+2\big)^2 \Bigl[-\frac{32}{3} S_1^2-\frac{32
    S_2}{3}\Bigr]}{(N-1) N^2 (N+1)^2 (N+2)}\Biggr] L_M
-\frac{32 P_{284}}{243 (N-1) N^5 (N+1)^5 (N+2)^4}
\nonumber\\&&
+\frac{32 P_{283} S_1}{81 (N-1) N^4 (N+1)^4 (N+2)^3}
-\frac{32 \big(N^2+N+2\big)^2 }{9 (N-1) N^2 (N+1)^2 (N+2)} L_M^3
\nonumber\\&&
+\frac{P_{281} \Bigl[-\frac{16}{27} S_1^2-\frac{16 S_2}{27}\Bigr]}{(N-1) N^3 (N+1)^3 (N+2)^2}
+\frac{\big(N^2+N+2\big)^2 \Bigl[\frac{80}{27} S_1^3
+\frac{80}{9} S_2 S_1+\frac{160 S_3}{27}+\frac{256 \zeta_3}{9}\Bigr]}
{(N-1) N^2 (N+1)^2 (N+2)}\Biggr\}\Biggr\},
\end{eqnarray}

}
with
{\small
\begin{eqnarray}
P_{280}&=&8 N^7+37 N^6+83 N^5+85 N^4+61 N^3+58 N^2-20 N-24
\\
P_{281}&=&40 N^7+185 N^6+430 N^5+521 N^4+452 N^3+404 N^2-16 N-96
\\
P_{282}&=&95 N^{10}+712 N^9+2379 N^8+4269 N^7+4763 N^6+4569 N^5+3309 N^4+200 N^3
\nonumber \\ &&
-808 N^2-48 N+144
\\
P_{283}&=&233 N^{10}+1744 N^9+5937 N^8+11454 N^7+14606 N^6+15396 N^5+12030
N^4+3272 N^3
\nonumber\\&&
-928 N^2-96 N+288
\\
P_{284}&=&1330 N^{13}+13931 N^{12}+66389 N^{11}+187681 N^{10}+354532 N^9+492456
N^8+532664 N^7
\nonumber\\&&
+423970 N^6+204541 N^5+34274 N^4-11704 N^3-3408 N^2-1008 N-864 
\end{eqnarray}

}
and
{\small
\begin{eqnarray}
\lefteqn{A_{qg,Q} =  \tfrac{1}{2}[1 + (-1)^N]} \nonumber\\ &&
\times \Biggl\{
\textcolor{blue}{a_s^3} \Biggl\{
\textcolor{blue}{C_F} \textcolor{blue}{N_F} \textcolor{blue}{T_F^2} \Biggl[
\Biggl[\frac{8 \big(N^2+N+2\big) P_{285}}{9 (N-1) N^3
    (N+1)^3 (N+2)^2}+\frac{8}{9} \tilde{\gamma}_{qg}^{0} S_1\Biggr] L_M^3
\nonumber \\ &&
+\Biggl[\frac{4 P_{291}}{9 (N-1) N^4 (N+1)^4 (N+2)^3}-\frac{32 \big(5 N^3+8 N^2+19
  N+6\big) S_1}{9 N^2 (N+1) (N+2)}+\tilde{\gamma}_{qg}^{0} \Bigl[-\frac{4}{3}
S_1^2-\frac{4 S_2}{3}\Bigr]\Biggr] L_M^2
\nonumber \\ &&
+\Biggl[
\frac{16 \big(10 N^3+13 N^2+29 N+6\big) S_1^2}{9 N^2 (N+1)
  (N+2)}-\frac{16 \big(103 N^4+257 N^3+594 N^2+524 N+120\big) S_1}{27 N^2
  (N+1)^2 (N+2)}
\nonumber \\ &&
+\frac{4 P_{293}}{27 (N-1) N^5 (N+1)^5 (N+2)^4}+\frac{16 \big(10 N^3+25 N^2+29
  N+6\big) S_2}{9 N^2 (N+1) (N+2)}
\nonumber \\ &&
+\tilde{\gamma}_{qg}^{0} \Bigl[\frac{4}{9} S_1^3+\frac{4}{3} S_2 S_1-\frac{16
  S_3}{9}\Bigr]\Biggr] L_M
+\frac{8 \big(215 N^4+481 N^3+930 N^2+748 N+120\big) S_1^2}{81 N^2 (N+1)^2
  (N+2)}
\nonumber \\ &&
-\frac{64}{9} \frac{
 \big(N^2+N+2\big) P_{285} \zeta_3}{(N-1) N^3 (N+1)^3
  (N+2)^2} 
+\frac{P_{295}}{243 (N-1) N^6 (N+1)^6 (N+2)^5}
\nonumber \\ &&
-\frac{16 \big(1523 N^5+5124 N^4+11200 N^3+14077 N^2+7930 N+1344\big) S_1}{243
  N^2 (N+1)^3 (N+2)}
\nonumber \\ &&
+\frac{8 \big(109 N^4+291 N^3+478 N^2+324 N+40\big) S_2}{27 N^2 (N+1)^2
  (N+2)}+\frac{\big(10 N^3+13 N^2+29 N+6\big) \Bigl[-\frac{16}{81}
  S_1^3-\frac{16}{27} S_2 S_1\Bigr]}{N^2 (N+1) (N+2)}
\nonumber \\ &&
+\frac{32 \big(5 N^3-16 N^2+N-6\big) S_3}{81 N^2 (N+1)
  (N+2)}+\tilde{\gamma}_{qg}^{0} \Bigl[-\frac{1}{27} S_1^4-\frac{2}{9} S_2
S_1^2-\frac{8}{27} S_3 S_1-\frac{64}{9} \zeta_3 S_1-\frac{1}{9} S_2^2+\frac{14
  S_4}{9}\Bigr]\Biggr] 
\nonumber \\ &&
+\textcolor{blue}{C_A} \textcolor{blue}{N_F} \textcolor{blue}{T_F^2}
\Biggl[
 L_M^3 \Biggl[
-\frac{64 \big(N^2+N+1\big) \big(N^2+N+2\big)}{9 (N-1) N^2
  (N+1)^2 (N+2)^2}-\frac{8}{9} \tilde{\gamma}_{qg}^{0} S_1\Biggr]
\nonumber \\ &&
+\Biggl[
\frac{8 P_{289}}{9 (N-1) N^2 (N+1)^3 (N+2)^3}+\frac{32 \big(5 N^4+20 N^3+47
  N^2+58 N+20\big) S_1}{9 N (N+1)^2 (N+2)^2}
\nonumber \\ &&
+\tilde{\gamma}_{qg}^{0} \Bigl[\frac{4}{3} S_1^2+\frac{4 S_2}{3}+\frac{8}{3}
S_{-2}\Bigr]
\Biggr] L_M^2
+\Biggl[-\frac{32 \big(5 N^4+20
  N^3+41 N^2+49 N+20\big) S_1^2}{9 N (N+1)^2 (N+2)^2}
\nonumber \\ &&
+\frac{16 P_{292}}{27 (N-1) N^4 (N+1)^4 (N+2)^4}
-\frac{32 \big(5 N^4+26 N^3+47 N^2+43 N+20\big) S_2}{9 N (N+1)^2 (N+2)^2}
\nonumber\\ &&
+\frac{16 P_{286} S_1}{27 N (N+1)^3 (N+2)^3}
-\frac{64 \big(5 N^2+8 N+10\big) S_{-2}}{9 N (N+1) (N+2)}
+\tilde{\gamma}_{qg}^{0} \Bigl[-\frac{4}{9} S_1^3+\frac{4}{3} S_2
S_1-\frac{8 S_3}{9}-\frac{16}{3} S_{-3}
\nonumber \\ &&
-\frac{16}{3} S_{2,1}\Bigr]\Biggr] L_M
-\frac{16 P_{287} S_1^2}{81 N (N+1)^3 (N+2)^3}+\frac{8 P_{294}}{243 (N-1) N^5
  (N+1)^5 (N+2)^5}
\nonumber \\ &&
+\frac{512}{9} \big(N^2+N+1\big) \big(N^2+N+2\big) \frac{\zeta_3}{(N-1) N^2
  (N+1)^2 (N+2)^2}+\frac{16 P_{290} S_1}{243 (N-1) N^2 (N+1)^4 (N+2)^4}
\nonumber \\ &&
-\frac{16 P_{288} S_2}{81 N (N+1)^3 (N+2)^3}+\frac{64 \big(5 N^4+38 N^3+59 N^2+31
  N+20\big) S_3}{81 N (N+1)^2 (N+2)^2}
\nonumber \\ &&
-\frac{32 \big(121 N^3+293 N^2+414 N+224\big) S_{-2}}{81 N (N+1)^2
  (N+2)}+\frac{128 \big(5 N^2+8 N+10\big) S_{-3}}{27 N (N+1) (N+2)}
\nonumber \\ &&
+\frac{\big(5 N^4+20 N^3+41 N^2+49 N+20\big) \Bigl[\frac{32}{81}
  S_1^3-\frac{32}{27} S_2 S_1+\frac{128}{27} S_{2,1}\Bigr]}{N (N+1)^2 (N+2)^2}
+\tilde{\gamma}_{qg}^{0} \Bigl[\frac{1}{27} S_1^4-\frac{2}{9} S_2 S_1^2
\nonumber \\ &&
+\Bigl[\frac{16}{9} S_{2,1}-\frac{40 S_3}{27}\Bigr] S_1+\frac{64}{9}
\zeta_3 S_1+\frac{1}{9} S_2^2+\frac{14 S_4}{9}+\frac{32}{9}
S_{-4}+\frac{32}{9} S_{3,1}
-\frac{16}{9} S_{2,1,1}\Bigr]\Biggr] \Biggr\}\Biggr\},
\end{eqnarray}

}
with the polynomials
{\small
\begin{eqnarray}
P_{285}&=&3 N^6+9 N^5-N^4-17 N^3-38 N^2-28 N-24
    \\
P_{286}&=&94 N^6+631 N^5+2106 N^4+4243 N^3+4878 N^2+2812 N+680
    \\
P_{287}&=&103 N^6+694 N^5+2148 N^4+3991 N^3+4494 N^2+2704 N+680
    \\
P_{288}&=&139 N^6+1093 N^5+3438 N^4+5776 N^3+5724 N^2+3220 N+752
    \\
P_{289}&=&9 N^8+54 N^7+56 N^6-182 N^5-717 N^4-1120 N^3-1012 N^2-672 N-160
 \\
P_{290}&=&1244 N^{10}+10557 N^9+40547 N^8+90323 N^7+114495 N^6+49344 N^5-69902
N^4
\nonumber \\ &&
-115200 N^3
-64352 N^2-11264 N+864
 \\
P_{291}&=&33 N^{11}+231 N^{10}+698 N^9+1290 N^8+1513 N^7+1463 N^6+2236 N^5+5096
N^4+7328 N^3
\nonumber \\ &&
+5456 N^2+3456 N+1152
\\
P_{292}&=&99 N^{12}+891 N^{11}+2902 N^{10}+3392 N^9-4300 N^8-20914 N^7-33059
N^6-28357 N^5
\nonumber \\ &&
-11406 N^4
+3840 N^3+7568 N^2+4176 N+864
\\
P_{293}&=&159 N^{14}+1590 N^{13}+7223 N^{12}+20982 N^{11}+43703 N^{10}+65162
N^9+62553 N^8+30282 N^7
\nonumber \\ &&
-28286 N^6-145968 N^5-257720 N^4-241760 N^3-158112 N^2-73728 N-17280
\\
P_{294}&=&3315 N^{15}+39780 N^{14}+194011 N^{13}+471164 N^{12}+416251
N^{11}-860568 N^{10}-3525799 N^9
\nonumber \\ &&
-6015120 N^8-6333994 N^7-4373672 N^6-1907512 N^5-499824 N^4-217952 N^3
\nonumber \\ &&
-264192 N^2
-160128 N-34560
\\
P_{295}&=&13923 N^{17}+180999 N^{16}+1064857 N^{15}+3812487 N^{14}+9348807
N^{13}+16391845 N^{12}
\nonumber \\ &&
+20248499 N^{11}+17070917 N^{10}+11536274 N^9+11303496 N^8+13846104
N^7+16104128 N^6
\nonumber \\ &&
+22643488 N^5+29337472 N^4+26395008 N^3+15388416 N^2+5612544 N+995328~. 
\end{eqnarray}

}

Next we present the OMEs, which are known except for the constant term in the unrenormalized 
massive 
OME at 3--loop order, $a_{ij}^{(3)}$. The matrix element $A_{Qq}^{\sf PS}$ is given by~: 
{\small
\begin{eqnarray}
\lefteqn{A_{Qq}^{\sf PS} =   \tfrac{1}{2}[1 + (-1)^N]}  \nonumber\\ &&
\times \Biggl\{ \textcolor{blue}{a_s^2} \textcolor{blue}{C_F} \textcolor{blue}{T_F} 
\Biggl\{-\frac{4 L_M^2 \big(N^2+N+2\big)^2}{(N-1) N^2 (N+1)^2 (N+2)}
-\frac{8 S_2 \big(N^2+N+2\big)^2}{(N-1) N^2 (N+1)^2 (N+2)}
\nonumber\\ &&
+\frac{4
  P_{312}}{(N-1) N^4 (N+1)^4 (N+2)^3}
-\frac{8 \big(N^2+5 N+2\big) \big(5 N^3+7 N^2+4 N+4\big) L_M}{(N-1) N^3 (N+1)^3 (N+2)^2}\Biggr\}
\nonumber \\&&
+
 \textcolor{blue}{a_s^3}
\Biggl\{
\textcolor{blue}{T_F} \textcolor{blue}{C_F^2} \Biggl[
\Biggl[
\frac{4 \big(N^2+N+2\big)^2 \big(3 N^2+3 N+2\big)}{3
    (N-1) N^3 (N+1)^3 (N+2)}-\frac{16 \big(N^2+N+2\big)^2 S_1}{3 (N-1) N^2
    (N+1)^2 (N+2)}\Biggr] L_M^3
\nonumber \\&&
+\Biggl[-\frac{8 \big(5 N^2+N-2\big) S_1 \big(N^2+N+2\big)^2}{(N-1) N^3 (N+1)^3
  (N+2)}+\frac{16 S_2 \big(N^2+N+2\big)^2}{(N-1) N^2 (N+1)^2 (N+2)}
\nonumber \\&&
+\frac{4 P_{297} \big(N^2+N+2\big)}{(N-1) N^4 (N+1)^4 (N+2)}\Biggr] L_M^2
+\Biggl[\frac{\Bigl[\frac{8}{3} S_1^3-24 S_2
  S_1-\frac{80 S_3}{3}+32 S_{2,1}+96 \zeta_3\Bigr] \big(N^2+N+2\big)^2}{(N-1)
  N^2 (N+1)^2 (N+2)}
\nonumber \\&&
-\frac{4 \big(5 N^3+4 N^2+9 N+6\big) S_1^2 \big(N^2+N+2\big)}{(N-1) N^2
  (N+1)^3 (N+2)}-\frac{4 P_{321}}{(N-1) N^5 (N+1)^5 (N+2)^3}
\nonumber \\&&
+\frac{8 P_{313} S_1}{(N-1) N^4 (N+1)^4 (N+2)^3}-\frac{4 P_{307} S_2}{(N-1) N^3
  (N+1)^3 (N+2)^2}\Biggr] L_M
- \frac{2 \big(N^2+N+2\big)\zeta_2 P_{304}}{(N-1) N^4 (N+1)^4 (N+2)} 
\nonumber \\&&
-\frac{4 \big(N^2+N+2\big) \big(N^4-5 N^3-32 N^2-18 N-4\big) S_1^2}{(N-1) N^3
  (N+1)^3 (N+2)}
+\frac{4 P_{323}}{(N-1) N^6 (N+1)^6 (N+2)^3}
\nonumber \\&&
-\frac{4}{3} 
\frac{\big(N^2+N+2\big)^2
\big(3 N^2+3 N+2\big) \zeta_3}{(N-1) N^3 (N+1)^3 (N+2)}
+ \frac{4 \big(N^2+N+2\big) \big(5 N^4+4 N^3+N^2-10 N-8\big)\zeta_2}{(N-1) N^3 (N+1)^3 (N+2)} S_1
\nonumber \\&&
+\frac{8 \big(N^2+N+2\big) \big(2 N^5-2 N^4-11 N^3-19 N^2-44 N-12\big)
  S_1}{(N-1) N^3 (N+1)^4 (N+2)}
-\frac{4 \big(N^2+N+2\big) P_{303} S_2}{(N-1) N^4 (N+1)^4 (N+2)}
\nonumber \\&&
+\frac{(3 N+2) \big(N^2+N+2\big) \Bigl[-\frac{8}{3} S_1^3-8 S_2 S_1\Bigr]}{(N-1) N^3 (N+1) (N+2)}
-\frac{8 \big(N^2+N+2\big) \big(3 N^4+48 N^3+43 N^2-22 N-8\big) S_3}{3 (N-1) N^3 (N+1)^3 (N+2)}
\nonumber \\&&
+\frac{32 \big(N^2-3 N-2\big) \big(N^2+N+2\big) S_{2,1}}{(N-1) N^3 (N+1)^2 (N+2)}
+\frac{\big(N^2+N+2\big)^2} {(N-1) N^2(N+1)^2 (N+2)}
\Bigl[\frac{2}{3} S_1^4+4 S_2 S_1^2
\nonumber \\&&
+\big(\frac{16 S_3}{3}+32 S_{2,1}\big) S_1+\frac{16}{3} \zeta_3 S_1+2 S_2^2-12 S_4+32 S_{3,1}
-64 S_{2,1,1}+\big(4 S_1^2-12 S_2\big) \zeta_2\Bigr]\Biggr] 
\nonumber \\&& 
+\textcolor{blue}{C_F} \textcolor{blue}{T_F^2} \Biggl[
-\frac{128 \big(N^2+N+2\big)^2 L_M^3}{9
  (N-1) N^2 (N+1)^2 (N+2)}
+\Biggl[\frac{32 \big(N^2+N+2\big)^2 S_1}{3 (N-1) N^2
  (N+1)^2 (N+2)}
\nonumber\\ &&
-\frac{32 P_{298}}{9 (N-1) N^3 (N+1)^2 (N+2)^2}\Biggr] L_M^2
+\Biggl[\frac{\big(-\frac{32}{3} S_1^2-32 S_2\big)
  \big(N^2+N+2\big)^2}{(N-1) N^2 (N+1)^2 (N+2)}
\nonumber \\&&
-\frac{64 P_{315}}{27 (N-1) N^4 (N+1)^4 (N+2)^3}+\frac{64 P_{309} S_1}{9 (N-1)
  N^3 (N+1)^3 (N+2)^2}\Biggr] L_M
\nonumber \\&&
+\frac{16 \big(N^2+N+2\big) \big(8 N^3+13 N^2+27 N+16\big) S_1^2}{9 (N-1) N^2
  (N+1)^3 (N+2)}+\frac{32}{9} \frac{P_{298}\zeta_2}{(N-1) N^3 (N+1)^2 (N+2)^2}
\nonumber \\&&
+\frac{32 P_{322}}{81 (N-1) N^5 (N+1)^5 (N+2)^4}-\frac{32 \big(N^2+N+2\big)
  \big(43 N^4+105 N^3+224 N^2+230 N+86\big) S_1}{27 (N-1) N^2 (N+1)^4 (N+2)}
\nonumber \\&&
+\frac{16 P_{310} S_2}{9 (N-1) N^3 (N+1)^3 (N+2)^2}+\frac{\big(N^2+N+2\big)^2
  \Bigl[-\frac{16}{9} S_1^3-\frac{16}{3} S_2 S_1-\frac{32}{3} \zeta_2
  S_1+\frac{160 S_3}{9}+\frac{128 \zeta_3}{9}\Bigr]}{(N-1) N^2 (N+1)^2
  (N+2)}
\Biggr]
\nonumber \\&&
+\textcolor{blue}{C_F} \textcolor{blue}{T_F^2} \textcolor{blue}{N_F} \Biggl[
-\frac{16}{9} \frac{P_{326} \zeta_2}{(N-1) N^3 (N+1)^3 (N+2)^2}
+L_M^2 \Biggl[\frac{32 P_{327}}{9 (N-1) N^3 (N+1)^3 (N+2)^2}
\N\\&&
-\frac{32 \big(N^2+N+2\big)^2 S_1}{3 (N-1) N^2 (N+1)^2 (N+2)}\Biggr] 
+ L_M \Biggl[
-\frac{32 P_{328}}{27 (N-1) N^4 (N+1)^4 (N+2)^3}
\N\\&&
+\frac{32 P_{326} S_1}{9 (N-1) N^3 (N+1)^3 (N+2)^2}
+\frac{\big(N^2+N+2\big)^2 \Bigl[
-\frac{16}{3} S_1^2
-\frac{80 S_2}{3}\Bigr]}{(N-1) N^2 (N+1)^2 (N+2)}\Biggr]
\N\\&&
-\frac{32 P_{329}}{3 (N-1) N^5 (N+1)^5 (N+2)^4}
+\frac{\big(N^2+N+2\big)^2 }{(N-1) N^2 (N+1)^2 (N+2)}
\Bigl[
\frac{16}{3} \zeta_2 S_1
+\frac{64 S_3}{3}
+\frac{32 \zeta_3}{9}
\Bigr]
\N\\&&
-\frac{32 \big(N^2+N+2\big)^2 L_M^3}{9 (N-1) N^2 (N+1)^2 (N+2)}
+\frac{64 \big(N^2+5 N+2\big) \big(5 N^3+7 N^2+4 N+4\big) S_2}{3 (N-1) N^3 (N+1)^3 (N+2)^2}
\Biggr]
\nonumber \\&&
+\textcolor{blue}{C_F} \textcolor{blue}{C_A} \textcolor{blue}{T_F}
\Biggl[
\frac{8 \big(N^2+N+2\big) \big(N^3+8 N^2+11 N+2\big) S_1^3}{3 (N-1)
  N^2 (N+1)^3 (N+2)^2}+\frac{4 \big(N^2+N+2\big) P_{296} S_1^2}{(N-1) N^2 (N+1)^4
  (N+2)^3}
\nonumber \\&&
-\frac{4}{3} \frac{\big(N^2+N+2\big) P_{301} \zeta_2}{(N-1)^2 N^3 (N+1)^3 (N+2)^2}
S_1
-\frac{8 \big(N^2+N+2\big) P_{311} S_1}{(N-1) N^2 (N+1)^5 (N+2)^4}
\nonumber \\&&
-\frac{8 \big(N^2+N+2\big) \big(3 N^3-12 N^2-27 N-2\big) S_2 S_1}{(N-1) N^2 (N+1)^3 (N+2)^2}
-\frac{8}{9} \frac{\big(N^2+N+2\big) P_{299} \zeta_3}{(N-1)^2 N^3 (N+1)^3 (N+2)^2}
\nonumber \\&&
+\frac{4}{9} \frac{P_{320} \zeta_2}{(N-1)^2 N^4 (N+1)^4 (N+2)^3}
+\frac{8 P_{325}}{3 (N-1)^2 N^6 (N+1)^6 (N+2)^5}
\nonumber \\&&
+L_M^3 \Biggl[
\frac{8 \big(11 N^4+22 N^3-23 N^2-34
  N-12\big) \big(N^2+N+2\big)^2}{9 (N-1)^2 N^3 (N+1)^3 (N+2)^2}
+\frac{16 S_1 \big(N^2+N+2\big)^2}{3 (N-1) N^2 (N+1)^2 (N+2)}\Biggr]
\nonumber \\&&
+\frac{4 P_{317} S_2}{3 (N-1)^2 N^4 (N+1)^4 (N+2)^3}
-\frac{16 \big(N^2+N+2\big) P_{300} S_3}{3 (N-1)^2 N^3 (N+1)^3 (N+2)^2}
\nonumber \\&&
+L_M^2 \Biggl[
\frac{\Bigl[16 S_2+32 S_{-2}\Bigr] \big(N^2+N+2\big)^2}{(N-1) N^2 (N+1)^2 (N+2)}
+\frac{8 P_{302} S_1
  \big(N^2+N+2\big)}{3 (N-1)^2 N^3 (N+1)^3 (N+2)^2}
\nonumber \\&&
-\frac{8 P_{319}}{9 (N-1)^2 N^4 (N+1)^4 (N+2)^3}\Biggr]
\nonumber \\&&
+\frac{\big(N^2+N+2\big) \big(N^4+2 N^3+7 N^2+22 N+20\big) \Bigl[32 (-1)^N
  S_{-2}+16 (-1)^N \zeta_2\Bigr]}{(N-1) N (N+1)^4 (N+2)^3}
\nonumber \\&&
+\frac{\big(N^2-N-4\big) \big(N^2+N+2\big)}
{(N-1) N (N+1)^3 (N+2)^2}
 \Bigl[-64 (-1)^N S_1 S_{-2}-32
  (-1)^N S_{-3}+64 S_{-2,1}
-32 (-1)^N S_1 \zeta_2
\nonumber \\&& 
-24 (-1)^N
  \zeta_3\Bigr]
+\frac{\big(N^2+N+2\big)^2 }{(N-1) N^2 (N+1)^2 (N+2)}
\Bigl[-\frac{2}{3} S_1^4-20 S_2 S_1^2-32 (-1)^N
  S_{-3} S_1+\big(64 S_{-2,1}
\nonumber\\ && 
-\frac{160 S_3}{3}\big) S_1-\frac{8}{3} \big(-7+9
  (-1)^N\big) \zeta_3 S_1-2 S_2^2
+S_{-2} \big(-32 (-1)^N S_1^2
\nonumber\\ &&
-32 (-1)^N S_2\big)-36 S_4-16 (-1)^N
  S_{-4}+16 S_{3,1}+32 S_{-2,2}+32 S_{-3,1}+16 S_{2,1,1}
\nonumber\\ &&
-64 S_{-2,1,1}+\big(-4
  \big(-3+4 (-1)^N\big) S_1^2-4 \big(-1+4 (-1)^N\big) S_2-8 \big(1+2
  (-1)^N\big) S_{-2}\big) \zeta_2\Bigr]
\nonumber \\&&
+L_M \Biggl[\frac{
  \big(N^2+N+2\big)^2}{(N-1) N^2 (N+1)^2 (N+2)}
\Bigl[-\frac{8}{3} S_1^3+40 S_2
  S_1+32 \big(1+(-1)^N\big) S_{-2} S_1
\nonumber\\ &&
 +16 (-1)^N S_{-3}-32 S_{2,1}+12 \big(-9+(-1)^N\big) \zeta_3\Bigr]
+\frac{8 P_{324}}{27 (N-1)^2 N^5 (N+1)^5 (N+2)^4}
\nonumber \\&&
+\frac{4 \big(17 N^4-6 N^3+41 N^2-16 N-12\big) S_1^2 \big(N^2+N+2\big)}{3
  (N-1)^2 N^3 (N+1)^2 (N+2)}+\frac{4 P_{305} S_2 \big(N^2+N+2\big)}{3 (N-1)^2
  N^3 (N+1)^3 (N+2)^2}
\nonumber \\&&
+\frac{8 \big(31 N^2+31 N+74\big) S_3 \big(N^2+N+2\big)}{3 (N-1) N^2 (N+1)^2
  (N+2)}+\frac{16 \big(7 N^2+7 N+10\big) S_{-3} \big(N^2+N+2\big)}{(N-1) N^2
  (N+1)^2 (N+2)}
\nonumber \\&&
-\frac{128 \big(N^2+N+1\big) S_{-2,1} \big(N^2+N+2\big)}{(N-1) N^2 (N+1)^2
  (N+2)}
+\frac{\big(N^2-N-4\big)
32 (-1)^N S_{-2}
\big(N^2+N+2\big)}{(N-1) N (N+1)^3 (N+2)^2}
\nonumber \\&&
-\frac{8 P_{316} S_1}{9 (N-1)^2 N^4 (N+1)^4 (N+2)^2}
+\frac{16 P_{306} S_{-2}}{(N-1) N^3 (N+1)^3 (N+2)^2}\Biggr]\Biggr]
+a_{Qq}^{{\sf PS},(3)}\Biggr\}\Biggr\},
\end{eqnarray}

}
with the polynomials
{\small
\begin{eqnarray}
P_{296}&=&N^6+6 N^5+7 N^4+4 N^3+18 N^2+16 N-8
\\
P_{297}&=&7 N^6+15 N^5+7 N^4-23 N^3-26 N^2-20 N-8
\\
P_{298}&=&8 N^6+29 N^5+84 N^4+193 N^3+162 N^2+124 N+24
\\
P_{299}&=&11 N^6+6 N^5+75 N^4+68 N^3-200 N^2-80 N-24
\\
P_{300}&=&11 N^6+29 N^5-7 N^4-25 N^3-56 N^2-72 N-24
\\
P_{301}&=&17 N^6+27 N^5+75 N^4+149 N^3-20 N^2-80 N-24
\\
P_{302}&=&17 N^6+51 N^5+51 N^4+89 N^3+40 N^2-80 N-24
\\
P_{303}&=&27 N^6+102 N^5+135 N^4+56 N^3-8 N^2-20 N-8
\\
P_{304}&=&38 N^6+108 N^5+151 N^4+106 N^3+21 N^2-28 N-12
\\
P_{305}&=&73 N^6+189 N^5+45 N^4+31 N^3-238 N^2-412 N-120
\\
P_{306}&=&2 N^7+14 N^6+37 N^5+102 N^4+155 N^3+158 N^2+132 N+40
\\
P_{307}&=&3 N^7-15 N^6-133 N^5-449 N^4-658 N^3-500 N^2-296 N-96
\\
P_{308}&=&8 N^7+37 N^6+68 N^5-11 N^4-86 N^3-56 N^2-104 N-48
\\
P_{309}&=&8 N^7+37 N^6+83 N^5+85 N^4+61 N^3+58 N^2-20 N-24
\\
P_{310}&=&8 N^7+37 N^6+158 N^5+565 N^4+796 N^3+628 N^2+400 N+96
\\
P_{311}&=&2 N^8+22 N^7+117 N^6+386 N^5+759 N^4+810 N^3+396 N^2+72 N+32
\\
P_{312}&=&N^{10}+8 N^9+29 N^8+49 N^7-11 N^6-131 N^5-161 N^4-160 N^3-168 N^2-80 N-16
\\
P_{313}&=&19 N^{10}+143 N^9+427 N^8+567 N^7+454 N^6+822 N^5+1560 N^4+1784
N^3+1488 N^2
\nonumber \\&&
+768 N
+192
\\
P_{314}&=&43 N^{10}+320 N^9+939 N^8+912 N^7-218 N^6-510 N^5-654 N^4-1232 N^3+16 N^2
\N\\ &&
+672 N+288
\\
P_{315}&=&43 N^{10}+320 N^9+1059 N^8+1914 N^7+2431 N^6+2874 N^5+2379 N^4+820 N^3+352 N^2\nonumber \\&&
+336 N+144
\\
P_{316}&=&136 N^{10}+647 N^9+1110 N^8-438 N^7-2555 N^6-2106 N^5-3105 N^4-3167 N^3+418 N^2\nonumber \\&&
+924 N+72
\\
P_{317}&=&3 N^{11}+66 N^{10}+104 N^9-1152 N^8-3801 N^7-2510 N^6+3318 N^5+8076 N^4+9608 N^3\nonumber \\&&
+6512 N^2+2432 N+384
\\
P_{318}&=&5 N^{11}+62 N^{10}+252 N^9+374 N^8+38 N^7-400 N^6-473 N^5-682 N^4-904 N^3-592 N^2\nonumber \\&&
-208 N-32
\\
P_{319}&=&118 N^{11}+793 N^{10}+2281 N^9+3402 N^8+2428 N^7+1457 N^6+1917 N^5+2476 N^4
\nonumber \\&&
+4392 N^3
+4976 N^2+2832 N+576
\\
P_{320}&=&127 N^{11}+820 N^{10}+2251 N^9+2196 N^8-1109 N^7-934 N^6+4491 N^5+9334 N^4
\nonumber \\&&
+12552 N^3
+9680 N^2+4656 N+864
\\
P_{321}&=&37 N^{12}+305 N^{11}+1107 N^{10}+2328 N^9+3520 N^8+5020 N^7+7642 N^6+10519 N^5
\nonumber \\&&
+10938 N^4
+8248 N^3+4656 N^2+1712 N+288
\\
P_{322}&=&248 N^{13}+2599 N^{12}+12793 N^{11}+39593 N^{10}+87182 N^9+148026 N^8+196942 N^7
\nonumber \\&&
+192416 N^6
+128195 N^5+63406 N^4+32344 N^3+15984 N^2+5616 N+864
\\
P_{323}&=&4 N^{14}+56 N^{13}+443 N^{12}+2139 N^{11}+6049 N^{10}+10762 N^9+13272 N^8+11692 N^7
\nonumber \\&&
+6106 N^6
+339 N^5-1254 N^4-72 N^3+496 N^2+240 N+32
\\
P_{324}&=&686 N^{14}+6560 N^{13}+25572 N^{12}+43489 N^{11}+9045 N^{10}-72944 N^9
-125240 N^8
\nonumber \\&&
-156761 N^7
-206883 N^6-241600 N^5-250212 N^4-225808 N^3-150864 N^2
\nonumber\\ &&
-56448 N-8640
\\
P_{325}&=&12 N^{17}+162 N^{16}+1030 N^{15}+4188 N^{14}+11527 N^{13}
+19051 N^{12}+11176 N^{11}-17182 N^{10}
\nonumber \\&&
-36527 N^9-27469 N^8-11770 N^7+5554 N^6+32640 N^5+46456 N^4+34528 N^3+14816 N^2\nonumber \\&&
+3584 N+384
   \\
P_{326}&=&8 N^7+37 N^6+68 N^5-11 N^4-86 N^3-56 N^2-104 N-48
    \\
P_{327}&=&8 N^7+37 N^6+83 N^5+85 N^4+61 N^3+58 N^2-20 N-24
 \\
P_{328}&=&43 N^{10}+320 N^9+939 N^8+912 N^7-218 N^6-510 N^5-654 N^4-1232 N^3
\N\\&&
+16 N^2+672 N+288
 \\
P_{329}&=&5 N^{11}+62 N^{10}+252 N^9+374 N^8+38 N^7-400 N^6-473 N^5-682 N^4
\N\\&&
-904 N^3-592 N^2-208 N-32~. 
\end{eqnarray}

}
The OME $A_{Qg}$ except for the term $a_{Qg}^{(3)}$ reads~:
{\small


}
where
{\small
\begin{eqnarray}
P_{330}&=&N^6-93 N^5-444 N^4-317 N^3+329 N^2+296 N+84
     \\
P_{331}&=&N^6-9 N^5-120 N^4-137 N^3+29 N^2+56 N+36
     \\
P_{332}&=&N^6+6 N^5+7 N^4+4 N^3+18 N^2+16 N-8
     \\
P_{333}&=&N^6+8 N^5+23 N^4+54 N^3+94 N^2+72 N+8
     \\
P_{334}&=&2 N^6+11 N^5+8 N^4-7 N^3+14 N^2+12 N-24
     \\
P_{335}&=&3 N^6+9 N^5-N^4-17 N^3-38 N^2-28 N-24
     \\
P_{336}&=&3 N^6+30 N^5+15 N^4-64 N^3-56 N^2-20 N-8
     \\
P_{337}&=&5 N^6+15 N^5+36 N^4+51 N^3+25 N^2+8 N+4
     \\
P_{338}&=&5 N^6+18 N^5+51 N^4+84 N^3+60 N^2+34 N+12
     \\
P_{339}&=&5 N^6+26 N^5+97 N^4+160 N^3+135 N^2+79 N+22
     \\
P_{340}&=&5 N^6+42 N^5+84 N^4+35 N^3+40 N^2+34 N+48
     \\
P_{341}&=&6 N^6+18 N^5+7 N^4-16 N^3-31 N^2-20 N-12
     \\
P_{342}&=&6 N^6+47 N^5+136 N^4+223 N^3+256 N^2+172 N+32
     \\
P_{343}&=&9 N^6+27 N^5-65 N^4-319 N^3-404 N^2-200 N-40
     \\
P_{344}&=&10 N^6-6 N^5-39 N^4-44 N^3-97 N^2+20 N+12
     \\
P_{345}&=&10 N^6+63 N^5+105 N^4+31 N^3+17 N^2+14 N+48
     \\
P_{346}&=&11 N^6-15 N^5-327 N^4-181 N^3+292 N^2-20 N-48
     \\
P_{347}&=&11 N^6+15 N^5-285 N^4-319 N^3-254 N^2-368 N-240
     \\
P_{348}&=&11 N^6+33 N^5-189 N^4-361 N^3-194 N^2-92 N-72
     \\
P_{349}&=&11 N^6+33 N^5-114 N^4-247 N^3-263 N^2-176 N-108
     \\
P_{350}&=&11 N^6+33 N^5-87 N^4-85 N^3+4 N^2-116 N-48
     \\
P_{351}&=&11 N^6+57 N^5-39 N^4-109 N^3-44 N^2-116 N-48
     \\
P_{352}&=&11 N^6+81 N^5+9 N^4-133 N^3-92 N^2-116 N-48
     \\
P_{353}&=&13 N^6+36 N^5+39 N^4+8 N^3-21 N^2-29 N-10
 \\
P_{354}&=&17 N^6+51 N^5+390 N^4+359 N^3-389 N^2-200 N-84
     \\
P_{355}&=&18 N^6+87 N^5+57 N^4-119 N^3-131 N^2-60 N-20
     \\
P_{356}&=&23 N^6+39 N^5+75 N^4+157 N^3+96 N^2+70 N+28
     \\
P_{357}&=&29 N^6+176 N^5+777 N^4+1820 N^3+1878 N^2+776 N+232
     \\
P_{358}&=&45 N^6+135 N^5-91 N^4-407 N^3-214 N^2+12 N-248
     \\
P_{359}&=&51 N^6+140 N^5+227 N^4+208 N^3-202 N^2-96 N-8
     \\
P_{360}&=&55 N^6+141 N^5-195 N^4-401 N^3-772 N^2-748 N-384
     \\
P_{361}&=&55 N^6+165 N^5-420 N^4-899 N^3-1561 N^2-1336 N-1188
     \\
P_{362}&=&57 N^6+297 N^5+519 N^4+399 N^3+92 N^2-68 N-16
     \\
P_{363}&=&67 N^6+93 N^5+351 N^4+259 N^3-1054 N^2-556 N-312
     \\
P_{364}&=&76 N^6+487 N^5+1692 N^4+3271 N^3+3186 N^2+1516 N+536
     \\
P_{365}&=&77 N^6+195 N^5+627 N^4+977 N^3-128 N^2-452 N+432
     \\
P_{366}&=&77 N^6+339 N^5-105 N^4-487 N^3-356 N^2-668 N-240
     \\
P_{367}&=&83 N^6+249 N^5-111 N^4-637 N^3-956 N^2-596 N-624
     \\
P_{368}&=&97 N^6+591 N^5+1311 N^4+229 N^3-712 N^2+308 N+192
     \\
P_{369}&=&N^8+24 N^7+62 N^6+8 N^5-123 N^4-128 N^3-108 N^2-72 N-48
     \\
P_{370}&=&N^8+427 N^7+1133 N^6+697 N^5-434 N^4-636 N^3-244 N^2-64 N-16
     \\
P_{371}&=&2 N^8+22 N^7+117 N^6+386 N^5+759 N^4+810 N^3+396 N^2+72 N+32
     \\
P_{372}&=&2 N^8+29 N^7+179 N^6+441 N^5+529 N^4+332 N^3+172 N^2+92 N+24
     \\
P_{373}&=&3 N^8+54 N^7+118 N^6-44 N^5-353 N^4-314 N^3-272 N^2-200 N-144
     \\
P_{374}&=&9 N^8+54 N^7+56 N^6-182 N^5-717 N^4-1120 N^3-1012 N^2-672 N-160
     \\
P_{375}&=&12 N^8+52 N^7+132 N^6+216 N^5+191 N^4+54 N^3-25 N^2-20 N-4
     \\
P_{376}&=&15 N^8+36 N^7+50 N^6-252 N^5-357 N^4+152 N^3-68 N^2+88 N+48
     \\
P_{377}&=&18 N^8+101 N^7+128 N^6+208 N^5+190 N^4-769 N^3-1200 N^2-212 N-48
     \\
P_{378}&=&33 N^8+132 N^7+350 N^6+636 N^5+685 N^4+528 N^3+292 N^2+128 N+32
     \\
P_{379}&=&121 N^8+370 N^7+924 N^6+358 N^5-381 N^4+184 N^3-1096 N^2-48 N+144
     \\
P_{380}&=&321 N^8+1674 N^7+2360 N^6-1378 N^5-6565 N^4-5992 N^3-1972 N^2+128 N-96
  \\
P_{381}&=&507 N^8+2190 N^7+3002 N^6+1692 N^5-681 N^4-2554 N^3-404 N^2+664 N+192
  \\
P_{382}&=&633 N^8+2532 N^7+5036 N^6+6174 N^5+4307 N^4+1182 N^3-176 N^2-184 N-48
     \\
P_{383}&=&N^9+6 N^8+15 N^7+25 N^6+36 N^5+85 N^4+128 N^3+104 N^2+64 N+16
     \\
P_{384}&=&N^9+21 N^8+85 N^7+105 N^6+42 N^5+290 N^4+600 N^3+456 N^2+256 N+64
     \\
P_{385}&=&4 N^9+53 N^8+193 N^7+233 N^6+87 N^5+554 N^4+1172 N^3+904 N^2+512 N+128
  \\
P_{386}&=&6 N^9+93 N^8+576 N^7+1296 N^6+586 N^5+359 N^4+2000 N^3+1996 N^2
\nonumber\\&&
+1488 N+384
  \\
P_{387}&=&25 N^9-43 N^8-424 N^7+462 N^6+4345 N^5+7513 N^4+6446 N^3+4020 N^2
\nonumber\\&&
+1944 N+480
  \\
P_{388}&=&36 N^9+156 N^8-115 N^7-1116 N^6-1251 N^5-78 N^4+300 N^3+84 N^2-128 N-48
  \\
P_{389}&=&40 N^9+273 N^8+635 N^7+613 N^6+119 N^5-2 N^4-314 N^3-668 N^2+24 N+144
  \\
P_{390}&=&45 N^9+270 N^8+724 N^7+1262 N^6+1731 N^5+2740 N^4+3484 N^3+2928 N^2
\nonumber\\ &&
+1696 N+384
  \\
P_{391}&=&66 N^9+534 N^8+1409 N^7+1080 N^6-933 N^5-1116 N^4+588 N^3+996 N^2
\nonumber\\&&
+736 N+240
  \\
P_{392}&=&69 N^9+366 N^8+1100 N^7+1894 N^6+2451 N^5+5276 N^4+7460 N^3+5352 N^2
\nonumber\\ &&
+3008 N+672
  \\
P_{393}&=&80 N^9+441 N^8+568 N^7-592 N^6-1202 N^5+2003 N^4+4106 N^3+3116 N^2
\nonumber\\ &&
+2712 N+864
  \\
P_{394}&=&94 N^9+597 N^8+1508 N^7+2086 N^6+1517 N^5+1381 N^4+2731 N^3+3802 N^2
\nonumber\\ &&
+2916 N+648
  \\
P_{395}&=&251 N^9+1586 N^8+4206 N^7+6764 N^6+4008 N^5-2242 N^4+13 N^3+7122 N^2
\nonumber\\ &&
+6156 N+1944
  \\
P_{396}&=&489 N^9+2934 N^8+7636 N^7+12206 N^6+6675 N^5-12692 N^4-24608
N^3-16272 N^2
 \nonumber \\ &&
-2864 N+2304
  \\
P_{397}&=&891 N^9+5751 N^8+15070 N^7+21430 N^6+37623 N^5+55339 N^4+44064
N^3+25144 N^2
\nonumber \\ &&
+9488 N+1776
 \\
P_{398}&=&10 N^{10}+62 N^9+407 N^8+1119 N^7+1405 N^6+889 N^5+240 N^4-90 N^3-114 N^2
\nonumber\\ &&
-48 N-8
 \\
P_{399}&=&36 N^{10}+456 N^9+2448 N^8+7171 N^7+12399 N^6+13213 N^5+8997 N^4+5000
N^3+2888 N^2
\nonumber \\ &&
+992 N+112
\\
P_{400}&=&37 N^{10}+392 N^9+2106 N^8+6514 N^7+9211 N^6+1258 N^5-9218 N^4-6116
N^3-72 N^2
\nonumber \\ &&
-752 N-192
\\
P_{401}&=&85 N^{10}+425 N^9+830 N^8+788 N^7-521 N^6-325 N^5+2238 N^4+2568
N^3+968 N^2
\nonumber \\ &&
-1296 N-576
\\
P_{402}&=&103 N^{10}+575 N^9+1124 N^8-334 N^7-1505 N^6+3755 N^5+4926 N^4+36
N^3-472 N^2
\nonumber \\ &&
-2160 N-864
\\
P_{403}&=&149 N^{10}+793 N^9+2368 N^8+5026 N^7+6853 N^6+6277 N^5+5062 N^4+3168
N^3+1296 N^2
\nonumber \\ &&
+400 N+96
\\
P_{404}&=&170 N^{10}+883 N^9+1897 N^8+2710 N^7-448 N^6-4745 N^5+561 N^4+5904
N^3+1132 N^2
\nonumber \\ &&
-2016 N-864
\\
P_{405}&=&170 N^{10}+1213 N^9+3091 N^8+2506 N^7-2692 N^6-3047 N^5-861 N^4-2352
N^3-5324 N^2
\nonumber \\ &&
-6240 N-2016
\\
P_{406}&=&436 N^{10}+3960 N^9+15787 N^8+36343 N^7+46431 N^6+17745 N^5-28270
N^4-33648 N^3
\nonumber \\ &&
-11056 N^2-1936 N+864
\\
P_{407}&=&3 N^{11}+42 N^{10}+144 N^9+74 N^8-459 N^7-1060 N^6-1152 N^5-1424
N^4-1688 N^3
\nonumber \\ &&
-1232 N^2
-736 N-192
\\
P_{408}&=&33 N^{11}+231 N^{10}+698 N^9+1290 N^8+1513 N^7+1463 N^6+2236 N^5+5096
N^4+7328 N^3
\nonumber \\ &&
+5456 N^2+3456 N+1152
\\
P_{409}&=&95 N^{11}+853 N^{10}+3599 N^9+9245 N^8+12320 N^7-282 N^6-23342
N^5-26920 N^4
\nonumber \\ &&
-10832 N^3
-1712 N^2-416 N-192
\\
P_{410}&=&129 N^{11}+903 N^{10}+2894 N^9+5730 N^8+6505 N^7+383 N^6-9464
N^5-13912 N^4-11680 N^3
\nonumber \\ &&
-6640 N^2-3648 N-1152
\\
P_{411}&=&243 N^{11}+1701 N^{10}+5378 N^9+10350 N^8+11479 N^7+1193 N^6-14684
N^5-20572 N^4
\nonumber \\ &&
-16288 N^3-8944 N^2-4992 N-1728
\\
P_{412}&=&333 N^{11}+2331 N^{10}+6556 N^9+9270 N^8+5081 N^7-6701 N^6-17554
N^5-20036 N^4
\nonumber \\ &&
-15680 N^3
-9200 N^2-5664 N-1728
\\
P_{413}&=&2 N^{12}+20 N^{11}+86 N^{10}+192 N^9+199 N^8-N^7-297 N^6-495 N^5-514
N^4-488 N^3
\nonumber \\ &&
-416 N^2
-176 N-32
\\
P_{414}&=&23 N^{12}+138 N^{11}-311 N^{10}-3148 N^9-7605 N^8-8462 N^7-4163
N^6+246 N^5+1540 N^4
\nonumber \\ &&
+1066 N^3+444 N^2+120 N+16
\\
P_{415}&=&111 N^{12}+1035 N^{11}+3634 N^{10}+5168 N^9-2662 N^8-21724 N^7-37157
N^6-34963 N^5
\nonumber \\ &&
-19122 N^4-4560 N^3+80 N^2+1008 N+288
\\
P_{416}&=&201 N^{12}+1845 N^{11}+6742 N^{10}+11990 N^9+7139 N^8-8917 N^7-15710
N^6-2110 N^5
\nonumber \\ &&
+16644 N^4+22080 N^3+12416 N^2+4128 N+576
\\
P_{417}&=&7299 N^{12}+53973 N^{11}+206656 N^{10}+532170 N^9+820775 N^8+650149
N^7+204230 N^6
\nonumber \\ &&
+189820 N^5+606016 N^4+664624 N^3+372192 N^2+143424 N+27648
\\
P_{418}&=&9 N^{13}+72 N^{12}+101 N^{11}-511 N^{10}-2325 N^9-4428 N^8-4619
N^7-3841 N^6-4462 N^5
\nonumber \\ &&
-6012 N^4-6992 N^3-5296 N^2-2592 N-576
\\
P_{419}&=&69 N^{13}+420 N^{12}+794 N^{11}-1501 N^{10}-11265 N^9-18414 N^8-4436
N^7-5017 N^6
\nonumber \\ &&
-41818 N^5
-65616 N^4-62960 N^3-39184 N^2-17184 N-3456
\\
P_{420}&=&296 N^{13}+2368 N^{12}+9908 N^{11}+22254 N^{10}+13564 N^9-31716
N^8-71723 N^7-71221 N^6
\nonumber \\ &&
-44369 N^5-33249 N^4-26584 N^3+4968 N^2+7344 N+432
\\
P_{421}&=&385 N^{14}+2567 N^{13}+6877 N^{12}+9235 N^{11}+5375 N^{10}-1207
N^9-3313 N^8+905 N^7
\nonumber \\ &&
+3876 N^6
+1676 N^5+256 N^4+1544 N^3+1616 N^2+736 N+192
\\
P_{422}&=&531 N^{14}+5454 N^{13}+25877 N^{12}+77604 N^{11}+159437 N^{10}+205070
N^9+82971 N^8
\nonumber \\ &&
-207408 N^7-490544 N^6-694320 N^5-735104 N^4-562304 N^3-355584 N^2-158976 N
\nonumber \\ &&
-34560\\
P_{423}&=&1773 N^{14}+18018 N^{13}+80795 N^{12}+214620 N^{11}+371423
N^{10}+398930 N^9+154773 N^8
\nonumber \\ &&
-228072 N^7-435356 N^6-492936 N^5-534656 N^4-453440 N^3-299712 N^2
\nonumber\\ &&
-144000 N-34560
\\
P_{424}&=&4 N^{15}+50 N^{14}+267 N^{13}+765 N^{12}+1183 N^{11}+682 N^{10}-826
N^9-1858 N^8-1116 N^7
\nonumber \\ &&
+457 N^6
+1500 N^5+2268 N^4+2400 N^3+1392 N^2+448 N+64
\\
P_{425}&=&26 N^{15}+314 N^{14}+1503 N^{13}+3222 N^{12}+2510 N^{11}+1996
N^{10}+15041 N^9+40728 N^8
\nonumber \\ &&
+54008 N^7+44956 N^6+31936 N^5+30416 N^4+29568 N^3+16704 N^2+5376 N+768
\\
P_{426}&=&28 N^{15}+335 N^{14}+1953 N^{13}+6497 N^{12}+11508 N^{11}+6624
N^{10}-11753 N^9-27541 N^8
\nonumber \\ &&
-33352 N^7-40915 N^6-40468 N^5-16628 N^4+7584 N^3+10416 N^2+4032 N+576
\\
P_{427}&=&435 N^{15}+5436 N^{14}+32317 N^{13}+119006 N^{12}+307057
N^{11}+620328 N^{10}+1065977 N^9
\nonumber \\ &&
+1575060 N^8+1889534 N^7+1704634 N^6+1113248 N^5+592440 N^4+328672 N^3
\nonumber \\ &&
+165984 N^2+59904 N+10368
\\
P_{428}&=&939 N^{16}+10527 N^{15}+47251 N^{14}+101719 N^{13}+66350
N^{12}-155710 N^{11}
\nonumber \\ &&
-322813 N^{10}-16829 N^9+702425 N^8+1332497 N^7+1596952 N^6+1640548 N^5
\nonumber \\ &&
+1506496 N^4+1099952 N^3+604032 N^2+211392 N+34560
\\
P_{429}&=&1623 N^{16}+20963 N^{15}+119399 N^{14}+394315 N^{13}+831483
N^{12}+1160715 N^{11}+1086519 N^{10}
\nonumber \\ &&
+712841 N^9+425270 N^8+337718 N^7+207634 N^6-73752 N^5-261272 N^4
\nonumber \\ &&
-200160 N^3-64672 N^2-4480 N+1408
\\
P_{430}&=&87 N^{17}+1099 N^{16}+6055 N^{15}+19019 N^{14}+37119 N^{13}+45159
N^{12}+29583 N^{11}-2639 N^{10}
\nonumber \\ &&
-30218 N^9-40778 N^8-39994 N^7-35844 N^6-30808
N^5-30384 N^4-28256 N^3
\nonumber \\ &&
-16064 N^2 -5248 N-768~. 
\end{eqnarray}

}

The operator matrix element $A_{gg,Q}$ except for the term $a_{gg,Q}^{(3)}$ is given by~:
{\small
\begin{eqnarray}
\lefteqn{A_{gg,Q} = \tfrac{1}{2}[1 + (-1)^N] \times}\nonumber\\
&&   \Biggl\{\textcolor{blue}{a_s}  \frac{4}{3} \textcolor{blue}{T_F} L_M
+ \textcolor{blue}{a_s^2}  \Biggl\{
\frac{16}{9} \textcolor{blue}{T_F^2} L_M^2
            + \textcolor{blue}{C_A T_F} 
\Biggl[
\Biggl[
\frac{16 \big(N^2+N+1\big)}{3 (N-1) N (N+1) (N+2)}-\frac{8
  S_1}{3}
\Biggr] 
L_M^2
\nonumber\\&&
+\Biggl[\frac{16 P_{433}}{9 (N-1) N^2 (N+1)^2 (N+2)}-\frac{80 S_1}{9}\Biggr] 
L_M
+\frac{2 P_{451}}{27 (N-1) N^3 (N+1)^3 (N+2)}
-\frac{4 (56 N+47) S_1}{27 (N+1)}\Biggr]
\nonumber\\&&
+ \textcolor{blue}{C_F T_F} \Bigg[\frac{4
  \big(N^2+N+2\big)^2 }{(N-1) N^2 (N+1)^2
  (N+2)} L_M^2
+\frac{4 P_{445}}{(N-1) N^3 (N+1)^3 (N+2)} L_M
\nonumber\\&&
-\frac{P_{461}}{(N-1) N^4 (N+1)^4 (N+2)}
\Biggr]
\Biggr\}
\nonumber\\
&&+ \textcolor{blue}{a_s^3} \Biggl\{
\Biggl[
\textcolor{blue}{T_F^3} \Biggl[
\frac{64}{27} L_M^3
-\frac{64 \zeta_3}{27}
\Biggr]
+ \textcolor{blue}{C_A T_F^2} 
\Biggl[
\Biggl[
\frac{448 \big(N^2+N+1\big)}{27 (N-1) N (N+1) (N+2)}
-\frac{224 S_1}{27}
\Biggr]
L_M^3
\nonumber\\&&
+\Biggl[
\frac{8 P_{441}}{27 (N-1) N^2 (N+1)^2 (N+2)}-\frac{640 S_1}{27}
\Biggr] 
L_M^2
+\Biggl[
-\frac{2 P_{454}}{27 (N-1) N^3 (N+1)^3 (N+2)}
\nonumber\\&&
-\frac{8 P_{440} S_1}{9 (N-1) N^2 (N+1)^2 (N+2)}\Biggr] L_M
+\frac{8 S_1^2}{3 (N+1)}
-\frac{4}{27} \frac{P_{442} \zeta_2}{(N-1) N^2 (N+1)^2 (N+2)} 
\nonumber\\&&
-\frac{8 P_{460}}{81 (N-1) N^4 (N+1)^4 (N+2)}
+\frac{16 \big(328 N^4+256 N^3-247 N^2-175 N+54\big) S_1}{81 (N-1) N (N+1)^2}
\nonumber\\&&
-\frac{448}{27} \frac{\big(N^2+N+1\big) \zeta_3}{(N-1) N (N+1) (N+2)}
-\frac{8 (2 N+1) S_2}{3 (N+1)}
+\frac{560}{27} S_1 \zeta_2
+\frac{224}{27} S_1 \zeta_3
\nonumber\\&&
+ \textcolor{blue}{N_F} 
\Biggl[
\Biggl[
\frac{128 \big(N^2+N+1\big)}{27 (N-1) N (N+1)
  (N+2)}-\frac{64 S_1}{27}\Biggr] L_M^3
+\Biggl[
-\frac{4 P_{457}}{81 (N-1) N^3 (N+1)^3 (N+2)}
\nonumber\\&&
-\frac{16 P_{443} S_1}{81 (N-1) N^2 (N+1)^2 (N+2)}\Biggr] L_M
+\frac{16 S_1^2}{9 (N+1)}
-\frac{4}{27} \frac{\zeta_2 P_{437}}{(N-1) N^2 (N+1)^2 (N+2)}
\nonumber\\&&
+\frac{32 P_{465}}{243 (N-1) N^4 (N+1)^4 (N+2)}
+\frac{32 \big(328 N^4+256 N^3-247 N^2-175 N+54\big) S_1}{243 (N-1) N (N+1)^2}
\nonumber\\&&
-\frac{128}{27} \frac{\big(N^2+N+1\big)\zeta_3}{(N-1) N (N+1) (N+2)}
-\frac{16 (2 N+1) S_2}{9 (N+1)}
+\frac{160}{27} S_1 \zeta_2
+\frac{64}{27} S_1 \zeta_3
\Biggr]
\Biggr] 
\nonumber\\&&
+ \textcolor{blue}{C_A^2 T_F} 
\Biggl[
\Biggl[
\frac{176 S_1}{27}
-\frac{352 \big(N^2+N+1\big)}{27 (N-1) N (N+1) (N+2)}
\Biggr] L_M^3
+\Biggl[
-\frac{2 P_{469}}{9 (N-1)^2 N^3 (N+1)^3 (N+2)^3}
\nonumber\\&&
-\frac{8 P_{452} S_1}{9 (N-1)^2 N^2 (N+1)^2 (N+2)^2}
+\frac{64}{3} S_{-2} S_1
+\frac{64}{3} S_1 S_2
+\frac{\big(N^2+N+1\big) \Bigl[-\frac{128}{3} S_2-\frac{128}{3} S_{-2}\Bigr]}{(N-1) N (N+1) (N+2)}
\nonumber\\&&
+\frac{32 S_3}{3}
+\frac{32}{3} S_{-3}-\frac{64}{3} S_{-2,1}
\Biggr] L_M^2
+\Biggl[
\frac{32}{3} S_{-2}^2
-\frac{16 P_{456} S_{-2}}{9 (N-1)^2 N^3 (N+1)^3 (N+2)^2}
\nonumber\\&&
+\frac{32 P_{439} S_1 S_{-2}}{9 (N-1) N^2 (N+1)^2 (N+2)}
+64 \frac{\big(2 N^4+4 N^3+7 N^2+5 N+6\big) \zeta_3}{(N-1) N^2 (N+1)^2 (N+2)}
\nonumber\\&&
+\frac{P_{479}}{81 (N-1)^2 N^4 (N+1)^4 (N+2)^3}
-\frac{4 P_{474} S_1}{81 (N-1)^2 N^4 (N+1)^4 (N+2)^2}
+ \Biggl[\frac{640 S_2}{9}-\frac{32 S_3}{3}\Biggr] S_1
\nonumber\\&&
+\frac{16 P_{432} S_2}{9 (N-1) N^2 (N+1)^2 (N+2)}
+\frac{8 \big(40 N^4+80 N^3+73 N^2+33 N+54\big) S_3}{9 N^2 (N+1)^2}
-64 S_1 \zeta_3
\nonumber\\&&
+\frac{P_{438} \big(\frac{16}{9} S_{-3}-\frac{32}{9} S_{-2,1}\big)}{(N-1) N^2 (N+1)^2 (N+2)}
\Biggr] L_M
-\frac{44 S_1^2}{9 (N+1)}
-\frac{8 P_{468}}{243 (N-1) N^4 (N+1)^4 (N+2)}
\nonumber\\&&
+\frac{4}{27} \frac{P_{470} \zeta_2}{(N-1)^2 N^3 (N+1)^3 (N+2)^3} 
-\frac{8 \big(2834 N^4+2042 N^3-1943 N^2-1151 N+594\big) S_1}{243 (N-1) N (N+1)^2}
\nonumber\\&&
+\frac{16}{27} \frac{P_{444} S_1 \zeta_2}{(N-1)^2 N^2 (N+1)^2 (N+2)^2}
+\frac{44 (2 N+1) S_2}{9 (N+1)}
+\Bigl[-\frac{32}{3} S_1 S_2-\frac{32}{3} S_{-2} S_1-\frac{16 S_3}{3}
\nonumber\\&&
-\frac{16}{3} S_{-3}+\frac{32}{3} S_{-2,1}\Bigr]\zeta_2
-\frac{176}{27} S_1 \zeta_3
+\frac{\big(N^2+N+1\big) \big(\big(\frac{64 S_2}{3}+\frac{64}{3} S_{-2}\big)
  \zeta_2+\frac{352 \zeta_3}{27}\big)}{(N-1) N (N+1) (N+2)}\Biggr] 
\nonumber\\
&&
+ \textcolor{blue}{C_F^2 T_F} 
\Biggl[
\Biggl[
\frac{16 \big(N^2+N+2\big)^2 S_1}{3 (N-1) N^2 (N+1)^2(N+2)}
-\frac{4 \big(N^2+N+2\big)^2 \big(3 N^2+3 N+2\big)}{3 (N-1) N^3
    (N+1)^3 (N+2)}
\Biggr] L_M^3
\nonumber\\&&
+\Biggl[\frac{8 \big(5 N^2+N-2\big) S_1 \big(N^2+N+2\big)^2}{(N-1) N^3 (N+1)^3
  (N+2)}-\frac{16 S_2 \big(N^2+N+2\big)^2}{(N-1) N^2 (N+1)^2 (N+2)}
\nonumber\\&&
-\frac{4 P_{436} \big(N^2+N+2\big)}{(N-1) N^4 (N+1)^4 (N+2)}
\Biggr] L_M^2
+\Biggl[
\frac{\big(-\frac{8}{3} S_1^3+24 S_2 S_1-32
  S_{2,1}\big) \big(N^2+N+2\big)^2}{(N-1) N^2 (N+1)^2 (N+2)}
\nonumber\\&&
+\frac{4 \big(5 N^3+4 N^2+9 N+6\big) S_1^2 \big(N^2+N+2\big)}{(N-1) N^2 (N+1)^3 (N+2)}
+\frac{16 \big(5 N^2+5 N-14\big) S_3 \big(N^2+N+2\big)}{3 (N-1) N^2 (N+1)^2 (N+2)}
\nonumber\\&&
-\frac{4 \big(17 N^4+48 N^3+69 N^2+10 N-8\big) S_2
  \big(N^2+N+2\big)}{(N-1) N^3 (N+1)^3 (N+2)}
-\frac{96 \zeta_3 \big(N^2+N+2\big)}{N^2 (N+1)^2}
\nonumber\\&&
+\frac{\big(-256 S_1 S_{-2}-128 S_{-3}+256 S_{-2,1}\big) \big(N^2+N+2\big)}{(N-1) N^2 (N+1)^2 (N+2)}
-\frac{2 P_{471}}{(N-1) N^5 (N+1)^5 (N+2)}
\nonumber\\&&
-\frac{8 P_{448} S_1}{(N-1) N^4 (N+1)^4 (N+2)}
-\frac{32 \big(N^6+3 N^5+N^4-3 N^3-26 N^2-24 N-16\big) S_{-2}}{(N-1) N^3
  (N+1)^3 (N+2)}\Biggr] L_M
\nonumber\\&&
+\frac{4 \big(N^2+N+2\big) \big(N^4-5 N^3-32 N^2-18 N-4\big) S_1^2}{(N-1) N^3 (N+1)^3 (N+2)}
-\frac{4}{3} \frac{P_{453} \zeta_3}{(N-1) N^3 (N+1)^3 (N+2)}
\nonumber\\&&
-2 \frac{P_{464} \zeta_2}{(N-1) N^4 (N+1)^4 (N+2)}
-\frac{8 \big(N^2+N+2\big) \big(2 N^5-2 N^4-11 N^3-19 N^2-44 N-12\big)
  S_1}{(N-1) N^3 (N+1)^4 (N+2)}
\nonumber\\&&
-\frac{P_{480}}{(N-1) N^6 (N+1)^6 (N+2)}
-4 \frac{\big(N^2+N+2\big) \big(5 N^4+4 N^3+N^2-10 N-8\big) \zeta_2}{(N-1) N^3
  (N+1)^3 (N+2)} S_1
\nonumber\\&&
+\frac{4 \big(N^2+N+2\big) P_{434} S_2}{(N-1) N^4 (N+1)^4 (N+2)}+\frac{(3 N+2)
  \big(N^2+N+2\big) \big(\frac{8}{3} S_1^3+8 S_2 S_1\big)}{(N-1) N^3 (N+1)
  (N+2)}
+128 \log (2) \zeta_2
\nonumber\\&&
+\frac{8 \big(N^2+N+2\big) \big(3 N^4+48 N^3+43 N^2-22 N-8\big) S_3}{3 (N-1)
  N^3 (N+1)^3 (N+2)}-\frac{32 \big(N^2-3 N-2\big) \big(N^2+N+2\big)
  S_{2,1}}{(N-1) N^3 (N+1)^2 (N+2)}
\nonumber\\&&
+\frac{\big(N^2+N+2\big)^2} 
{(N-1) N^2 (N+1)^2 (N+2)}
\Bigg[-\frac{2}{3} S_1^4-4 S_2
  S_1^2+\Bigl[-\frac{16}{3} S_3-32 S_{2,1}\Bigr] S_1-\frac{16}{3} \zeta_3 S_1-2
  S_2^2+12 S_4
\nonumber\\ && 
-32 S_{3,1}
+64 S_{2,1,1}+\Bigl[12 S_2-4 S_1^2\Bigr]
  \zeta_2\Bigr] \Biggr]
\nonumber\\&&
+ \textcolor{blue}{C_F T_F^2} 
\Biggl[
\frac{80 \big(N^2+N+2\big)^2}{9 (N-1) N^2 (N+1)^2 (N+2)}
L_M^3
+\Biggl[\frac{32 \big(N^2+N+2\big)^2 S_1}{3 (N-1) N^2 (N+1)^2 (N+2)}
\nonumber\\&&
+\frac{8 P_{449}}{9 (N-1) N^3 (N+1)^3 (N+2)}
\Biggr] L_M^2
+\Biggl[\frac{\big(\frac{16}{3} S_1^2-16 S_2\big)
  \big(N^2+N+2\big)^2}{(N-1) N^2 (N+1)^2 (N+2)}
\nonumber\\&&
-\frac{8 P_{467}}{27 (N-1) N^4 (N+1)^4 (N+2)}+\frac{32 P_{435} S_1}{9 (N-1) N^3
  (N+1)^3 (N+2)}\Biggr] L_M
\nonumber\\&&
-\frac{8}{9} \frac{P_{450} \zeta_2}{(N-1) N^3 (N+1)^3 (N+2)}
+\frac{2 P_{475}}{9 (N-1) N^5 (N+1)^5 (N+2)}
\nonumber\\&&
+ \textcolor{blue}{N_F} 
\Biggl[
\frac{64 \big(N^2+N+2\big)^2}{9(N-1) N^2 (N+1)^2 (N+2)}
L_M^3
+\Biggl[\frac{\Bigl[16 S_1^2-\frac{80 S_2}{3}\Bigr]
  \big(N^2+N+2\big)^2}{(N-1) N^2 (N+1)^2 (N+2)}
\nonumber\\&&
-\frac{4 P_{466}}{9 (N-1) N^4
  (N+1)^4 (N+2)}
-\frac{32 \big(5 N^5+52 N^4+109 N^3+90 N^2+48 N+16\big) S_1}{3 (N-1) N^3
  (N+1)^3 (N+2)}\Biggr] L_M
\nonumber\\&&
+\frac{4}{9} \frac{P_{455} \zeta_2}{(N-1) N^3 (N+1)^3 (N+2)}
-\frac{16 \big(N^2+N+2\big) \big(8 N^3+13 N^2+27 N+16\big)
  \big(S_1^2+ S_2 \big)}{9 (N-1) N^2 (N+1)^3 (N+2)}
\nonumber\\&&
+\frac{2 P_{476}}{81 (N-1) N^5 (N+1)^5 (N+2)}
+\frac{32 \big(N^2+N+2\big) \big(43 N^4+105 N^3+224 N^2+230 N+86\big) S_1}{27 (N-1) N^2 (N+1)^4 (N+2)}
\nonumber\\&&
\nonumber\\&&
+\frac{\big(N^2+N+2\big)^2}{(N-1) N^2 (N+1)^2 (N+2)}
\Bigr[\frac{16}{9} S_1^3+\frac{16}{3} S_2 S_1+\frac{16}{3} \zeta_2 S_1+\frac{32 S_3}{9}-\frac{64
    \zeta_3}{9}\Bigl]
\Biggr] 
\nonumber\\ &&
+ \frac{\big(N^2+N+2\big)^2
  \big(-\frac{16}{3} S_1 \zeta_2-\frac{80 \zeta_3}{9}\big)}{(N-1) N^2 (N+1)^2
  (N+2)}\Biggr]
\nonumber\\&&
+ \textcolor{blue}{C_A C_F T_F} 
\Biggl[
-\frac{8 \big(N^2+N+2\big) \big(N^3+8 N^2+11 N+2\big) S_1^3}{3 (N-1) N^2 
(N+1)^3 (N+2)^2}
-\frac{4 \big(N^2+N+2\big) P_{431} S_1^2}{(N-1) N^2 (N+1)^4 (N+2)^3}
\nonumber\\&&
-\frac{4}{3} \frac{P_{463} \zeta_2}{(N-1)^2 N^3 (N+1)^3 (N+2)^2} S_1
-\frac{2
  P_{473} S_1}{9 (N-1) N^2 (N+1)^5 (N+2)^4}
\nonumber\\&&
+\frac{8 \big(N^2+N+2\big) \big(3 N^3-12 N^2-27 N-2\big) S_2 S_1}{(N-1) N^2
  (N+1)^3 (N+2)^2}+\frac{8}{9} \frac{\zeta_3}{(N-1)^2 N^3 (N+1)^3 (N+2)^2}
P_{462}
\nonumber\\&&
+\frac{4}{9} \frac{P_{477}\zeta_2}{(N-1)^2 N^4 (N+1)^4 (N+2)^3}
-\frac{P_{482}}{18 (N-1)^2 N^6 (N+1)^6 (N+2)^5}
\nonumber\\&&
+L_M^3 
\Biggl[
-\frac{8 \big(11 N^4+22 N^3-23 N^2-34
  N-12\big) \big(N^2+N+2\big)^2}{9 (N-1)^2 N^3 (N+1)^3 (N+2)^2}
-\frac{16 S_1
  \big(N^2+N+2\big)^2}{3 (N-1) N^2 (N+1)^2 (N+2)}
\Biggr]
\nonumber\\&&
-\frac{4 \big(N^2+N+2\big) P_{458} S_2}{(N-1)^2 N^4 (N+1)^4 (N+2)^3}-\frac{64
  \big(N^2+N+2\big) \big(N^5+10 N^4+9 N^3+3 N^2+7 N+6\big) S_3}{3 (N-1)^2 N^3
  (N+1)^3 (N+2)^2}
\nonumber\\&&
+L_M^2 
\Biggl[
\frac{\Bigl[-16 S_2-32 S_{-2}\Bigr]
  \big(N^2+N+2\big)^2}{(N-1) N^2 (N+1)^2 (N+2)}-\frac{2 P_{478}}{9 (N-1)^2 N^4
  (N+1)^4 (N+2)^3}
\nonumber\\&&
-\frac{8 P_{459} S_1}{3 (N-1)^2 N^3 (N+1)^3 (N+2)^2}\Biggr]-64 \log (2) \zeta_2
\nonumber\\&&
+\frac{\big(N^2+N+2\big) \big(N^4+2 N^3+7 N^2+22 N+20\big) \Bigl[-32 (-1)^N
  S_{-2}-16 (-1)^N \zeta_2\Bigr]}{(N-1) N (N+1)^4 (N+2)^3}
\nonumber\\&&
+\frac{\big(N^2-N-4\big) \big(N^2+N+2\big)}{(N-1) N (N+1)^3 (N+2)^2}
\Bigl[64 (-1)^N S_1 S_{-2}+32 (-1)^N
  S_{-3}-64 S_{-2,1}
+32 (-1)^N S_1 \zeta_2
\nonumber\\&&
+24 (-1)^N \zeta_3\Bigr]
+\frac{\big(N^2+N+2\big)^2} {(N-1) N^2 (N+1)^2 (N+2)}
\Bigl[\frac{2}{3} S_1^4+20 S_2 S_1^2+32 (-1)^N
  S_{-3} S_1
\nonumber\\ && 
+\Bigl[\frac{160 S_3}{3}-64 S_{-2,1}\Bigr] S_1
+\frac{8}{3} \big(-7+9
  (-1)^N\big) \zeta_3 S_1+2 S_2^2
+S_{-2} \Bigl[32 (-1)^N S_1^2+32 (-1)^N S_2\Bigr]
\nonumber\\ && 
+36 S_4+16 (-1)^N
  S_{-4}-16 S_{3,1}-32 S_{-2,2}-32 S_{-3,1}-16 S_{2,1,1}
+64 S_{-2,1,1}
+\Bigl[
4 \big(-3+4 (-1)^N\big) S_1^2
\nonumber\\ &&
+4 \big(-1+4 (-1)^N\big) S_2
+8 \big(1+2 (-1)^N\big) S_{-2}
\Bigr] \zeta_2\Bigr]
+L_M 
\Biggl[
\frac{\big(N^2+N+2\big)^2}{(N-1) N^2 (N+1)^2 (N+2)}
\Bigl[
\frac{8}{3} S_1^3
-40 S_2 S_1
\nonumber\\&&
-32 (-1)^N S_{-2} S_1
-16 (-1)^N S_{-3}
+32 S_{2,1}
-12 (-1)^N \zeta_3
\Bigr] 
-\frac{16 \big(5 N^2+5 N-26\big) S_3 \big(N^2+N+2\big)}{3 (N-1) N^2 (N+1)^2 (N+2)}
\nonumber\\&&
-\frac{4 \big(17 N^4-6 N^3+41 N^2-16 N-12\big) S_1^2 \big(N^2+N+2\big)}{3 (N-1)^2 N^3 (N+1)^2 (N+2)}
+\frac{96 \big(N^2+N+4\big) S_{-3} \big(N^2+N+2\big)}{(N-1) N^2 (N+1)^2 (N+2)}
\nonumber\\&&
-\frac{32 \big(N^2+N+14\big) S_{-2,1} \big(N^2+N+2\big)}{(N-1) N^2 (N+1)^2 (N+2)}
-\frac{\big(N^2-N-4\big)
32 (-1)^N S_{-2}
\big(N^2+N+2\big)}{(N-1) N (N+1)^3 (N+2)^2}
\nonumber\\&&
+\frac{8 P_{481}}{27 (N-1)^2 N^5 (N+1)^5 (N+2)^4}
- \frac{4 \big(N^2+N+10\big) \big(5
N^2+5 N+18\big) \zeta_3}{(N-1) N^2 (N+1)^2 (N+2)}
\nonumber\\
&&-\frac{8 P_{472} S_1}{9 (N-1)^2 N^4 (N+1)^4 (N+2)^2}+\frac{64 \big(N^4+2
    N^3+7 N^2+6 N+16\big) S_{-2} S_1}{(N-1) N^2 (N+1)^2 (N+2)}
\nonumber\\&&
-\frac{4 P_{446} S_2}{(N-1)^2 N^3 (N+1)^3 (N+2)^2}-\frac{32 P_{447}
  S_{-2}}{(N-1)^2 N^3 (N+1)^3 (N+2)^2}+64 S_1 \zeta_3
\Biggr] 
\Biggr] 
\Biggr] 
 + a_{gg,Q}^{(3)}
\Biggr\} 
\Biggr\}~,
\end{eqnarray}

}
with
{\small
\begin{eqnarray}
P_{431}&=&N^6+6 N^5+7 N^4+4 N^3+18 N^2+16 N-8
 \\
P_{432}&=&3 N^6+9 N^5-113 N^4-241 N^3-274 N^2-152 N-24
   \\
P_{433}&=&3 N^6+9 N^5+22 N^4+29 N^3+41 N^2+28 N+6
    \\
P_{434}&=&3 N^6+30 N^5+15 N^4-64 N^3-56 N^2-20 N-8
    \\
P_{435}&=&4 N^6+3 N^5-50 N^4-129 N^3-100 N^2-56 N-24
    \\
P_{436}&=&7 N^6+15 N^5+7 N^4-23 N^3-26 N^2-20 N-8
    \\
P_{437}&=&9 N^6+27 N^5+161 N^4+277 N^3+358 N^2+224 N+48
     \\
P_{438}&=&20 N^6+60 N^5+11 N^4-78 N^3-121 N^2-72 N-108
     \\
P_{439}&=&20 N^6+60 N^5+11 N^4-78 N^3-85 N^2-36 N-108
     \\
P_{440}&=&40 N^6+114 N^5+19 N^4-132 N^3-147 N^2-70 N-32
     \\
P_{441}&=&63 N^6+189 N^5+367 N^4+419 N^3+626 N^2+448 N+96
     \\
P_{442}&=&99 N^6+297 N^5+631 N^4+767 N^3+1118 N^2+784 N+168
     \\
P_{443}&=&136 N^6+390 N^5+19 N^4-552 N^3-947 N^2-630 N-288
     \\
P_{444}&=&N^8+4 N^7+2 N^6+64 N^5+173 N^4+292 N^3+256 N^2-72 N-72
       \\
P_{445}&=&N^8+4 N^7+8 N^6+6 N^5-3 N^4-22 N^3-10 N^2-8 N-8
       \\
P_{446}&=&3 N^8-14 N^7-164 N^6-454 N^5-527 N^4-204 N^3-112 N^2+80 N+48
       \\
P_{447}&=&3 N^8+10 N^7+13 N^6+N^5+28 N^4+81 N^3+4 N^2-12 N-32
       \\
P_{448}&=&3 N^8+23 N^7+51 N^6+95 N^5+142 N^4+158 N^3+56 N^2-32 N-16
       \\
P_{449}&=&15 N^8+60 N^7+76 N^6-18 N^5-275 N^4-546 N^3-400 N^2-224 N-96
       \\
P_{450}&=&15 N^8+60 N^7+86 N^6+12 N^5-166 N^4-378 N^3-245 N^2-148 N-84
    \\
P_{451}&=&15 N^8+60 N^7+572 N^6+1470 N^5+2135 N^4+1794 N^3+722 N^2-24 N-72
     \\
P_{452}&=&23 N^8+92 N^7+46 N^6-88 N^5+79 N^4+476 N^3+428 N^2-96 N-96
     \\
P_{453}&=&24 N^8+96 N^7+93 N^6-57 N^5-143 N^4-79 N^3-34 N^2-20 N-8
  \\
P_{454}&=&27 N^8+108 N^7-1440 N^6-4554 N^5-5931 N^4-3762 N^3-256 N^2+1184 N+480
\\
P_{455}&=&63 N^8+252 N^7+196 N^6-258 N^5-551 N^4-282 N^3-220 N^2-80 N+48
    \\
P_{456}&=&131 N^8+524 N^7+691 N^6+239 N^5-848 N^4-1483 N^3-586 N^2+108 N+360
  \\
P_{457}&=&297 N^8+1188 N^7+640 N^6-2094 N^5-1193 N^4+2874 N^3+5008 N^2+3360 N+864
     \\
P_{458}&=&N^9+21 N^8+85 N^7+105 N^6+42 N^5+290 N^4+600 N^3+456 N^2+256 N+64
     \\
P_{459}&=&3 N^{10}+15 N^9+35 N^8+50 N^7+91 N^6+233 N^5+255 N^4+150 N^3-24 N^2
\nonumber\\&&
-184 N-48
  \\
P_{460}&=&3 N^{10}+15 N^9+3316 N^8+12778 N^7+22951 N^6+23815 N^5+14212 N^4+3556
N^3
\nonumber\\&&
-30 N^2+288 N+216 
 \\
P_{461}&=&15 N^{10}+75 N^9+112 N^8+14 N^7-61 N^6+107 N^5+170 N^4+36 N^3-36 N^2
\nonumber\\&&
-32 N-16
  \\
P_{462}&=&18 N^{10}+90 N^9+119 N^8-91 N^7-167 N^6+101 N^5+162 N^4-72 N^3-504 N^2
\nonumber  \\ &&
-184 N-48
  \\
P_{463}&=&30 N^{10}+150 N^9+163 N^8-224 N^7-586 N^6-368 N^5-39 N^4-78 N^3+144 N^2
\nonumber\\ &&
+184 N+48
  \\
P_{464}&=&40 N^{10}+200 N^9+282 N^8-66 N^7-615 N^6-753 N^5-509 N^4-205 N^3-2 N^2
\nonumber\\ &&
+68 N+24
  \\
P_{465}&=&63 N^{10}+315 N^9-1142 N^8-6260 N^7-11927 N^6-12359 N^5-7235 N^4-1778 N^3
\nonumber\\&&
+15 N^2-144 N-108
  \\
P_{466}&=&67 N^{10}+335 N^9+368 N^8-762 N^7-3349 N^6-6669 N^5-8310 N^4-7656 N^3
\nonumber\\&&
-4648 N^2-1600 N-288
     \\
P_{467}&=&219 N^{10}+1095 N^9+1640 N^8-82 N^7-2467 N^6-2947 N^5-3242 N^4-4326N^3
\nonumber\\&&
-3466 N^2-1488 N-360
  \\
P_{468}&=&693 N^{10}+3465 N^9-11014 N^8-62668 N^7-120361 N^6-125113 N^5-73393 N^4
\nonumber\\&&
-18010 N^3+165 N^2-1584 N-1188
  \\
P_{469}&=&3 N^{11}+21 N^{10}-124 N^9-1014 N^8-2185 N^7-2099 N^6-934 N^5-2060
N^4-4632 N^3
\nonumber\\&&
-4256 N^2
-2688 N-768
  \\
P_{470}&=&27 N^{11}+189 N^{10}+631 N^9+1356 N^8+2155 N^7+2207 N^6+211 N^5-4984
N^4-8400 N^3
\nonumber\\&&
-5824 N^2-2544 N-576
  \\
P_{471}&=&N^{12}+6 N^{11}-5 N^{10}-80 N^9-379 N^8-846 N^7-1057 N^6-786 N^5+84
N^4+490 N^3
\nonumber\\&&
+324 N^2
+152 N+48
  \\
P_{472}&=&15 N^{12}+90 N^{11}+80 N^{10}-452 N^9-1401 N^8-2298 N^7-5002 N^6-6516
N^5-1116 N^4
\nonumber\\&&
+2960 N^3
+3464 N^2+2400 N+864
  \\
P_{473}&=&233 N^{12}+2724 N^{11}+13349 N^{10}+34680 N^9+46703 N^8+12096 N^7-69461 N^6
\nonumber\\&&
-137724 N^5-141176 N^4-91776 N^3-34832 N^2-6336 N-2304
  \\
P_{474}&=&310 N^{12}+2058 N^{11}+5939 N^{10}+17235 N^9+44700 N^8+93240 N^7+140861 N^6
\nonumber\\&&
+113169 N^5+14578 N^4-40374 N^3-33372 N^2-12312 N-3888
  \\
P_{475}&=&391 N^{12}+2346 N^{11}+4795 N^{10}+2758 N^9-2243 N^8+1150 N^7+7713 N^6
\nonumber\\&&
+4546 N^5-792 N^4+1224 N^2+864 N+288
  \\
P_{476}&=&1593 N^{12}+9558 N^{11}+15013 N^{10}-8758 N^9-62269 N^8-82318 N^7-79041 N^6
\nonumber\\&&
-90898 N^5-70928 N^4-15872 N^3+7344 N^2+5184 N+1728
  \\
P_{477}&=&15 N^{13}+120 N^{12}+530 N^{11}+1562 N^{10}+2042 N^9-1680 N^8-9220 N^7-12524 N^6
\nonumber\\&&
-7911 N^5-5230 N^4-5880 N^3-3344 N^2-2544 N-864
  \\
P_{478}&=&33 N^{13}+264 N^{12}+479 N^{11}-1366 N^{10}-8809 N^9-23124 N^8-34351
N^7-26198 N^6
\nonumber\\&&
-3624 N^5 +5240 N^4-2496 N^3-7232 N^2-7104 N-2304
  \\
P_{479}&=&2493 N^{13}+19944 N^{12}+79295 N^{11}+208394 N^{10}+375431 N^9+531516
N^8+623697 N^7
\nonumber\\&&
+733338 N^6+963340 N^5+1047352 N^4+895648 N^3+559488 N^2+222336 N+41472
  \\
P_{480}&=&39 N^{14}+273 N^{13}+741 N^{12}+1025 N^{11}+1343 N^{10}+3479 N^9+6707 N^8+6555 N^7
\nonumber\\&&
+2258 N^6-1520 N^5-1944 N^4-532 N^3+280 N^2+208 N+32 
\\
P_{481}&=&276 N^{16}+3036 N^{15}+13660 N^{14}+30172 N^{13}+22446 N^{12}-50653
N^{11}-171627 N^{10}
\nonumber\\&&
-246412 N^9
-204934 N^8-83791 N^7+28263 N^6+43144 N^5-33372 N^4-82640 N^3
\nonumber\\&&
-79152 N^2-47232 N -12096
\\
P_{482}&=&3135 N^{19}+43890 N^{18}+257636 N^{17}+794084 N^{16}+1224418
N^{15}+244448 N^{14}
\nonumber\\&&
-2371724 N^{13}-3594388 N^{12}-792201 N^{11}+2719198 N^{10}+2284064 N^9-85568 N^8
\nonumber\\&&
-227344 N^7 +952768 N^6 +1160704 N^5+807552 N^4+574464 N^3+305664 N^2
\nonumber\\&&
+104448 N+18432~. 
\end{eqnarray}

}
\section{The asymptotic Heavy Flavor Wilson Coefficients contributing to \boldmath $F_2(x,Q^2)$ in 
$z$-space}
\label{app:B}

\vspace*{1mm}
\noindent
The representation of the Wilson coefficients in momentum fraction or $z$-space of the 
contributions being known at present can be obtained in terms of harmonic polylogarithms 
\cite{HPL} over the alphabet
\begin{eqnarray}
\mathfrak{A} = \left\{ \frac{dz}{z}, \frac{dz}{1-z}, \frac{dz}{1+z} \right\} \equiv \left\{f_0(z), f_1(z), f_{-1}(z) 
\right\}
\end{eqnarray}
as iterated integrals 
\begin{eqnarray}
\label{eq:HPL}
H_{b,\vec{a}}(z) = \int_0^z f_b(y) H_{\vec{a}}(y),~~~f_b  \in \mathfrak{A},~~~~H_\emptyset = 1, H_{\tiny {\underbrace{0, 
..., 
0}_{k}}}(z) := \frac{1}{k!} \ln^k(z).
\end{eqnarray}
For brevity we will drop the argument $z$ of the harmonic polylogarithms in the following.
As a shorthand notation we introduce the Mellin-inversion of (\ref{eq:gqgtil})
\begin{eqnarray}
\gamma_{qg} = -4 \left[z^2 + (1-z)^2\right]~,
\end{eqnarray}
where the Mellin transform is defined by
\begin{eqnarray}
\Mvec[f(x)](N) = \int_0^1 dx~x^{N-1}~f(x)~.
\end{eqnarray}

The Wilson coefficients $L_{q,2}^{\sf PS}$ and $L_{g,2}^{\sf S}$ are given in $z$-space by~: 
{\small
\begin{eqnarray}
\lefteqn{L_{q,2}^{\sf PS} =} \nonumber\\ &&
\textcolor{blue}{a_s^3} 
\Biggl\{
\textcolor{blue}{C_F}
\textcolor{blue}{N_F} \textcolor{blue}{T_F^2}
\Biggl[
\Biggl[
\frac{64}{9} (z+1) H_0
-\frac{32 (z-1) \big(4 z^2+7 z+4\big)}{27 z}\Biggr] L_Q^3
+\Bigl[
-\frac{64}{3} (z+1) H_0^2
\N\\&&
+\frac{64}{9} \big(4 z^2-11 z-8\big) H_0
+\frac{32 (z-1) \big(10 z^2+33 z-2\big)}{9 z}\Bigr] L_Q^2
+ L_Q \Biggl[
\frac{64}{3} H_0^3 (z+1)
\N\\&&
+\frac{\Bigl[\frac{256}{9} H_{0,-1}
-\frac{256}{9} H_{-1} H_0\Bigr] (z+1)^3}{z}
-\frac{32}{9} \big(12 z^2-59 z-29\big) H_0^2
\N\\&&
+\Biggl[\frac{64}{3} (z+1) H_0
-\frac{32 (z-1) \big(4 z^2+7 z+4\big)}{9 z}\Biggr] L_M^2
-\frac{64 (z-1) \big(304 z^2+811 z+124\big)}{81 z}
\N\\&&
-\frac{64}{27} \big(60 z^2-155 z-233\big) H_0
-\frac{256 \big(3 z^2+1\big) \zeta_2}{9 z}
+L_M \Biggl[
-\frac{64 (z-1) \big(38 z^2+47 z+20\big)}{27 z}
\N\\&&
+\frac{128}{9} \big(2 z^2+11 z+8\big) H_0
+\frac{64 (z-1) \big(4 z^2+7 z+4\big) H_1}{9 z}
+(z+1) \Bigl[\frac{128 \zeta_2}{3}
-\frac{128}{3} H_{0,1}\Bigr]\Biggr]\Biggr]
\N\\&&
+L_M^3 \Biggl[\frac{32 (z-1) \big(4 z^2+7 z+4\big)}{27 z}
-\frac{64}{9} (z+1) H_0\Biggr]
-\frac{32}{81} \big(57 z^2+367 z+295\big) H_0^2
\N\\&&
-\frac{16 (z-1) \big(118 z^2-107 z+118\big) H_1^2}{81 z}+ L_M^2 \Biggl[
-\frac{64}{3} (z+1) H_0^2
+\frac{64}{9} \big(4 z^2-11 z-8\big) H_0
\N\\&&
+\frac{32 (z-1) \big(10 z^2+33 z-2\big)}{9 z}\Biggr]
+\frac{64 (z-1) \big(3821 z^2+6698 z+500\big)}{729 z}
\N\\&&
-\frac{32}{27} \big(68 z^3+53 z^2+5 z-32\big) \frac{\zeta_3}{z}
-\frac{32}{243} \big(48 z^2+5087 z+2783\big) H_0
\N\\&&
-\frac{32 (z-1) \big(256 z^2+521 z-14\big) H_1}{81 z}
-\frac{32 (z-1) \big(38 z^2+47 z+20\big) H_0 H_1}{27 z}
\N\\&&
-\frac{64 \big(42 z^3-227 z^2-74 z+30\big) H_{0,1}}{81 z}
+\frac{64}{9} \big(2 z^2+11 z+8\big) H_0 H_{0,1}
\N\\&&
-\frac{64}{27} \big(6 z^2+4 z-5\big) H_{0,0,1}
+\frac{128 \big(9 z^3+7 z^2+7 z+3\big) H_{0,1,1}}{27 z}
\N\\&&
+\frac{32}{81} \big(198 z^2-427 z-229\big) \zeta_2
-\frac{64}{27} \big(6 z^2+62 z+53\big) H_0 \zeta_2
+\frac{(z-1) \big(4 z^2+7 z+4\big)}{z} \Bigl[\frac{64}{81} H_1^3
\N\\&&
+\frac{16}{9} H_0 H_1^2
-\frac{32}{9} \zeta_2 H_1\Bigr]
+L_M \Biggl[\frac{256 \zeta_2 z^2}{3}
-\frac{16}{9} \big(8 z^2+73 z+61\big) H_0^2
\N\\&&
+\frac{64}{27} \big(28 z^2-221 z-107\big) H_0+(z+1) \Bigl[
-\frac{32}{9} H_0^3
+\frac{128}{3} H_{0,1} H_0
-\frac{256}{3} \zeta_2 H_0
+\frac{128}{3} H_{0,1,1}
\N\\&&
-\frac{128 \zeta_3}{3}\Bigr]
+\frac{32 (z-1) \big(832 z^2+1213 z+76\big)}{81 z}
+\frac{128 (z-1) \big(4 z^2-26 z+13\big) H_1}{27 z}
\N\\&&
+\frac{(z-1) \big(4 z^2+7 z+4\big) \Bigl[
-\frac{32}{9} H_1^2
-\frac{64}{9} H_0 H_1\Bigr]}{z}
-\frac{64 \big(8 z^3-3 z^2+3 z+4\big) H_{0,1}}{9 z}\Biggr]
\N\\&&
+(z+1) \Bigl[
-\frac{8}{27} H_0^4
-\frac{464}{81} H_0^3
-\frac{64}{3} H_{0,1,1} H_0
+\frac{832}{9} \zeta_3 H_0
-\frac{224 \zeta_2^2}{15}
+\frac{128}{9} H_{0,0,0,1}
\N\\&&
-\frac{64}{9} H_{0,0,1,1}
-\frac{256}{9} H_{0,1,1,1}+\big(\frac{64}{3} H_{0,1}
-\frac{64}{9} H_0^2\big) \zeta_2\Bigr]
\Biggr]
+\textcolor{blue}{N_F} \hat{\tilde{C}}_{2,q}^{{\sf PS},(3)}({N_F})\Biggr\}
\end{eqnarray}

}
and
{\small
\begin{eqnarray}
\lefteqn{L_{g,2}^{\sf S} =} \nonumber\\ &&
\textcolor{blue}{a_s^2} 
\textcolor{blue}{N_F} \textcolor{blue}{T_F^2} \Biggl\{\Biggl[{\gamma}_{qg}^{0} \Bigl[\frac{4 H_0}{3}
+\frac{4 H_1}{3}\Bigr]
-\frac{16}{3} \big(8 z^2-8 z+1\big)\Biggr] L_M
-\frac{4}{3} {\gamma}_{qg}^{0} L_Q L_M\Biggr\}
\N\\&&
+\textcolor{blue}{a_s^3} \Biggl\{
\textcolor{blue}{N_F}  \textcolor{blue}{T_F^3} \Biggl[
\big({\gamma}_{qg}^{0} \Bigl[\frac{16 H_0}{9}
+\frac{16 H_1}{9}\Bigr]
-\frac{64}{9} \big(8 z^2-8 z+1\big)\big) L_M^2
-\frac{16}{9} {\gamma}_{qg}^{0} L_Q L_M^2\Biggr]
\N\\&&
+\textcolor{blue}{C_A} \textcolor{blue}{N_F} \textcolor{blue}{T_F^2} 
\Biggl[
-\frac{4}{27} (30 z-13) H_0^4
-\frac{8}{81} \big(32 z^2+628 z-169\big) H_0^3
-\frac{16}{81} \big(532 z^2+2586 z-193\big) H_0^2
\N\\&&
-\frac{64}{27} \big(7 z^2+7 z+5\big) H_{-1} H_0^2
+\frac{16}{3} (2 z-1) \zeta_2 H_0^2
+\frac{8}{243} \big(4641 z^2-67330 z+3473\big) H_0
\N\\&&
+\frac{32}{27} \big(59 z^2-50 z+25\big) H_1 H_0
+\frac{64}{27} \big(14 z^2+11 z+10\big) H_{0,-1} H_0
-\frac{64}{27} \big(7 z^2-4 z+5\big) H_{0,1} H_0
\N\\&&
+\frac{16}{9} (62 z-19) \zeta_2 H_0
+\frac{32}{9} (38 z+5) \zeta_3 H_0
+ L_M^3 \Biggl[\frac{16 (z-1) \big(31 z^2+7 z+4\big)}{27 z}
-\frac{32}{9} (4 z+1) H_0
\N\\&&
-\frac{8}{9} {\gamma}_{qg}^{0} H_1\Biggr]
+\frac{16}{81} \big(172 z^2-163 z+56\big) H_1^2
-\frac{64}{45} (8 z+3) \zeta_2^2
\N\\&&
+\frac{8 \big(276317 z^3-271875 z^2+11280 z-6182\big)}{729 z}
-\frac{16}{27} \big(248 z^3-438 z^2+33 z-32\big) \frac{\zeta_3}{z}
\N\\&&
+\frac{8 \big(28805 z^3-28460 z^2+4612 z-1596\big) H_1}{243 z}+L_Q^3 \Biggl[
-\frac{16 (z-1) \big(31 z^2+7 z+4\big)}{27 z}
\N\\&&
+\frac{32}{9} (4 z+1) H_0
+\frac{8}{9} {\gamma}_{qg}^{0} H_1\Biggr]
+\big(200 z^2+191 z+112\big) \Bigl[\frac{32}{81} H_{0,-1}
-\frac{32}{81} H_{-1} H_0\Bigr]
\N\\&&
-\frac{32}{27} \big(98 z^2+347 z+16\big) H_{0,1}
-\frac{128}{27} \big(7 z^2+4 z+5\big) H_{0,0,-1}
+\frac{16}{27} \big(28 z^2-202 z+77\big) H_{0,0,1}
\N\\&&
+\frac{32}{9} \big(14 z^2-15 z+10\big) H_{0,1,1}+\big(2 z^2+2 z+1\big) \Bigl[
-\frac{64}{27} H_{-1} H_0^3
+\frac{64}{9} H_{0,-1} H_0^2
-\frac{128}{9} H_{0,0,-1} H_0
\N\\&&
+\frac{128}{9} H_{0,0,0,-1}\Bigr]
-\frac{64}{3} z^2 H_{0,0,0,1}
+\frac{32}{81} \big(117 z^2+1000 z-27\big) \zeta_2
+L_M^2 \Biggl[
-\frac{16}{3} (8 z+1) H_0^2
\N\\&&
+\frac{16}{9} \big(49 z^2-136 z-13\big) H_0
+\frac{8 \big(1048 z^3-894 z^2-87 z-40\big)}{27 z}
+\frac{16 \big(87 z^3-80 z^2+13 z-4\big) H_1}{9 z}
\N\\&&
+{\gamma}_{qg}^{0} \Bigl[
-\frac{4}{3} H_1^2
-\frac{8}{3} H_0 H_1\Bigr]+\big(2 z^2+2 z+1\big) \Bigl[\frac{32}{3} H_{0,-1}
-\frac{32}{3} H_{-1} H_0\Bigr]
-\frac{32}{3} (4 z+1) H_{0,1}
\N\\&&
-\frac{64}{3} (z-2) z \zeta_2\Biggr]
+L_Q^2 \Biggl[
-\frac{16}{3} (8 z+1) H_0^2
+\frac{16}{9} \big(49 z^2-136 z-13\big) H_0
\N\\&&
+\frac{8 \big(1048 z^3-894 z^2-87 z-40\big)}{27 z}
+\frac{16 \big(87 z^3-80 z^2+13 z-4\big) H_1}{9 z}
+{\gamma}_{qg}^{0} \Bigl[
-\frac{4}{3} H_1^2
\N\\&&
-\frac{8}{3} H_0 H_1\Bigr]+\big(2 z^2+2 z+1\big) \Bigl[\frac{32}{3} H_{0,-1}
-\frac{32}{3} H_{-1} H_0\Bigr]
-\frac{32}{3} (4 z+1) H_{0,1}
\N\\&&
+\Biggl[
-\frac{16 (z-1) \big(31 z^2+7 z+4\big)}{9 z}
+\frac{32}{3} (4 z+1) H_0
+\frac{8}{3} {\gamma}_{qg}^{0} H_1\Biggr] L_M
-\frac{64}{3} (z-2) z \zeta_2\Biggr]
\N\\&&
+\big(7 z^2-7 z+5\big) \Bigl[
\frac{32}{81} H_1^3
+\frac{32}{27} H_0 H_1^2
+\frac{32}{27} H_0^2 H_1
-\frac{128}{27} H_{0,1} H_1
+\frac{64}{27} \zeta_2 H_1\Bigr]
\N\\&&
+L_Q \Biggl[
\frac{448}{9} z H_0^3
-\frac{16}{9} \big(63 z^2-296 z-10\big) H_0^2
+\frac{32}{3} \big(5 z^2+4 z+2\big) H_{-1} H_0^2
\N\\&&
-\frac{32}{3} \big(3 z^2-4 z+2\big) H_1 H_0^2
-\frac{64}{27} \big(325 z^2-832 z-31\big) H_0
-\frac{64}{3} \big(3 z^2+4 z+2\big) H_{0,-1} H_0
\N\\&&
+\frac{64}{3} \big(z^2+2 z+2\big) H_{0,1} H_0
+\frac{128}{3} (z-5) z \zeta_2 H_0+\Biggl[
-\frac{16 (z-1) \big(31 z^2+7 z+4\big)}{9 z}
\N\\&&
+\frac{32}{3} (4 z+1) H_0
+\frac{8}{3} {\gamma}_{qg}^{0} H_1\Biggr] L_M^2
-\frac{8 \big(25657 z^3-23556 z^2-969 z-412\big)}{81 z}
\N
\\
&&
+\frac{32}{9} \big(105 z^3-184 z^2+3 z-4\big) \frac{\zeta_2}{z}
-\frac{32 \big(1032 z^3-964 z^2+65 z-20\big) H_1}{27 z}
\N\\&&
+\frac{\big(87 z^3-80 z^2+13 z-4\big) \Bigl[
-\frac{16}{9} H_1^2
-\frac{32}{9} H_0 H_1\Bigr]}{z}
+\frac{\big(17 z^3+7 z^2-4 z-2\big) \Bigl[\frac{64}{9} H_{-1} H_0
-\frac{64}{9} H_{0,-1}\Bigr]}{z}
\N\\&&
-\frac{64 \big(9 z^3-59 z^2-5 z+2\big) H_{0,1}}{9 z}
+{\gamma}_{qg}^{0} \Bigl[\frac{8}{9} H_1^3
+\frac{16}{3} H_0 H_1^2
-\frac{32}{3} H_{0,1} H_1\Bigr]+(2 z+1) \Bigl[\frac{32}{3} H_0 H_{-1}^2
\N\\&&
-\frac{64}{3} H_{0,-1} H_{-1}
+\frac{64}{3} H_{0,-1,-1}\Bigr]
+\frac{64}{3} \big(z^2+4 z+2\big) H_{0,0,-1}
-\frac{64}{3} \big(z^2-2 z+2\big) H_{0,0,1}
\N\\&&
+\big(2 z^2+2 z+1\big) \Bigl[\frac{64}{3} H_{-1} H_{0,1}
-\frac{64}{3} H_{0,-1,1}
-\frac{64}{3} H_{0,1,-1}\Bigr]
-\frac{128}{3} \big(3 z^2-5 z+1\big) H_{0,1,1}
\N\\&&
-\frac{32}{3} \big(4 z^2+2 z+1\big) H_{-1} \zeta_2
-\frac{32}{3} (2 z-1) H_1 \zeta_2
+L_M \Biggl[
-\frac{32}{3} (4 z+3) H_0^2
\N\\&&
+\frac{32}{9} \big(101 z^2-8 z+13\big) H_0
-\frac{32 \big(8 z^3-14 z^2+11 z-8\big)}{3 z}
+\frac{32 \big(47 z^3-40 z^2-7 z-4\big) H_1}{9 z}
\N\\&&
+{\gamma}_{qg}^{0} \Bigl[
-\frac{8}{3} H_1^2
-\frac{16}{3} H_0 H_1\Bigr]+\big(2 z^2+2 z+1\big) \Bigl[\frac{64}{3} H_{0,-1}
-\frac{64}{3} H_{-1} H_0\Bigr]
-\frac{64}{3} (4 z+1) H_{0,1}
\N\\&&
-\frac{128}{3} (z-2) z \zeta_2\Biggr]
+\frac{32}{3} \big(6 z^2-22 z+3\big) \zeta_3\Biggr]+L_M \Biggl[
\frac{160}{9} H_0^3
-\frac{16}{9} \big(131 z^2+32 z+13\big) H_0^2
\N\\&&
-\frac{32}{3} (z-1)^2 H_1 H_0^2
+\frac{8}{3} {\gamma}_{qg}^{0} H_1^2 H_0
-\frac{16}{27} \big(880 z^2+1486 z-353\big) H_0
-\frac{32 \big(47 z^3-40 z^2-7 z-4\big) H_1 H_0}{9 z}
\N\\&&
-\frac{64}{3} \big(z^2-4 z-1\big) H_{0,1} H_0
+\frac{128}{3} \big(z^2-3 z-1\big) \zeta_2 H_0
-\frac{16 \big(35 z^3-28 z^2-7 z-4\big) H_1^2}{9 z}
\N\\&&
+\frac{8 \big(4933 z^3-5084 z^2+113 z+92\big)}{27 z}
+\frac{32}{9} \big(117 z^3-32 z^2+9 z-4\big) \frac{\zeta_2}{z}
\N\\&&
+\frac{16 \big(404 z^3-476 z^2+271 z-144\big) H_1}{27 z}
+\frac{\big(3 z^3-7 z^2-14 z-2\big) \Bigl[\frac{64}{9} H_{-1} H_0
-\frac{64}{9} H_{0,-1}\Bigr]}{z}
\N\\&&
-\frac{64 \big(35 z^3+11 z^2+8 z+2\big) H_{0,1}}{9 z}
+(2 z+1) \Bigl[\frac{32}{3} H_0 H_{-1}^2
-\frac{64}{3} H_{0,-1} H_{-1}
+\frac{64}{3} H_{0,-1,-1}\Bigr]
\N\\&&
+z^2 \Bigl[\frac{32}{3} H_{-1} H_0^2
+\frac{64}{3} H_{0,-1} H_0-64 H_{0,0,-1}\Bigr]
+\frac{64}{3} \big(z^2-4 z+1\big) H_{0,0,1}
+\big(2 z^2+2 z+1\big) \Bigl[\frac{64}{3} H_{-1} H_{0,1}
\N\\&&
-\frac{64}{3} H_{0,-1,1}
-\frac{64}{3} H_{0,1,-1}\Bigr]
+\frac{64}{3} (4 z+1) H_{0,1,1}
-\frac{32}{3} \big(4 z^2+2 z+1\big) H_{-1} \zeta_2
+\frac{32}{3} \big(4 z^2-6 z+3\big) H_1 \zeta_2
\N\\&&
+\frac{32}{3} \big(6 z^2-2 z-1\big) \zeta_3\Biggr]+{\gamma}_{qg}^{0} \Bigl[\frac{1}{27} H_1^4
-\frac{8}{9} H_{0,1} H_1^2
-\frac{4}{9} H_0^3 H_1+\Bigl[\frac{16}{9} H_{0,0,1}
+\frac{16}{9} H_{0,1,1}\Bigr] H_1
\N\\&&
+\frac{16}{3} \zeta_3 H_1
-\frac{8}{9} H_{0,1}^2+H_0^2 \Bigl[\frac{4}{3} H_{0,1}
-\frac{2}{9} H_1^2\Bigr]+H_0 \Bigl[\frac{4}{27} H_1^3
-\frac{8}{3} H_{0,0,1}
+\frac{8}{9} H_{0,1,1}\Bigr]
-\frac{8}{9} H_{0,0,1,1}
\N\\&&
-\frac{8}{9} H_{0,1,1,1}+\Bigl[\frac{4}{9} H_1^2
-\frac{8}{9} H_0 H_1
+\frac{8}{9} H_{0,1}\Bigr] \zeta_2\Bigr]\Biggr] 
+\textcolor{blue}{C_F} \textcolor{blue}{N_F} \textcolor{blue}{T_F^2}\Biggl[
-\frac{2}{27} \big(56 z^2+448 z-179\big) H_0^4
\N\\&&
+\frac{4}{81} \big(288 z^2-6524 z+3259\big) H_0^3
-\frac{4}{81} \big(4096 z^2+23771 z-21328\big) H_0^2+8 (12 z-5) \zeta_2 H_0^2
\N\\&&
+\frac{56}{243} \big(1491 z^2-4715 z+17578\big) H_0
-\frac{448}{81} \big(z^2-z+2\big) H_1 H_0+112 (5 z-2) \zeta_2 H_0
\N\\&&
+\frac{16}{9} \big(96 z^2+92 z-109\big) \zeta_3 H_0+\Biggl[
-\frac{16}{3} (2 z-1) H_0^2
-\frac{32}{9} \big(6 z^2-z-4\big) H_0
\N\\&&
+\frac{8 \big(124 z^3-258 z^2+159 z-16\big)}{27 z}
+\frac{8}{9} {\gamma}_{qg}^{0} H_1\Biggr] L_M^3
+\frac{8}{81} \big(364 z^2-373 z+224\big) H_1^2
\N
\\
&&
-\frac{32}{45} \big(4 z^2-112 z+47\big) \zeta_2^2
+\frac{2540132 z^3-7301946 z^2+4812411 z+34912}{729 z}
\N\\&&
-\frac{16}{27} \big(468 z^3-1949 z^2+994 z-64\big) \frac{\zeta_3}{z}
+\frac{4 \big(43898 z^3-106070 z^2+58429 z+1296\big) H_1}{243 z}
\N\\&&
+L_Q^3 \Biggl[\frac{16}{3} (2 z-1) H_0^2
+\frac{16}{9} (4 z-11) H_0
-\frac{16 \big(62 z^3-147 z^2+84 z-8\big)}{27 z}
+\frac{16}{9} {\gamma}_{qg}^{0} H_1\Biggr]
\N\\&&
-\frac{16}{81} \big(580 z^2+797 z-3196\big) H_{0,1}+\big(16 z^2-16 z+5\big) \Bigl[\frac{16}{27} H_0^2 H_1
-\frac{32}{27} H_0 H_{0,1}\Bigr]
\N\\&&
+\frac{16}{27} \big(32 z^2-977 z+388\big) H_{0,0,1}+\big(7 z^2-7 z+5\big) \Bigl[
-\frac{32}{81} H_1^3
-\frac{64}{27} H_{0,1,1}\Bigr]
\N\\&&
-\frac{16}{9} \big(8 z^2+100 z-41\big) H_{0,0,0,1}
+\frac{16}{81} \big(608 z^2+769 z-3140\big) \zeta_2
+{\gamma}_{qg}^{0} \Bigl[
-\frac{1}{27} H_1^4
-\frac{8}{27} H_0^3 H_1
\N\\&&
-\frac{64}{9} \zeta_3 H_1
+\frac{8}{9} H_0^2 H_{0,1}
-\frac{16}{9} H_0 H_{0,0,1}
-\frac{8}{9} H_{0,1,1,1}\Bigr]
+L_Q^2 \Biggl[
-\frac{8}{3} \big(8 z^2+56 z-31\big) H_0^2
\N\\&&
+\frac{8}{9} \big(244 z^2-236 z+571\big) H_0
+\frac{4 \big(2156 z^3-7632 z^2+4977 z+256\big)}{27 z}
\N\\&&
+\frac{16 \big(130 z^3-215 z^2+112 z-8\big) H_1}{9 z}+{\gamma}_{qg}^{0} \Bigl[-4 H_1^2
-\frac{32}{3} H_0 H_1\Bigr]
-\frac{16}{3} \big(12 z^2-8 z-5\big) H_{0,1}
\N\\&&
+\Bigl[-16 (2 z-1) H_0^2
-\frac{32}{3} \big(6 z^2-z-4\big) H_0
+\frac{8 \big(124 z^3-258 z^2+159 z-16\big)}{9 z}
+\frac{8}{3} {\gamma}_{qg}^{0} H_1\Bigr] L_M
\N\\&&
-\frac{16}{3} \big(4 z^2-8 z+13\big) \zeta_2
+(2 z-1) \Bigl[-16 H_0^3+32 \zeta_2 H_0-32 H_{0,0,1}+32 \zeta_3\Bigr]\Biggr]
\N\\&&
+L_M^2 \Biggl[-8 \big(8 z^2+16 z-9\big) H_0^2
+\frac{8}{9} \big(92 z^2-160 z+521\big) H_0
-\frac{4}{3} {\gamma}_{qg}^{0} H_1^2
\N\\&&
+\frac{4 \big(2372 z^3-7140 z^2+4611 z+256\big)}{27 z}
+\frac{16 \big(54 z^3-127 z^2+71 z-8\big) H_1}{9 z}
+\big(4 z^2+4 z-11\big) \Bigl[\frac{16 \zeta_2}{3}
\N\\&&
-\frac{16}{3} H_{0,1}\Bigr]+(2 z-1) \Bigl[-16 H_0^3
+32 \zeta_2 H_0-32 H_{0,0,1}+32 \zeta_3\Bigr]\Biggr]
+L_Q \Biggl[\frac{64}{3} \big(2 z^2+15 z-5\big) H_0^3
\N\\&&
-\frac{16}{45} \big(72 z^3+560 z^2-2660 z+2975\big) H_0^2
-\frac{64}{3} \big(5 z^2-4 z+2\big) H_1 H_0^2
\N\\&&
-\frac{16 \big(1444 z^3-5632 z^2+9213 z+4\big) H_0}{45 z}
-\frac{32 \big(198 z^3-283 z^2+140 z-8\big) H_1 H_0}{9 z}
\N\\&&
+\frac{128}{3} (z-2) (3 z-2) H_{0,-1} H_0
+\frac{32}{3} \big(16 z^2-8 z-5\big) H_{0,1} H_0
-\frac{32}{3} \big(4 z^2+84 z-33\big) \zeta_2 H_0
\N\\&&
-\frac{16 \big(168 z^3-253 z^2+131 z-8\big) H_1^2}{9 z}
+\Biggl[16 (2 z-1) H_0^2
+\frac{16}{3} \big(8 z^2-9\big) H_0
\N\\&&
-\frac{16 (z-1) \big(62 z^2-73 z+8\big)}{9 z}\Biggr] L_M^2
-\frac{4 \big(269954 z^3-828996 z^2+567861 z-11744\big)}{405 z}
\N\\&&
+\frac{16}{45} \big(144 z^4+1600 z^3-1400 z^2+3045 z-80\big) \frac{\zeta_2}{z}
-\frac{8 \big(3080 z^3-8448 z^2+5247 z+256\big) H_1}{27 z}
\N\\&&
+\frac{\big(36 z^5-155 z^4+40 z^3+225 z^2-20 z+1\big) \Bigl[\frac{64}{45} H_{-1} H_0
-\frac{64}{45} H_{0,-1}\Bigr]}{z^2}
\N\\&&
+\frac{16 \big(76 z^3-254 z^2-329 z-16\big) H_{0,1}}{9 z}
+{\gamma}_{qg}^{0} \Bigl[\frac{8}{3} H_1^3+8 H_0 H_1^2
+\frac{32}{3} H_{0,1} H_1\Bigr]
\N\\&&
-\frac{128}{3} \big(7 z^2-14 z+9\big) H_{0,0,-1}
-\frac{32}{3} \big(8 z^2-60 z+15\big) H_{0,0,1}
+\frac{32}{3} \big(24 z^2-20 z+1\big) H_{0,1,1}
\N
\\
&&
+\frac{64}{3} \big(8 z^2-6 z+3\big) H_1 \zeta_2
+(z+1)^2 \Bigl[
-\frac{128}{3} H_0 H_{-1}^2+\big(\frac{64}{3} H_0^2
+\frac{256}{3} H_{0,-1}\big) H_{-1}
-\frac{128}{3} \zeta_2 H_{-1}
\N\\&&
-\frac{256}{3} H_{0,-1,-1}\Bigr]
+L_M \Biggl[\frac{16}{3} \big(32 z^2+44 z-25\big) H_0^2
-\frac{32}{3} \big(8 z^2-21 z+89\big) H_0
\N\\&&
-\frac{16 \big(1006 z^3-3333 z^2+2208 z+128\big)}{27 z}
-\frac{16 \big(64 z^3-198 z^2+141 z-16\big) H_1}{9 z}
+{\gamma}_{qg}^{0} \Bigl[
-\frac{16}{3} H_1^2
\N\\&&
-\frac{32}{3} H_0 H_1\Bigr]
+\frac{64}{3} (z-1) (4 z+5) H_{0,1}
-\frac{64}{3} \big(8 z^2-3 z-3\big) \zeta_2
+(2 z-1) \Bigl[32 H_0^3-64 \zeta_2 H_0+64 H_{0,0,1}
\N\\&&
-64 \zeta_3\Bigr]\Biggr]
+\frac{128}{3} \big(4 z^2-21 z+11\big) \zeta_3+(2 z-1) \Bigl[16 H_0^4-96 \zeta_2 H_0^2+64 H_{0,0,1} H_0-128 \zeta_3 H_0
-\frac{32 \zeta_2^2}{5}
\N\\&&
+64 H_{0,0,1,1}\Bigr]\Biggr]
+L_M \Biggl[
-\frac{64}{9} \big(14 z^2+41 z-13\big) H_0^3
-\frac{16}{45} \big(72 z^3+360 z^2+2390 z-2835\big) H_0^2
\N\\&&
-\frac{64}{3} z^2 H_1 H_0^2
-\frac{8 \big(524 z^3+29468 z^2-50797 z+24\big) H_0}{135 z}
+\frac{16 \big(28 z^3-162 z^2+123 z-16\big) H_1 H_0}{9 z}
\N\\&&
-\frac{128}{3} \big(z^2-4 z+2\big) H_{0,-1} H_0-64 (z-1) (2 z+1) H_{0,1} H_0
+\frac{32}{3} \big(36 z^2+64 z-23\big) \zeta_2 H_0
\N\\&&
+\frac{8 \big(44 z^3-178 z^2+143 z-16\big) H_1^2}{9 z}
+\frac{4 \big(259856 z^3-763164 z^2+514989 z-11456\big)}{405 z}
\N\\&&
+\frac{32}{45} \big(72 z^4-70 z^3+345 z^2-1340 z+40\big) \frac{\zeta_2}{z}
+\frac{16 \big(794 z^3-3157 z^2+2114 z+128\big) H_1}{27 z}
\N\\&&
+\frac{\big(36 z^5+155 z^4+40 z^3-45 z^2+20 z+1\big) \Bigl[\frac{64}{45} H_{-1} H_0
-\frac{64}{45} H_{0,-1}\Bigr]}{z^2}
+\frac{16 \big(56 z^2+413 z+16\big) H_{0,1}}{9 z}
\N\\&&
+{\gamma}_{qg}^{0} \Bigl[\frac{8}{3} H_1^3
+\frac{16}{3} H_0 H_1^2
+\frac{16}{3} H_{0,1} H_1\Bigr]
+\frac{128}{3} \big(z^2-10 z+3\big) H_{0,0,-1}
-\frac{32}{3} \big(8 z^2+52 z-11\big) H_{0,0,1}
\N\\&&
-\frac{64}{3} \big(2 z^2+3 z-6\big) H_{0,1,1}
+\frac{64}{3} \big(4 z^2-2 z+1\big) H_1 \zeta_2+(z+1)^2 \Bigl[
-\frac{128}{3} H_0 H_{-1}^2+\big(\frac{64}{3} H_0^2
\N\\&&
+\frac{256}{3} H_{0,-1}\big) H_{-1}
-\frac{128}{3} \zeta_2 H_{-1}
-\frac{256}{3} H_{0,-1,-1}\Bigr]
+\frac{32}{3} \big(24 z^2+82 z-27\big) \zeta_3
\N\\&&
+(2 z-1) \Bigl[-16 H_0^4+96 \zeta_2 H_0^2-64 H_{0,0,1} H_0+128 \zeta_3 H_0
+\frac{32 \zeta_2^2}{5}-64 H_{0,0,1,1}\Bigr]\Biggr]
\N\\&&
+(2 z-1) \Bigl[
-\frac{4}{3} H_0^5
+\frac{16}{3} \zeta_2 H_0^3
+\frac{176}{3} \zeta_3 H_0^2
+\frac{64}{5} \zeta_2^2 H_0
-32 H_{0,0,0,0,1}
\N\\&&
+32 \zeta_5\Bigr]\Biggr]
+\textcolor{blue}{N_F} \hat{\tilde{C}}_{2,g}^{{\sf S},(3)}({N_F})\Biggr\}
\end{eqnarray}

}
The pure-singlet Wilson coefficient $H_{q,2}^{\sf PS}$ reads~:
{\small


}
The Wilson coefficient $H_{g,2}^{\sf S}$ is given by~:
{\small
%

}

\noindent
Numerical implementations of the harmonic polylogarithms were given in Refs.~\cite{NUMHPL}.
\section{The Longitudinal Wilson Coefficients in \boldmath $z$-space}
\label{app:C}

\vspace*{1mm}
\noindent
The Wilson coefficients for the longitudinal structure function $F_L(x,Q^2)$ in the asymptotic
region in $z$-space are presented in the following. $L_{q,L}^{\sf PS}$ and $L_{g,L}^{\sf S}$
read~: 
{\small
\begin{eqnarray}
\lefteqn{L_{q,L}^{\sf PS} =} \nonumber\\ &&
\textcolor{blue}{a_s^3} \Biggl\{
\textcolor{blue}{C_F} \textcolor{blue}{N_F} \textcolor{blue}{T_F^2} \Biggl[
L_M \Biggl[z \Bigl[\frac{256}{3} H_{0,1}
-\frac{256 \zeta_2}{3}\Bigr]
-\frac{256 (z-1) \big(2 z^2+2 z-1\big) H_1}{9 z}
\N\\&&
-\frac{256}{9} z (2 z+11) H_0
+\frac{256 (z-1) \big(19 z^2+16 z-5\big)}{27 z}\Biggr]
+z (2 z+11) \Biggl[
\frac{128}{9} H_{0,1}
-\frac{128 \zeta_2}{9}\Biggr]
\N\\&&
+z \Biggl[
\frac{128 \zeta_3}{3}
-\frac{128}{3} H_{0,1,1}\Biggr]
+L_Q^2 \Biggl[
\frac{128 (z-1) \big(2 z^2+2 z-1\big)}{9 z}
-\frac{128}{3} z H_0\Biggr]
\N\\&&
+L_Q \Biggl[
-\frac{256}{9} \big(4 z^2-8 z-3\big) H_0
+\frac{256}{3} z H_0^2
-\frac{256 (z-1) \big(3 z^2+6 z-2\big)}{9 z}\Biggr]
\N\\&&
+\Biggl[
\frac{128 (z-1) \big(2 z^2+2 z-1\big)}{9 z}-\frac{128}{3} z H_0\Biggr] L_M^2
+\frac{64 (z-1) \big(2 z^2+2 z-1\big) H_1^2}{9 z}
\nonumber\\ &&
-\frac{128 (z-1) \big(19 z^2+16 z-5\big) H_1}{27 z}
-\frac{128}{27} z (19 z+67) H_0+\frac{256 (z-1) \big(55 z^2+43 z-14\big)}{81 z}\Biggr]
\nonumber\\ &&
+\textcolor{blue}{N_F} \hat{\tilde{C}}_{L,q}^{{\sf PS},(3)}({N_F})\Biggr\}~, 
\end{eqnarray}

}
{\small
\begin{eqnarray}
\lefteqn{L_{g,L}^{\sf S} =} \nonumber\\ &&
-\frac{64}{3} \textcolor{blue}{a_s^2} \textcolor{blue}{N_F} \textcolor{blue}{T_F^2} (z-1) z L_M
+\textcolor{blue}{a_s^3} \Biggl\{
-\textcolor{blue}{N_F} \textcolor{blue}{T_F^3}\frac{256}{9}  (z-1) z L_M^2
\N\\&&
+\textcolor{blue}{C_A} \textcolor{blue}{T_F^2} \textcolor{blue}{N_F} \Biggl[
\Biggl[
\frac{64 (z-1) \big(17 z^2+2 z-1\big)}{9 z}
-\frac{256}{3} z H_0
+\frac{128}{3} (z-1) z H_1\Biggr] L_Q^2
\N\\&&
+\Biggl[
-\frac{64 (z-1) \big(461 z^2+11 z-25\big)}{27 z}
-\frac{128}{9} z (26 z-59) H_0
\N\\&&
-\frac{128 (z-1) \big(39 z^2+2 z-1\big) H_1}{9 z}+(z-1) z \Biggl[
-\frac{128}{3} H_1^2
-\frac{256}{3} H_0 H_1\Biggr]
\N\\&&
+z (z+1) \Bigl[\frac{256}{3} H_{-1} H_0
-\frac{256}{3} H_{0,-1}\Bigr]+z \Bigl[\frac{512}{3} H_0^2
+\frac{512}{3} H_{0,1}\Bigr]
\N\\&&
+\Biggl[\frac{128 (z-1) \big(17 z^2+2 z-1\big)}{9 z}
-\frac{512}{3} z H_0
+\frac{256}{3} (z-1) z H_1\Biggr] L_M
\N\\&&
+\frac{256}{3} (z-2) z \zeta_2\Biggr] L_Q
-\frac{32}{9} (28 z-3) H_0^2
+\Biggl[\frac{64 (z-1) \big(17 z^2+2 z-1\big)}{9 z}
-\frac{256}{3} z H_0
\N\\&&
+\frac{128}{3} (z-1) z H_1\Biggr] L_M^2
+\frac{32 (z-1) \big(2714 z^2-106 z-139\big)}{81 z}
-\frac{64}{27} \big(110 z^2+277 z-33\big) H_0
\N\\&&
+\frac{4160}{27} (z-1) z H_1+z \Bigl[
-\frac{64}{9} H_0^3
-\frac{64}{3} H_{0,1}
+\frac{64 \zeta_2}{3}\Bigr]
+L_M \Bigl[
\frac{64 (z-1) \big(68 z^2+z-7\big)}{9 z}
\N\\&&
-\frac{128}{9} (4 z-1) (13 z+6) H_0
-\frac{128 (z-1) \big(19 z^2+2 z-1\big) H_1}{9 z}
+(z-1) z \Bigl[
-\frac{128}{3} H_1^2
\N\\&&
-\frac{256}{3} H_0 H_1\Bigr]+z (z+1) \Bigl[\frac{256}{3} H_{-1} H_0
-\frac{256}{3} H_{0,-1}\Bigr]+z \Bigl[\frac{256}{3} H_0^2
\N\\&&
+\frac{512}{3} H_{0,1}\Bigr]
+\frac{256}{3} (z-2) z \zeta_2\Biggr]\Biggr] 
+\textcolor{blue}{C_F} \textcolor{blue}{T_F^2} \textcolor{blue}{N_F} \Biggl[
-\frac{16}{3} z H_0^4
-\frac{32}{3} (7 z-1) H_0^3-32 (19 z-3) H_0^2
\N\\&&
-64 \big(6 z^2+7 z-8\big) H_0+\Biggl[-64 z H_0^2
-\frac{64}{3} \big(4 z^2+3 z-3\big) H_0
\N\\&&
+\frac{128 (z-1) \big(17 z^2-10 z-1\big)}{9 z}\Biggr] L_M^2
+\frac{16 (z-1) \big(343 z^2-242 z+4\big)}{3 z}
\N\\&&
+L_Q^2 \Biggl[-64 z H_0^2
-\frac{64}{3} (z+1) (4 z-3) H_0
+\frac{64 (z-1) \big(28 z^2-23 z-2\big)}{9 z}\Biggr]
\N\\&&
+L_Q \Biggl[
-\frac{128 (25 z+2) H_1 (z-1)^2}{9 z}
-\frac{32 \big(2474 z^2-4897 z+44\big) (z-1)}{135 z}
\N\\&&
-\frac{64}{45} \big(12 z^3-180 z^2-265 z+90\big) H_0^2
-\frac{64 \big(354 z^3-397 z^2+388 z+4\big) H_0}{45 z}
\N\\&&
+\frac{(z+1) \big(6 z^4-6 z^3+z^2-z+1\big) \big(\frac{256}{45} H_{-1} H_0
-\frac{256}{45} H_{0,-1}\big)}{z^2}
+\frac{128}{3} (z+1) (4 z-3) H_{0,1}
\N\\&&
+\Biggl[128 z H_0^2
+\frac{128}{3} (2 z-1) (2 z+3) H_0
-\frac{64 (z-1) \big(74 z^2-37 z-4\big)}{9 z}\Biggr] L_M
\N\\&&
+\frac{128}{45} \big(12 z^3-60 z^2-25 z+45\big) \zeta_2
+z \Bigl[128 H_0^3-256 \zeta_2 H_0+256 H_{0,0,1}-256 \zeta_3\Bigr]\Biggr]
\N\\&&
+L_M \Biggl[
-\frac{64}{45} \big(12 z^3+180 z^2+335 z-90\big) H_0^2
+\frac{64 \big(456 z^3-708 z^2+347 z-4\big) H_0}{45 z}
\N\\&&
+\frac{64 (z-1) \big(2368 z^2-2189 z-2\big)}{135 z}
+\frac{64 (z-1) \big(80 z^2-37 z-4\big) H_1}{9 z}
\N\\&&
+\frac{(z+1) \big(6 z^4-6 z^3+z^2-z+1\big) \Bigl[\frac{256}{45} H_{-1} H_0
-\frac{256}{45} H_{0,-1}\Bigr]}{z^2}
-\frac{128}{3} (2 z-1) (2 z+3) H_{0,1}
\N\\&&
+\frac{128}{45} \big(12 z^3+60 z^2+50 z-45\big) \zeta_2
+z \Bigl[-128 H_0^3+256 \zeta_2 H_0-256 H_{0,0,1}+256 \zeta_3\Bigr]
\Biggr]\Biggr] 
\N\\&&
+\textcolor{blue}{N_F} \hat{\tilde{C}}_{L,g}^{{\sf S},(3)}({N_F}) \Biggr\}~.
\end{eqnarray}

}
\noindent
The flavor non-singlet Wilson coefficient is given by~: 
{\small
\begin{eqnarray}
\lefteqn{L_{q,L}^{\sf NS} =}  \nonumber\\ &&
 \textcolor{blue}{a_s^2} \textcolor{blue}{C_F} \textcolor{blue}{T_F} \Biggl\{
\frac{16 L_Q z}{3}
+\Bigl[
-\frac{32}{3} H_0
-\frac{16 H_1}{3}\Bigr] z
-\frac{8}{9} (25 z-6)\Biggr\}
\N\\&&
+\textcolor{blue}{a_s^3} \Biggl\{
\textcolor{blue}{C_F^2}\textcolor{blue}{T_F} \Biggl[
\Biggl[8 (z+2)
+z \Bigl[-16 H_0-32 H_1\Bigr]\Biggr] L_Q^2
+\Biggl[
\frac{16}{15} z \big(24 z^2-5\big) H_0^2
+\frac{80}{9} (5 z-6) H_1
\N\\&&
+\frac{16 \big(144 z^3-227 z^2-72 z-96\big) H_0}{45 z}
+\frac{32 \big(72 z^3-223 z^2-77 z+48\big)}{45 z}
\N\\&&
+\frac{\big(3 z^5-5 z^3-10 z^2-2\big) \Bigl[\frac{256}{15} H_{0,-1}-\frac{256}{15} H_{-1} H_0\Bigr]}{z^2}
-\frac{32}{5} z \big(8 z^2+5\big) \zeta_2
+z \Bigl[-\frac{256}{3} H_0 H_{-1}^2
\N\\&&
+\Bigl[\frac{128}{3} H_0^2+\frac{512}{3} H_{0,-1}\Bigr] H_{-1}
+\frac{128}{3} H_1^2-\frac{128}{3} H_0^2 H_1+32 H_{0,1}
+H_0 \Bigl[\frac{256 H_1}{3}-\frac{256}{3} H_{0,-1}
\N\\&&
+\frac{256}{3} H_{0,1}\Bigr]
-\frac{512}{3} H_{0,-1,-1}+\frac{256}{3} H_{0,0,-1}
-\frac{256}{3} H_{0,0,1}+\Bigl[\frac{256 H_1}{3}
-\frac{256}{3} H_{-1}\Bigr] \zeta_2
+\frac{512 \zeta_3}{3}\Bigr]\Biggr] L_Q
\N\\&&
+\frac{8}{9} (z+3) H_0^2
+\Biggl[\frac{8 (z+2)}{3}
+z \Bigl[-\frac{16}{3} H_0-\frac{32 H_1}{3}\Bigr]\Biggr] L_M^2
-\frac{2}{27} (653 z-872)
+\frac{16}{27} (11 z+42) H_0
\N\\&&
+z \Bigl[
-\frac{8}{9} H_0^3
-\frac{16}{3} H_1 H_0^2+\Bigl[\frac{32}{3} H_{
0,1}-\frac{160 H_1}{9}\Bigr] H_0
-\frac{896 H_1}{27}
+\frac{160}{9} H_{0,1}
-\frac{32}{3} H_{0,0,1}\Biggr]
\N\\&&
+\Biggl[
-\frac{8}{9} (53 z-56)
+\frac{32}{9} (z+3) H_0
+z \Biggl[
-\frac{16}{3} H_0^2
-\frac{64}{3} H_1 H_0
-\frac{320 H_1}{9}
+\frac{64}{3} H_{0,1}\Bigr]\Biggr] L_M\Biggr] 
\N\\&&
+\Biggl[
\Biggl[
\frac{64 z L_Q^2}{9}
+\Biggl[z \Bigl[
-\frac{256}{9} H_0-\frac{128 H_1}{9}\Bigr]
-\frac{64}{27} (25 z-6)\Biggr] L_Q
+\textcolor{blue}{N_F} \Biggl[
\frac{128 z L_Q^2}{9}
+\Biggl[
z \Bigl[-\frac{512}{9} H_0
\N\\&&
-\frac{256 H_1}{9}\Bigr]
-\frac{128}{27} (25 z-6)\Biggr] L_Q\Biggr]\Biggr] \textcolor{blue}{T_F^2}
+\textcolor{blue}{C_A} \Biggl[
L_Q \Biggl[
-\frac{16 \big(216 z^3-3329 z^2+624 z+144\big)}{135 z}
\N\\&&
-\frac{64 \big(18 z^3-149 z^2+6 z-12\big) H_0}{45 z}
+\frac{\big(3 z^5-5 z^3-10 z^2-2\big) 
\Bigl[\frac{128}{15} H_{-1} H_0-\frac{128}{15} H_{0,-1}\Bigr]}{z^2}
\N\\&&
+z \big(3 z^2-5\big) \Bigl[\frac{128 \zeta_2}{15}
-\frac{64}{15} H_0^2\Bigr]
+z \Bigl[\frac{128}{3} H_0 H_{-1}^2
+\Bigl[-\frac{64}{3} H_0^2-\frac{256}{3} H_{0,-1}\Bigr] H_{-1}
+\frac{64}{3} H_0^2 H_1
\N\\&&
+\frac{1088 H_1}{9}
+H_0 \Bigl[\frac{128}{3} H_{0,-1}
-\frac{128}{3} H_{0,1}\Bigr]
+\frac{256}{3} H_{0,-1,-1}
-\frac{128}{3} H_{0,0,-1}
+\frac{128}{3} H_{0,0,1}
+\Bigl[
\frac{128}{3} H_{-1}
\N\\&&
-\frac{128 H_1}{3}\Bigr] \zeta_2
-\frac{256 \zeta_3}{3}\Bigr]\Biggr]-\frac{352 L_Q^2 z}{9}\Biggr] 
\textcolor{blue}{T_F}\Biggr] \textcolor{blue}{C_F}
+\hat{C}_{L,q}^{{\sf NS},(3)}({N_F})\Biggr\}~. 
 \end{eqnarray}

}
The pure-singlet Wilson coefficient $H_{q,L}^{\sf PS}$ reads~:
{\small
 \begin{eqnarray}
\lefteqn{H_{q,L}^{\sf PS} =} \nonumber\\ &&
\textcolor{blue}{a_s^2} \textcolor{blue}{C_F} \textcolor{blue}{T_F} \Biggl\{\frac{32 (z-1) \big(10 z^2-2 z+1\big)}{9 z}
 -32 (z+1) (2 z-1) H_0
 -\frac{32 (z-1) \big(2 z^2+2 z-1\big) H_1}{3 z}
  \N\\&&
 +L_Q \Biggl[\frac{32 (z-1) \big(2 z^2+2 z-1\big)}{3 z}
 -32 z H_0\Biggr]
 +z \Bigl[32 H_0^2+32 H_{0,1} -32 \zeta_2\Bigr]\Biggr\}
 \N\\&&
+\textcolor{blue}{a_s^3}\Biggl\{ \textcolor{blue}{C_F^2}
\textcolor{blue}{T_F} \Biggl[
-\frac{8}{3} (5 z+2) H_0^3
-\frac{8}{3} \big(8 z^2+3\big) H_0^2
+\frac{16}{9} \big(160 z^2+93 z-39\big) H_0
\N\\&&
+\Biggl[16 z H_0^2
-16 (z+2) H_0
-\frac{16 (z-1) \big(4 z^2-11 z-2\big)}{3 z}\Biggr] L_M^2
-\frac{32 (z-1) \big(440 z^2-91 z-28\big)}{27 z}
\N\\&&
+\frac{(z-1) \big(4 z^2-11 z-2\big) \Bigl[\frac{32}{3} H_0 H_1-\frac{32}{3} H_{0,1}\Bigr]}{z}
+\Biggl[
-\frac{32}{3} z H_0^3+16 (5 z+2) H_0^2
\N\\&&
+\frac{32}{3} \big(8 z^2+18 z+3\big) H_0
-\frac{32 (z-1) \big(80 z^2+17 z-10\big)}{9 z}\Biggr] L_M
+L_Q^2 \Biggl[
-\frac{64}{3} \big(2 z^2-3\big) H_0
\N\\&&
+\frac{32 (z-1) \big(2 z^2-9 z-1\big)}{3 z}
-\frac{64 (z-1) \big(2 z^2+2 z-1\big) H_1}{3 z}
+z \Bigl[64 H_{0,1}-64 \zeta_2\Bigr]\Biggr]
\N\\&&
+(z+2) \Bigl[32 H_0 H_{0,1}-64 H_{0,0,1}+64 \zeta_3\Bigr]
+z \Bigl[\frac{4}{3} H_0^4-64 H_{0,0,1} H_0-128 \zeta_3 H_0
-\frac{384 \zeta_2^2}{5}
\N\\&&
+192 H_{0,0,0,1}\Bigr]
+L_Q \Biggl[
-\frac{32 (z-1) \big(86 z^2-33 z+6\big)}{45 z}
-\frac{32 \big(56 z^3-813 z^2+142 z+16\big) H_0}{45 z}
\N\\&&
+\frac{64 (z-1) \big(4 z^2+55 z-5\big) H_1}{9 z}
+\frac{(z-1) \big(2 z^2+2 z-1\big)
 \Bigl[\frac{64}{3} H_1^2+\frac{128}{3} H_0 H_1\Bigr]}{z}
 \N\\&&
 +\frac{(z+1) \big(6 z^4-6 z^3+11 z^2+4 z-4\big) \Bigl[\frac{128}{45} H_{0,-1}
 -\frac{128}{45} H_{-1} H_0\Bigr]}{z^2}
 \N\\&&
 +\frac{128 \big(4 z^3-6 z^2-3 z-1\big) H_{0,1}}{3 z}
 +\big(6 z^3+90 z^2-85 z-90\big) \Bigl[\frac{64}{45} H_0^2
 -\frac{128 \zeta_2}{45}\Bigr]
 \N\\&&
 +z \Bigl[
 -\frac{64}{9} H_0^3+\Bigl[\frac{128}{3} H_{0,-1}-128 H_{0,1}\Bigr] H_0
 +\frac{896}{3} \zeta_2 H_0-\frac{256}{3} H_{0,0,-1}-128 H_{0,1,1}
 +192 \zeta_3\Bigr]\Biggr]\Biggr]
 \N\\&&
  +\textcolor{blue}{C_F} \textcolor{blue}{T_F^2}\textcolor{blue}{N_F} \Biggl[\Biggl[
 \frac{128 (z-1) \big(2 z^2+2 z-1\big)}{9 z}
 -\frac{128}{3} z H_0\Biggr] L_Q^2+
 \Biggl[
 \frac{256}{3} z H_0^2
 \N\\&&
 -\frac{256}{9} \big(4 z^2-8 z-3\big) H_0
 -\frac{256 (z-1) \big(3 z^2+6 z-2\big)}{9 z}\Biggr] L_Q\Biggr]
  \N\\&&
 +\textcolor{blue}{C_F}  \textcolor{blue}{T_F^2} \Biggl[\Biggl[
 \frac{128 (z-1) \big(2 z^2+2 z-1\big)}{9 z}
 -\frac{128}{3} z H_0\Biggr] L_Q^2
 +\Biggl[
 \frac{256}{3} z H_0^2
 -\frac{256}{9} \big(4 z^2-8 z-3\big) H_0
  \N\\&&
 -\frac{256 (z-1) \big(3 z^2+6 z-2\big)}{9 z}\Biggr] L_Q
 +\frac{64 (z-1) \big(2 z^2+2 z-1\big) H_1^2}{9 z}
 +\Biggl[
 \frac{128 (z-1) \big(2 z^2+2 z-1\big)}{9 z}
 \N\\&&
 -\frac{128}{3} z H_0\Biggr] L_M^2
 +\frac{256 (z-1) \big(55 z^2+43 z-14\big)}{81 z}
 -\frac{128}{27} z (19 z+67) H_0
 \N\\&&
 -\frac{128 (z-1) \big(19 z^2+16 z-5\big) H_1}{27 z}
 +L_M \Biggl[
 \frac{256 (z-1) \big(19 z^2+16 z-5\big)}{27 z}
 -\frac{256}{9} z (2 z+11) H_0
 \N\\&&
 -\frac{256 (z-1) \big(2 z^2+2 z-1\big) H_1}{9 z}
 +z \Bigl[
 \frac{256}{3} H_{0,1}
 -\frac{256 \zeta_2}{3}\Bigr]\Biggr]
 +z (2 z+11) \Bigl[
 \frac{128}{9} H_{0,1}
 -\frac{128 \zeta_2}{9}\Bigr]
 \N\\&&
 +z \Bigl[\frac{128 \zeta_3}{3}-\frac{128}{3} H_{0,1,1}\Bigr]\Biggr] 
 +\textcolor{blue}{C_F} \textcolor{blue}{C_A} \textcolor{blue}{T_F} \Biggl[\Biggl[
 -\frac{16 (z-1) \big(46 z^2-z-21\big)}{3 z}
 +\frac{64 \big(7 z^2-3 z-1\big) H_0}{3 z}
 \N\\&&
 -\frac{64 (z-1) \big(2 z^2+2 z-1\big) H_1}{3 z}
 +z \Bigl[64 H_0^2+64 H_{0,1}-64 \zeta_2\Bigr]\Biggr] L_Q^2
 +\Biggl[
 -\frac{32}{3} (19 z-12) H_0^2
 \N\\&&
 +\frac{32 \big(422 z^3-137 z^2-114 z+4\big) H_0}{9 z}
 -\frac{32 (z-1) \big(670 z^2-245 z+46\big)}{27 z}
 \N\\&&
 +\frac{32 (z-1) \big(106 z^2-23 z-65\big) H_1}{9 z}
 +\frac{(z-1) \big(2 z^2+2 z-1\big) \Bigl[\frac{128}{3} H_1^2+\frac{256}{3} H_0 H_1\Bigr]}{z}
 \N\\&&
 +\frac{(z+1) \big(2 z^2-2 z-1\big) \Bigl[\frac{256}{3} H_{0,-1}-\frac{256}{3} H_{-1} H_0\Bigr]}{z}
 -64 (z-4) H_{0,1}-\frac{64}{3} z (8 z-3) \zeta_2
 \N\\&&
 +z \Bigl[-\frac{256}{3} H_0^3+\Bigl[-256 H_{0,-1}-256 H_{0,1}\Bigr] H_0
 +256 \zeta_2 H_0+512 H_{0,0,-1}-256 H_{0,1,1}
 \N\\&&
 -128 \zeta_3\Bigr]\Biggr] L_Q\Biggr] 
 +\tilde{C}_{L,q}^{{\sf PS},(3)}({N_F}+1)\Biggr\}~.
\end{eqnarray}

}
\noindent
Finally, the gluonic Wilson coefficient is given by~:
{\small
\begin{eqnarray}
\lefteqn{H_{g,L}^{\sf S} =} \nonumber\\ &&
-16 \textcolor{blue}{T_F} (z-1) z \textcolor{blue}{a_s}
+ \textcolor{blue}{a_s^2} \Biggl\{
-\frac{64}{3} (z-1) z L_M \textcolor{blue}{T_F^2}+\textcolor{blue}{C_A}\textcolor{blue}{T_F}\Biggl[
-\frac{32 (z-1) \big(53 z^2+2 z-1\big)}{9 z}
\N\\&&
-32 \big(13 z^2-8 z-1\big) H_0
-\frac{32 (z-1) \big(29 z^2+2 z-1\big) H_1}{3 z}+L_Q \Biggl[
\frac{32 (z-1) \big(17 z^2+2 z-1\big)}{3 z}
\N\\&&
-128 z H_0+64 (z-1) z H_1\Biggr]
+(z-1) z \Bigl[-32 H_1^2-64 H_0 H_1\Bigr]
+z (z+1) \Bigl[64 H_{-1} H_0-64 H_{0,-1}\Bigr]
\N\\&&
+z \Bigl[96 H_0^2+128 H_{0,1}\Bigr]
+64 (z-2) z \zeta_2\Biggr] 
+\textcolor{blue}{C_F} \textcolor{blue}{T_F}\Biggl[
-\frac{64}{15} z \big(3 z^2+5\big) H_0^2
\N\\&&
+\frac{16 \big(36 z^3-78 z^2-13 z-4\big) H_0}{15 z}
+\frac{32 (z-1) \big(63 z^2+6 z-2\big)}{15 z}
\N\\&&
+L_Q \Biggl[32 z H_0-16 (z-1) (2 z+1)\Biggr]
+16 (z-1) (4 z+1) H_1
\N\\&&
+\frac{(z+1) \big(6 z^4-6 z^3+z^2-z+1\big) }{z^2}\Bigl[
\frac{64}{15} H_{-1} H_0
-\frac{64}{15} H_{0,-1}\Bigr]
-32 z H_{0,1}
\N\\&&
+\Biggl[16 (z-1) (2 z+1)-32 z H_0\Biggr] L_M
+\frac{32}{15} z \big(12 z^2+5\big) \zeta_2\Biggr] \Biggr\}
\N\\&&
+\textcolor{blue}{a_s^3}\Biggl\{
-\textcolor{blue}{T_F^3} \frac{256}{9} (z-1) z L_M^2 
+\textcolor{blue}{C_A} \textcolor{blue}{T_F^2} \Biggl[
\Biggl[
\frac{64 (z-1) \big(17 z^2+2 z-1\big)}{9 z}
-\frac{256}{3} z H_0
\N\\&&
+\frac{128}{3} (z-1) z H_1\Biggr] L_Q^2+\Biggl[
-\frac{64 (z-1) \big(461 z^2+11 z-25\big)}{27 z}
-\frac{128}{9} z (26 z-59) H_0
\N\\&&
-\frac{128 (z-1) \big(39 z^2+2 z-1\big) H_1}{9 z}+(z-1) z \Bigl[
-\frac{128}{3} H_1^2
-\frac{256}{3} H_0 H_1\Bigr]
\N\\&&
+z (z+1) \Bigl[\frac{256}{3} H_{-1} H_0
-\frac{256}{3} H_{0,-1}\Bigr]
+z \Bigl[\frac{512}{3} H_0^2
+\frac{512}{3} H_{0,1}\Bigr]+\Biggl[\frac{128 (z-1) \big(17 z^2+2 z-1\big)}{9 z}
\N\\&&
-\frac{512}{3} z H_0
+\frac{256}{3} (z-1) z H_1\Biggr] L_M
+\frac{256}{3} (z-2) z \zeta_2\Biggr] L_Q
-\frac{32}{9} (28 z-3) H_0^2
\N\\&&
+\Biggl[
\frac{64 (z-1) \big(17 z^2+2 z-1\big)}{9 z}
-\frac{256}{3} z H_0
+\frac{128}{3} (z-1) z H_1\Biggr] L_M^2
\N\\&&
+\frac{32 (z-1) \big(2714 z^2-106 z-139\big)}{81 z}
-\frac{64}{27} \big(110 z^2+277 z-33\big) H_0
+\frac{4160}{27} (z-1) z H_1
\N\\&&
+z \Bigl[
-\frac{64}{9} H_0^3
-\frac{64}{3} H_{0,1}
+\frac{64 \zeta_2}{3}\Bigr]+L_M \Biggl[\frac{64 (z-1) \big(68 z^2+z-7\big)}{9 z}
-\frac{128}{9} (4 z-1) (13 z+6) H_0
\N\\&&
-\frac{128 (z-1) \big(19 z^2+2 z-1\big) H_1}{9 z}+(z-1) z \Bigl[
-\frac{128}{3} H_1^2
-\frac{256}{3} H_0 H_1\Bigr]
\N\\&&
+z (z+1) \Bigl[\frac{256}{3} H_{-1} H_0
-\frac{256}{3} H_{0,-1}\Bigr]+z \Bigl[\frac{256}{3} H_0^2
+\frac{512}{3} H_{0,1}\Bigr]
+\frac{256}{3} (z-2) z \zeta_2\Biggr] \Biggr]
\N\\&&
+\textcolor{blue}{C_A} \textcolor{blue}{T_F^2}  \textcolor{blue}{N_F} \Biggl[\Biggl[
\frac{64 (z-1) \big(17 z^2+2 z-1\big)}{9 z}
-\frac{256}{3} z H_0
+\frac{128}{3} (z-1) z H_1\Biggr] L_Q^2
\N\\&&
+\Biggl[
-\frac{64 (z-1) \big(461 z^2+11 z-25\big)}{27 z}
-\frac{128}{9} z (26 z-59) H_0
-\frac{128 (z-1) \big(39 z^2+2 z-1\big) H_1}{9 z}
\N\\&&
+(z-1) z \Bigl[
-\frac{128}{3} H_1^2
-\frac{256}{3} H_0 H_1\Bigr]
+z (z+1) \Bigl[\frac{256}{3} H_{-1} H_0
-\frac{256}{3} H_{0,-1}\Bigr]+z \Bigl[\frac{512}{3} H_0^2
+\frac{512}{3} H_{0,1}\Bigr]
\N\\&&
+\frac{256}{3} (z-2) z \zeta_2\Biggr] L_Q\Biggr]
\N\\&&
+\textcolor{blue}{C_A^2}  \textcolor{blue}{T_F}\Biggl[\Biggl[
-\frac{16 (z-1) \big(1033 z^2-26 z-65\big)}{9 z}
-\frac{64 \big(6 z^3-47 z^2+3 z+1\big) H_0}{3 z}
\N\\&&
-\frac{32 (z-1) \big(79 z^2+8 z-4\big) H_1}{3 z}
+(z-1) z \Bigl[-128 H_1^2-128 H_0 H_1\Bigr]
+128 z (z+3) H_{0,1}
\N\\&&
+z \Bigl[256 H_0^2-512 \zeta_2\Bigr]\Biggr] L_Q^2
+\Biggl[
\frac{64}{3} \big(18 z^2-91 z+6\big) H_0^2
+\frac{32 \big(2713 z^3-1405 z^2-60 z+4\big) H_0}{9 z}
\N\\&&
+\frac{64 (z-1) \big(137 z^2+12 z-6\big) H_1 H_0}{3 z}
+128 z (3 z-5) H_{0,-1} H_0-128 z (z+11) H_{0,1} H_0
\N\\&&
+\frac{32 (z-1) \big(161 z^2+12 z-6\big) H_1^2}{3 z}
+\frac{32 (z-1) \big(680 z^2-60 z-13\big)}{9 z}
\N\\&&
+\frac{32 (z-1) \big(1919 z^2+30 z-93\big) H_1}{9 z}
+\frac{(z+1) \big(79 z^2-8 z-4\big) \Bigl[\frac{64}{3} H_{0,-1}
-\frac{64}{3} H_{-1} H_0\Bigr]}{z}
\N\\&&
-\frac{128 \big(z^3+53 z^2-6 z+1\big) H_{0,1}}{3 z}-128 z (3 z-13) H_{0,0,-1}-128 (z-3) z H_{0,0,1}
\N\\&&
-256 z (z+5) H_{0,1,1}
-\frac{64}{3} \big(135 z^2-160 z-6\big) \zeta_2+z (z+1) \Bigl[256 H_0 H_{-1}^2
+\Bigl[-192 H_0^2
\N\\&&
-512 H_{0,-1}-256 H_{0,1}\Bigr] H_{-1}
+512 \zeta_2 H_{-1}+512 H_{0,-1,-1}+256 H_{0,-1,1}+256 H_{0,1,-1}\Bigr]
\N\\&&
+(z-1) z \Bigl[128 H_1^3+384 H_0 H_1^2+192 H_0^2 H_1-512 \zeta_2 H_1\Bigr]
+z \Bigl[
-\frac{1024}{3} H_0^3
+1792 \zeta_2 H_0-128 \zeta_3\Bigr]\Biggr] L_Q\Biggr]
\N\\&&
+\textcolor{blue}{C_F^2} \textcolor{blue}{T_F} \Biggl[
-\frac{8}{3} \big(4 z^2-4 z-1\big) H_0^3
-4 \big(20 z^2-11 z-1\big) H_0^2
+8 \big(24 z^2+37 z-7\big) H_0
\N\\&&
-16 (z-1) (10 z-1) H_1 H_0
+32 \big(2 z^2+5 z-2\big) H_{0,1} H_0
+48 (z-1) z H_1^2-16 (z-1) (20 z+3)
\N\\&&
+32 (z-1) (6 z-1) H_1
+16 \big(16 z^2-24 z+3\big) H_{0,1}-32 \big(2 z^2+17 z-3\big) H_{0,0,1}
\N\\&&
+(z-1) (2 z+1) \Bigl[
\frac{16}{3} H_1^3-16 H_0^2 H_1
+32 H_{0,1,1}\Bigr]
-16 (z-2) (6 z-1) \zeta_2
+L_Q^2 \Biggl[
32 (2 z+1) H_1 (z-1)
\N\\&&
+24 (z-1)
+16 (2 z-1) (2 z+1) H_0
+z \Bigl[-16 H_0^2-64 H_{0,1}+64 \zeta_2\Bigr]\Biggr]
\N\\&&
+L_M^2 \Biggl[
32 (2 z+1) H_1 (z-1)+24 (z-1)
+16 (2 z-1) (2 z+1) H_0
+z \Bigl[
-16 H_0^2
\N\\&&
-64 H_{0,1}+64 \zeta_2\Bigr]\Biggr]
-32 \big(2 z^2-19 z+1\big) \zeta_3
+z \Biggl[
\frac{4}{3} H_0^4+32 H_{0,1} H_0^2-192 H_{0,0,1} H_0+192 \zeta_2 H_0
\N\\&&
-64 \zeta_3 H_0
-\frac{608 \zeta_2^2}{5}+384 H_{0,0,0,1}-64 H_{0,0,1,1}-64 H_{0,1,1,1}\Bigr]
+L_M \Biggl[
\frac{32}{15} \big(24 z^3+90 z^2-95 z-15\big) H_0^2
\N\\&&
+\frac{32 \big(78 z^3+141 z^2-34 z+8\big) H_0}{15 z}
+128 \big(2 z^2-3 z-1\big) H_{0,1} H_0
-\frac{8 (z-1) (6 z+1) (153 z-32)}{15 z}
\N\\&&
+16 (z-1) (4 z-3) H_1
+\frac{(z+1) \big(12 z^4+3 z^3-73 z^2-2 z+2\big) 
\Bigl[\frac{128}{15} H_{0,-1}
-\frac{128}{15} H_{-1} H_0\Bigr]}{z^2}
\N\\&&
+32 (6 z+1) H_{0,1}-64 \big(4 z^2-5 z-2\big) H_{0,0,1}
-\frac{32}{15} \big(48 z^3+120 z^2-250 z-45\big) \zeta_2
\N\\&&
+(z+1) (2 z-1) \Bigl[128 H_0 H_{-1}^2
+\Bigl[-64 H_0^2-256 H_{0,-1}\Bigr] H_{-1}
+128 \zeta_2 H_{-1}-128 H_0 H_{0,-1}
\N\\&&
+256 H_{0,-1,-1}+384 H_{0,0,-1}\Bigr]
+(z-1) (2 z+1) \Bigl[-64 H_1 H_0^2+128 H_1 H_0
+64 H_1^2+128 H_1 \zeta_2\Bigr]
\N\\&&
+z \Bigl[\frac{32}{3} H_0^3+\Bigl[128 H_{0,1}-128 H_{0,-1}\Bigr] H_0^2
+\Bigl[512 H_{0,-1,-1}+512 H_{0,0,-1}-512 H_{0,0,1}\Bigr] H_0
\N\\&&
-256 H_{0,-1}^2
+\frac{768 \zeta_2^2}{5}-256 H_{0,1,1}-768 H_{0,0,0,-1}+768 H_{0,0,0,1}
+\Bigl[64 H_0+256 H_{0,-1}-256 H_{0,1}\Bigr] \zeta_2
\N\\&&
+\Bigl[-512 H_0-576\Bigr] \zeta_3\Bigr]\Biggr]
+L_Q \Biggl[
-\frac{32}{15} \big(24 z^3+90 z^2-95 z-15\big) H_0^2
\N\\&&
-\frac{32 \big(78 z^3+141 z^2-34 z+8\big) H_0}{15 z}
-128 \big(2 z^2-3 z-1\big) H_{0,1} H_0
+\frac{8 (z-1) (6 z+1) (153 z-32)}{15 z}
\N\\&&
-16 (z-1) (4 z-3) H_1
+\frac{(z+1) \big(12 z^4+3 z^3-73 z^2-2 z+2\big) \Bigl[\frac{128}{15} H_{-1} H_0
-\frac{128}{15} H_{0,-1}\Bigr]}{z^2}
\N\\&&
-32 (6 z+1) H_{0,1}+64 \big(4 z^2-5 z-2\big) H_{0,0,1}
+L_M \Biggl[-64 (2 z+1) H_1 (z-1)-48 (z-1)
\N\\&&
-32 (2 z-1) (2 z+1) H_0
+z \Bigl[32 H_0^2+128 H_{0,1}-128 \zeta_2\Bigr]\Biggr]
+\frac{32}{15} \big(48 z^3+120 z^2-250 z-45\big) \zeta_2
\N\\&&
+(z+1) (2 z-1) \Bigl[-128 H_0 H_{-1}^2
+\Bigl[64 H_0^2+256 H_{0,-1}\Bigr] H_{-1}
-128 \zeta_2 H_{-1}+128 H_0 H_{0,-1}
\N\\&&
-256 H_{0,-1,-1}-384 H_{0,0,-1}\Bigr]
+(z-1) (2 z+1) \Bigl[64 H_1 H_0^2-128 H_1 H_0-64 H_1^2-128 H_1 \zeta_2\Bigr]
\N\\&&
+z \Bigl[
-\frac{32}{3} H_0^3+\Bigl[128 H_{0,-1}-128 H_{0,1}\Bigr] H_0^2
+\Bigl[-512 H_{0,-1,-1}-512 H_{0,0,-1}+512 H_{0,0,1}\Bigr] H_0+256 H_{0,-1}^2
\N\\&&
-\frac{768 \zeta_2^2}{5}+256 H_{0,1,1}+768 H_{0,0,0,-1}
-768 H_{0,0,0,1}+\Bigl[-64 H_0-256 H_{0,-1}+256 H_{0,1}\Bigr] \zeta_2
\N\\&&
+\Bigl[512 H_0+576\Bigr] \zeta_3\Bigr]\Bigr]\Biggr] 
+\textcolor{blue}{C_F} \textcolor{blue}{T_F^2} \Biggl[
-\frac{16}{3} z H_0^4
-\frac{32}{3} (7 z-1) H_0^3
\N\\&&
-32 (19 z-3) H_0^2-64 \big(6 z^2+7 z-8\big) H_0+\Biggl[-64 z H_0^2
-\frac{64}{3} \big(4 z^2+5 z-3\big) H_0
\N\\&&
+\frac{64 (z-1) \big(40 z^2-17 z-2\big)}{9 z}\Biggr] L_M^2
+\frac{16 (z-1) \big(343 z^2-242 z+4\big)}{3 z}+L_Q^2 \Biggl[-64 z H_0^2
\N\\&&
-\frac{64}{3} (z+1) (4 z-3) H_0
+\frac{64 (z-1) \big(28 z^2-23 z-2\big)}{9 z}\Biggr]
+L_Q \Biggl[
-\frac{128 (25 z+2) H_1 (z-1)^2}{9 z}
\N\\&&
-\frac{32 \big(2474 z^2-4897 z+44\big) (z-1)}{135 z}
-\frac{64}{45} \big(12 z^3-180 z^2-265 z+90\big) H_0^2
\N\\&&
-\frac{64 \big(354 z^3-397 z^2+388 z+4\big) H_0}{45 z}
+\frac{(z+1) \big(6 z^4-6 z^3+z^2-z+1\big) }{z^2}
\Bigl[\frac{256}{45} H_{-1} H_0
\N\\&&
-\frac{256}{45} H_{0,-1}\Bigr]
+\frac{128}{3} (z+1) (4 z-3) H_{0,1}+\Biggl[128 z H_0^2
+\frac{128}{3} \big(4 z^2+3 z-3\big) H_0
\N\\&&
-\frac{256 (z-1) \big(17 z^2-10 z-1\big)}{9 z}\Biggr] L_M
+\frac{128}{45} \big(12 z^3-60 z^2-25 z+45\big) \zeta_2
+z \Bigl[128 H_0^3
\N\\&&
-256 \zeta_2 H_0+256 H_{0,0,1}-256 \zeta_3\Bigr]\Biggr]
+L_M \Biggl[
-\frac{64}{45} \big(12 z^3+180 z^2+305 z-90\big) H_0^2
\N\\&&
+\frac{64 \big(426 z^3-553 z^2+362 z-4\big) H_0}{45 z}
+\frac{32 (z-1) \big(3716 z^2-4753 z-4\big)}{135 z}
\N\\&&
+\frac{128 (z-1) \big(37 z^2-20 z-2\big) H_1}{9 z}+\frac{(z+1) \big(6 z^4-6 z^3+z^2-z+1\big) \Bigl[\frac{256}{45} H_{-1} H_0
-\frac{256}{45} H_{0,-1}\Bigr]}{z^2}
\N\\&&
-\frac{128}{3} \big(4 z^2+3 z-3\big) H_{0,1}
+\frac{128}{45} \big(12 z^3+60 z^2+35 z-45\big) \zeta_2
\N\\&&
+z \Bigl[-128 H_0^3+256 \zeta_2 H_0-256 H_{0,0,1}
+256 \zeta_3\Bigr]\Biggr]\Biggr] 
+\textcolor{blue}{C_F} \textcolor{blue}{N_F} \textcolor{blue}{T_F^2} \Biggl[\Biggl[-64 z H_0^2
\N\\&&
-\frac{64}{3} (z+1) (4 z-3) H_0
+\frac{64 (z-1) \big(28 z^2-23 z-2\big)}{9 z}\Biggr] L_Q^2+\Biggl[
-\frac{128 (25 z+2) H_1 (z-1)^2}{9 z}
\N\\&&
-\frac{32 \big(2474 z^2-4897 z+44\big) (z-1)}{135 z}
-\frac{64}{45} \big(12 z^3-180 z^2-265 z+90\big) H_0^2
\N\\&&
-\frac{64 \big(354 z^3-397 z^2+388 z+4\big) H_0}{45 z}
+\frac{(z+1) \big(6 z^4-6 z^3+z^2-z+1\big) }{z^2}\Bigl[\frac{256}{45} H_{-1} H_0
\N\\&&
-\frac{256}{45} H_{0,-1}\Bigr]
+\frac{128}{3} (z+1) (4 z-3) H_{0,1}+\Biggl[\frac{64}{3} (z-1) (2 z+1)
-\frac{128}{3} z H_0\Biggr] L_M
\N\\&&
+\frac{128}{45} \big(12 z^3-60 z^2-25 z+45\big) \zeta_2
+z \Bigl[128 H_0^3-256 \zeta_2 H_0+256 H_{0,0,1}-256 \zeta_3\Bigr]\Biggr] L_Q
\N\\&&
+L_M \Biggl[
-\frac{32}{9} (z-1) (68 z+25)
-\frac{64}{9} \big(6 z^2-31 z-3\big) H_0
-\frac{64}{3} (z-1) (2 z+1) H_1
\N\\&&
+z \Bigl[\frac{128}{3} H_0^2
+\frac{128}{3} H_{0,1}
-\frac{128 \zeta_2}{3}\Bigr]\Biggr]\Biggr]
+\textcolor{blue}{C_A} \textcolor{blue}{C_F} \textcolor{blue}{T_F}\Biggl[
-\frac{16}{3} (3 z+1) H_0^3
\N\\&&
-\frac{8}{3} \big(11 z^2-18 z+3\big) H_0^2
+\frac{16}{9} \big(772 z^2+480 z-39\big) H_0
+\frac{16 (z-1) \big(77 z^2-25 z-4\big) H_1 H_0}{3 z}
\N\\&&
-32 (7 z-2) H_{0,1} H_0-8 (z-1) (9 z-1) H_1^2
-\frac{32 (z-1) \big(2168 z^2-91 z-28\big)}{27 z}
\N\\&&
-16 (z-1) (16 z-1) H_1
+(z-1) (2 z+1) \Bigl[16 H_0 H_1^2
-\frac{16}{3} H_1^3\Bigr]
\N\\&&
+z (z+1) \Bigl[192 H_{0,-1}-192 H_{-1} H_0\Bigr]
\N\\&&
-\frac{16 \big(68 z^3-117 z^2+21 z+4\big) H_{0,1}}{3 z}
-32 \big(2 z^2-2 z-1\big) H_{0,1,1}+L_M^2 \Biggl[
\N\\&&
-\frac{16 (z-1) \big(43 z^2-11 z-2\big)}{3 z}
+32 (3 z-1) H_0-32 (z-1) (2 z+1) H_1
\N\\&&
+z \Bigl[64 H_0^2+64 H_{0,1}-64 \zeta_2\Bigr]\Biggr]
-16 z (3 z+17) \zeta_2
+(z+1) (2 z-1) \Bigl[32 H_0 H_{-1}^2
\N\\&&
+\Bigl[-16 H_0^2-64 H_{0,-1}\Bigr] H_{-1}
+32 \zeta_2 H_{-1}+32 H_0 H_{0,-1}+64 H_{0,-1,-1}-32 H_{0,0,-1}\Bigr]
\N\\&&
+L_Q^2 \Biggl[
\frac{16 (z-1) \big(65 z^2-2\big)}{3 z}
-\frac{32}{3} (20 z-3) H_0
+32 (z-1) (2 z+1) H_1
+z \Bigl[-64 H_0^2
\N\\&&
-64 H_{0,1}+64 \zeta_2\Bigr]\Biggr]
+(15 z-4) \Bigl[
32 H_{0,0,1}-32 \zeta_3\Bigr]
+z \Bigl[
\frac{8}{3} H_0^4-32 H_{0,-1} H_0^2
+\Bigl[128 H_{0,-1,-1}
\N\\&&
+128 H_{0,0,-1}-256 H_{0,0,1}-64 H_{0,1,1}\Bigr] H_0
-448 \zeta_3 H_0-64 H_{0,-1}^2
-\frac{1472 \zeta_2^2}{5}-192 H_{0,0,0,-1}
\N\\&&
+768 H_{0,0,0,1}+128 H_{0,0,1,1}+64 H_{0,1,1,1}+\Bigl[64 H_{0,-1}-32 H_0\Bigr] \zeta_2\Biggr]
+L_Q \Biggl[
\frac{128}{45} z \big(6 z^2+35\big) H_0^3
\N\\&&
+\frac{64 \big(84 z^4-9 z^3+272 z^2-48 z+6\big) H_0^2}{45 z}
-\frac{32 (z+1) \big(24 z^4+6 z^3-11 z^2-4 z+4\big) H_{-1} H_0^2}{15 z^2}
\N\\&&
-32 (z-1) (2 z+1) H_1 H_0^2
-\frac{16 \big(4668 z^3-5233 z^2-130 z-64\big) H_0}{45 z}
-64 (z-1) (4 z+1) H_1 H_0
\N\\&&
-\frac{64 \big(30 z^4-35 z^3-15 z^2-4\big) H_{0,-1} H_0}{15 z^2}
+64 (z+1) (2 z-1) H_{0,1} H_0
-\frac{256}{5} z \big(4 z^2+5\big) \zeta_2 H_0
\N\\&&
-32 (z-1) (6 z+1) H_1^2
-\frac{8 (z-1) \big(11062 z^2+1335 z-168\big)}{45 z}
\N\\&&
-\frac{64}{45} \big(168 z^4-108 z^3+343 z^2-6 z+24\big) \frac{\zeta_2}{z}
+\frac{64}{5} (z+1) \big(12 z^4-2 z^3-3 z^2-2 z+2\big) \frac{\zeta_2}{z^2} H_{-1}
\N\\&&
-\frac{64 (z-1) \big(371 z^2+37 z-14\big) H_1}{15 z}
-\frac{64}{15} (z-1) \big(12 z^4-18 z^3-13 z^2+2 z+2\big) \frac{\zeta_2}{z^2} H_1
\N\\&&
+\frac{(z+1) \big(42 z^4-69 z^3-35 z^2-4 z+7\big) \Bigl[\frac{256}{45} H_{0,-1}
-\frac{256}{45} H_{-1} H_0\Bigr]}{z^2}
\N\\&&
+\frac{64 \big(24 z^3+208 z^2-17 z+4\big) H_{0,1}}{15 z}
+\frac{(z+1) \big(12 z^4+18 z^3-13 z^2-2 z+2\big)}{z^2}  \Bigl[\frac{64}{15} H_0 H_{-1}^2
\N\\&&
-\frac{128}{15} H_{0,-1} H_{-1}
+\frac{128}{15} H_{0,-1,-1}\Bigr]
+\frac{64 \big(24 z^5+90 z^4-75 z^3-45 z^2-4\big) H_{0,0,-1}}{15 z^2}
\N\\&&
+\frac{64}{15} \big(24 z^3-30 z^2+55 z+15\big) H_{0,0,1}+\frac{(z+1) \big(6 z^4-6 z^3+z^2-z+1\big) }{z^2}
\Bigl[
-\frac{256}{15} H_{-1} H_{0,1}
\N\\&&
+\frac{256}{15} H_{0,-1,1}
+\frac{256}{15} H_{0,1,-1}\Bigr]
+\Biggl[\frac{352}{3} z H_0
-\frac{176}{3} (z-1) (2 z+1)\Biggr] L_M
-\frac{256}{3} z \big(3 z^2+2\big) \zeta_3
\N\\&&
+z \Bigl[\Bigl[64 H_{0,1}-64 H_{0,-1}\Bigr] H_0^2
+\Bigl[256 H_{0,-1,-1}+256 H_{0,0,-1}-256 H_{0,0,1}\Bigr] H_0
-256 \zeta_3 H_0
\N\\&&
-128 H_{0,-1}^2
+\frac{384 \zeta_2^2}{5}+128 H_{0,1,1}
-384 H_{0,0,0,-1}+384 H_{0,0,0,1}
+\Bigl[128 H_{0,-1}-128 H_{0,1}\Bigr] \zeta_2\Bigr]\Biggr]
\N\\&&
+L_M \Biggl[
-\frac{32}{5} \big(4 z^3+5 z^2-10 z-5\big) H_0^2
+\frac{16 \big(2226 z^3-43 z^2-63 z-24\big) H_0}{45 z}
\N\\&&
-\frac{8 (z-1) \big(3758 z^2-1299 z-152\big)}{45 z}
-\frac{16}{3} (z-1) (2 z-23) H_1
\N\\&&
+\frac{(z+1) \big(12 z^4-27 z^3-58 z^2-2 z+2\big) }{z^2}
\Bigl[\frac{64}{15} H_{-1} H_0
-\frac{64}{15} H_{0,-1}\Bigr]
-64 \big(6 z^2-z-3\big) H_{0,0,-1}
\N\\&&
+\frac{32}{15} z \big(24 z^2-85\big) \zeta_2
+(z+1) (2 z-1) \Bigl[-64 H_0 H_{-1}^2
+\Bigl[32 H_0^2+128 H_{0,-1}\Bigr] H_{-1}
-64 \zeta_2 H_{-1}
\N\\&&
-128 H_{0,-1,-1}\Bigr]
+(z-1) (2 z+1) \Bigl[32 H_1 H_0^
2+\Bigl[64 H_{0,-1}-64 H_{0,1}\Bigr] H_0-32 H_1^2+64 H_{0,0,1}
\N\\&&
-64 H_1 \zeta_2\Bigr]+z \Bigl[
-\frac{128}{3} H_0^3+\Bigl[64 H_{0,-1}-64 H_{0,1}\Bigr] H_0^2+\Bigl[-256 H_{0,-1,-1}-256 H_{0,0,-1}
\N\\&&
+256 H_{0,0,1}\Bigr] H_0
+256 \zeta_3 H_0+128 H_{0,-1}^2
-\frac{384 \zeta_2^2}{5}
-\frac{544}{3} H_{0,1}+128 H_{0,1,1}+384 H_{0,0,0,-1}
\N\\&&
-384 H_{0,0,0,1}+\Bigl[128 H_{0,1}-128 H_{0,-1}\Bigr] \zeta_2\Bigr]
\Biggr] \Biggr]
+\tilde{C}_{L,g}^{{\sf S},(3)}({N_F}+1)
\Biggr\}~. 
\end{eqnarray}

}
\section{The OMEs in \boldmath $z$-Space}
\label{app:D}

\vspace*{1mm}
\noindent
In the following, we present the massive operator matrix elements in $z$-space. They are given 
by~:
{\small
\begin{eqnarray}
\lefteqn{A_{qq,Q}^{\sf PS} = }&&
\nonumber \\ &&
\textcolor{blue}{a_s^3} \Biggl\{
\textcolor{blue}{C_F} \textcolor{blue}{N_F} \textcolor{blue}{T_F^2} 
\Biggl[
L_M^2 \Biggl[(z+1)
\Bigl[\frac{64}{3} H_{0,1}-\frac{32}{3} H_0^2-\frac{64 \zeta_2}{3}\Bigr]+\frac{32}{9} \big(4 z^2-7 z-13\big) H_0
\nonumber\\&&
-\frac{32 (z-1) \big(4 z^2+7 z+4\big) H_1}{9 z}-\frac{32}{3} (z-1) (2z-5)\Biggr] +
L_M \Biggl[\big(4 z^2-7 z-13\big)\Bigl[\frac{64 \zeta_2}{9}-\frac{64}{9} H_{0,1}\Bigr]
\nonumber\\&&
+(z+1) \Bigl[\frac{128}{3} H_{0,0,1}-\frac{128}{3} H_{0,1,1}-\frac{128}{3} H_0 \zeta_2-\frac{32}{9} H_0^3-\frac{464}{9} H_0^2\Bigr]+\frac{32 (z-1) \big(4 z^2+7 z+4\big) H_1^2}{9 z}
\nonumber\\&&
+\frac{128}{27} \big(19 z^2-16 z-40\big) H_0+\frac{64}{3} (z-1) (2 z-5)H_1-\frac{32 (z-1) \big(80 z^2-511 z-136\big)}{81 z}\Biggr]
\nonumber\\&&
+\big(57 z^2-131 z-203\big) \Bigl[\frac{64 \zeta_2}{81}-\frac{64}{81}H_{0,1}\Bigr]+(z+1) \Bigl[\frac{1856}{27} H_{0,0,1}+\frac{128}{9}H_{0,0,0,1}-\frac{256}{9} H_{0,0,1,1}
\nonumber\\&&
+\frac{320}{9} H_{0,1,1,1}+H_0^2 \Bigl[-\frac{64}{9} \zeta_2-\frac{5312}{81}\Bigr]+H_0 \Bigl[\frac{640 \zeta_3}{9}-\frac{1856  \zeta_2}{27}\Bigr]-\frac{8}{27} H_0^4-\frac{464}{81} H_0^3 
\nonumber\\&&
-\frac{256}{15} \zeta_2^2\Bigr]+\frac{64}{27} \big(6 z^2-25 z-34\big)H_{0,1,1}+
 L_M^3 \Biggl[\frac{32 (z-1) \big(4 z^2+7 z+4\big)}{27 z}-\frac{64}{9} (z+1)H_0\Biggr]
\nonumber\\&&
-\frac{80 (z-1) \big(4 z^2+7 z+4\big) H_1^3}{81 z}-\frac{16 (z-1) \big(34 z^2-227 z-20\big) H_1^2}{81 z}
\nonumber\\&&
-\frac{64 (z-1) \big(10 z^2+213 z+64\big) H_1}{81 z}
+\frac{32}{243} \big(660 z^2-1577 z-2351\big) H_0
\nonumber\\&&
-\frac{64 \big(22 z^3+16 z^2-17 z-16\big) \zeta_3}{27 z}
-\frac{64 (z-1) \big(139 z^2-3701 z-752\big)}{729 z}
\Biggr] 
\Biggr\}~,
\end{eqnarray}

}
{\small
\begin{eqnarray}
\lefteqn{A_{qg,Q} =} &&
\nonumber \\ &&
\textcolor{blue}{a_s^3} 
\Biggl\{
\textcolor{blue}{C_A} \textcolor{blue}{T_F^2} \textcolor{blue}{N_F} \Biggl[-\frac{8}{27} \big(4 z^2+16 z-5\big) H_0^4+\frac{32}{81}  z (14 z+29) H_0^3
\nonumber \\ &&
-\frac{16}{81} \big(44 z^2+243 z-56\big) H_0^3+\frac{16}{81}  z (200 z+347) H_0^2-\frac{16}{81} \big(402 z^2+1472 z-205\big) H_0^2
\nonumber \\ &&
-\frac{64}{27}  \big(7 z^2+7 z+5\big) H_{-1} H_0^2+\frac{8}{243} \big(5772 z^2-27934 z-451\big) H_0+z \Bigl[\frac{96}{9}
\zeta_2
\nonumber \\ &&
+\frac{11392 }{81}\Bigr] H_0+\frac{256}{9} (4 z+1) \zeta_3
H_0-\frac{16}{9} \big(4 z^2-7 z-1\big) H_1 H_0
\nonumber \\ &&
+\frac{64}{27}  \big(14 z^2+11 z+10\big) H_{0,-1} H_0-\frac{32}{27}
\big(14 z^2-17 z+10\big) H_{0,1} H_0
\nonumber \\ &&
+ L_M^3 \Biggl[\frac{16 (z-1) \big(31 z^2+7
  z+4\big)}{27 z}
-\frac{32}{9} (4 z+1) H_0-\frac{8}{9} {\gamma}_{qg}^{0} H_1\Biggr]-\frac{16}{81} \big(218 z^2-200 z+85\big)
H_1^2
\nonumber \\ &&
-\frac{11392}{81}  (z-1) z
+\frac{4 \big(173275 z^3-157668 z^2+21651 z-17368\big)}{729 z}+\frac{32}{81}
\big(109 z^2+47 z+47\big) \zeta_2
\nonumber \\ &&
-\frac{16}{81}  \big(254 z^2+137 z+112\big) \zeta_2-\frac{32
  \big(145 z^3-123 z^2+3 z-16\big) \zeta_3}{27 z}
\nonumber \\ &&
+z (z+1) \Bigl[\frac{32}{27}  H_0^4+\frac{0}{3}
\zeta_2 H_0^2\Bigr]+\frac{32 \big(1184 z^3-1067 z^2+487 z+18\big) H_1}{243
  z}
\nonumber \\ &&
+\big(200 z^2+191 z+112\big) \Bigl[\frac{32}{81}  H_{0,-1}-\frac{32}{81}
 H_{-1} H_0\Bigr]
\nonumber \\ &&
+L_M^2 \Biggl[-\frac{64}{3} z H_0^2+\frac{32}{9}
\big(9 z^2-20 z-5\big) H_0+\frac{8 \big(205 z^3-168 z^2+42 z-52\big)}{27 z}
\nonumber \\ &&
+\frac{32}{9} \big(4 z^2-4 z+5\big) H_1+{\gamma}_{qg}^{0}
\big(\frac{4}{3} H_1^2-\frac{4 \zeta_2}{3}\big)+\big(2 z^2+2 z+1\big)
\Bigl[-\frac{16}{3}  \zeta_2-\frac{32}{3}  H_{-1} H_0
\nonumber \\ &&
+\frac{32}{3}  H_{0,-1}\Bigr]\Biggr]
+\big(7 z^2-7 z+5\big)
\Bigl[\frac{32}{81} H_1^3+\frac{32}{27} H_0 H_1^2+\frac{32}{27} H_0^2
H_1+\Bigl[\frac{64 \zeta_2}{27}-\frac{128}{27} H_{0,1}\Bigr] H_1\Bigr]
\nonumber \\ &&
+\frac{16}{27} \big(24 z^2-134 z+3\big) H_{0,1}+\big(7 z^2+4 z+5\big)
\Bigl[\frac{32}{9}  \zeta_3-\frac{128}{27}  H_{0,0,-1}\Bigr]
\nonumber \\ &&
+\frac{32}{27} \big(14 z^2-35 z+10\big) H_{0,0,1}
+L_M \Biggl[-\frac{16}{9} (10 z-3) H_0^3-\frac{8}{9}
\big(16 z^2+128 z-3\big) H_0^2
\nonumber \\ &&
+\frac{16}{27} \big(238 z^2-646 z+5\big) H_0-\frac{32}{9} \big(7 z^2-7
z+5\big) H_1^2+\frac{8 \big(3791 z^3-3318 z^2+465 z-344\big)}{81 z}
\nonumber \\ &&
+\frac{32}{9} \big(7 z^2-z+5\big) \zeta_2+\frac{16}{27} \big(158 z^2-149
z+85\big) H_1+\big(7 z^2+7 z+5\big) \Bigl[-\frac{32}{9}  \zeta_2
\nonumber \\ &&
-\frac{64}{9}  H_{-1} H_0+\frac{64}{9}  H_{0,-1}\Bigr]-\frac{64}{3}
z H_{0,1}+\big(2 z^2+2 z+1\big) \Bigl[-\frac{32}{3}  H_{-1} H_0^2
\nonumber \\ &&
+\frac{64}{3}  H_{0,-1} H_0+16  \zeta_3-\frac{64}{3} 
H_{0,0,-1}\Bigr]+{\gamma}_{qg}^{0} \Bigl[-\frac{4}{9} H_1^3-\frac{4}{3}
H_0^2 H_1+\Bigl[\frac{16}{3} H_{0,1}
\nonumber \\ &&
-\frac{8 \zeta_2}{3}\Bigr] H_1+4 \zeta_3+H_0 \Bigl[\frac{8}{3}
H_{0,1}-\frac{4}{3} H_1^2\Bigr]-\frac{8}{3} H_{0,0,1}-8
H_{0,1,1}\Bigr]\Biggr]
+\frac{32}{9} \big(14 z^2-9 z+10\big) H_{0,1,1}
\nonumber \\ &&
+\big(2 z^2+2 z+1\big) \Bigl[-\frac{64}{27}  H_{-1} H_0^3+\frac{64}{9}
 H_{0,-1} H_0^2-\frac{128}{9}  H_{0,0,-1} H_0-\frac{224}{45}
 \zeta_2^2
\nonumber \\ &&
+\frac{128}{9}  H_{0,0,0,-1}\Bigr]+{\gamma}_{qg}^{0}
\Bigl[\frac{1}{27} H_1^4+\Bigl[\frac{4 \zeta_2}{9}-\frac{8}{9} H_{0,1}\Bigr]
H_1^2-\frac{4}{9} H_0^3 H_1+\Bigl[\frac{16 \zeta_3}{3}+\frac{16}{9}
H_{0,0,1}
\nonumber \\
&&+\frac{16}{9} H_{0,1,1}\Bigr] H_1-\frac{56}{45} \zeta_2^2-\frac{8}{9}
  H_{0,1}^2+\frac{8}{9} H_{0,1} \zeta_2+H_0^2 \Bigl[\frac{4}{3}
  H_{0,1}-\frac{2}{9} H_1^2\Bigr]+H_0 \Bigl[\frac{4}{27} H_1^3-\frac{8}{9}
  \zeta_2 H_1
\nonumber \\ &&
-\frac{8}{3} H_{0,0,1}+\frac{8}{9} H_{0,1,1}\Bigr]+\frac{8}{3}
H_{0,0,0,1}-\frac{8}{9} H_{0,0,1,1}-\frac{8}{9} H_{0,1,1,1}\Bigr]\Biggr]
\N\\ &&
+\textcolor{blue}{C_F}  \textcolor{blue}{T_F^2} \textcolor{blue}{N_F} \Biggl[
-\frac{4}{27} \big(28 z^2+116 z-67\big) H_0^4
+\frac{4}{81} \big(288 z^2-2582 z+2341\big) H_0^3
\nonumber\\ &&
-\frac{4}{81} \big(2152
z^2+5141 z-13876\big) H_0^2+\frac{4}{243} \big(14880 z^2+73472 z
\nonumber \\ &&
+154967\big) H_0+\frac{256}{9} \big(6 z^2-z-4\big) \zeta_3
H_0-\frac{448}{81} \big(z^2-z+2\big) H_1 H_0
+
L_M^3
\Biggl[-\frac{16}{3} (2 z-1) H_0^2
\nonumber \\ &&
-\frac{32}{9} \big(6 z^2-z-4\big) H_0+\frac{8 \big(124 z^3-258 z^2+159
  z-16\big)}{27 z}+\frac{8}{9} {\gamma}_{qg}^{0} H_1\Biggr]
\nonumber \\ &&
+\frac{8}{81} \big(364 z^2-373 z+224\big) H_1^2-\frac{179524 z^3+2535258
  z^2-2713863 z-42688}{729 z}
\nonumber \\ &&
-\frac{448}{81} \big(13 z^2-13 z+8\big) \zeta_2-\frac{64 \big(117 z^3-251
  z^2+154 z-16\big) \zeta_3}{27 z}+(2 z-1) \Bigl[\frac{128}{3} H_0^2
\zeta_3
\nonumber \\ &&
-\frac{16}{15} H_0^5\Bigr]-\frac{16}{243} \big(2188 z^2-2278 z+1613\big)
H_1
+L_M^2 \Biggl[-\frac{32}{3} (2 z-1)
H_0^3
\nonumber \\ &&
-\frac{16}{3} (2 z+3) (4 z-3) H_0^2
+\frac{8}{9} \big(160 z^2+146 z+305\big) H_0
\nonumber\\ &&
-\frac{4 \big(1000 z^3+1356
  z^2-2247 z-208\big)}{27 z}-\frac{32}{9} \big(4 z^2-4 z+5\big) H_1
\nonumber \\ &&
+{\gamma}_{qg}^{0} \Bigl[-\frac{4}{3} H_1^2+\frac{8 \zeta_2}{3}-\frac{8}{3} H_{0,1}\Bigr]\Biggr]
+\big(16 z^2-16 z+5\big)
\Bigl[\frac{16}{27} H_1 H_0^2-\frac{32}{27} H_{0,1} H_0+\frac{32}{27}
H_{0,0,1}\Bigr]
\nonumber \\ &&
+\big(7 z^2-7 z+5\big) \Bigl[-\frac{32}{81} H_1^3+\frac{896}{81}
H_{0,1}-\frac{64}{27} H_{0,1,1}\Bigr]
+
L_M
\Biggl[-\frac{20}{3} (2 z-1) H_0^4
\nonumber \\ &&
-\frac{16}{9} \big(14 z^2+37 z-26\big) H_0^3+\frac{4}{9} \big(136 z^2-174
z+909\big) H_0^2-\frac{8}{27} \big(32 z^2-2393 z-4145\big) H_0
\nonumber \\ &&
+\frac{64}{3} (z-1) z H_1 H_0-\frac{4 \big(3556 z^3+33342 z^2-38175
  z+800\big)}{81 z}-\frac{16}{27} \big(140 z^2-149 z+112\big) H_1
\nonumber \\ &&
+\big(7 z^2-7 z+5\big) \Bigl[\frac{32}{9} H_1^2-\frac{64 \zeta_2}{9}\Bigr]+\frac{64}{9} \big(4 z^2-4 z+5\big)
H_{0,1}+{\gamma}_{qg}^{0} \Bigl[\frac{4}{9} H_1^3-\frac{4}{3} H_0^2
H_1-\frac{8 \zeta_3}{3}
\nonumber \\ &&
+\frac{8}{3} H_0 H_{0,1}-\frac{8}{3} H_{0,0,1}+\frac{8}{3}
H_{0,1,1}\Bigr]\Biggr]
+{\gamma}_{qg}^{0} \Bigl[-\frac{1}{27}
H_1^4-\frac{8}{27} H_0^3 H_1-\frac{64}{9} \zeta_3 H_1+\frac{16}{45}
\zeta_2^2
\nonumber \\ &&
+\frac{8}{9} H_0^2 H_{0,1}-\frac{16}{9} H_0 H_{0,0,1}+\frac{16}{9}
H_{0,0,0,1}-\frac{8}{9} H_{0,1,1,1}\Bigr]\Biggr]
\Biggr\}~, 
\end{eqnarray}

}

{\small
\begin{eqnarray}
\lefteqn{A_{Qq}^{\sf PS} =} &&
\nonumber \\ &&
\textcolor{blue}{a_s^2} \Biggl\{
\textcolor{blue}{C_F} \textcolor{blue}{T_F} 
\Biggl[\frac{2}{3} \big(8 z^2+15 z+3\big) H_0^2-\frac{8}{9} \big(56z^2+33 z+21\big) H_0
\nonumber\\&&
+ L_M^2 \Biggl[\frac{4 (z-1) \big(4 z^2+7 z+4\big)}{3 z}-8 (z+1) H_0\Biggr]+\frac{4 (z-1) \big(400 z^2+121 z+112\big)}{27 z}
\nonumber\\&&
+\frac{(z-1) \big(4 z^2+7 z+4\big) \Bigl[\frac{8}{3} H_{0,1}-\frac{8}{3} H_0  H_1\Bigr]}{z}+(z+1) \Bigl[-\frac{4}{3} H_0^3+16 H_{0,1} H_0+32 \zeta_3-32
H_{0,0,1}\Bigr]
\nonumber\\&&
+L_M \Biggl[8 (z+1) H_0^2-\frac{8}{3} \big(8 z^2+15 z+3\big) H_0+\frac{16 (z-1)  \big(28 z^2+z+10\big)}{9 z}\Biggr] \Biggr]
\Biggr\}
\nonumber\\&&
+ \textcolor{blue}{a_s^3} \Biggl\{
a_{Qq}^{{\sf PS},(3)}
+\textcolor{blue}{C_F^2} \textcolor{blue}{T_F}
\Biggl[-\frac{2}{9} \big(4  z^2-3 z+3\big) H_0^4-\frac{2}{9} \big(40 z^2+149 z+115\big) H_0^3
\nonumber\\&&
+\frac{2}{3} \big(160 z^2+191 z-117\big) H_0^2-\frac{8}{3} \big(4 z^2-3 z-3\big) \zeta_2 H_0^2-\frac{4 (z-1) \big(20  z^2+41 z-4\big) H_1 H_0^2}{3 z}
\nonumber\\&&
+\frac{8 \big(4 z^3+27 z^2+3 z-4\big) H_{0,1} H_0^2}{3 z}-\frac{4}{3} \big(400z^2-135 z+222\big) H_0-\frac{2}{3} \big(80 z^2+469 z+221\big) \zeta_2 H_0
\nonumber\\&&
+\frac{16}{9} \big(44 z^2+51 z-18\big) \zeta_3 H_0-\frac{8 (z-1) \big(16  z^2-43 z+66\big) H_1 H_0}{3 z}
\nonumber\\&&
+\frac{16 (z-1) \big(10 z^2+11 z-2\big) H_{0,1} H_0}{3 z}+\frac{16 \big(4  z^3-33 z^2-15 z+4\big) H_{0,0,1} H_0}{3 z}
\nonumber\\&&
-\frac{8 (z-1) \big(6 z^2-z-6\big)
  H_1^3}{9 z}
+\frac{8}{15} \big(188 z^2-27 z-105\big) \zeta_2^2-\frac{4 (z-1) \big(24  z^2-13 z+17\big) H_1^2}{3 z}
\nonumber\\&&
+\frac{64 (z+1)^2 (2 z-1) H_{0,1}^2}{3 z}
+\frac{4 (z-1) (2 z+1) (352 z+233)}{9 z}+\frac{4 \big(84 z^3+79 z^2+75  z-60\big) \zeta_2}{3 z}
\nonumber\\&&
-\frac{4 (z-1) \big(40 z^2+33 z+4\big) H_1 \zeta_2}{3 z}-\frac{8 \big(12  z^3+15 z^2+9 z+8\big) H_{0,1} \zeta_2}{3 z}
\nonumber\\&&
+\frac{4}{9} \big(72 z^2-809 z-145\big) \zeta_3-\frac{4 (z-1) \big(80  z^2-181 z-9\big) H_1}{3 z}
+L_M^3 \Biggl[\frac{16 \big(4 z^2+7 z+4\big) H_1  (z-1)}{9 z}
\nonumber\\&&
+\frac{92 (z-1)}{9}+\frac{16}{9} z (4 z+3) H_0
+(z+1) \Bigl[-\frac{8}{3} H_0^2+\frac{32 \zeta_2}{3}-\frac{32}{3}H_{0,1}\Bigr]\Biggr]
\nonumber\\&&
-\frac{8 \big(8 z^3+100 z^2-85 z+66\big) H_{0,1}}{3 z}
-\frac{32}{3} (z-1) H_1H_{0,1}
-\frac{8 \big(20 z^3-95 z^2-43 z+4\big) H_{0,0,1}}{3 z}
\nonumber\\&&
+L_M^2 \Biggl[\frac{8}{3} \big(4 z^2-9 z-3\big)H_0^2+\frac{8}{3} (z+1) (32 z-31) H_0
+\frac{16 (z-1) \big(4 z^2+7 z+4\big) H_1 H_0}{3 z}
\nonumber\\&&
+\frac{4 (z-1) \big(32  z^2+81 z+12\big)}{3 z}+32 (3 z+2) \zeta_2
+\frac{8 (z-1) \big(32 z^2+35 z+8\big) H_1}{3 z}
\nonumber\\&&
-\frac{16 \big(4 z^3+21 z^2+9  z-4\big) H_{0,1}}{3 z}
+(z+1) \Bigl[-\frac{16}{3} H_0^3+\Bigl[32 \zeta_2-32 H_{0,1}\Bigr] H_0-32 \zeta_3+32H_{0,0,1}\Bigr]\Biggr]
\nonumber\\&&
+\frac{(z-1) \big(4 z^2+7 z+4\big)}{z}
\Bigl[-\frac{2}{9} H_1^4+\Bigl[\frac{16}{3}  H_{0,1}-\frac{20 \zeta_2}{3}\Bigr] H_1^2-\frac{8}{9} H_0^3  H_1
\nonumber\\&&
+\Bigl[\frac{176 \zeta_3}{9}-\frac{64}{3} H_{0,0,1}-\frac{64}{3}  H_{0,1,1}\Bigr] H_1
+H_0 \Bigl[H_1 \Bigl[\frac{32}{3} H_{0,1}-\frac{16    \zeta_2}{3}\Bigr]-\frac{64}{3} H_{0,1,1}\Bigr]\Bigr]
\nonumber\\&&
-\frac{8}{3} \big(12 z^2-23 z-22\big) H_{0,1,1}-\frac{16 \big(20 z^3-21 z^2-33
  z+4\big) H_{0,0,0,1}}{3 z}-\frac{32 \big(3 z^2+15 z+8\big) H_{0,0,1,1}}{3 z}
\nonumber\\&&
+\frac{16 \big(20 z^3+15 z^2-27 z-24\big) H_{0,1,1,1}}{3 z}
+L_M \Biggl[-\frac{32}{3} \big(4 z^2+z+1\big)
H_0^3
\nonumber\\&&
+\frac{2}{3} \big(136 z^2-111 z+213\big) H_0^2
+\frac{8}{27} \big(242 z^2-3984 z-633\big) H_0
+\frac{32}{3} (2 z+3) (8 z-3) \zeta_2 H_0
\nonumber\\&&
+\frac{8 (z-1) \big(140 z^2-127 z+104\big) H_1 H_0}{9 z}
-32 \big(4 z^2+3 z+1\big) H_{0,1} H_0
\nonumber\\&&
+\frac{4 (z-1) \big(28 z^2+21 z+4\big)
  H_1^2}{3 z}
+\frac{4 (z-1) \big(3204 z^2+1625 z+180\big)}{27 z}
-\frac{16}{9} \big(74 z^2+18 z+297\big) \zeta_2
\nonumber \\ &&
-\frac{16 \big(12 z^3-z^2+z-8\big) \zeta_3}{z}
+\frac{16 (z-1) \big(229 z^2-1175
  z-239\big) H_1}{27 z}
\nonumber\\&&
+\frac{8 \big(8 z^3+303 z^2+363 z+104\big) H_{0,1}}{9 z}+\frac{(z-1) \big(4
    z^2+7 z+4\big) \Bigl[-\frac{8}{9} H_1^3-\frac{16}{3} H_0 H_1^2+\frac{32}{3}
    H_{0,1} H_1\Bigr]}{z}
\nonumber\\&&
+\frac{32}{3} \big(8 z^2+15\big) H_{0,0,1}-\frac{16 \big(12 z^3+27 z^2+3
  z-8\big) H_{0,1,1}}{3 z}+(z+1) \Bigl[6 H_0^4-96 \zeta_2 H_0^2
\nonumber\\&&
+\Bigl[192 \zeta_3+96 H_{0,0,1}+64 H_{0,1,1}\Bigr] H_0+\frac{288}{5}
\zeta_2^2-32 H_{0,1}^2-96 H_{0,0,0,1}+32 H_{0,1,1,1}\Bigr]\Biggr]
\nonumber\\&&
+(z+1) \Bigl[\frac{2}{15} H_0^5+\Bigl[4 \zeta_2+\frac{16}{3} H_{0,1}\Bigr]
H_0^3+\Bigl[-\frac{88}{3} \zeta_3-48 H_{0,0,1}\Bigr] H_0^2
+\Bigl[-\frac{448}{5} \zeta_2^2+32 H_{0,1} \zeta_2
\nonumber\\&&
-32 H_{0,1}^2
+160H_{0,0,0,1}+128 H_{0,0,1,1}\Bigr] H_0
+32 H_{0,0,1} \zeta_2+80 H_{0,1,1} \zeta_2-\frac{80}{3} \zeta_2
\zeta_3+160 \zeta_5
\nonumber\\&&
+H_{0,1} \Bigl[-\frac{352}{3} \zeta_3+128 H_{0,0,1}-64 H_{0,1,1}\Bigr]-192
H_{0,0,0,0,1}-768 H_{0,0,0,1,1}-320 H_{0,0,1,0,1}
\nonumber\\&&
+416 H_{0,0,1,1,1}+192 H_{0,1,0,1,1}+32 H_{0,1,1,1,1}\Bigr]\Biggr]
+
\textcolor{blue}{C_F}
\textcolor{blue}{T_F^2}
\Biggl[
\frac{16}{27} \big(8 z^2+15 z+3\big) H_0^3
\nonumber\\&&
-\frac{32}{27} \big(56z^2+33 z+21\big) H_0^2
+\frac{32}{81} \big(1020 z^2+697 z+607\big) H_0+\frac{32}{9} \big(12 z^2+37z+19\big) \zeta_2 H_0
\nonumber\\&&
+\frac{128 (z-1) \big(28 z^2+z+10\big) H_1 H_0}{27 z}
-\frac{128 \big(2 z^3+6 z^2+3 z+2\big) H_{0,1} H_0}{9 z}
\nonumber\\&&
+
L_M^3
\Biggl[\frac{128 (z-1) \big(4 z^2+7 z+4\big)}{27 z}-\frac{256}{9} (z+1) H_0\Biggr] 
\nonumber\\&&
-\frac{16 (z-1) \big(38 z^2+47 z+20\big) H_1^2}{27 z}-\frac{64 (z-1) \big(1781  z^2+539 z+656\big)}{243 z}
\nonumber\\&&
-\frac{32 \big(56 z^3-179 z^2-95 z-40\big) \zeta_2}{27 z}-\frac{64  \big(62 z^3+129 z^2+36 z-8\big) \zeta_3}{27 z}
\nonumber\\&&
+\frac{128 (z-1) \big(55 z^2+64 z+28\big) H_1}{81 z}+\frac{(z-1) \big(4 z^2+7  z+4\big) \Bigl[\frac{16}{27} H_1^3-\frac{32}{9} H_0^2 H_1+\frac{32}{9}  \zeta_2 H_1\Bigr]}{z}
\nonumber\\&&
-\frac{64 \big(75 z^3+13 z^2+61 z-20\big) H_{0,1}}{27 z}
+L_M^2 \Biggl[\frac{32 (z-1) \big(94 z^2+49 z+40\big)}{27  z}
\nonumber\\&&
-\frac{32}{9} \big(12 z^2+37 z+19\big) H_0
-\frac{32 (z-1) \big(4 z^2+7 z+4\big) H_1}{9 z}+(z+1) \Bigl[\frac{32}{3}H_0^2-\frac{64 \zeta_2}{3}+\frac{64}{3} H_{0,1}\Bigr]\Biggr]
\nonumber\\&&
+\frac{64 \big(12 z^3+27 z^2+9 z+4\big) H_{0,0,1}}{9 z}
+L_M \Biggl[\frac{256 (z-1) \big(40 z^2+79  z+31\big)}{81 z}
\nonumber\\&&
-\frac{64}{27} \big(18 z^2+65 z+101\big) H_0+\big(4 z^2-7 z-13\big) \Bigl[\frac{32}{9} H_0^2+\frac{64 \zeta_2}{9}\Bigr]+\frac{64}{3} (z-1) (2 z-5) H_1
\nonumber\\&&
+\frac{(z-1) \big(4 z^2+7 z+4\big) \Bigl[\frac{32}{9} H_1^2-\frac{64}{9} H_0
  H_1\Bigr]}{z}+\frac{128 \big(5 z^2+5 z-2\big) H_{0,1}}{9 z}
\nonumber\\&&
+(z+1) \Bigl[-\frac{64}{9} H_0^3+\Bigl[\frac{128}{3} H_{0,1}-\frac{128
  \zeta_2}{3}\Bigr] H_0+\frac{256 \zeta_3}{3}-\frac{128}{3}
H_{0,0,1}-\frac{128}{3} H_{0,1,1}\Bigr]\Biggr]
\nonumber\\&&
+\frac{64}{9} \big(2 z^2+11 z+8\big) H_{0,1,1}+(z+1) \Bigl[-\frac{8}{9}
H_0^4+\Bigl[\frac{64}{3} H_{0,1}-\frac{32 \zeta_2}{3}\Bigr] H_0^2
\nonumber\\&&
+\Bigl[\frac{1024 \zeta_3}{9}-\frac{128}{3} H_{0,0,1}\Bigr]
H_0+\frac{448}{15} \zeta_2^2-\frac{64}{3} H_{0,1} \zeta_2-\frac{64}{3} H_{0,1,1,1}\Bigr]
\Biggr]
\nonumber\\&&
+\textcolor{blue}{C_F} \textcolor{blue}{T_F^2} \textcolor{blue}{N_F} \Biggl[
L_M^2 \Biggl[
(z+1) \Bigl[
-\frac{64}{3} H_{0,1}
+\frac{32}{3} H_0^2
+\frac{64 \zeta_2}{3}
\Bigr]
-\frac{32}{9} \big(4 z^2-7 z-13\big) H_0
\N\\&&
+\frac{32 (z-1) \big(4 z^2+7 z+4\big) H_1}{9 z}
+\frac{32}{3} (z-1) (2 z-5)\Biggr]
+L_M \Biggl[
(z+1) \Bigl[
\frac{128}{3} H_0 H_{0,1}
\N\\&&
-\frac{128}{3} H_{0,0,1}
-\frac{64}{3} H_{0,1,1}
-\frac{128}{3} \zeta_2 H_0
-\frac{64}{9} H_0^3
+64 \zeta_3
\Bigr]
-\frac{64 \big(2 z^3+z^2-2 z+4\big) H_{0,1}}{9 z}
\N\\&&
+\frac{32}{9} \big(4 z^2-7 z-13\big) H_0^2
+\frac{64}{27} \big(z^2+2 z-58\big) H_0
\N\\&&
+\frac{32 (z-1) \big(74 z^2-43 z+20\big) H_1}{27 z}
+\frac{(z-1) \big(4 z^2+7 z+4\big) \Bigl[
\frac{16}{9} H_1^2
-\frac{64}{9} H_0 H_1
\Bigr]}{z}
\N\\&&
+\frac{64}{9} \zeta_2 \big(6 z^2+4 z-5\big)
+\frac{128 (z-1) \big(25 z^2+94 z+34\big)}{81 z}
\Biggr]
+(z+1) \Bigl[
\zeta_2 \Bigl[
\frac{32}{3} H_{0,1}
-\frac{32}{3} H_0^2\Bigr]
\N\\&&
+\frac{64}{3} H_0^2 H_{0,1}
-\frac{128}{3} H_0 H_{0,0,1}
+\frac{832}{9} \zeta_3 H_0
-\frac{8}{9} H_0^4
-\frac{32 \zeta_2^2}{3}\Bigr]
+\frac{(z-1) \big(28 z^2+z+10\big) }{z}
\Bigl[
\N\\&&
\frac{128}{27} H_0 H_1
-\frac{128}{27} H_{0,1}\Bigr]
-\frac{128 \big(2 z^3+6 z^2+3 z+2\big) H_0 H_{0,1}}{9 z}
+\frac{64 \big(12 z^3+27 z^2+9 z+4\big) H_{0,0,1}}{9 z}
\N\\&&
-\frac{16}{27} (z-1) \big(74 z^2-43 z+20\big) \frac{\zeta_2}{z}
-\frac{32}{27} \big(100 z^3+183 z^2+33 z-4\big) \frac{\zeta_3}{z}
\N\\&&
+ L_M^3 \Biggl[
\frac{32 (z-1) \big(4 z^2+7 z+4\big)}{27 z}
-\frac{64}{9} (z+1) H_0\Biggr]
+\frac{32}{9} \zeta_2 \big(6 z^2+4 z-5\big) H_0
\N\\&&
+\frac{(z-1) \big(4 z^2+7 z+4\big) \Bigl[
-\frac{16}{9} \zeta_2 H_1
-\frac{32}{9} H_1 H_0^2\Bigr]}{z}
+\frac{16}{27} \big(8 z^2+15 z+3\big) H_0^3
\N\\&&
-\frac{32}{27} \big(56 z^2+33 z+21\big) H_0^2
+\frac{32}{81} \big(800 z^2-57 z+111\big) H_0
\N\\&&
-\frac{64 (z-1) \big(1156 z^2-203 z+328\big)}{243 z}\Biggr]
+\textcolor{blue}{C_A} 
\textcolor{blue}{C_F}
\textcolor{blue}{T_F}
\Biggl[-\frac{2}{9} (4 z-17) H_0^4-\frac{4}{9} \big(36 z^2+47 z+36\big)H_0^3
\nonumber\\&&
-\frac{8}{3} (z+3) \zeta_2 H_0^3
+\frac{64}{3}  z^2 H_0^2+\frac{4}{27} \big(1988 z^2-681 z+855\big)H_0^2+8  (z-1) (2 z+1) \zeta_2 H_0^2
\nonumber\\&&
-\frac{8}{3} (2 z+5) (3 z-4) \zeta_2 H_0^2-\frac{16}{3} (20 z-13)\zeta_3 H_0^2+\frac{8 (z-1) \big(122 z^2-19 z+113\big) H_1 H_0^2}{9 z}
\nonumber\\&&
-\frac{8 \big(19 z^2+19 z+8\big) H_{0,1} H_0^2}{3 z}+\frac{16}{5} (9 z-4)\zeta_2^2 H_0+\frac{16}{5}  (29 z-1) \zeta_2^2 H_0
\nonumber\\&&
-\frac{16 (z-1) \big(19 z^2+16 z+10\big) H_1^2 H_0}{9 z}-\frac{64}{9} \big(37 z^2+6\big) H_0
\nonumber\\&&
-\frac{4 \big(48876 z^3+9339 z^2+16218 z+2624\big) H_0}{81 z}-\frac{8}{9} \big(152 z^2-39 z+60\big) \zeta_2 H_0
\nonumber\\&&
-\frac{4 \big(170 z^3+199 z^2+175 z+80\big) \zeta_2 H_0}{9 z}-\frac{16 \big(24 z^3-31 z^2+215 z+4\big) \zeta_3 H_0}{9 z}
\nonumber\\&&
-\frac{32 (z-1) \big(733 z^2-62 z+301\big) H_1 H_0}{27 z}-\frac{32 \big(19  z^3-24 z^2-6 z+10\big) H_{0,-1} H_0}{9 z}
\nonumber\\&&
+\frac{16 \big(18 z^3+119 z^2-2 z+51\big) H_{0,1} H_0}{3 z}-\frac{32 \big(4  z^3-23 z^2-2 z-8\big) H_{0,0,1} H_0}{3 z}
\nonumber\\&&
+\frac{8 (z-1) (2 z+1) (14 z+1) H_1^3}{27 z}-\frac{8 \big(96 z^3-427 z^2+134  z-148\big) \zeta_2^2}{15 z}
\nonumber\\&&
+\frac{8  \big(116 z^3-87 z^2-3 z+4\big) \zeta_2^2}{15 z}+\frac{4  (z-1) \big(616 z^2+313 z+355\big) H_1^2}{27 z}
\nonumber\\&&
+\frac{4 (z-1) \big(75516 z^2-7654 z+23205\big)}{81 z}-\frac{8}{9} \big(9 z^2+185 z+38\big) \zeta_2
\nonumber \\
&&
+\frac{8 \big(1868 z^3-1164 z^2+1344 z-515\big) \zeta_2}{27 z}+\frac{4  (z-1) \big(154 z^2+163 z+46\big) H_1 \zeta_2}{9 z}
\nonumber\\&&
-\frac{8}{3} (23 z+14) H_{0,1} \zeta_2-\frac{8  \big(247 z^3-9  z^2+18 z+50\big) \zeta_3}{9 z}
\nonumber\\&&
+\frac{8 \big(1015 z^3+1149 z^2+705 z+126\big) \zeta_3}{9  z}-\frac{256}{3} (z-2) \zeta_2 \zeta_3+8  (25 z-9)
\zeta_2 \zeta_3
\nonumber\\&&
+8  (3 z+5) \zeta_5+8 (67 z-35) \zeta_5-\frac{64  (z-1)  \big(37 z^2+16\big) H_1}{9 z}
\nonumber\\&&
+\frac{4 (z-1) \big(7828 z^2+2755 z+4075\big) H_1}{81 z}+\frac{(z-1)  \big(z^2+1\big) \Bigl[-\frac{128}{3}  H_1^2-\frac{128}{3}  H_0  H_1\Bigr]}{z}
\nonumber\\&&
+\frac{(z+1) \big(182 z^2-122 z+47\big) \Bigl[\frac{32}{27} H_{-1}  H_0-\frac{32}{27} H_{0,-1}\Bigr]}{z}+\frac{64  \big(6 z^3+19 z^2+10  z-6\big) H_{0,1}}{9 z}
\nonumber\\&&
+\frac{8 \big(2820 z^3-3849 z^2+1128 z-1204\big) H_{0,1}}{27 z}
+L_M^3
\Biggl[\frac{16}{3} (2 z-1) H_0^2+\frac{16 \big(8  z^2+11 z+4\big) H_0}{9 z}
\nonumber\\&&
-\frac{8 (z-1) \big(44 z^2-z+44\big)}{9 z}-\frac{16 (z-1) \big(4 z^2+7  z+4\big) H_1}{9 z}+(z+1) \Bigl[\frac{32}{3} H_{0,1}-\frac{32 \zeta_2}{3}\Bigr]\Biggr]
\nonumber\\&&
+\frac{32 \big(19 z^3-51 z^2-6 z+10\big) H_{0,0,-1}}{9 z}+\frac{64}{9} \big(19 z^2-15 z-6\big) H_{0,0,1}
\nonumber\\&&
-\frac{16 \big(306 z^3+561 z^2+144 z+193\big) H_{0,0,1}}{9 z}
+L_M^2
\Biggl[-\frac{8}{3} \big(4 z^2-25 z+23\big) H_0^2
\nonumber\\&&
+\frac{8}{3}  \big(4 z^2-9 z+6\big) H_0^2+\frac{16}{9}  \big(13z^2-30 z+24\big) H_0+\frac{8 \big(246 z^3+163 z^2+91 z+40\big) H_0}{9 z}
\nonumber\\&&
+\frac{16 (z-1) \big(4 z^2+7 z+4\big) H_1 H_0}{3 z}
-\frac{8  (z-1)  \big(35 z^2-82 z+89\big)}{27 z}
\nonumber\\&&
-\frac{8 (z-1) \big(1829 z^2-403  z+605\big)}{27 z}
-\frac{8 (z-1) \big(104 z^2+119 z+32\big) H_1}{9 z}
\nonumber\\&&
+\frac{(z+1) \big(4 z^2-7
  z+4\big) \Bigl[-\frac{16}{3}  \zeta_2-\frac{32}{3}  H_{-1}  H_0+\frac{32}{3}  H_{0,-1}\Bigr]}{z}
+(10 z+7) \Bigl[\frac{32}{3} H_{0,1}-\frac{32 \zeta_2}{3}\Bigr]
\nonumber\\&&
+(z-1)\Bigl[-\frac{32}{3} H_0^3+\Bigl[-32  \zeta_2-64  H_{0,-1}\Bigr]H_0
-96  \zeta_3+128  H_{0,0,-1}\Bigr]
\nonumber\\&&
+(z+1) \Bigl[-32 \zeta_3-32 H_0 H_{0,1}+32 H_{0,0,1}\Bigr]\Biggr]
+\frac{(z+1) \big(19 z^2-16 z+10\big)}{z}
 \Bigl[-\frac{32}{9} H_0  H_{-1}^2
\nonumber\\&&
+\Bigl[\frac{16}{9} H_0^2-\frac{32}{9}  \zeta_2+\frac{64}{9} H_{0,-1}
\Bigr] H_{-1}-\frac{64}{9} H_{0,-1,-1}
\Bigr]
+\frac{8 \big(56 z^3-201 z^2-162 z-40\big) H_{0,1,1}}{9 z}
\nonumber\\&&
+\frac{(z-1) \big(4 z^2+7 z+4\big)}{z}
 \Bigl[\frac{2}{9} H_1^4
+\frac{4}{3} H_0^2  H_1^2
+\frac{4}{3} \zeta_2 H_1^2+\Bigl[-\frac{80}{9} \zeta_3+\frac{80}{3} H_{0,0,1}+\frac{16}{3} H_{0,1,1}\Bigr] H_1
\nonumber\\&&
-\frac{16}{3}  H_{0,1}^2
+H_0 \Bigl[-\frac{16}{9} H_1^3+\Bigl[-\frac{8}{3} \zeta_2-16  H_{0,1}\Bigr] H_1+\frac{80}{3} H_{0,1,1}\Bigr]\Bigr]
\nonumber\\&&
+\frac{16 \big(32 z^3-87 z^2+45 z-24\big) H_{0,0,0,1}}{3 z}+(2 z-1) \Bigl[128 H_0 H_{0,0,0,1}-32 H_0^2 H_{0,0,1}\Bigr]
\nonumber\\&&
-\frac{16 \big(36 z^3+6 z^2-15 z-20\big) H_{0,0,1,1}}{3 z}+z (4 z-3) \Bigl[H_0\Bigl[\frac{32}{3}  \zeta_3+\frac{32}{3}  H_{0,0,1}\Bigr]
\nonumber\\&&
-\frac{128}{3}  H_{0,0,0,1}+\frac{64}{3}  H_{0,0,1,1}\Bigr]
+\frac{(z+1) \big(4 z^2-7 z+4\big) }{z}
\Bigl[\frac{32}{9} H_0  H_{-1}^3
\nonumber\\&&
+\Bigl[-\frac{8}{3} H_0^2-\frac{-16}{3}  \zeta_2-\frac{32}{3} H_{0,-1}
\Bigr] H_{-1}^2+\Bigl[-\frac{8}{9} H_0^3+\Bigl[\frac{24}{3} \zeta_2
+\frac{32}{3} H_{0,-1}-\frac{32}{3}   H_{0,1}\Bigr] H_0
\nonumber\\&&
-16 \zeta_3+\frac{64}{3}  H_{0,-1,-1}-\frac{32}{3}  H_{0,0,-1}
+\frac{64}{3} H_{0,0,1}
\Bigr] H_{-1}
-\frac{24}{3} H_{0,-1} \zeta_2+\frac{8}{3} H_0^2 H_{0,-1}
\nonumber\\&&
+H_0 \Bigl[-\frac{32}{3} H_{0,-1,-1}+\frac{32}{3}   H_{0,-1,1}-\frac{16}{3} H_{0,0,-1}
+\frac{32}{3}   H_{0,1,-1}\Bigr]-\frac{64}{3} H_{0,-1,-1,-1}
\nonumber\\&&
-\frac{32}{3} H_{0,-1,0,1}
+\frac{32}{3}  H_{0,0,-1,-1}
-\frac{64}{3} H_{0,0,-1,1}
+\frac{16}{3}  H_{0,0,0,-1}
-\frac{64}{3} H_{0,0,1,-1}
\Bigr]
\nonumber \\
&&+L_M
\Biggl[\frac{4}{3} (4 z-5) H_0^4-\frac{8}{9}  (35 z-46) H_0^3-\frac{4}{9} \big(606 z^2-346 z+377\big) H_0^2
\nonumber\\&&
+\frac{8  (z+1) \big(16 z^2-19 z+16\big) H_{-1} H_0^2}{3 z}-\frac{4
  (z-1) \big(8 z^2+17 z+8\big) H_1 H_0^2}{z}
\nonumber\\&&
+\frac{128}{3}  z^2 H_0+\frac{16 \big(3657 z^3+2093 z^2+2330 z+224\big)
  H_0}{27 z}-\frac{16}{3} \big(16 z^2-19 z-25\big) \zeta_2 H_0
\nonumber\\&&
+\frac{8}{3}  \big(40 z^2-51 z+9\big) \zeta_2 H_0+8  (9 z-25)
\zeta_3 H_0-8 (73 z+11) \zeta_3 H_0
\nonumber\\&&
+\frac{16 (z-1) \big(203 z^2+47 z+140\big) H_1 H_0}{9 z}+\frac{64 
  \big(2 z^3-9 z^2+3 z-4\big) H_{0,-1} H_0}{3 z}
\nonumber\\&&
+\frac{8 \big(16 z^3-41 z^2-77 z-40\big) H_{0,1} H_0}{3 z}-32  (5 z-1)
H_{0,0,-1} H_0-32 (11 z+5) H_{0,0,1} H_0
\nonumber\\&&
-\frac{16}{5}  (7 z+27) \zeta_2^2-\frac{24}{5} (65 z+11) \zeta_2^2-\frac{4 (z-1) \big(20 z^2+21 z+2\big) H_1^2}{3 z}
\nonumber\\&&
-\frac{8 (z-1) \big(11542 z^2+399 z+4036\big)}{27 z}-\frac{32  (z+1)
  \big(53 z^2-14 z+26\big) \zeta_2}{9 z}
\nonumber\\&&
+\frac{8 \big(80 z^3-157 z^2+521 z-64\big) \zeta_2}{9 z}-\frac{16 
  (z+1) \big(4 z^2-z+4\big) H_{-1} \zeta_2}{z}
\nonumber\\&&
+\frac{4 \big(44 z^3-81 z^2-213 z-180\big) \zeta_3}{3 z}+\frac{4 
  \big(164 z^3-231 z^2+81 z-12\big) \zeta_3}{3 z}
\nonumber\\&&
+\frac{128  (z-1) \big(z^2+1\big) H_1}{3 z}-\frac{8 (z-1) \big(258
  z^2-559 z-138\big) H_1}{9 z}
\nonumber\\&&
+\frac{(z+1) \big(19 z^2-16 z+10\big)}{z} \Bigl[\frac{32}{9} H_{0,-1}-\frac{32}{9}
  H_{-1} H_0\Bigr]
\nonumber\\&&
+\frac{(z+1) \big(53 z^2-2 z+26\big)}{z} \Bigl[\frac{64}{9} 
  H_{0,-1}-\frac{64}{9}  H_{-1} H_0\Bigr]
\nonumber\\&&
+\frac{(z-1) \big(4 z^2+7 z+4\big)}{z} \Bigl[\frac{8}{9} H_1^3+8 H_0
  H_1^2+\Bigl[\frac{16 \zeta_2}{3}-\frac{32}{3} H_{0,1}\Bigr] H_1\Bigr]
\nonumber\\&&
-\frac{8 \big(274 z^3-253 z^2+611 z-200\big) H_{0,1}}{9 z}-\frac{16 
  \big(32 z^3-75 z^2+21 z-16\big) H_{0,0,-1}}{3 z}
\nonumber\\&&
-\frac{32}{3}  z (4 z-3) H_{0,0,1}+\frac{32 \big(8 z^2+17 z+14\big)
  H_{0,0,1}}{3 z}
\nonumber\\&&
+\frac{(z+1) \big(4 z^2-7 z+4\big) }{z}
\Bigl[\frac{32}{3} H_0  H_{-1}^2
+\Bigl[-\frac{8}{3} H_0^2+16 \zeta_2-\frac{64}{3} H_{0,-1}
\nonumber\\&&
-\frac{32}{3} H_{0,1}\Bigr] H_{-1}+\frac{16}{3}
  H_0 H_{0,-1}+\frac{64}{3} H_{0,-1,-1}
+\frac{32}{3}  H_{0,-1,1}-\frac{16}{3} H_{0,0,-1}+\frac{32}{3} H_{0,1,-1}\Bigr]
\nonumber\\&&
+\frac{(z+1) \big(8 z^2-5 z+8\big) \Bigl[\frac{32}{3}  H_{-1}
  H_{0,1}-\frac{32}{3}  H_{0,-1,1}-\frac{32}{3} 
  H_{0,1,-1}\Bigr]}{z}
\nonumber\\&&
+\frac{16 \big(8 z^3+23 z^2+5 z-8\big) H_{0,1,1}}{3 z}+(z-1)
\Bigl[\Bigl[8 \zeta_2+32 H_{0,-1}\Bigr]
H_0^2
\nonumber\\&&
+\Bigl[128 H_{0,-1,-1}+64 H_{0,0,-1}\Bigr] H_0-64 H_{0,-1}^2+64 H_{0,-1} \zeta_2
\nonumber\\&&
-96 H_{0,0,0,-1}\Bigr]-64  z
H_{0,0,0,1}+160 (5 z+1) H_{0,0,0,1}
\nonumber\\&&
+(z+1) \Bigl[56 H_{0,1} H_0^2-96 H_{0,1,1} H_0+32 H_{0,1}^2+H_{0,1} \Bigl[-32
\zeta_2-\frac{128 }{3}\Bigr]
\nonumber\\&&
+192  H_{0,0,0,-1}+32 H_{0,0,1,1}-32 H_{0,1,1,1}\Bigr]+\frac{128 
  (z-1)}{3 z}\Biggr]
\nonumber \\ &&
-\frac{32 (z-1) (z+2) (2 z+1) H_{0,1,1,1}}{3 z}
-64 (3 z-2) H_{0,0,0,0,1}+32 (23 z+27) H_{0,0,0,1,1}
\nonumber\\&&
+z \Bigl[\frac{32}{3}  \zeta_2 H_0^3+32  \zeta_3 H_0^2
+64  H_{0,0,0,1} H_0-320  H_{0,0,0,0,1}+128  H_{0,0,0,1,1}\Bigr]
\nonumber\\&&
+(z-1) \Bigl[\frac{4}{15} H_0^5-\frac{16}{3} H_{0,-1} H_0^3
+\Bigl[32 H_{0,0,-1}-32 H_{0,-1,-1}\Bigr] H_0^2
\nonumber\\&&
+\Bigl[32 H_{0,-1}^2
+16 \big(-2+5
\big) \zeta_2 H_{0,-1}
+128 H_{0,-1,-1,-1}
-64  H_{0,-1,0,1}-96 H_{0,0,0,-1}\Bigr] H_0
\nonumber\\&&
+64
H_{0,-1,-1} \zeta_2
-96 H_{0,0,-1} \zeta_2+H_{0,-1} \Bigl[48 \big(-5+3
\big) \zeta_3
-128 H_{0,-1,-1}-64 H_{0,0,-1}
+128 H_{0,0,1}\Bigr]
\nonumber\\&&
+256 H_{0,-1,0,-1,-1}
+512 H_{0,0,-1,-1,-1}+64
H_{0,0,-1,0,-1}
-128 H_{0,0,-1,0,1}+192 H_{0,0,0,-1,-1}
\nonumber\\&&
-384 H_{0,0,0,-1,1}+128 H_{0,0,0,0,-1}
-384 H_{0,0,0,1,-1}-128 H_{0,0,1,0,-1}\Bigr]
\nonumber\\&&
+(z+1) \Bigl[-16 H_{0,1,1} H_0^2+\Bigl[48 H_{0,1}^2+\Bigl[16 \zeta_2+\frac{128 }{3}\Bigr] H_{0,1}-128 H_{0,0,1,1}+64 H_{0,1,1,1}\Bigr] H_0
\nonumber\\&&
+H_{0,1,1} \Bigl[\frac{256 }{3}-16 \zeta_2\Bigr]-16 H_{0,0,1}
\zeta_2+H_{0,1} \Bigl[\frac{160 \zeta_3}{3}-160 H_{0,0,1}\Bigr]+288
H_{0,0,1,0,1}
\nonumber\\&&
-128 H_{0,0,1,1,1}-32 H_{0,1,0,1,1}-32 H_{0,1,1,1,1}\Bigr]-\frac{640 
  (z-1)}{9 z}\Biggr]
 \Biggr\}~,
\end{eqnarray}

}
{\small

+16 \big(80 z^2+22 z-1\big) H_{0,0,0,0,1}+\big(20 z^2+14z+7\big) \Bigl[32 H_{0,0,0,-1,1}+32 H_{0,0,0,1,-1}\Bigr]
\nonumber \\ &&
+\big(4 z^2+10 z+5\big)\Bigl[-32  H_{0,0,0,-1,1}-32  H_{0,0,0,1,-1}\Bigr]+128  z (9z+10) H_{0,0,0,1,1}
\nonumber \\ &&
-16 \big(60 z^2+374 z+129\big) H_{0,0,0,1,1}+\big(5 z^2+4z+2\big) \Bigl[64 H_{0,0,1,0,-1}-64 H_{0,-1} H_{0,0,1}\Bigr]
\nonumber \\ &&
+\big(4 z^2+6z+3\big) \Bigl[32  H_{0,-1} H_{0,0,1}+64 H_0 H_{0,-1,-1,-1}+64 H_{0,-1,-1,0,1}-32  H_{0,0,1,0,-1}\Bigr]
\nonumber \\ &&
-16 (110 z+53) H_{0,0,1,0,1}+32\big(104 z^2-8 z+55\big) H_{0,0,1,1,1}+z (z+1) \Bigl[-64  H_{0,0,1}H_0^2
\nonumber \\ &&
-128 H_{-1} H_{0,1} H_0+\Bigl[128 H_{0,-1,1}+128 H_{0,1,-1}-384 H_{0,0,1,1}\Bigr] H_0+384  H_{0,0,1} \zeta_2
\nonumber \\ &&
-128 H_{0,1,1}+128  H_{0,-1,0,1}-768  H_{0,0,1,1,1}\Bigr]+16 \big(72z^2-26 z+49\big) H_{0,1,0,1,1}
\nonumber \\ &&
+\big(2 z^2+2 z+1\big) \Bigl[16 H_0H_{-1}^4+\Bigl[-\frac{64}{3} H_0^2-\frac{-160}{3}\zeta_2-64 H_{0,-1}
\nonumber \\ &&
-\frac{64}{3} H_{0,1}\Bigr]H_{-1}^3+\Bigl[\frac{8}{3} H_0^3+\Bigl[-56 \zeta_2+64H_{0,-1}
-32 H_{0,1}\Bigr] H_0-160\zeta_3+192 H_{0,-1,-1}
\nonumber \\ &&
+64 H_{0,-1,1}+96 H_{0,0,1}+64 H_{0,1,-1}
\Bigr] H_{-1}^2+\Bigl[\frac{4}{3} H_0^4+\Bigl[-28  \zeta_2+16H_{0,-1}+32  H_{0,1}\Bigr] H_0^2
\nonumber \\ &&
+\Bigl[64\zeta_3-128 -128 H_{0,-1,-1}+64 H_{0,-1,1}-96H_{0,0,-1}
-64 H_{0,0,1}+64 H_{0,1,-1}+64 H_{0,1,1}\Bigr] H_0
\nonumber \\ &&
+\frac{296}{5} \zeta_2^2
+48 H_{0,-1} \zeta_2-96H_{0,1} \zeta_2-384 H_{0,-1,-1,-1}
-128H_{0,-1,-1,1}-128  H_{0,-1,0,1}-128H_{0,-1,1,-1}
\nonumber \\ &&
-192  H_{0,0,-1,1}
+160H_{0,0,0,-1}+32 H_{0,0,0,1}
-192  H_{0,0,1,-1}-128 H_{0,0,1,1}-128 H_{0,1,-1,-1}
\Bigr] H_{-1}
\nonumber \\ &&
+96H_{0,-1,1} \zeta_2
+96 H_{0,1,-1} \zeta_2+128 H_{0,-1}
+H_0^2 \Bigl[-32  H_{0,-1,1}-32  H_{0,1,-1}\Bigr]
+H_0\Bigl[-64 H_{0,-1,-1,1}
\nonumber \\ &&
+64 H_{0,-1,0,1}
-64 H_{0,-1,1,-1}
-64 H_{0,-1,1,1}+96H_{0,0,-1,-1}+64 H_{0,0,-1,1}
+64 H_{0,0,1,-1}-64H_{0,1,-1,-1}
\nonumber \\ &&
-64 H_{0,1,-1,1}
-64H_{0,1,1,-1}\Bigr]
+384 H_{0,-1,-1,-1,-1}
+128H_{0,-1,-1,-1,1}
\nonumber \\ &&
+128 H_{0,-1,-1,1,-1}+128  H_{0,-1,0,-1,1}
+128 H_{0,-1,0,1,-1}+128 H_{0,-1,1,-1,-1}
+192 H_{0,0,-1,-1,1}
\nonumber \\ &&
-64  H_{0,0,-1,0,1}+192  H_{0,0,-1,1,-1}+128 H_{0,0,-1,1,1}
+192  H_{0,0,1,-1,-1}
\nonumber \\ &&
+128  H_{0,0,1,-1,1}+128 H_{0,0,1,1,-1}+128 H_{0,1,-1,-1,-1}
\Bigr]
+16 \big(20 z^2-14 z+13\big) H_{0,1,1,1,1}\Biggr]\Biggr\}~.
\end{eqnarray}

}

\noindent
The OME $A_{gg,Q}(z)$ as a diagonal element in the singlet-gluon matrix has distribution-valued 
(+, $\delta(1-z))$ and regular (reg) contributions~: 
\begin{eqnarray} 
A_{gg,Q}(z) = \left[A_{gg,Q,+}(z)\right]_+ + A_{gg,Q,\rm reg}(z) + C_{gg,Q} \delta(1-z),
\end{eqnarray}
with 
\begin{eqnarray} 
\int_0^1 dz f(z) \left[A_{gg,Q}(z)\right]_+ &=& \int_0^1 dz [f(z)-f(1)] A_{gg,Q,+}(z)
\\
\int_0^1 dz C_{gg,Q} \delta(1-z) &=& C_{gg,Q}~.
\end{eqnarray}
The different parts are given by~:
%
%
{\small
\begin{eqnarray}
\lefteqn{A_{gg,Q,+} =} \nonumber \\ 
&&
\textcolor{blue}{a_s^2} \frac{1}{z-1} \Biggl\{
 \textcolor{blue}{C_A} \textcolor{blue}{T_F} \Biggl[- \frac{8}{3} L_M^2 - \frac{80}{9} L_M 
-\frac{224}{27}\Biggr]
\Biggr\}
\nonumber \\ &&
+
\textcolor{blue}{a_s^3} \frac{1}{z-1} \Biggl\{
\textcolor{blue}{C_A^2} \textcolor{blue}{T_F} 
\Biggl[L_M 
\Biggl[\Bigl[H_0 \Bigl[-\frac{64}{3} H_{0,-1}+\frac{32}{3} H_{0,1}-\frac{640}{9} H_1
-\frac{16}{3}\Bigr]+\frac{128}{3} H_{0,0,-1}-\frac{64}{3} H_{0,0,1}
\nonumber \\ &&
+\Bigl[-\frac{16}{3} H_1-\frac{160}{9}\Bigr] H_0^2+\frac{320 \zeta_2}{9}
-\frac{256 \zeta_3}{3}-\frac{1240}{81}\Bigr]\Biggr]
+ L_M^2\Biggl[\Bigl[-\frac{16}{3} H_0^2-\frac{64}{3} H_1 H_0+\frac{32 \zeta_2}{3}
-\frac{184}{9}\Bigr]\Biggr]
\nonumber \\ &&
+\Bigl[\zeta_2 \Bigl[\frac{8}{3} H_0^2+\frac{32}{3} H_1 H_0+\frac{16}{27}\Bigr]
-\frac{88 H_0}{9}-\frac{16 \zeta_2^2}{3}-\frac{176 \zeta_3}{27}-\frac{22672}{243}\Bigr]
+ L_M^3 \frac{176}{27}\Biggr]
\nonumber \\ &&
+\textcolor{blue}{C_A} \textcolor{blue}{C_F} \textcolor{blue}{T_F} \Biggl[ L_M \Bigl[64 \zeta_3
-\frac{40}{3}\Bigr]
- 8 L_M^2 + \Bigl[-40 \zeta_2-\frac{466}{9}\Bigr]\Biggr]
\nonumber \\ &&
+\textcolor{blue}{C_A} \textcolor{blue}{T_F^2} 
\Biggl[
\Bigl[\frac{16 H_0}{3}+\frac{560 \zeta_2}{27}+\frac{224 \zeta_3}{27}
+\frac{5248}{81}\Bigr]- L_M^3 \frac{224}{27}- L_M^2 \frac{640}{27}- L_M \frac{320}{9}\Biggr]
\nonumber \\ &&
+\textcolor{blue}{C_A} \textcolor{blue}{T_F^2}
\textcolor{blue}{N_F} \Biggl[\Bigl[\frac{32 H_0}{9}+\frac{160 \zeta_2}{27}
+\frac{64 \zeta_3}{27}+\frac{10496}{243}\Bigr]- L_M^3 \frac{64}{27}- L_M \frac{2176}{81}\Biggr]
+ a_{gg,Q,(+)}^{(3)}
\Biggr\}~,
\end{eqnarray}

}
{\small
\begin{eqnarray}
\lefteqn{C_{gg,Q} =} \nonumber\\ && 
\frac{4}{3} \textcolor{blue}{a_s} \textcolor{blue}{T_F} L_M
+\textcolor{blue}{a_s^2} \Biggl\{
\textcolor{blue}{C_A} \textcolor{blue}{T_F} \Biggl[
\frac{16}{3} L_M+\frac{10}{9}\Biggr]+\textcolor{blue}{C_F} \textcolor{blue}{T_F} \Biggl[4 L_M-15\Biggr]+\frac{16}{9} \textcolor{blue}{T_F^2} L_M^2\Biggr\}
\N\\&&
+\textcolor{blue}{a_s^3} \Biggl\{
\textcolor{blue}{C_A^2} \textcolor{blue}{T_F} 
\Biggl[
\Biggl[\frac{16 \zeta_3}{3}-\frac{2}{3}\Biggr] L_M^2
+\Biggl[
\frac{16 \zeta_2^2}{3}
+\frac{160 \zeta_3}{9}
+\frac{277}{9}\Biggr] L_M
+\zeta_2 \big(4-\frac{8 \zeta_3}{3}\big)-\frac{616}{27} \Biggr]
\N\\&&
+\textcolor{blue}{C_F} \textcolor{blue}{C_A} \textcolor{blue}{T_F} \Biggl\{
-\frac{22}{3} L_M^2+\frac{736}{9} L_M
+\frac{20 \zeta_2}{3}+16 \zeta_3-\frac{1045}{6}
-64 \zeta_2 \log (2)\Biggr\}
\N\\&&
+\textcolor{blue}{C_F} \textcolor{blue}{T_F^2} \textcolor{blue}{N_F} \Biggl\{28 \zeta_2+\frac{118}{3} -\frac{268}{9} L_M\Biggr\}
+\textcolor{blue}{C_F} \textcolor{blue}{T_F^2} \Biggl[\frac{40}{3} L_M^2-\frac{584}{9} L_M+\frac{782}{9}-\frac{40 \zeta_2}{3}\Biggr]
\N\\&&
+\textcolor{blue}{C_A} \textcolor{blue}{T_F^2} \textcolor{blue}{N_F} \Biggl[\frac{224}{27}-\frac{4 \zeta_2}{3}-\frac{44}{3} L_M\Biggr]
+\textcolor{blue}{C_A} \textcolor{blue}{T_F^2} \Biggl[
\frac{56}{3} L_M^2-2 L_M
-\frac{44 \zeta_2}{3}-\frac{8}{27}\Biggr]
\N\\&&
+\textcolor{blue}{C_F^2} \textcolor{blue}{T_F} \Biggl[
-2 L_M+-80 \zeta_2-32 \zeta_3-39+128 \zeta_2 \log (2)\Biggr]
+\textcolor{blue}{T_F^3} \Biggl[\frac{64}{27} L_M^3-\frac{64 \zeta_3}{27}\Biggr]
\N\\&&
+ a_{gg,Q,\delta}^{(3)}
\Biggr\},
\end{eqnarray}

}
\normalsize
and
%
%
{\small


}

\vspace*{5mm}
\noindent
{\bf Acknowledgment.}~We would like to thank J.~Ablinger, S.~Alekhin, S~Moch and C.~Schneider for 
discussions. This 
work has been supported in part by DFG Sonderforschungsbereich Transregio 9, Computergest\"utzte 
Theoretische Teilchenphysik, by the EU Networks {\sf LHCPHENOnet} PITN-GA-2010-264564 and 
{\sf HIGGSTOOLS} PITN-GA-2012-316704, and by FP7 ERC Starting Grant 257638 PAGAP.
%
%

\end{document}